\documentclass[twocolumn, twocolappendix]{aastex631}

\begin{document}

\title{The High-Redshift Clusters Occupied by Bent Radio AGN (COBRA) Survey: Investigating the Role of Environment on Bent Radio AGNs using LOFAR}

\author[0000-0001-5160-6713]{Emmet Golden-Marx}
\affiliation{Department of Astronomy, Tsinghua University, Beijing 100084, China}

\correspondingauthor{Emmet Golden-Marx}
\email{emmetgm@mail.tsinghua.edu.cn}

\author[0000-0001-9793-5416]{E. Moravec}
\affiliation{Green Bank Observatory, P.O. Box 2, Green Bank, WV 24944 USA}

\author[0000-0001-9495-7759]{L. Shen}
\affiliation{Department of Physics and Astronomy, Texas A$\&$M University, College Station, TX, 77843-44242 USA}
\affiliation{George P. and Cynthia Woods Mitchell Institute for Fundamental Physics and Astronomy, Texas A$\&$M University, College Station, TX, 77843-4242 USA}

\author[0000-0001-8467-6478]{Z. Cai}
\affiliation{Department of Astronomy, Tsinghua University, Beijing 100084, China}

\author[0000-0002-0485-6573]{E.\,L. Blanton}
\affiliation{Department of Astronomy and The Institute for Astrophysical Research, Boston University, 725 Commonwealth Avenue, Boston, MA 02215, USA}

\author[0000-0002-7326-5793]{M.\,L. Gendron-Marsolais}
\affiliation{Instituto de Astrofísica de Andalucía (IAA-CSIC), Glorieta de la Astronomía, E-18008 Granada, Spain}

\author[0000-0001-8887-2257]{H.\,J.\,A. R\"{o}ttgering}
\affiliation{Leiden Observatory, Leiden University, Niels Bohrweg 2, NL-2300 RA Leiden, the Netherlands}

\author[0000-0002-0587-1660]{R.\,J. van Weeren}
\affiliation{Leiden Observatory, Leiden University, Niels Bohrweg 2, NL-2300 RA Leiden, the Netherlands}

\author{V. Buiten}
\affiliation{Leiden Observatory, Leiden University, Niels Bohrweg 2, NL-2300 RA Leiden, the Netherlands}

\author[0000-0001-9578-6111]{R.\,D.\,P. Grumitt}
\affiliation{Department of Astronomy, Tsinghua University, Beijing 100084, China}

\author[0000-0002-6394-045X]{J. Golden-Marx}
\affiliation{Department of Astronomy, School of Physics and Astronomy, Shanghai Jiao Tong University, Shanghai 200240, China}

\author[0000-0001-8109-806X]{S. Pinjarkar}
\affiliation{Centre for Astrophysics Research, University of Hertfordshire, College Lane, Hatfield AL10 9AB, UK}

\author[0000-0002-7300-9239]{H. Tang}
\affiliation{Department of Astronomy, Tsinghua University, Beijing 100084, China}

\begin{abstract}
Bent radio AGN morphology depends on the density of the surrounding gas.  However, bent sources are found inside and outside clusters, raising the question of how environment impacts bent AGN morphology.  We analyze new LOw-Frequency Array Two-metre Sky Survey (LoTSS) Data Release II observations of 20 bent AGNs in clusters and 15 not in clusters from the high-$z$ Clusters Occupied by Bent Radio AGN (COBRA) survey (0.35 $<$ $z$ $<$ 2.35).  We measure the impact of environment on size, lobe symmetry, and radio luminosity.  We find that the most asymmetric radio lobes lie outside of clusters and we uncover a tentative correlation between the total projected physical area and cluster overdensity.  Additionally, we, for the first time, present spectral index measurements of a large sample of high-$z$ bent sources using LoTSS and Very Large Array Faint Images of the Radio Sky at Twenty-centimeters (VLA FIRST) observations.  We find that the median spectral index for the cluster sample is -0.76 $\pm$ 0.01, while the median spectral index for the non-cluster sample median is -0.81 $\pm$ 0.02.  Furthermore, 13 of 20 cluster bent AGNs have flat cores ($\alpha$ $\geq$ -0.6) compared to 4 of 15 of non-clusters, indicating a key environmental signature.  Beyond core spectral index, bent AGNs inside and outside clusters are remarkably similar.  We conclude that the non-cluster sample may be more representative of bent AGNs at large offsets from the cluster center ($>$ 1.2\,Mpc) or bent AGNs in weaker groups rather than the field. 
\end{abstract}


\section{Introduction}\label{sect:intro}
Galaxy clusters are the largest gravitationally-bound structures in the universe.  Since galaxy clusters host large populations of similarly aged galaxies, clusters offer a unique lens to study galaxy evolution.  Such studies require a large sample of similar clusters across a range of redshifts.  While there are many well-studied low-$z$ clusters that have been observed across the entire electromagnetic spectrum \citep[e.g.,][]{Eisenhardt2007}, as we probe higher redshifts, the number of well-studied clusters dwindles.  However, large optical and infrared surveys performed with telescopes like the $Spitzer$ Space Telescope and the \textit{Wide-field Infrared Survey Explorer} (WISE), have led to a recent boom in the number of confirmed high-$z$ clusters and cluster candidates \citep[e.g.,][]{Muzzin2009,Wilson2009,Wylezalek2013,Gonzalez2015,Paterno-Mahler2017,Gonzalez2019}.  Recently, the Massive and Distant Clusters of WISE Survey (MaDCoWS) identified thousands of cluster candidates at $z$ $\approx$ 1 \citep[e.g.,][]{Gonzalez2019}.  While surveys like MaDCoWS are excellent cluster finders, they preferentially identify massive, Coma cluster-like progenitors.  However, as shown in low-$z$ cluster surveys \citep[e.g.,][]{Miller2005,Rykoff2014}, there are far more low-mass clusters and groups.  As such, with the onset of large data sets and new, higher resolution observations coming from JWST, LSST, and DESI, we find a conundrum in how to best identify potential lower mass clusters at high-$z$ where projection effects and large observational foreground and background contamination can cause false detections.  

One strong tracer for finding high-$z$ galaxy clusters and their progenitors are radio active galactic nuclei (AGNs) \citep[e.g.,][]{Minkowski1960,Blanton2003,Wylezalek2013,Wylezalek2014,Cooke2015,Cooke2016,Noirot2016,Noirot2018,Moravec2019,Moravec2020,Shen2021}.  Specifically, the Clusters Around Radio Loud AGN (CARLA) survey found that $\approx$ 92$\%$ of their sample of radio loud AGNs were in overdense systems and that 55$\%$ were in clusters \citep{Wylezalek2013}.  Based on the density distribution of massive clusters/protoclusters and radio loud AGNs, \citet{Hatch2014} found that approximately all protoclusters host radio-loud AGNs at 1.3 $<$ $z$ $<$ 3.2, making them an ideal tracer.  Importantly, radio AGNs are also found in clusters across a range of masses and cluster dynamical/evolutionary states \citep[e.g.,][]{Blanton2001,Wing2011,Cooke2015,Cai2017,Garon2019}. 

Despite radio AGNs being a very strong tracer, not every radio AGN is in a rich cluster \citep[e.g.,][]{Blanton2000,Wing2011,Wylezalek2013,Paterno-Mahler2017}.  To better trace large scale cluster environments with radio AGNs, the morphology of the AGNs should be considered.  The standard classification scheme of radio AGNs is the Fanaroff-Riley (FR) system \citep{Fanaroff1974}, which divides radio AGNs into FRI and FRII sources based on their radio luminosity (L$_{1.44\,GHz}$ $\approx$ 10$^{25}$\,WHz$^{-1}$) and the location of the brightest portion of the core and lobes.  FRIs have brighter cores and lobes that fade at the edges, while FRIIs have large bright lobes farther from the AGN core \citep{Fanaroff1974}.   FRIs are commonly in clusters at both low and high redshift \citep[e.g.,][]{Longair1979,Prestage1988,Hill1991,Ledlow1996,Zirbel1997,Miller1999,Stocke1999,Wing2011,Gendre2013,Fujita2016,Shen2017}, while very powerful FRIIs are more commonly found in rich environments at high redshift \citep{Best2000}, although they have been found in clusters at low-$z$ \citep[e.g.,][]{Wing2011}.  Despite the distinct morphological divide between FRIs and FRIIs, recent analysis in \citet{Mingo2019} suggests that this difference in morphology does not follow the difference in the radio luminosity.  

One unique type of radio AGN commonly found in clusters of all masses are bent, double-lobed radio sources \citep[e.g.,][]{Blanton2000,Sakelliou2000,Blanton2001,Wing2011,Wing2013,Paterno-Mahler2017,Silverstein2018,Garon2019,Golden-Marx2019,Golden-Marx2021,deVos2021,Vardoulaki2021,Morris2022}.  The radio luminosity of bent radio sources typically straddles the FRI/FRII border \citet{Blanton2000,Blanton2001} and their unique bent morphology implies the presence of a dense gaseous medium (like the intracluster medium [ICM]), which is necessary to create the ram pressure to bend the radio lobes \citep[e.g.,][]{Owen1976,ODonoghue1993,Hardcastle2005,Morsony2013}.

Low-$z$ bent radio AGNs are preferentially found in rich cluster environments \citep[e.g.,][]{Wing2011,Morris2022}, and in galaxy clusters between 40 - 80\,$\%$ of the time depending on the richness, about a factor of 2$\times$ higher than straight radio AGNs \citep{Wing2011}.  Outside of galaxy clusters and groups, low-$z$ bent radio AGNs have been found in large-scale filaments between galaxy clusters \citep[e.g.,][]{Edwards2010} and fossil groups, neither of which would necessarily be characterized by large overdensities of red sequence galaxies.  The study of the environments of bent radio AGNs was extended out to high redshift using the Clusters Occupied by Bent Radio AGN (COBRA) survey \citep{Blanton2015,Paterno-Mahler2017,Golden-Marx2019,Golden-Marx2021}.  Similar to their low-$z$ counterparts, high-$z$ bent radio AGNs have been found in rich, red sequence selected galaxy cluster candidates as well as poorer and underdense environments \citep{Golden-Marx2019}.  

Bent radio AGN morphology is dependent on environment, which leads to the question of whether one can one differentiate the environment of bent radio AGNs at high-$z$ based solely on radio characteristics?  If so, does this allow us to better determine which bent radio AGNs to target for cluster searches, thus yielding a potential method of identifying new, lower mass, high-$z$ galaxy clusters and groups? 

In this paper, we take advantage of newly available radio observations from the LOw-Frequency ARray (LOFAR) Two-metre Sky Survey (LoTSS) Data Release II \citep{Shimwell2022} to attempt to answer these questions.  We present two samples of bent radio AGNs previously identified in the high-$z$ COBRA survey, one in red sequence cluster environments and one in non-cluster environments.  We compare their radio properties, including the projected physical size, projected physical area, radio luminosity, and spectral indices. 

The order of the paper is as follows.  We introduce the COBRA survey in Section~\ref{sect:COBRA} and our cluster and non-cluster sample in Section~\ref{sect:Cluster-NonCluster}.  We discuss our archival radio observations and the measurements of the radio source properties in Section~\ref{sect:RadioObs} through Section~\ref{sect:SpectralIndex}.  We compare the properties of bent radio AGNs in cluster candidates to those outside of clusters in Section~\ref{sect:AGNcluster}.  Lastly, we compare our findings to those of similar work probing the environment and properties of radio AGNs in Section~\ref{sect:discussion}.  Throughout this work, we adopt a flat $\Lambda$CDM cosmology, using H$_{o}$ = 70km\,s$^{-1}$\,Mpc$^{-1}$, $\Omega_{m}$ = 0.3, and $\Omega_{\Lambda}$=0.7.  Unless noted, all distances are given as projected distances and all magnitudes are given as AB magnitudes.  In this paper, we refer to high-$z$ as the redshift range of the bent radio AGNs in this study, 0.35 $<$ $z$ $<$ 2.35, or similar redshifts (with the majority of the sources being at 0.5 $<$ $z$ $<$ 1.5).  For the remainder of the paper, we refer to \citet{Golden-Marx2019} as GM19 and \citet{Golden-Marx2021} as GM21.

\section{Data}\label{sect:data}
In order to better characterize the properties of high-$z$ bent radio AGNs as a function of environment, we need a high-$z$ cluster sample and a high-$z$ non-cluster sample.  For our analysis, we take advantage of bent radio AGNs identified as part of the high-$z$ COBRA survey.  To do this analysis, we use publicly available data from the Very Large Array (VLA) Faint Images of the Radio Sky at Twenty-centimeters (FIRST) \citep{Becker1995} and LoTSS DR2 \citep[e.g.,][]{Shimwell2017,Shimwell2019,Shimwell2022}. Our principal focus are the newly available LoTSS DR2 \citep{Shimwell2022} data described in Section~\ref{sect:RadioObs}.

\subsection{The High-z COBRA Survey}\label{sect:COBRA}
The high-$z$ COBRA survey is a sample of 646 bent, double-lobed radio sources selected from the FIRST survey at 0.35 $<$ $z$ $<$ 3.0 (with the majority of AGNs having redshifts between 0.5 $<$ $z$ $<$ 1.5).  These bent radio AGNs were originally identified as part of the sample of radio sources studied in \citet{Wing2011}.  Specifically, \citet{Wing2011} built on the visually selected sample of 384 bent radio AGNs from \citet{Blanton2000} to identify potential bent radio AGNs in the $\approx$ 10,000 square degree FIRST survey.  Unlike the visual-bent selected sample, \citet{Wing2011} used pattern recognition software from \citet{Proctor2006} to identify 1546 three component radio sources in what they refer to as the auto-bent sample.  Because FIRST was designed to cover the same portion of the sky as the Palomar Sky Survey, which is the same area of the sky observed by the Sloan Digital Sky Survey (SDSS), \citet{Wing2011} cross-matched the radio sources with SDSS DR7 to identify low-$z$ host galaxies.  As not every bent AGN had a host galaxy detected above the magnitude threshold (m$_{r}$ = 22.0~\,mag) used in \citet{Wing2011}, the 653 remaining sources became the initial high-$z$ COBRA sample \citep[e.g.,][]{Blanton2015,Paterno-Mahler2017}.  Of these 653 sources, 646 were successfully observed with the IRAC 3.6\,$\mu$m band on $Spitzer$ (PI Blanton).  

Of these 646 sources, 41 are spectroscopically-confirmed quasars from SDSS.  Using this sample, \citet{Paterno-Mahler2017} cross-matched the 3.6\,$\mu$m images with radio observations from FIRST to identify potential host galaxies.  \citet{Paterno-Mahler2017} then measured a single-band IR overdensity in both 1$\arcmin$ and 2$\arcmin$ regions centered on the radio AGN and found that $\approx$ 82$\%$ of bent radio sources are in overdense environments, but only $\approx$ 29$\%$ are found in rich cluster candidates, defined by a minimum of a 2$\sigma$ overdensity relative to a background field.

To further determine if these bent radio AGNs are in cluster candidates, GM19 analyzed the population of red sequence galaxies surrounding each radio AGN in a subset of the COBRA sample.  To estimate the color of the surrounding galaxies, GM19 introduced $i$-band observations of 90 of the fields hosting bent radio AGNs taken using the 4.3\,m Lowell Discovery Telescope (LDT).  These fields were chosen based on either a particularly clear bent radio morphology, the radio sources being identified as quasars in SDSS, or the overdensity measurements from \citet{Paterno-Mahler2017}.  Of these 90 fields, 38 were also observed in $r$-band on the LDT (see Table 1 in GM19 for the observed fields and exposure times).  

All cluster candidate redshift estimates are based on the identification of the host galaxy.  For the bent radio AGNs without spectroscopic redshifts, GM19 estimated a photometric redshift by comparing the colors of the host galaxies to spectral energy distribution (SED) models of quiescent galaxies across the redshift range 0 $<$ $z$ $<$ 3.0 generated from EzGal \citep{Mancone2012}, with an estimated redshift error of $\pm$ 0.1 (see GM19 for the input parameters of the models used; for the radio AGNs in this paper, the redshift, error, and source of that redshift are specified in Table~\ref{tb:RadioProp}).  Of the 90 radio AGNs observed in the optical, GM19 estimate photometric redshifts or find an SDSS spectroscopic redshift for 77 AGNs.  Although only four of the sources with spectroscopic redshifts are not quasars (and thus have a color typical of a red sequence galaxy), GM19 found strong agreement between their photometric redshift estimates using optical and IR photometry and sources with SDSS photometric or spectroscopic redshifts.  

Using the available optical and IR observations, GM19 measured the color of the host galaxy identified in \citet{Paterno-Mahler2017} ($i - [3.6]$, $[3.6] - [4.5]$, and/or $r - i$) and compared it to the color of the surrounding galaxies.  To identify potential red sequence galaxy cluster candidates, GM19 used a red sequence width of $\pm$0.15 from the color of the host in the $i - [3.6]$ or $r - i$ color, in agreement with the 3$\sigma$ red sequence scatter found in spectroscopic studies at $z$ $\approx$ 1.0 \citep[e.g.,][]{Blakeslee2003,Mei2006,Mei2009,Snyder2012,Lemaux2012,Cerulo2016}.  For the higher redshift sources observed primarily with $Spitzer$ photometry, GM19 used a slight modification to the well-studied Spitzer color cut (\citealp{Papovich2008}; $[3.6] - [4.5]$ $>$ -0.15), which has been used to identify numerous high-$z$ cluster and protocluster candidates \citep[e.g.,][]{Wylezalek2013}.

To estimate which fields host galaxy cluster candidates, GM19 used two distinct methods.  First, they compared the number of red sequence galaxies within a 1$\arcmin$ region centered on the radio AGN to the average number in a background field.  Second, to account for all bent radio AGNs not being at the centers of cluster \citep[e.g.,][]{Sakelliou2000}, GM19 measured the surface density of red sequence galaxies within the field of view (FOV) and chose the peak of that distribution as the new cluster center.  We use these red sequence surface density measurements and cluster centers in this paper.  

Because cluster galaxies, especially at higher redshift, are not always red, GM19 also measured a combined overdensity, which accounts for the overdensity of red sequence galaxies, galaxies bluer than the red sequence, and galaxies redder than the red sequence.  To scale this measurement, GM19 used observations with more statistically robust photometric redshifts from the Observations of Redshift Evolution in Large-Scale Environments Survey (ORELSE) \citep[e.g.,][]{Lubin2009,Hung2019}, to measure the fraction of galaxies at a given redshift in the ORELSE survey that are on the red sequence or redder/bluer.  For both overdensity measurements, the detection threshold is either a m*+1 magnitude limit (if $z$ $<$ 1.0), where the magnitude of an m* galaxy is estimated via EzGal, or the magnitude limit of the $Spitzer$ observations (21.4~\,mag). To prevent single galaxy detections from yielding a cluster candidate at the highest redshifts, GM19 require both a 2$\sigma$ detection when centered on either the radio AGN or the estimated cluster center and that each cluster candidate have at least three red sequence galaxies within the 1$\arcmin$ search region.  From the sample with redshifts, GM19 identify 39 high-$z$ bent radio AGNs in galaxy cluster candidates at 0.35 $<$ $z$ $<$ 2.2 (see GM19 for the red sequence and combined overdensities).

\subsubsection{The High-z Cluster and Non-Cluster Sample}\label{sect:Cluster-NonCluster}
  
Although GM19 identified 39 red sequence galaxy cluster candidates, after further analysis of the bent radio AGNs, GM21 removed three radio AGNs from the sample due to uncertain morphology.  As such, the initial high-$z$ sample in this work consists of the 36 cluster candidates in GM21.  To create a uniformly observed and analyzed comparison sample of high-$z$ bent radio AGNs that are not in clusters, we return to the sample from GM19.  Since certain characteristics depend on redshift (e.g., radio luminosity, projected physical size, projected physical area), we only select AGNs that either have a spectroscopically confirmed redshift from SDSS, a photometric redshift from SDSS, or an estimated redshift from GM19.  As mentioned above, 39 bent radio AGNs with optical imaging and redshift estimates are found in cluster candidate environments, while 38 are not.  Thus, using this sample, we select bent AGNs with no evidence of being in a rich, red sequence cluster environments.  These sources have the same available photometry as the cluster sample.  Although GM19 does not identify any of these sources as cluster candidates based on the surrounding distribution of red sequence galaxies, we explore whether these radio AGNs are truly representative of a field sample in Section~\ref{sect:Size-Asymmetry-Environment}. 

\subsection{Radio Observations}\label{sect:RadioObs}
While much of the pioneering work on large samples of bent radio AGNs in clusters was done using FIRST \citep[e.g.,][]{Blanton2000,Blanton2001,Wing2011,Wing2013}, we are entering the LOFAR era of radio astronomy and understanding the similarities in radio properties is vital to further characterizing bent radio AGNs.  We chose LOFAR due to its sensitivity and because lower frequency emission is more sensitive to older, less energetic synchrotron emitting electrons.  This less energetic synchrotron emission may be more impacted by the cluster environment, making LOFAR the ideal tool for our analysis.  Specifically, we use the publicly available data from the LoTSS DR2 \citep{Shimwell2022} observed in the 120-160\,MHz range (with a median frequency of 144\,MHz). 

\begin{figure}
\begin{center}
\includegraphics[scale=0.6,trim={0.3in 0.1in 0.0in 0.3in},clip=true]{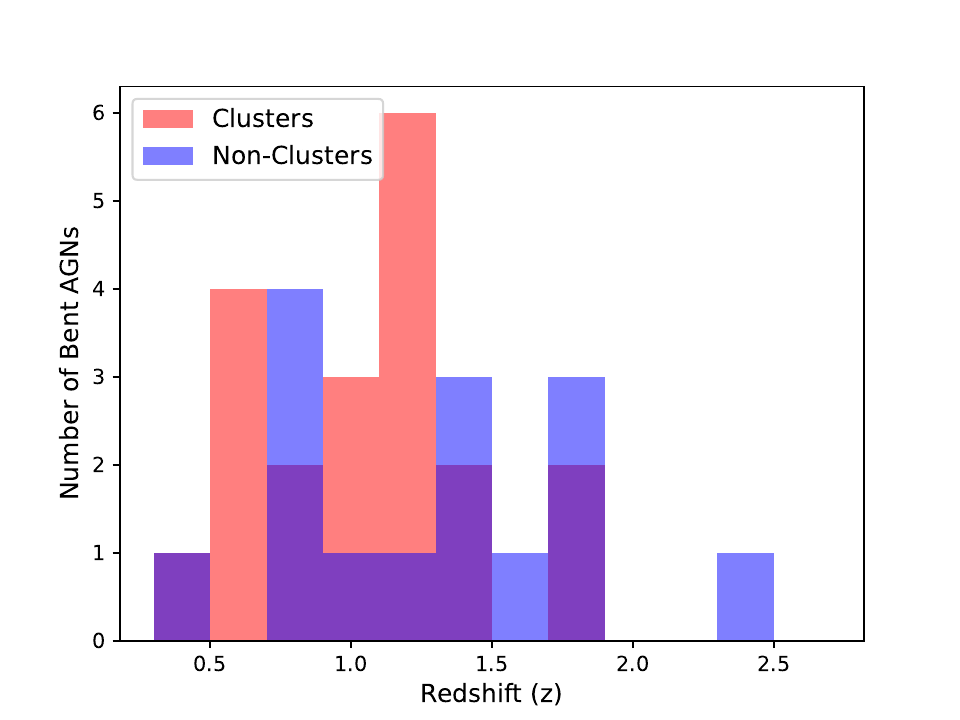}
\caption{The redshift distribution of bent radio AGNs in the cluster (red) and non-cluster (blue) sample.  While we have more clusters at 0.5$<$ $z$ $<$ 1.5 than non-clusters, each sample spans a similar redshift range (with the exception of the highest redshift non-cluster source).}
\label{Fig:ClusterNon-z}
\end{center}
\end{figure}

\begin{figure*}
\begin{center}
\includegraphics[scale=0.14,trim={0.0in 0.0in 0.0in 0.0in},clip=true]{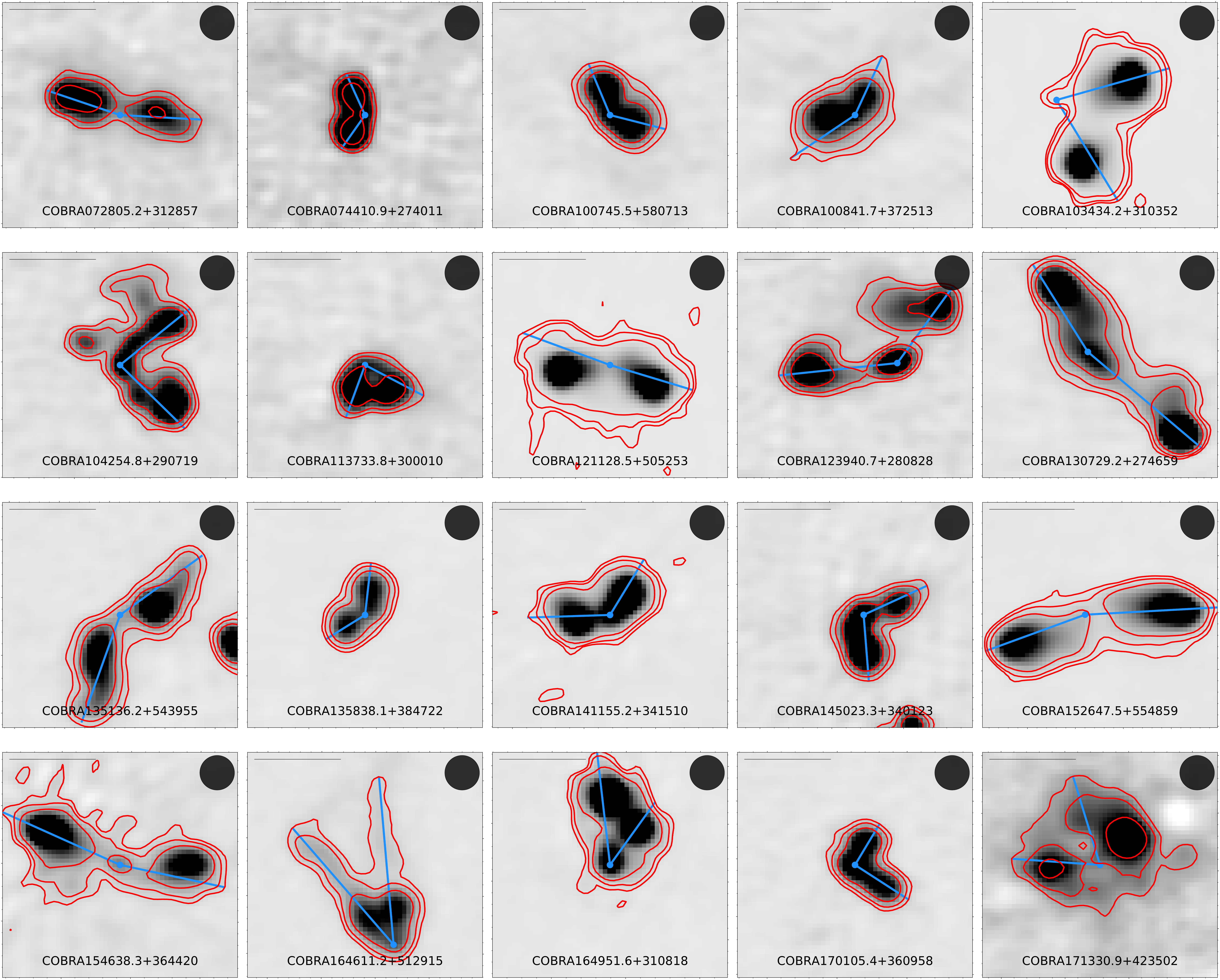}

\caption{Cutouts of the 20 bent, double-lobed radio sources in the COBRA cluster candidate sample observed by LoTSS DR2.  While most are 1$\farcm$3$\times$1$\farcm$3, some are slightly larger to encompass the extent of the entire radio source.  The three red contours show 10$\sigma$, 20$\sigma$, and 50$\sigma$ contours based on the measurement of the rms noise in each image.  The blue lines show our estimate of the projected physical size of each radio lobe and the filled blue circle shows the location of the AGN's core identified from both the optical/IR host and the FIRST images (see Figure~\ref{Fig:AllClustersIR} for the optical/IR host galaxies).  The black circle shows the beam size and the black line shows 0$\farcm$5.}
\label{Fig:AllClusters}
\end{center}
\end{figure*}

\begin{figure*}
\begin{center}
\includegraphics[scale=0.14,trim={0.0in 0.0in 0.0in 0.0in},clip=true]{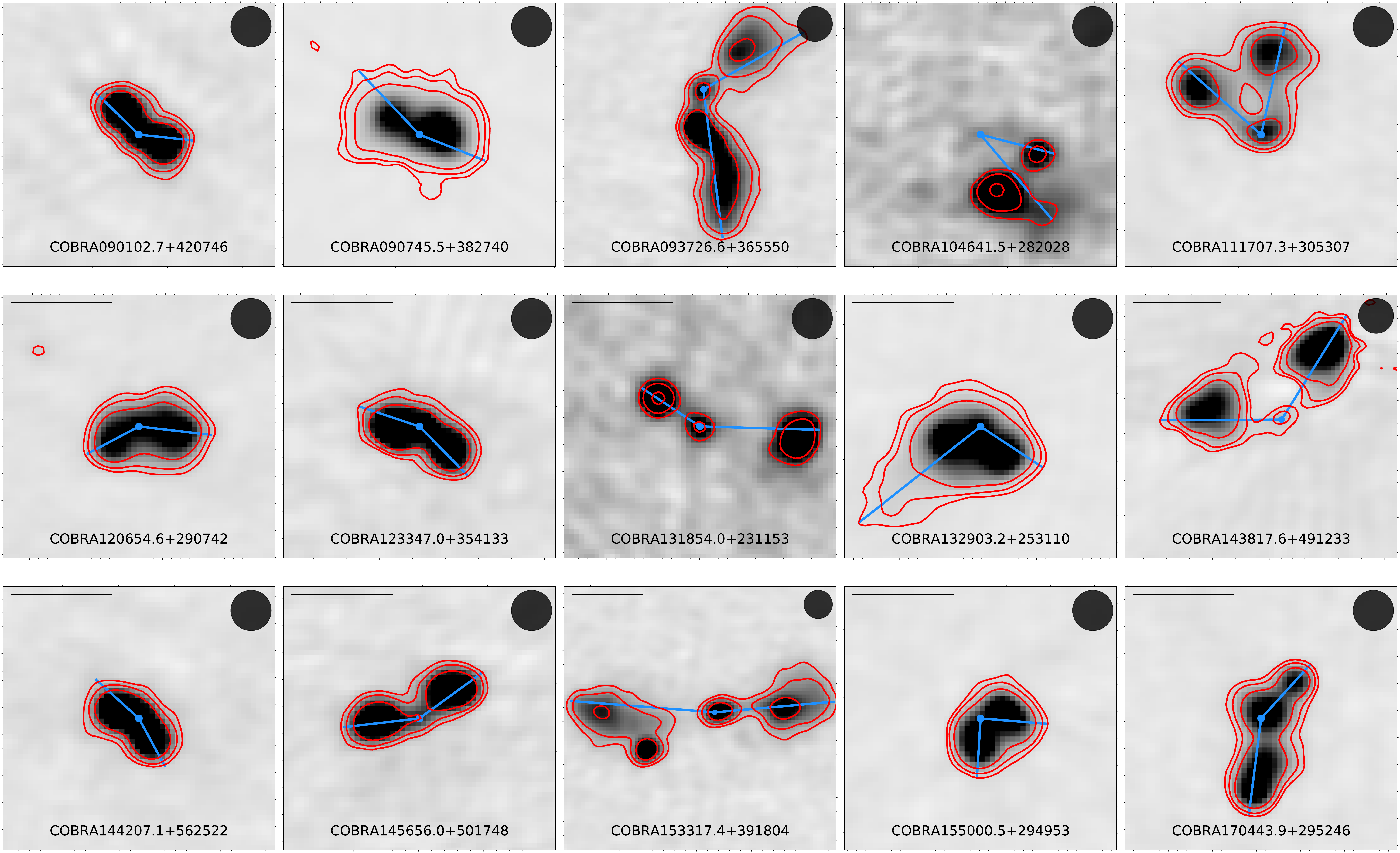}

\caption{Cutouts of the 15 bent, double-lobed radio sources in the non-cluster sample observed by LoTSS DR2.  While most of the cutouts are 1$\farcm$3$\times$1$\farcm$3, some of the images are slightly larger due to the larger angular size of the radio AGNs.  All contours, lines, and circles are the same as in Figure~\ref{Fig:AllClusters} (the location of the core for COBRA104641.5+282028 comes from the optical/IR host and the FIRST contours; see Figure~\ref{Fig:AllNonClustersIR} for the host galaxies).}
\label{Fig:AllNonClusters}
\end{center}
\end{figure*}

Of the 36 COBRA cluster candidates and 38 COBRA non-cluster candidates, only four sources were observed with LoTSS DR1 \citep[e.g.,][]{Shimwell2017,Shimwell2019}.  However, 37 COBRA sources from both samples were observed as part of LoTSS DR2.  In total, 21 are in high-$z$ cluster candidates and 16 are in the non-cluster sample.  Of the 21 high-$z$ cluster candidates, we remove one source where we do not have a well identified host galaxy (See Figures~\ref{Fig:AllClusters} and \ref{Fig:AllNonClusters} for the LoTSS images and Figures~\ref{Fig:AllClustersIR} and \ref{Fig:AllNonClustersIR} for $Spitzer$ images with host galaxy identification).  Additionally, of the 16 non-cluster sources, we remove one radio AGN that is not actually a bent radio AGN, but two overlapping radio sources.  For the remainder of this paper, the high-$z$ cluster sample (non-cluster sample) is these 20 (15) bent radio AGNs (see Table~\ref{tb:RadioProp} for the full list of bent radio AGNs in each sample).  

The goal of this paper is to study the impact of environment on bent AGNs, not redshift evolution.  While a true analysis of redshift evolution is beyond the scope of this paper due to the small sample, we show the redshift distribution of each sample in Figure~\ref{Fig:ClusterNon-z}.  As shown in Figure~\ref{Fig:ClusterNon-z}, despite there being more bent radio AGNs in the cluster sample at 0.5 $<$ z $<$ 1.5 than in the non-cluster sample, both samples include radio AGNs that cover a similar redshift range.  However, there is likely some redshift evolution influencing the observed properties, especially the environment.  Specifically, the density of radio AGNs in clusters increases with redshift \citep[e.g.,][]{Galametz2009,Mo2018,Mo2020} and the cluster environment has greater impact on the opening angle of bent AGNs due to the ICM density decreasing as a function of increasing redshift \citep{Vardoulaki2021}.  Additionally, the luminosity functions of low and high-luminosity radio AGNs peak at different redshifts \citep{Hardcastle2020}. Given the similar redshift distributions, we don't expect redshift evolution to be dramatically different between the two samples.  

Of note, the highest redshift source in the non-cluster sample is at z $=$ 2.346, while the highest redshift source in the cluster sample is at z $=$ 1.818 (both are SDSS spectroscopically-confirmed quasars).  This difference in redshift is not necessarily an evolutionary factor, where bent AGNs are only found in clusters at $z$ $<$ 2.0.  GM19 identified one cluster candidate at $z$ $=$ 2.18, which to date has not been observed by LoTSS.  Additionally, we find a slightly higher fraction of the bent radio AGNs in the non-cluster sample are SDSS-identified quasars (6 of 15 [40.0$\%$] vs 4 of 20 [25.0$\%$]) (See Table~\ref{tb:RadioProp}).  Because quasars are typically more energetic than other radio AGNs and have been shown to have slightly different properties in terms of their locations within high-$z$ clusters \citep[e.g.,][]{Mo2018}, this difference may impact our analysis. 

\subsubsection{Radio Source Projected Physical Size}\label{sect:RadioSize}

We aim to determine if the size and asymmetry of bent radio AGNs is impacted by the cluster environment. We estimate the projected physical size of each bent radio AGN in a similar manner to \citet{Moravec2019,Moravec2020}, who estimated the largest physical extent by identifying the two most distant points along 4$\sigma$ contours of their radio observations.  Since the ``C" shape of bent radio AGNs means that the two lobes might be closer to one another than to the core, we treat each lobe separately and measure the largest projected physical extent from the radio source core (identified from FIRST/$Spitzer$ in \citealp{Paterno-Mahler2017} and GM19) to each lobe and combine these two distances to get our measurement of the projected physical size (see Figures~\ref{Fig:AllClusters} and \ref{Fig:AllNonClusters} for examples showing the rays tracing the projected physical size and see Table\,\ref{tb:RadioProp} for our measurement of the projected physical size and the angular sizes of each lobe).  For bent radio AGNs with less complicated morphology, this is approximately the total extent of the radio source if it were straight.  However, for the more complex sources, this measurement tends to underestimate the true physical extent.  

We determine our detection criteria by measuring the rms noise in each LoTSS image.  Although we only show 1$\farcm$3 $\times$ 1$\farcm$3 cutouts in Figures~\,\ref{Fig:AllClusters} and \ref{Fig:AllNonClusters}, we did our analysis using 5$\arcmin$ $\times$ 5$\arcmin$ cutouts centered on the radio AGN.  We estimate the rms by analyzing a 50\,pixel $\times$ 50\,pixel region in each LoTSS cutout ($\approx$ 1$\farcm$25 $\times$ 1$\farcm$25).  This region is near the radio source and devoid of other bright radio sources.  We report the rms values of each field in Table~\ref{tb:RadioProp-2}.  Based on the appearance of radio artifacts, as well as the high level of detection in the LoTSS data, we use a 10$\sigma$ threshold for the minimum detection for each bent radio AGN.  This value is based on the rms noise of each image to account for slight differences in the local background.  

As size measurements are highly dependent on the detection threshold, we estimate the error in our projected physical sizes by measuring the difference in the size of each radio lobe if we instead use 9$\sigma$ or 11$\sigma$ as our minimum threshold.  To more thoroughly account for this error, we do not extend the ray between the location of the maximum projected distance at 10$\sigma$ to the 9$\sigma$ and 11$\sigma$ contours.  Rather, we remeasure the location of the greatest physical extent at 9$\sigma$ and 11$\sigma$ and measure the projected physical size at those contour values.  We then add this measurement error in quadrature with our estimate of our observational error, which we assume is half of the beam radius (3$\arcsec$ for LoTSS) to account for the total error.  To convert our projected angular size into a projected physical size, we multiply by the angular diameter distance, which we calculate using \verb|astropy| \citep[][]{astropy:2013,astropy:2018}.  We estimate the error in the angular diameter distance following GM21 and measure the difference in the angular diameter distance based on the error in the redshift (see Table~\ref{tb:RadioProp}).  We then combine these values with the measurement error to get the total error in the projected physical size.  Because our combined error is larger than the 0$\farcs$2 astrometric uncertainty reported in \citet{Shimwell2022}, we do not account for this additional source of error. 

To verify the validity of these measurements, we examine each source by eye.  The extent of most sources are well estimated by our measurement of the projected physical size.  However, a few sources have greatest physical sizes that are impacted by small features in the 10$\sigma$ contour creating rays that do not trace the structure as determined by eye (e.g., COBRA104254.8+290719, COBRA154638+364420).  For example, COBRA104254.8+290719 has an additional radio feature to the left of the upper lobe, which was initially identified as the location of the greatest radio extent.  While these features are real (and the feature for COBRA104254.8+290719 appears, though much fainter, in the FIRST image as well), we ignore these features when measuring the projected physical size and use the greatest extent from the main lobe as shown in Figure~\ref{Fig:AllClusters}.  Additionally, for COBRA104541+282028 (see Figure~\ref{Fig:AllNonClusters}), the radio contours trace two separate lobes with no identified core.  For this source, we estimate the size as the position from the core, which is the location of the host galaxy identified in \citet{Paterno-Mahler2017}, to the farthest extent of the radio lobes.  Thus, the physical extent extends beyond the actual size traced by the radio lobes, which may add an additional source of error.

Although we treat the projected physical size as an intrinsic property of the radio source, it is highly dependent on the redshift and our measurement of corresponding noise levels.  While we use a 10$\sigma$ clipping for each radio AGN, the surface brightness of the radio source is dependent on the redshift and thus, the level of that 10$\sigma$ clip relative to the radio lobes, is also impacted.  Since the majority of our sources have lobes with brighter edges, they are not as impacted as sources such as COBRA164611.2+512915 and COBRA171330.9+423502 in Figure~\ref{Fig:AllClusters} and COBRA090745.5+382740 and COBRA132903.2+253110 in Figure~\ref{Fig:AllNonClusters}, which have clearly visible faint tails.  To determine the effect on the system as a whole, we examine the total angular size as a function of redshift using a Spearman Test.  For the cluster sample, we find a weak correlation with redshift and no evidence to reject the null hypothesis (r$_s$ = -0.25, $p$ = 0.29).  For the non-cluster sample, we find a moderate trend, with no evidence to reject the null hypothesis (r$_s$ = -0.42, $p$ = 0.12).  Given the smaller sample size and that the highest redshift source (one of the smallest sources) is in the non-cluster sample we remove this source to determine if it is an outlier.  In this case, we find a weak trend with no evidence to reject the null hypothesis (r$_s$ = -0.310, $p$ = 0.281).  While our sample is not statistically impacted by the redshift dependence of the measurement, it may slightly bias some of our results and should be taken as another potential source of error when we discuss the impacts of physical size and the asymmetry of radio lobes as a function of environment.

\subsubsection{Radio Source Projected Physical Area}\label{sect:RadioArea}
Although the projected physical size is a common measurement of radio AGNs \citep[e.g.,][]{Moravec2019,Moravec2020,Shen2020}, the physical size is only a one-dimensional measurement.  The cluster environment might limit the degree of collimation within each radio lobe, which would not necessarily be accounted for in a length measurement.  As such, we estimate the projected physical area of each radio lobe to account for the total volume of the lobes.  We measure the projected physical area by measuring the area enclosed within each 10$\sigma$ contour using Green's Theorem, which transforms the line integral over the curve of our contour into a double integral over the region within the contour.  We convert this to the projected physical area by multiplying the area within the contour by the angular diameter distance squared (see Table\,\ref{tb:RadioProp} for the total projected physical area measurements).  
\begin{figure}
\begin{center}
\includegraphics[scale=0.4,trim={1.098in 0.53in 0.12in 0.093in},clip=true]{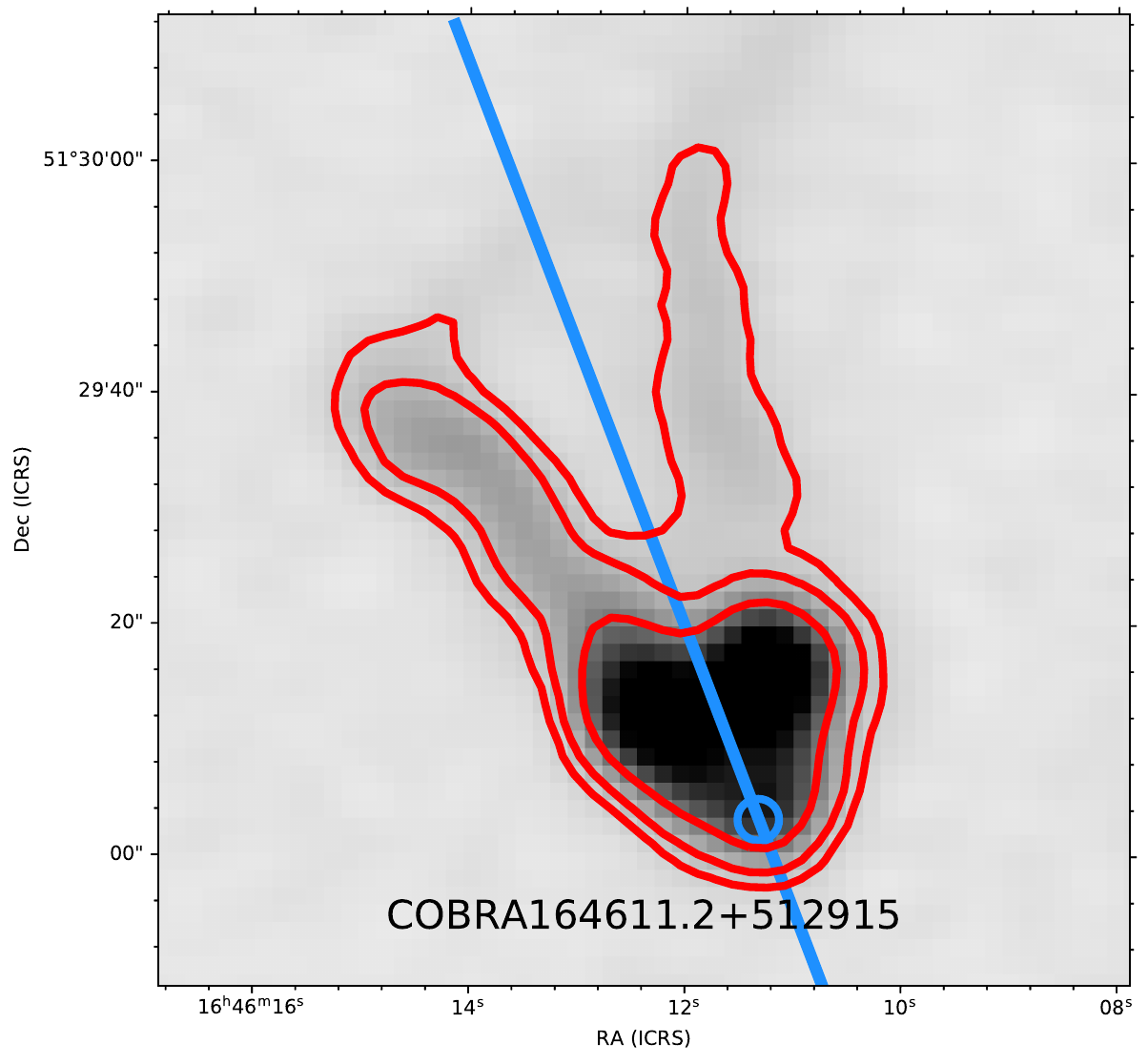}

\caption{Radio contours for COBRA164611.2+512915.  Here we show our fiducial divide of the radio source (the blue line) through the radio core that we use to divide the radio AGNs into lobe 1 and lobe 2.  As in Figure~\ref{Fig:AllClusters}, the blue circle shows the AGN core and the red contours show 10$\sigma$, 20$\sigma$, and 50$\sigma$ values.}
\label{Fig:AreaDivide}
\end{center}
\end{figure}

As we are interested in probing asymmetry between the bent lobes of each AGN, we estimate a fiducial divide centered on the radio core and extending to the 10$\sigma$ contour to separate the radio lobes (see Figure~\ref{Fig:AreaDivide}).  While this allows us to measure lobes separately, we do not treat the central core region as a separate component (this region becomes evident in some sources when we look at the spectral index - see Section~\ref{sect:SpectralIndex}), but instead include it equally as part of the area of each lobe. Since the core region is relatively symmetric, we believe this is the best solution in estimating lobe asymmetries.  

Unlike the projected physical size, for the projected physical area, we only measure the extent within the 10$\sigma$ contours of the main lobes (for some of the radio sources, such as COBRA104254.8+290719, there are additional associated components, which we do not include).  As discussed for our measurement of the projected physical size, there is one example where no emission is detected at the core at the 10$\sigma$ level and multiple sources that re detected as separate components (See Figure~\ref{Fig:AllClusters}).  While our measurement of the largest projected physical size accounts for the distance regardless of whether or not the entire source is contained within the 10$\sigma$ contour (including the core), our measurement of projected physical area only includes the area of the components, not any space connecting the core to the lobes.  Thus, our measurements of the projected area may be a slight underestimate, as there should be a radio jet connecting the regions. 

We estimate the error in this measurement following the same approach as in Section~\ref{sect:RadioSize}.  Using 9$\sigma$ and 11$\sigma$ contours, we measure the area of each lobe and compare it to the 10$\sigma$ value, which we similarly combine in quadrature with half of the area of the radio beam.  We then combine this error with the error measurement for the angular diameter distance to account for the total error.  

Similar to the projected physical size, the projected physical area is subject to the redshift dependence of surface brightness relative to our 10$\sigma$ flux density threshold and is not an intrinsic property of the radio source.  Again, we do a Spearman test to compare the angular area as a function of redshift and find a very weak correlation and no evidence to reject the null hypothesis within the cluster sample (r$_s$ = -0.18, $p$ = 0.43).  We again find a moderate correlation among the sources in the non-cluster sample, with no evidence to reject the null hypothesis (r$_s$ = -0.44, $p$ = 0.10).  Like the projected physical size, the slight redshift dependence in the non-cluster sample may be due to the highest redshift source being one of the smallest sources.  When we do the Spearman test without this source, we find a weaker trend with no evidence to reject the null hypothesis (r$_s$ = -0.36, $p$ = 0.20).  While we see no statistical trends with redshift, it is possible that our measurements of the projected physical area and the asymmetry of the radio lobes, are impacted by this redshift effect.  As such, this should be treated as an additional source of error in our analysis.    

\subsubsection{Radio Source Opening Angle}\label{sect:RadioAngle}
\begin{figure*}
\begin{center}
\includegraphics[scale=0.25,trim={1.1in 0.54in 0.1in 0.1in},clip=true]{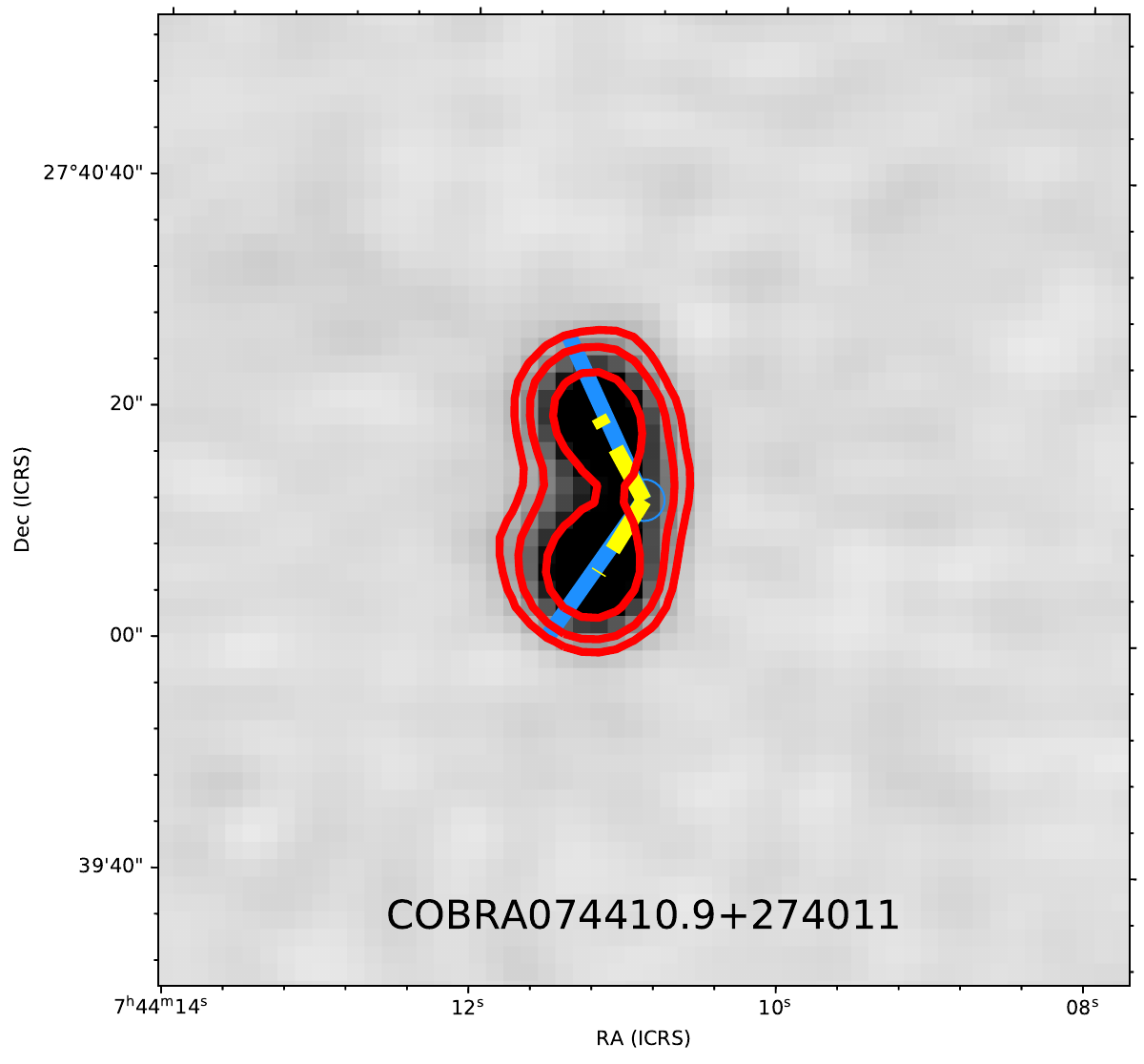}
\includegraphics[scale=0.25,trim={1.1in 0.54in 0.1in 0.1in},clip=true]{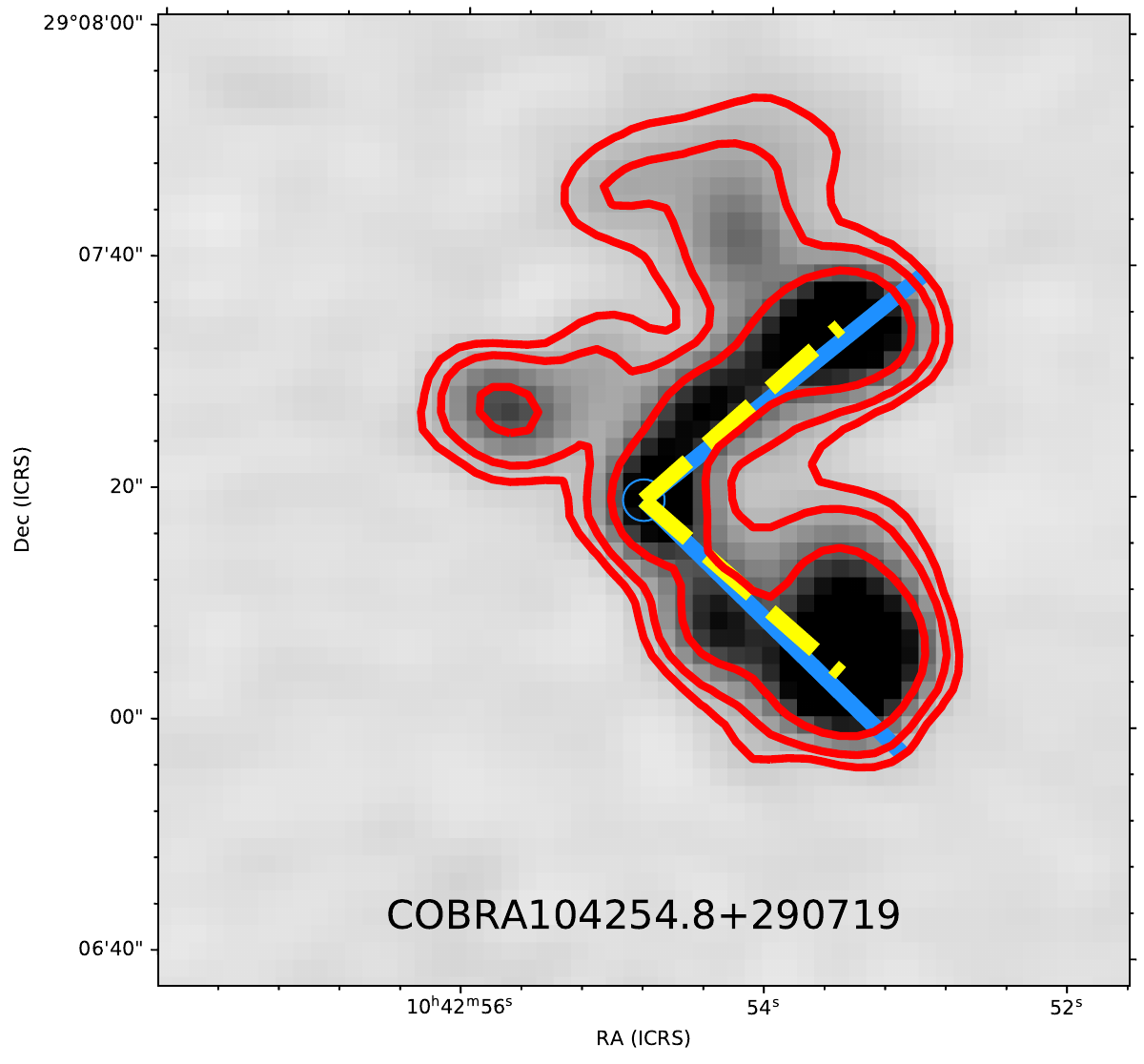}
\includegraphics[scale=0.25,trim={1.1in 0.54in 0.1in 0.1in},clip=true]{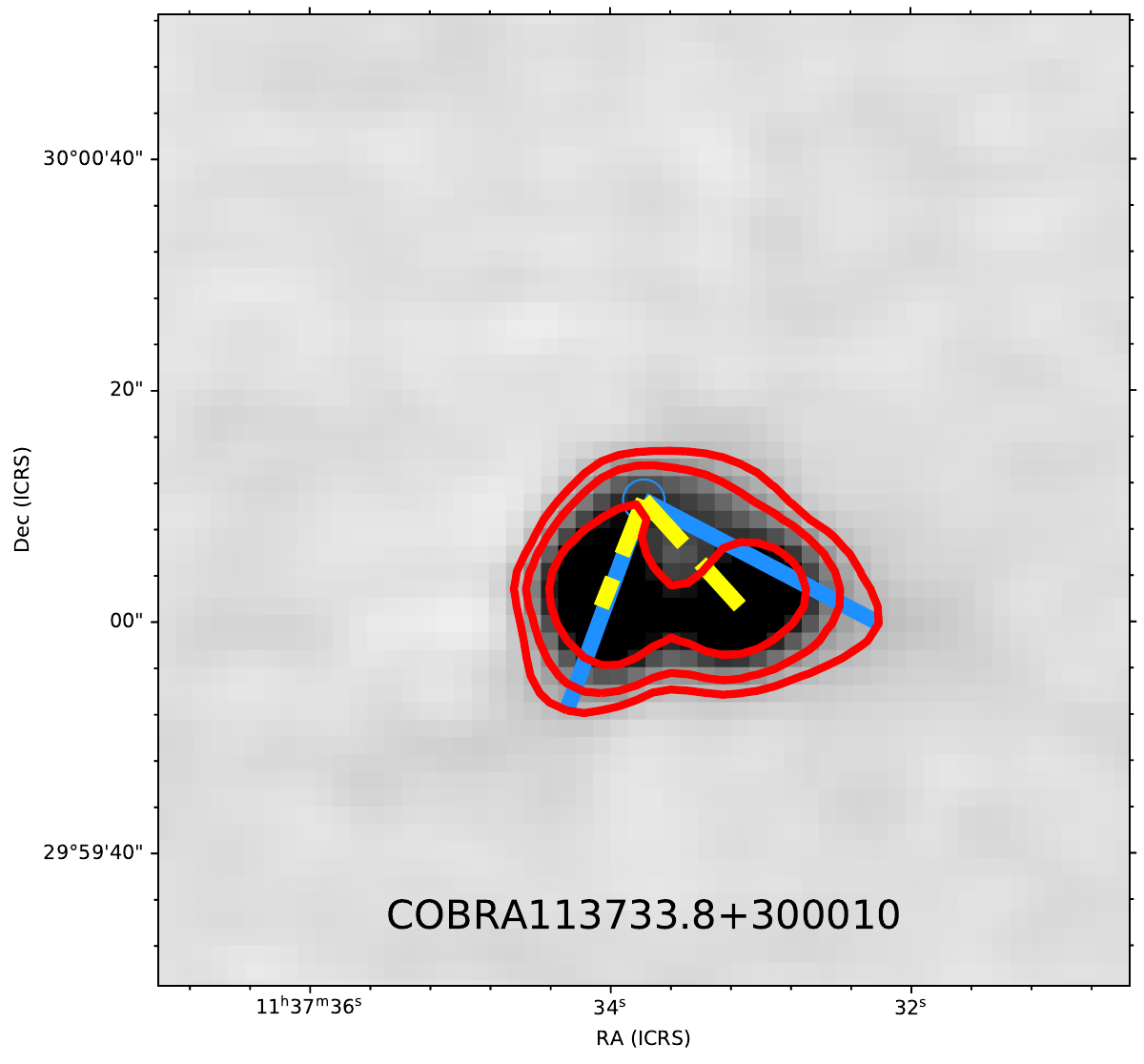}
\includegraphics[scale=0.25,trim={1.1in 0.54in 0.1in 0.1in},clip=true]{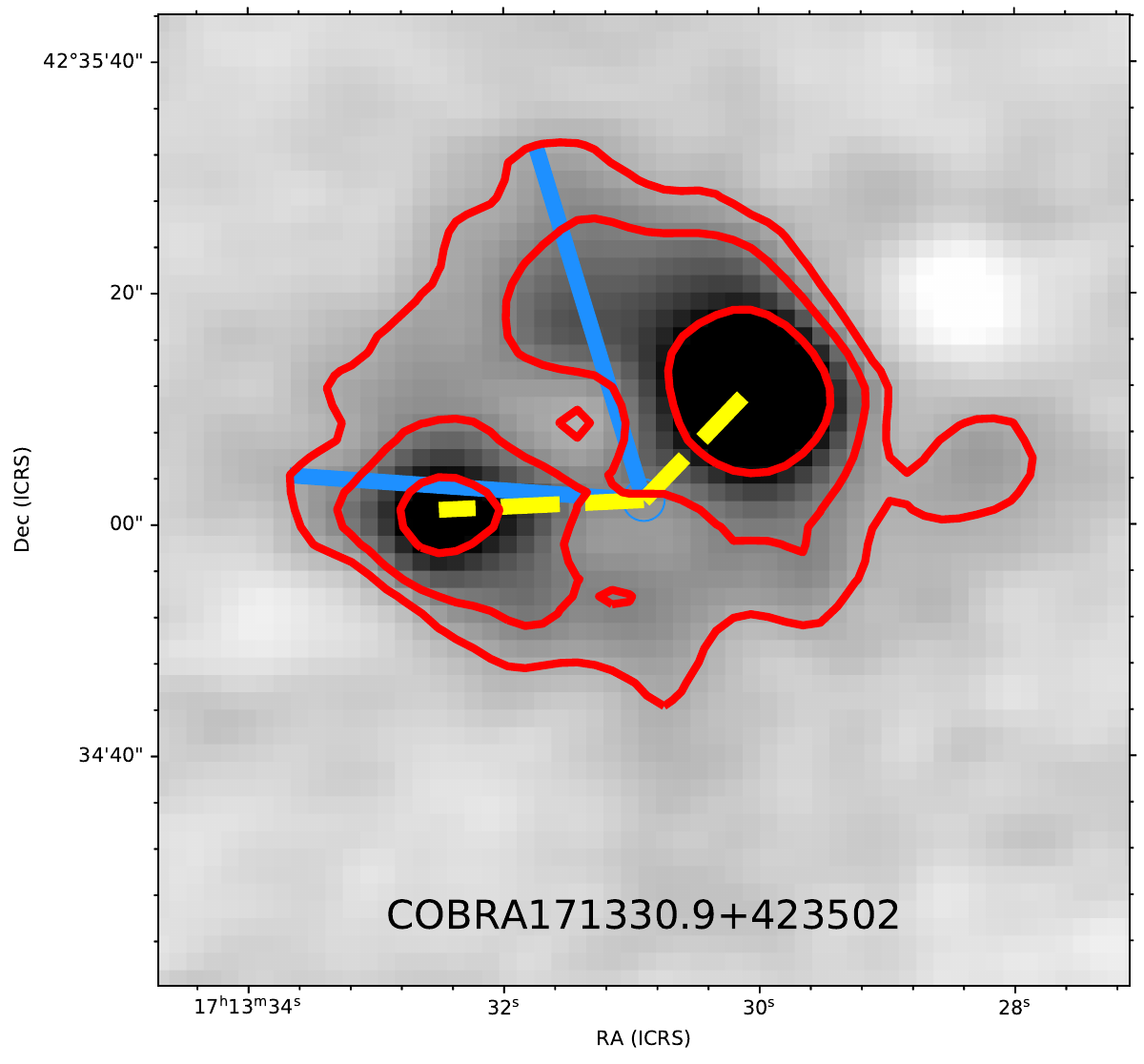}

\caption{Examples of our measurement of the opening angle for the LoTSS images.  The red contours are the same contours as in Figure~\ref{Fig:AllClusters}.  The yellow dashed lines show L$_{1}$ and L$_{2}$ from Equation~\ref{equation:Arccos}, while the blue lines show our estimate of the opening angle using the largest physical extent of each radio source.  In some cases, the opening angles are nearly identical, while most slightly differ.  However, some give a dramatically different value, as is the case for COBRA171330.9+423502, in the far right panel.}
\label{Fig:OpeningAngle-NEWVersion-SHOW}
\end{center}
\end{figure*}

Although our measurements of the projected physical size of each radio lobe yield a first order estimate of the opening angle of each bent radio AGN, there are a number of sources, including COBRA171330.9+423502, where the projected physical extents do not trace the peak regions of flux and thus yield an inaccurate representation of the opening angle.  To better capture the opening angle for all bent radio AGNs, we measure the angle between the radio source core (as identified in GM21) and the brightest region in each lobe (see Figure~\ref{Fig:OpeningAngle-NEWVersion-SHOW}).  

We quantify this opening angle, following the convention in GM21: 
\begin{equation}\label{equation:Arccos}
    \rm{Opening\,\,Angle} = \rm{arccos}\left(\frac{(H)^2 - (L_1)^2 - (L_2)^2}{-2(L_1)(L_2)}\right),
\end{equation}
where H is the ray connecting the two brightest pixels in each lobe, and L$_{1}$ and L$_{2}$ are the rays between the core and the brightest pixel in each lobe respectively.  Although we define lobe 1 and lobe 2 and treat them separately throughout the majority of this paper, they are arbitrarily defined, meaning any trends between properties of lobe 1 and lobe 2 simply show a degree of asymmetry, not necessarily a defining characteristic of each lobe.  Similar to GM21, we estimate the error in the opening angle using the error in the relative position of each component.  

\begin{figure}
\begin{center}
\includegraphics[scale=0.7,trim={0.0in 0.0in 0.0in 0.0in},clip=true]{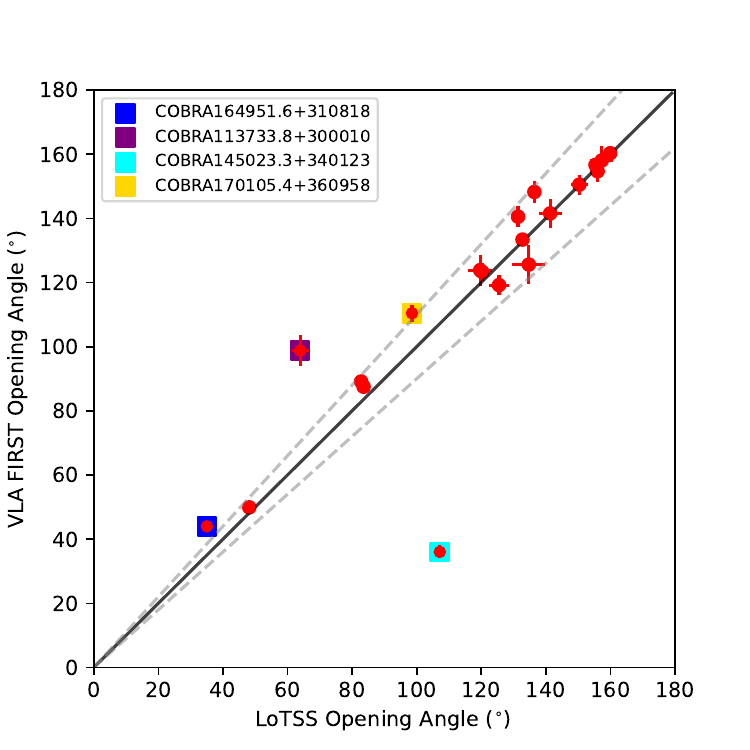}

\caption{A comparison of the opening angle measurements from FIRST (GM21) with the opening angle measurements from LoTSS DR2 for the cluster sample.  The black lines shows a one-to-one trend, while the two grey dashed lines highlight a 10$\%$ difference between the opening angles.  We overlay colored squares on all bent AGNs where the opening angles differ by greater than 10$\%$.  For the majority of our sources, these measurements are very similar.}
\label{Fig:OpeningAngle-COMP}
\end{center}
\end{figure}

As shown in Figure~\ref{Fig:OpeningAngle-COMP}, by comparing the opening angle measurements from GM21 from FIRST with the opening angle measurements from LoTSS (see Table\,\ref{tb:RadioProp} for the full list of opening angles), we find that there is relative agreement between the values, with all but four sources having a measurement difference of less than 10$\%$. We individually examine these four sources with the largest difference in the opening angle.  For COBRA164951.6+310818, the major offset is due to an error misidentifying the radio source core (and thus the opening angle) in GM21.  Similarly, for COBRA145023.3+340123, the difference comes from differences in the identified lobes.  As shown in Figure~\ref{Fig:3band}, in the FIRST image, the component to the west of the core, shown as a separate component in FIRST, is much fainter and was not included in the FIRST catalog.  Thus, we instead identified the second lobe as what appears to be a secondary lobe feature in the eastern lobe in the LoTSS image.  Although we note that the western component in COBRA145023.3+340123 does appear to be a radio lobe, the lack of inclusion in the FIRST catalog is intriguing.  While other radio AGNs like COBRA164611.2+512915 have additional radio emission extending beyond the brightest components of the main lobes, no other source has a previously unreported feature.  As such, we caution that this new lobe may be a co-spatial source and requires further analysis to accurately trace the opening angle of this radio source.  For the third source, COBRA113733.8+300010, which is in the only spectroscopically confirmed cluster in our sample \citep{Blanton2003}, the difference is due to the highly curved nature of the lobes, which create differences based on the location of the peak brightness in the lobes.  For the fourth source, COBRA170105.4+360958, the offset is slight ($\approx$ 12$\%$), and again likely due to small differences in the location of the peak brightness.  

\subsubsection{Radio Luminosity}\label{sect:RadioPower}
Beyond the visual morphology, we are also interested in probing the differences in the total radio luminosity of each radio source to see if any differences exist among the populations inside and outside of clusters.  To measure the radio luminosity, we follow \citet{Moravec2019,Moravec2020} and GM21 and use Equation~\ref{Eq:RadioPower},
\begin{equation}\label{Eq:RadioPower}
    L_{144\,MHz} = 4\pi\,D^{2}_{L}S_{144\,MHz}\,(1+z)^{\alpha - 1},
\end{equation}
where L$_{144\,MHz}$ is the radio luminosity (W\,Hz$^{-1}$) at 144\,MHz, the median frequency of the LoTSS survey, D$_{L}$ is the luminosity distance (Mpc), calculated with Astropy using the redshift estimates from GM19, $\alpha$ is the spectral index calculated by measuring the integrated spectral index of the entire source (see Section~\ref{sect:SpectralIndex}), and S$_{144\,MHz}$ is the integrated radio flux density (mJy) of each radio source.  Although \citet{Shimwell2022} report flux densities for these radio sources, to more accurately capture the flux density of the radio sources for our analysis, we use the CASA tool \verb|imstat| to meausure the flux density across the radio source (above the 10$\sigma$ threshold, as shown in the radio contours in Figures~\ref{Fig:AllClusters} and \ref{Fig:AllNonClusters}). 

For our bent radio AGNs in cluster candidates, we find that the radio luminosities range from 1.7$\times$10$^{25}$\,WHz$^{-1}$ to 4.1$\times$10$^{27}$\,WHz$^{-1}$ (see Table\,\ref{tb:RadioProp-2} for the full list of radio luminosities).  Compared to the measurements to the radio luminosities from GM21, we find that the three most powerful radio sources are three of the four most powerful radio sources reported in that paper (the fourth brightest bent AGN has not been observed as part of LoTSS).  Similarly, we find that for the bent radio AGNs in the non-cluster sample, the radio luminosities range from 0.8$\times$10$^{25}$\,WHz$^{-1}$ to 3.3$\times$10$^{27}$\,WHz$^{-1}$.  Based purely on the radio luminosity, we do not see any major differences between the two samples.    

\subsection{Spectral Index}\label{sect:SpectralIndex}
To more completely probe the energetics of each bent radio AGN, we measure the integrated spectral index of the radio emission across the radio source.  Because the SED of an AGN is dominated by synchrotron emission, it can be estimated by a smooth power law of the form $\nu^{\alpha}$.  The value of $\alpha$, the spectral index, is typically of the order $-$0.7 to $-$0.8 for extended radio sources \citep[e.g.,][]{Kellermann1988,Sarazin1988,Condon1992,Peterson1997,Lin2007,Miley2008} and can be calculated by measuring the flux density in different wavebands.

\begin{figure*}
\begin{center}
\includegraphics[scale=0.25,trim={1.1in 0.54in 0.10in 0.1in},clip=true]{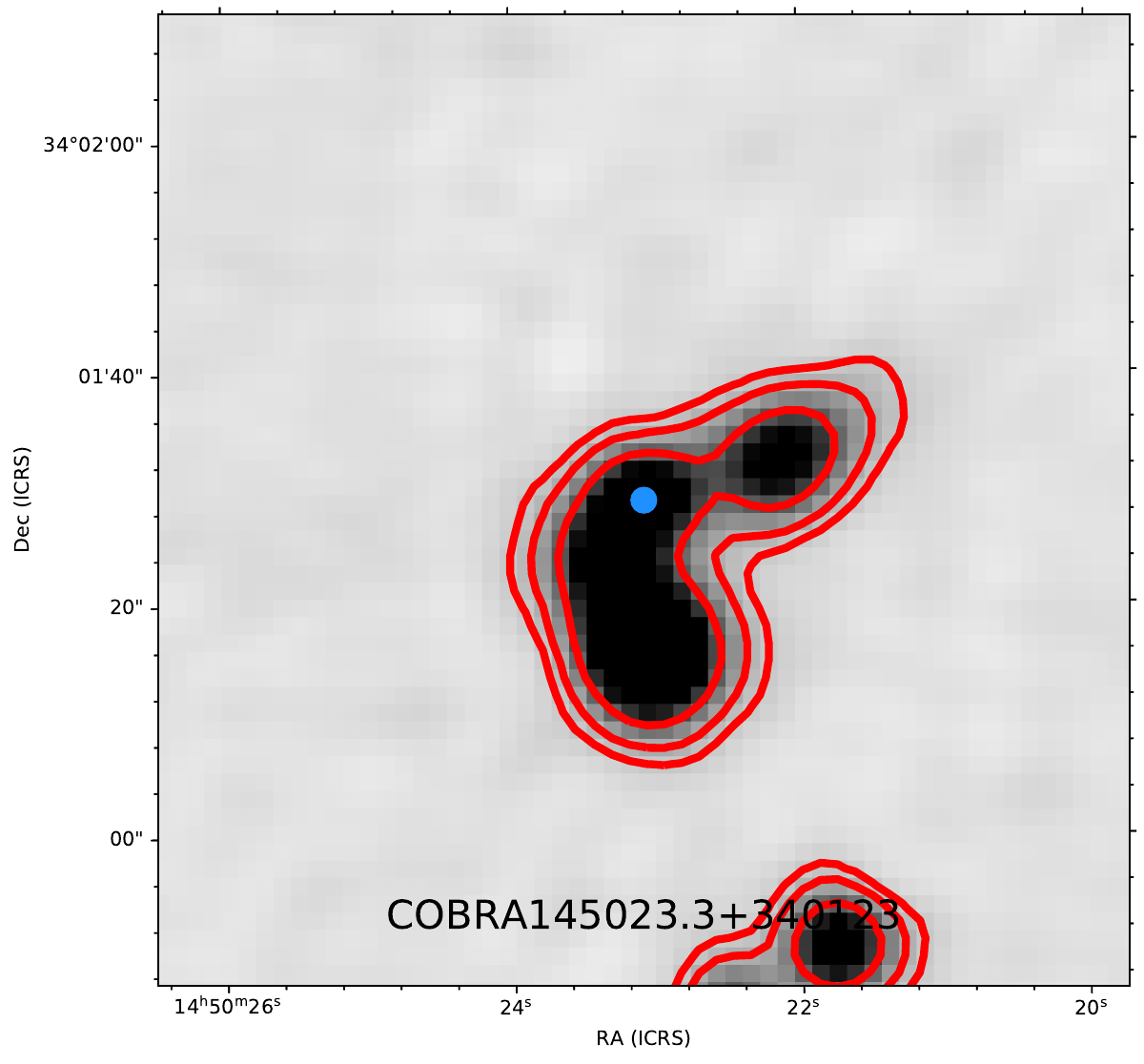}
\includegraphics[scale=0.25,trim={1.1in 0.54in 0.10in 0.1in},clip=true]{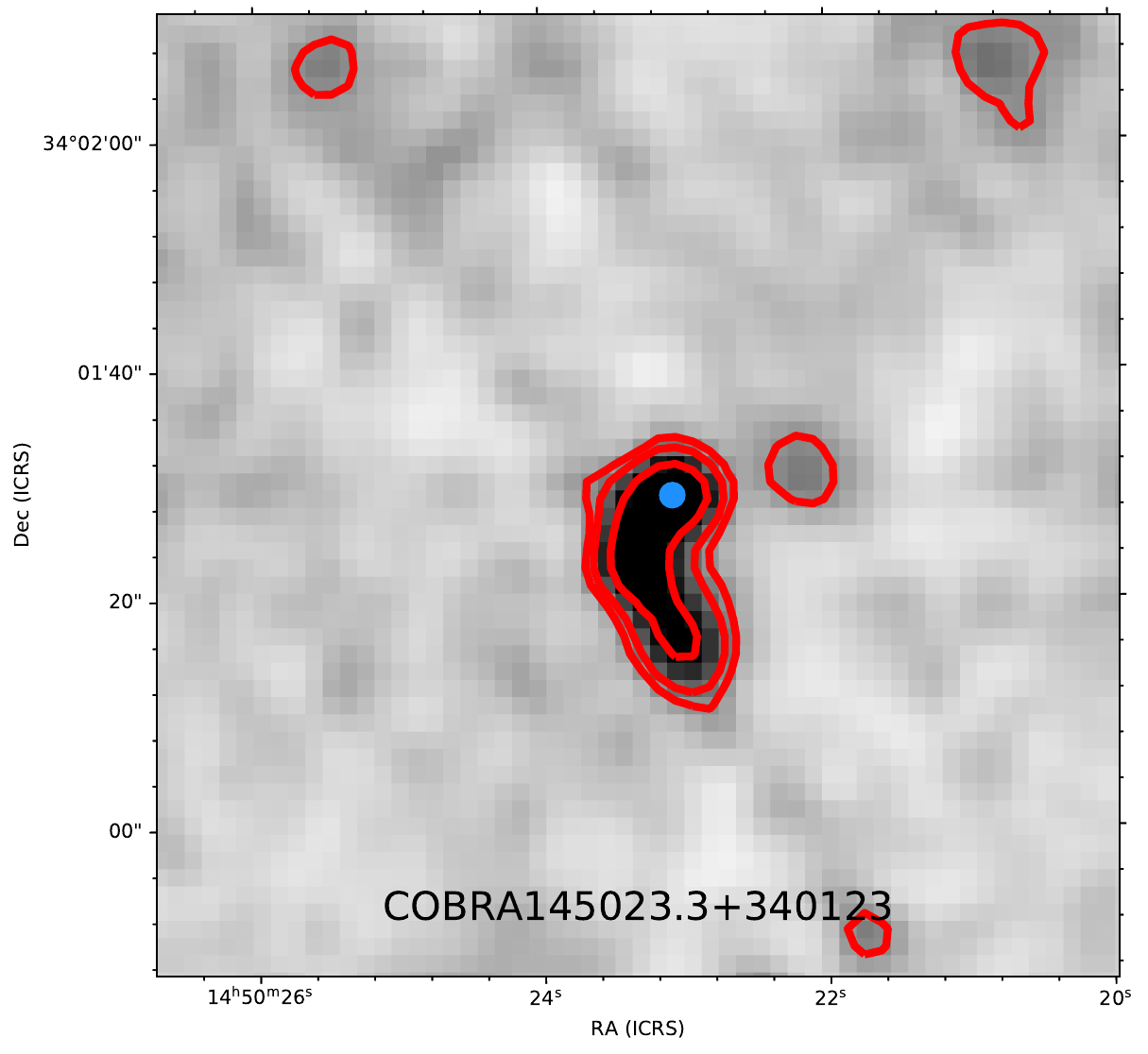}
\includegraphics[scale=0.25,trim={1.1in 0.54in 0.10in 0.1in},clip=true]{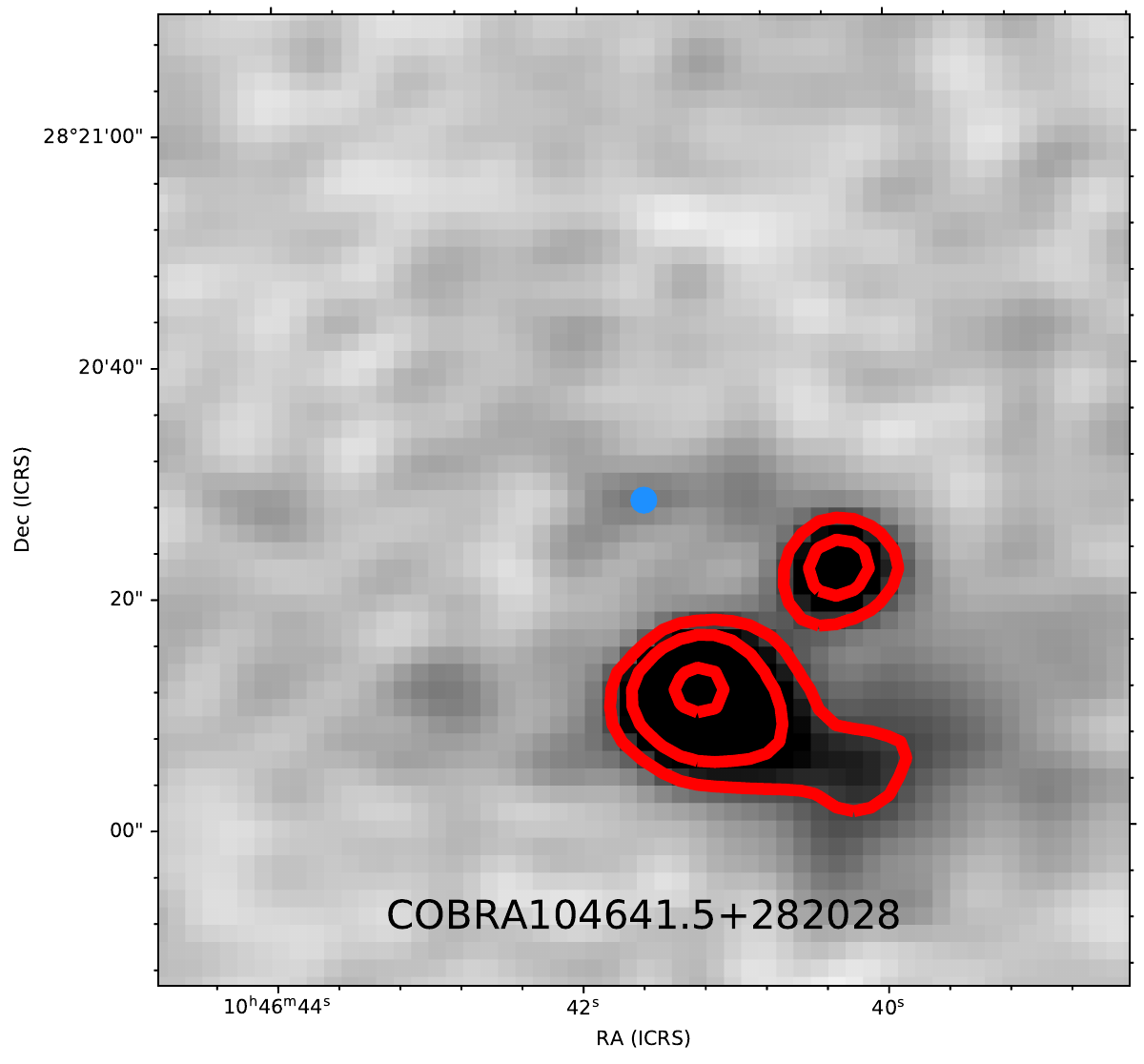}
\includegraphics[scale=0.25,trim={1.1in 0.54in 0.10in 0.1in},clip=true]{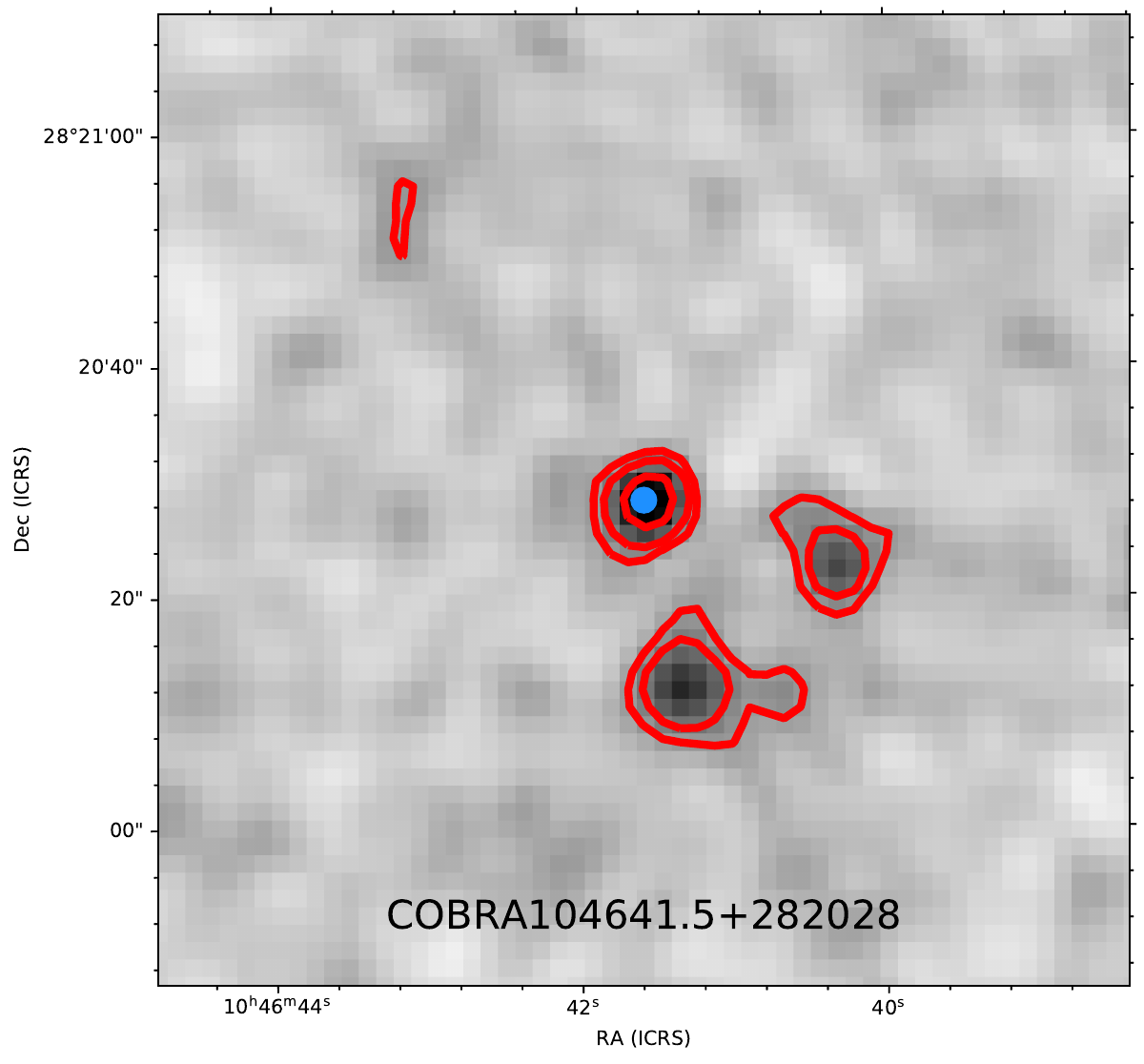}

\caption{Examples of the two bands used to create spectral index maps.  The left two images show LoTSS and FIRST images of COBRA145023.3+340123, while the right two images are from COBRA104641.5+282028.  In the LoTSS images, the red contours show the 10$\sigma$, 20$\sigma$, and 50$\sigma$ contours, while in the FIRST images, the red contours show 3$\sigma$, 5$\sigma$, and 10$\sigma$.  The blue circle identifies the location of the core.  Each FIRST image is smoothed and regridded to match LoTSS (6$\arcsec$ resolution and 1$\farcs$5 pixels).  We specifically highlight two sources with slight differences in the components.  For COBRA145023.3+340123, the west component is not as strongly identified in FIRST, while for COBRA104641+282028, the core is not detected at the 10$\sigma$ level in LoTSS.}

\label{Fig:3band}
\end{center}
\end{figure*}

\citet{Shimwell2022} measured a spectral index for sources identified in the LoTSS DR2 catalog by dividing the wide LOFAR band into two bands, and noted that multiple factors led to a large degree of uncertainty, with error in the value of the order $\gtrapprox$ $0.2$.  To better probe the energetics of the core and lobes, we measure the spectral index by combining the LoTSS data with publicly available data from FIRST \citep{Becker1995} (the same data used in GM19 and GM21).  Using these two bands, we cover a frequency range of 144\,MHz (LoTSS) to 1.44\,GHz (FIRST).  To accurately compare each bent radio AGN in each waveband, we use the CASA tool \verb|imsmooth| \citep{CASA} to smooth the FIRST images to the angular resolution of LoTSS (6$\arcsec$; the angular resolution of FIRST is 5$\farcs$5).  We then use the CASA tool \verb|imregrid| to rescale the FIRST images to the pixel scale of LoTSS (1$\farcs$5 per pixel; the pixel scale of FIRST is 1$\farcs$8 per pixel).  Using these recalibrated FIRST images, we measure the rms noise for the FIRST images using a 50\,pixel $\times$ 50\,pixel region as done for our previous LoTSS analysis.  

While our radio AGNs are very bright relative to the background in the LoTSS images, the same radio AGNs are not as bright relative to the background in the FIRST images (see Figure~\ref{Fig:3band}).  To determine the spectral index, we measure the ratio of the flux densities in the LoTSS and FIRST images.  To best account for the flux density in the same spatial region of the radio source, we measure the flux density within the 3$\sigma$ contours based on the FIRST images for both the FIRST and LoTSS observations.  As shown in Figure~\ref{Fig:3band}, the 3$\sigma$ FIRST contours generally fall within the 10$\sigma$ LoTSS contours.  However, for a small number of sources, the 3$\sigma$ FIRST contours extend outside of the LoTSS contours.  For those sources, we measure the flux within the 3$\sigma$ FIRST contours in each band to best account for the flux of the radio source.  As these regions are generally only slightly offset from the LoTSS 10$\sigma$ contours, the FIRST contours should accurately probe the spectral index.  The exception is the core in COBRA104641.5+282028, which is detected at the 3$\sigma$ level in FIRST and not at the 10$\sigma$ level LoTSS (this field is flagged in Table~\ref{tb:RadioProp-2}). 

To calculate the spectral index, we first write the flux density, F, as: 
\begin{equation}\label{eq:1}
 \textnormal{F} = A\nu\,^\alpha,   
\end{equation} where A is the amplitude, $\nu$ is the rest-frame central frequency of the band, and $\alpha$ is the spectral index.  We calculate the spectral index by taking the ratio of the two flux densities, noting that the amplitude is the same for the same radio source.  Thus, the spectral index has the form:
\begin{equation}\label{eq:2}
    \alpha = \frac{\textnormal{log}(\frac{F1}{F2})}{\textnormal{log}({\frac{\nu1}{\nu2}})},
\end{equation} where F1 is the flux density (mJy) in LoTSS, F2 is the flux density (mJy) in FIRST, $\nu$1 is the frequency of LoTSS, and $\nu$2 is the frequency of FIRST.

We measure the error in the spectral index from the error in the radio flux density at each frequency following the methodology to \citet{DiGennaro2018a}.  For the LoTSS and FIRST images, we follow \citet{Shimwell2022}, and assume a 10$\%$ error in these flux density measurements.  We add this error in quadrature with the error due to the rms noise (see Table~\ref{tb:RadioProp-2}) to estimate the total error in the flux density.  We then sum these errors in quadrature following Equation 1 in \citet{DiGennaro2018a} to get the error in the spectral index.

\section{A Comparison of the Physical Properties of Bent AGNs Inside and Outside of Cluster candidates}\label{sect:AGNcluster}

\subsection{Comparing Size \& Asymmetry in Radio Lobes}\label{sect:Asymmetry}
\begin{figure}
\begin{center}
\includegraphics[scale=0.7,trim={0.0in 0.0in 0.0in 0.4in},clip=true]{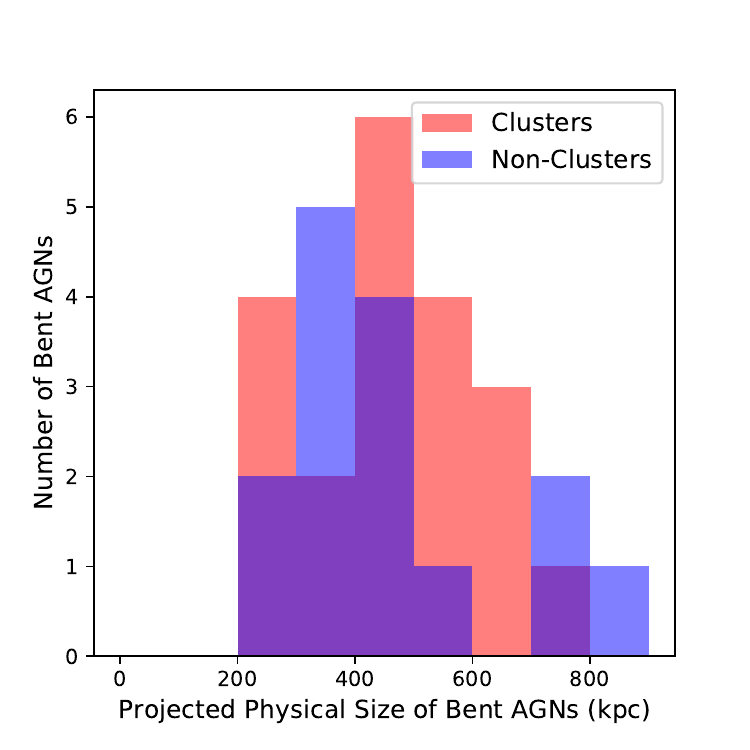}
\includegraphics[scale=0.7,trim={0.0in 0.0in 0.0in 0.4in},clip=true]{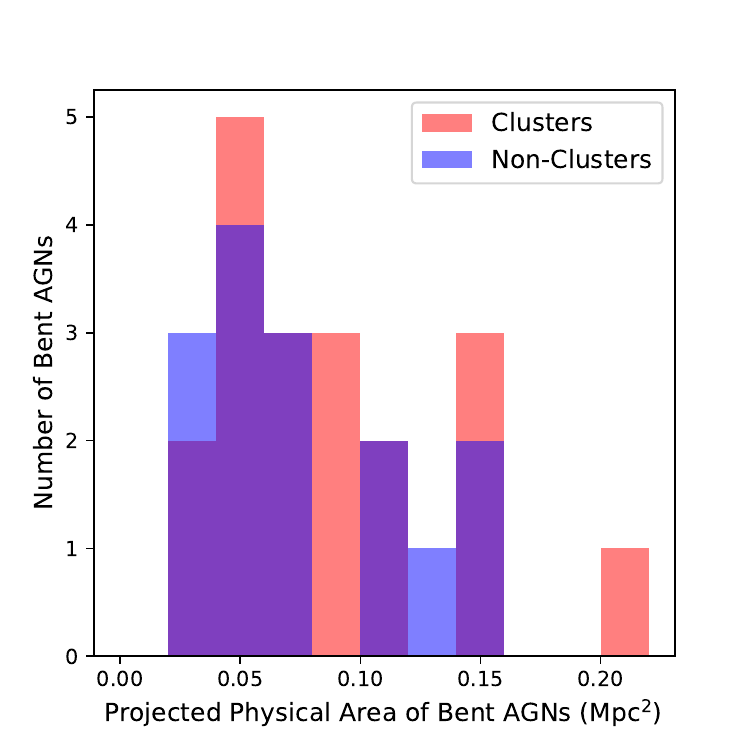}

\caption{The projected physical size and area distributions of the bent radio AGNs in the cluster and non-cluster sample.  We see that each sample spans a similar range of sizes and areas.}
\label{Fig:SizeHist}
\end{center}
\end{figure}

To understand the impact of environment on bent radio AGN morphology beyond the opening angle \citep[e.g.,][]{Hardcastle2005,Morsony2013,Garon2019,Vardoulaki2021}, we compare the distribution of radio source sizes and areas within each sample.  As shown in Figure~\ref{Fig:SizeHist} and  Table\,\ref{tb:RadioProp}, bent radio AGNs in the cluster sample span a similar range of projected physical sizes as their non-cluster counterparts (256\,kpc to 724\,kpc for the cluster sample and 275\,kpc to 838\,kpc in the non-cluster sample).  Similarly, we find relative agreement between the projected physical areas of bent radio AGNs in the cluster sample and their non-cluster counterparts (0.029\,Mpc$^{2}$ to 0.229\,Mpc$^{2}$ for the cluster sample and 0.027\,Mpc$^{2}$ to 0.157\,Mpc$^{2}$ for the non-cluster sample). 

\begin{figure*}
\begin{center}
\includegraphics[scale=0.7,trim={0in 0in 0in 0in},clip=true]{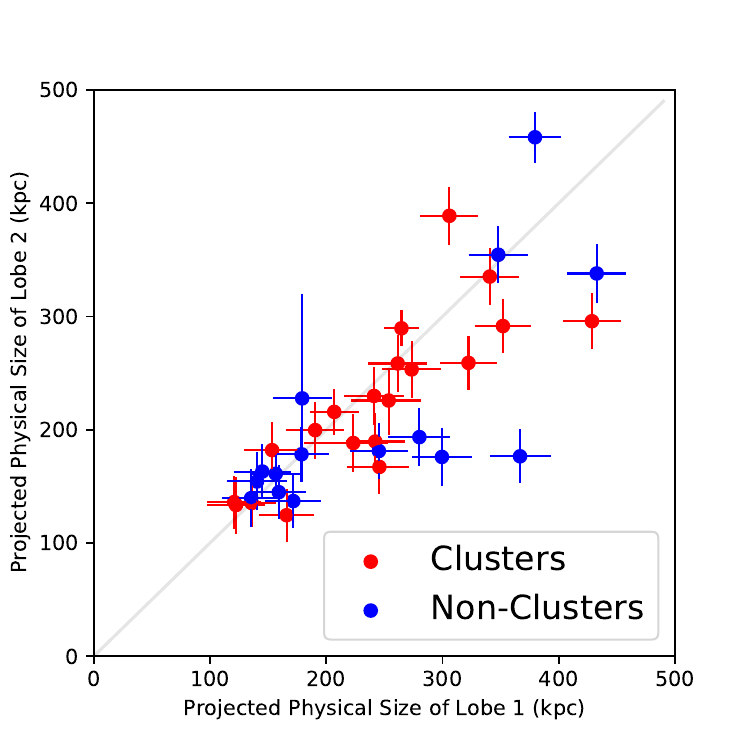}
\includegraphics[scale=0.7,trim={0in 0in 0in 0in},clip=true]{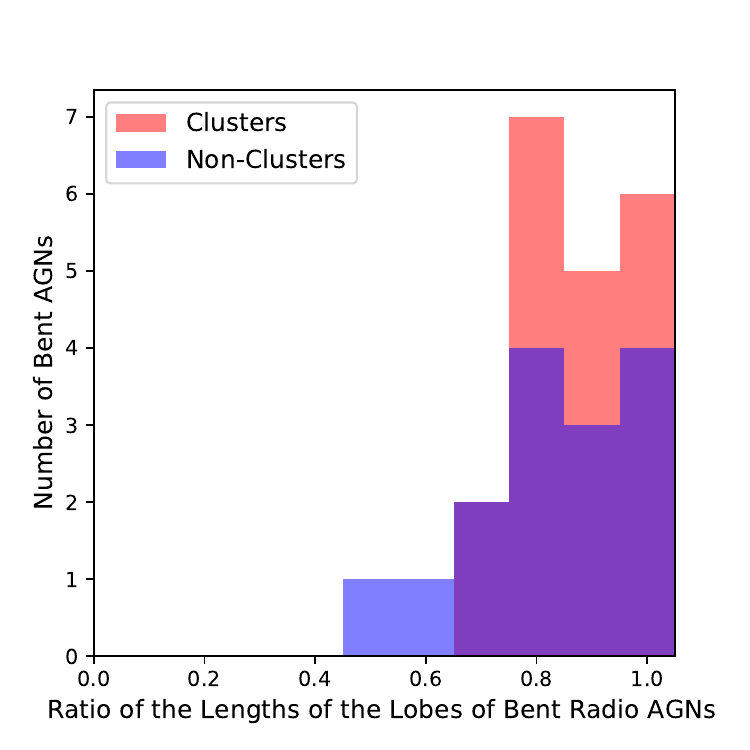}
\includegraphics[scale=0.675,trim={0in 0in 0in 0in},clip=true]{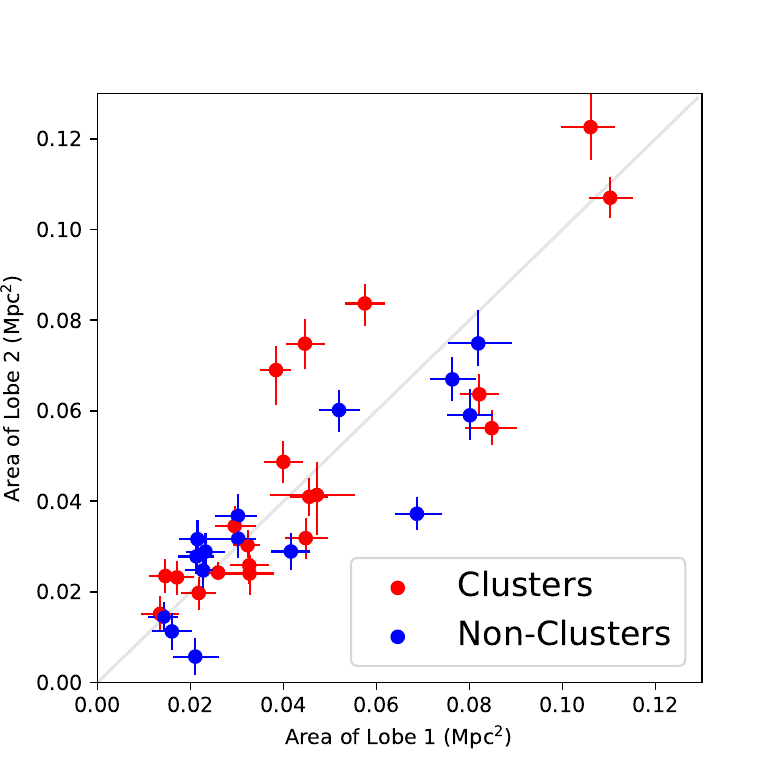}
\includegraphics[scale=0.7,trim={0in 0in 0in 0in},clip=true]{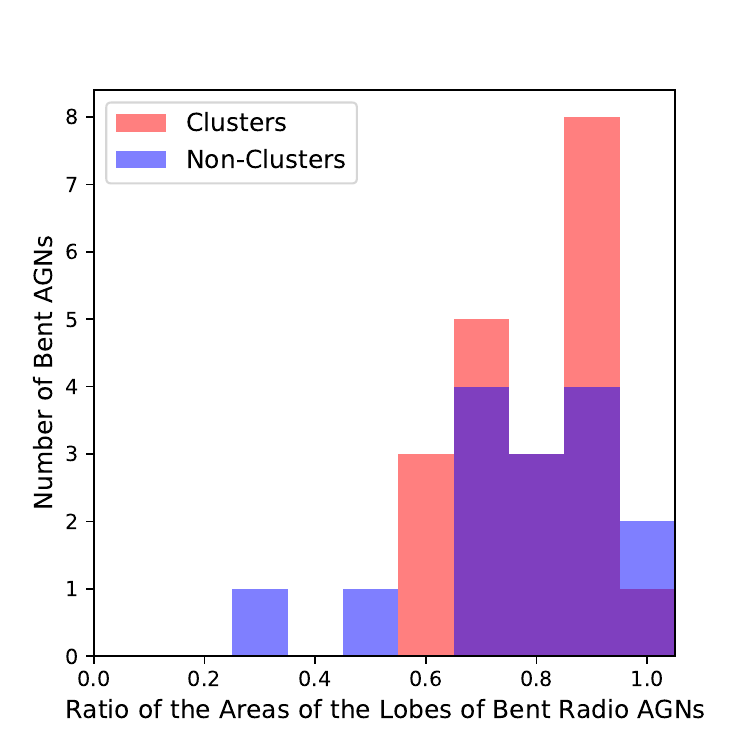}
\caption{Probes of the symmetry of the lobes of bent radio AGNs.  The top panels shows the projected physical size (kpc) of each lobe when compared to the opposite lobe.  The left-hand side is a direct comparison of the projected physical size, while the right-hand side shows a histogram of the ratios of the lengths of each lobe.  For the ratio, we compare smallest lobe length and the largest lobe length regardless of which lobe is lobe 1 or lobe 2.  Red points are bent AGNs in the cluster candidate sample.  Blue points are bent AGNs in the non-cluster sample.  A grey dashed line in each of the left hand panels shows a one-to-one trend.  The bottom panels similarly show the projected physical area (Mpc$^{2}$) of each radio lobe when compared to its opposite lobe.  It follows the same conventions as the top panel with the left side showing a direct comparison and the right side showing a histogram of the ratio of the areas of each lobe.  Both plots comparing the radio lobes indicate that most of the lobes have similar projected physical sizes and areas when compared to their opposite lobe.  Although no overall trend with the total projected physical size or area is found, when we compare the ratio of the projected physical sizes and projected physical areas, the two most asymmetric sources are in non-cluster environments.  }
\label{Fig:Asymmetry-comparison}
\end{center}
\end{figure*}

To probe differences in environment, we compare the projected physical size and projected physical area of each lobe to one another (see Figure~\ref{Fig:Asymmetry-comparison}).  We find that the majority of bent radio AGNs appear relatively symmetric ($>$ 80$\%$), regardless of their large-scale environment.  To avoid bias from size differences in larger radio AGNs, the right-hand panels of Figure~\ref{Fig:Asymmetry-comparison} show histograms of the ratios of the projected physical size and projected physical area of each radio lobe.  Although we treat lobe 1 and lobe 2 separately, these ratios show the smallest lobe over the largest lobe, regardless of which is lobe 1.  Each right-hand plot confirms that the majority of sources have lobes with similar sizes and areas.  Although both samples show sources with a similar degree of symmetry in terms of the projected physical size and projected physical area, as highlighted in Figure~\ref{Fig:Asymmetry-comparison}, the two sources where the ratios are the lowest are in the non-cluster sample.  Given the similarities, it is unclear if the environment or radio luminosity play any role in the asymmetry or size of bent AGNs (see Section~\ref{sect:Size-Asymmetry-Environment}), or if instead, the environments of these two samples is more similar than the analysis in GM19 suggests (see Section~\ref{sect:discussion}). 

As discussed in GM21, all measurements in regards to the size, area, and opening angles of bent radio AGNs are subject to projection effects.  Since it is difficult to quantify all possible viewing orientations of bent radio AGNs, we note that differences in both the projected physical size and the projected physical area might be a measure of the impact of viewing angle on these measurements rather than differences in the environment.  Thus, sources with slight differences in the projected physical size and projected physical area of the radio lobes may be indicative of similarly sized radio lobes viewed at differing orientations rather than slightly different environments or the effects of cluster weather and buoyancy.  

For the remainder of the paper, our analysis of the size of bent radio AGNs will focus on the projected physical area.  Although both measurements show a similar degree of symmetry, there are a number of sources (see Figures~\ref{Fig:AllClusters} and \ref{Fig:AllNonClusters}) where the rays estimating the projected physical size do not encapsulate the overall shape of the radio AGN and would require a more complex measurement to account for the unique morphology.  As such, a more complex measurement of the projected physical size, one that accounts for the radio flux density, frequency, sensitivity, and resolution to better trace the complete shape of the radio source and the redshift to trace the flux density cutoff is needed to accurately map the length of bent radio AGNs.  While imperfect, our measurement of the projected physical area better accounts for these factors.

\subsection{Differences in Radio Luminosity}\label{sect:RadioPower-Comp}

\begin{figure}
\begin{center}
\includegraphics[scale=0.6,trim={0.0in 0.0in 0.0in 0.6in},clip=true]{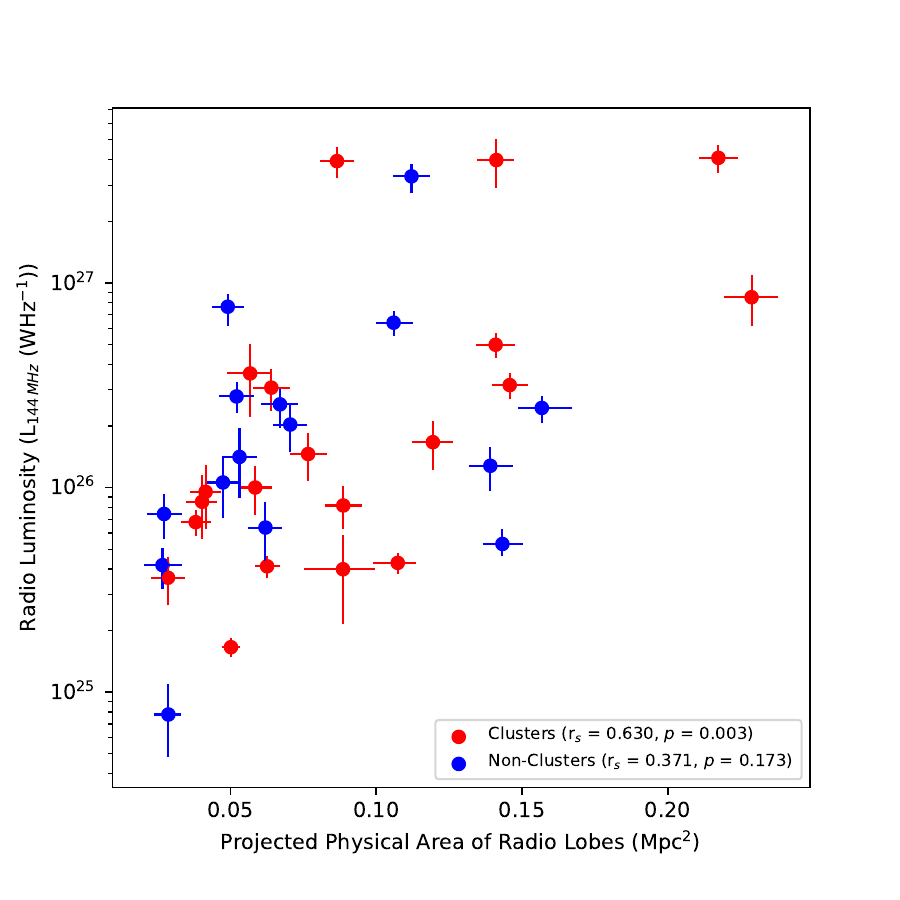}

\caption{The radio luminosity of bent radio AGNs (L$_{.144\,GHz}$) as a function of the total projected physical area of each bent radio AGN.  The cluster candidates are shown in red and the non-cluster candidates are shown in blue.  We find a moderate trend with strong evidence to reject the null hypothesis for the cluster sample, but only a moderate trend with no evidence to reject the null hypothesis for the non-cluster sample.}
\label{Fig:Area-RadioPower}
\end{center}
\end{figure}

As discussed in Section~\ref{sect:RadioPower}, we measure the radio luminosity of each bent radio AGN using the flux density within our 10$\sigma$ LoTSS contours.  To look for differences between the radio luminosity of these two populations, we plot the radio luminosity (L$_{.144GHz}$) as a function of the projected physical area of each radio source.  As shown in Figure~\ref{Fig:Area-RadioPower}, we see a similar relationship between these populations.  Statistically, the cluster sample (shown in red) shows a moderate trend with very strong evidence to reject the null hypothesis (r$_{s}$ = 0.630, $p$ = 0.003), with the most powerful radio AGNs being the largest. However, the non-cluster sample (shown in blue) shows a weak to moderate trend with no evidence to reject the null hypothesis (r$_{s}$ = 0.371, $p$ = 0.173). The difference in the strength and statistical robustness of the correlations may imply that the cluster environment plays a minor role in shaping the trend between radio AGN size and luminosity, likely with respect to the overall size of the radio AGN.  Alternatively, it may be indicative of the smaller non-cluster sample size and the lack of the most luminous radio AGNs in that sample ($>$ 10$^{27}$W\,Hz$^{-1}$) weakening the overall trend (see Section~\ref{sect:radioPower-environment} for further discussion).

To examine whether the difference in the strength of the statistical trends between the projected physical area and radio luminosity imply differences between the populations of bent AGNs, we run a Kolmogorov-Smirnov (KS) test to compare the two quantities.  In both cases, we find strong evidence that these samples are drawn from the same population ($p$ = 0.834 for the radio luminosity and $p$ = 0.750 for the projected physical areas).  As the only underlying difference between these samples is the measurement of the environment from GM19, this is further evidence that either the role of the environment is marginal and only affects the correlation between the two values or that even though GM19 measured differences in the overdensity of red sequence galaxies, these sources are not in entirely different environments.  Specifically, some of these non-cluster systems, which have positive overdensities (but below the 2$\sigma$ threshold), could be similar to the population of bent radio AGNs that are offset from clusters but still in local overdensities \citep{Garon2019}, and thus may be examples of poorer galaxy groups \citep[e.g.,][]{Morsony2013,Vardoulaki2021}.

\subsection{Verification of the Opening Angle - Richness Correlation in Clusters }\label{OpeningAngle-Comp}
\begin{figure}
\begin{center}
\includegraphics[scale=0.55,trim={0.2in 0.0in 0.0in 0.6in},clip=true]{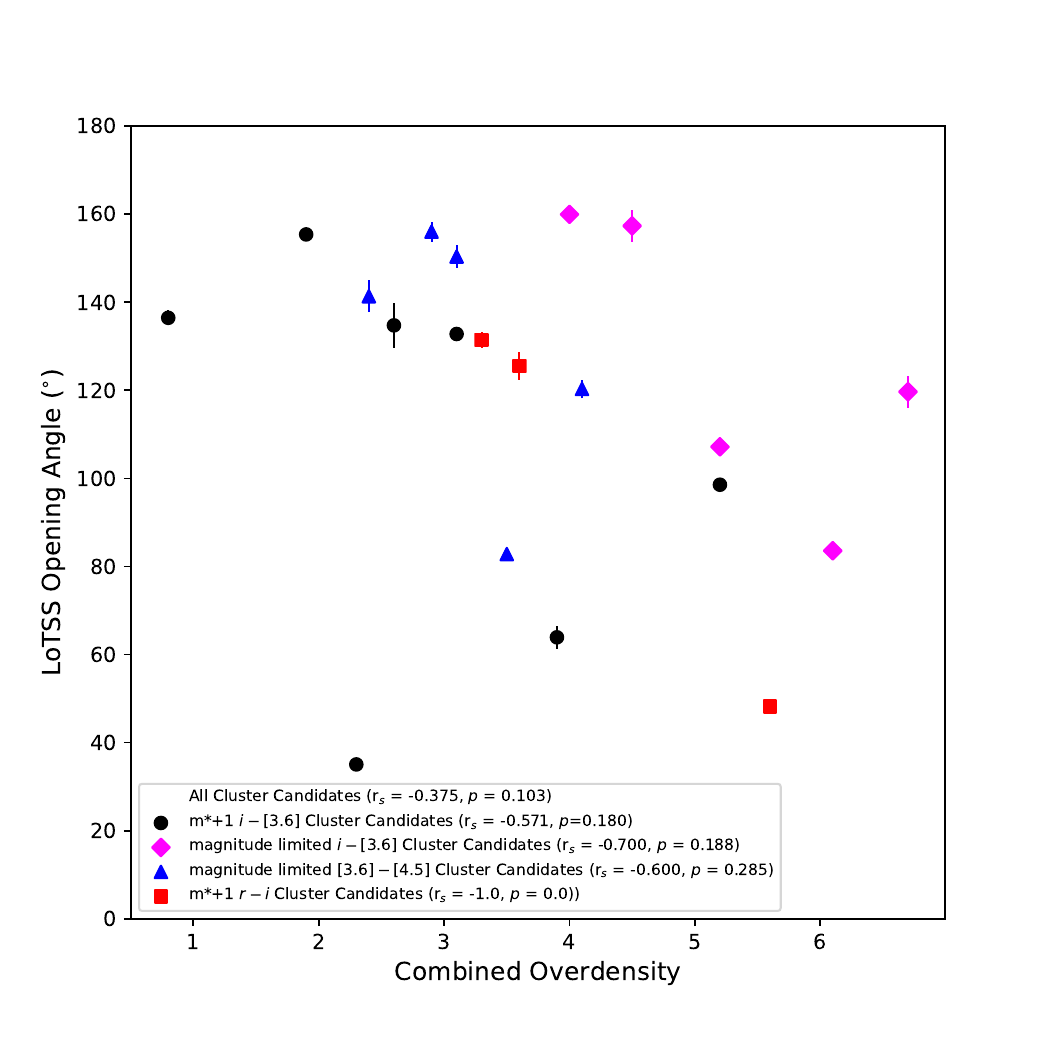}
\includegraphics[scale=0.55,trim={0.2in 0.0in 0.0in 0.6in},clip=true]{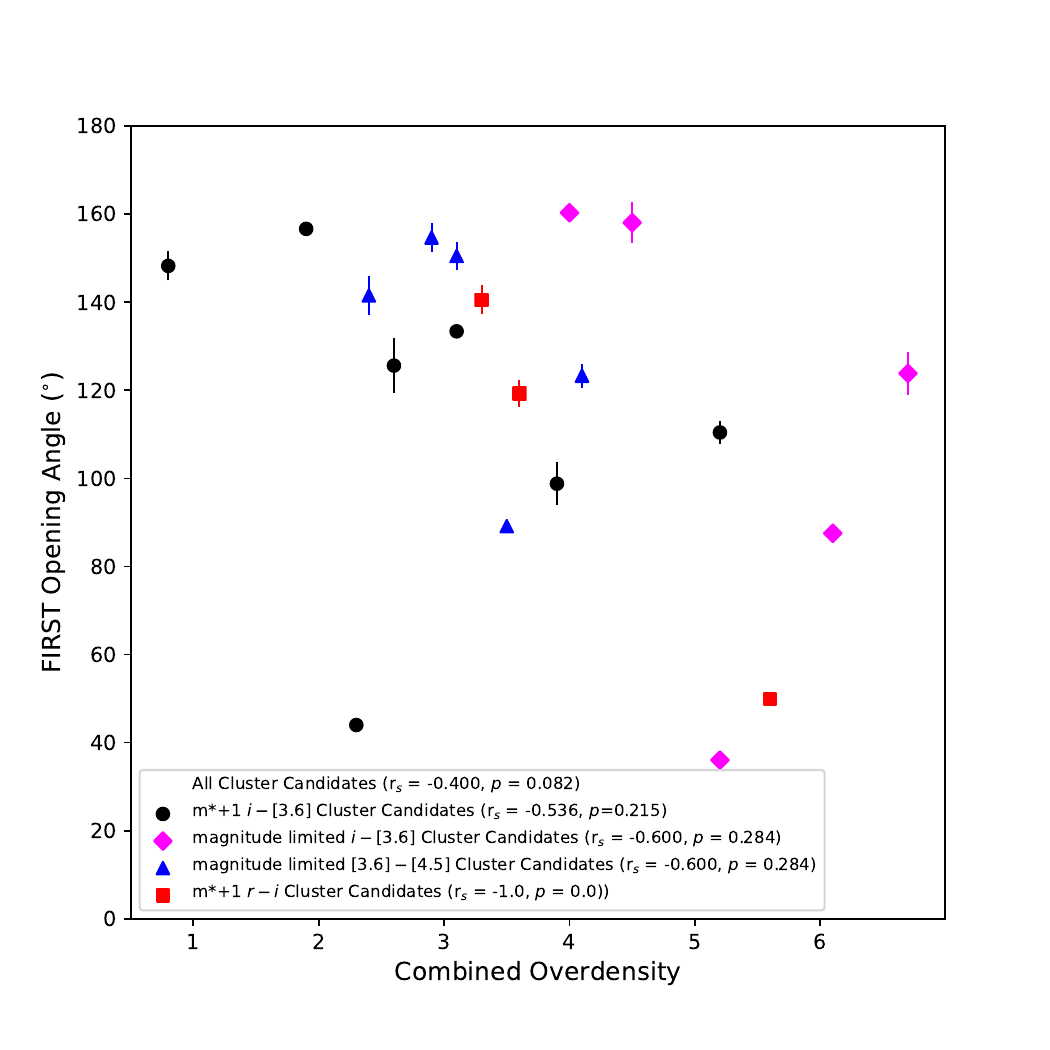}
\caption{The LoTSS opening angle as a function of the combined overdensity (top panel) and the FIRST opening angle as a function of the combined overdensity (bottom panel).  Unlike the trend shown in GM21, we find slightly weaker trends between the samples using the LoTSS images and very weak to no statistical evidence to reject the null hypothesis.  However, we do find a moderate trend with weak evidence to reject the null hypothesis in the FIRST sample.}
\label{Fig:OpeningOverdensity}
\end{center}
\end{figure}

Using the larger COBRA high-$z$ sample, GM21 found that richer galaxy clusters tend to host narrower bent AGNs. Assuming similar jet properties and the lack of projection effects, this would require a denser ICM for a greater degree of bending \citep[e.g.,][]{Hardcastle2005,Morsony2013}.  We compare the opening angles measured with LoTSS to the combined overdensity defined in GM19 to determine the impact of the frequency and angular resolution on the observations on this correlation (see Figure~\ref{Fig:OpeningOverdensity}).

Unlike the statistically robust trends identified in GM21, in our analysis in Figure~\ref{Fig:OpeningOverdensity}, the correlation it is not as statistically robust.  Using a Spearman test, our trend for the m*+1 $i - [3.6]$ sample shows a moderate trend, but little to no evidence to reject the null hypothesis (r$_{s}$ = $-$0.571, $p$ = 0.180).  As highlighted in Section~\ref{sect:RadioObs}, although the LoTSS observations are excellent at tracing the overall larger scale of the radio emission of a bent radio AGN, the lower frequency means that LoTSS data is less effective at tracing the most energetic regions of radio emission.  As the opening angle is estimated by measuring the curvature based on the brightest regions of the lobes and the core, LoTSS data may introduce noise, which would impact our results.  

To better estimate the biases in our measurements, we re-compare the opening angles measured in FIRST from GM21 (we correct for the misidentified opening angle in GM21, which appears as an outlier to the trend) to the same combined overdensities (see the bottom panel of Figure~\ref{Fig:OpeningAngle-COMP}).  Unlike the larger sample in GM21, we find similar results in terms of both the strength and statistical robustness of the trend based on the Spearman test.  However, for the FIRST measurements, we find that for the entire cluster candidate sample, and none of the subsamples, we see a moderate correlation with weak evidence to reject the null hypothesis (r$_s$ = -0.400, $p$ = 0.082).  As these values imply only a slightly stronger and statistically robust correlation than the LoTSS values, it would appear that the smaller sample size in this analysis may have a more dramatic effect than sensitivity, angular resolution, and wavelength coverage.

Although we do not find a statistical correlation, we considered the non-cluster sample and looked at the median value of the opening angle and combined overdensity (for those with a combined overdensity measurement) and found that the median values were located at the upper left portion of Figure~\ref{Fig:OpeningOverdensity}.  This adds further evidence that the opening angle is caused in part due to the environment \citep[e.g.,][]{Hardcastle2005,Morsony2013,Garon2019}. 

\subsection{Characterizing the Spectral Index of Bent Radio AGNs}\label{sect:SI-Cluster-Comp}

Because of the angular resolution of LoTSS relative to the overall size of the bent sources, we only analyze the spectral index of the entire radio source, similar to \citet{Mahony2016}, \citet{deGasparin2018}, and \citet{Dabhade2020}, or the core/lobe regions, similar to \citet{DiGennaro-o} (see Figure~\ref{Fig:SI-Hist} and Table~\ref{tb:RadioProp-2}).  While we previously measured the asymmetry of radio lobes in Section~\ref{sect:Asymmetry}, the coverage of the spectral index is not uniform across the two lobes for all of our sources, so we look at the lobes as a combined unit relative to the core as to not bias our results. Although a detailed accounting of the spectral index across the radio source is beyond the scope of this paper, we do present spectral index maps and error maps in the Appendix (see Figures~\ref{Fig:SI-clusters}, \ref{Fig:SI-nonclusters}, \ref{Fig:SI-clustersERROR}, and \ref{Fig:SI-nonclustersERROR}).

\subsubsection{Total Spectral Index}
\begin{figure}
\begin{center}
\includegraphics[scale=0.55,trim={0.2in 0.in 0.4in 0.4in},clip=true]{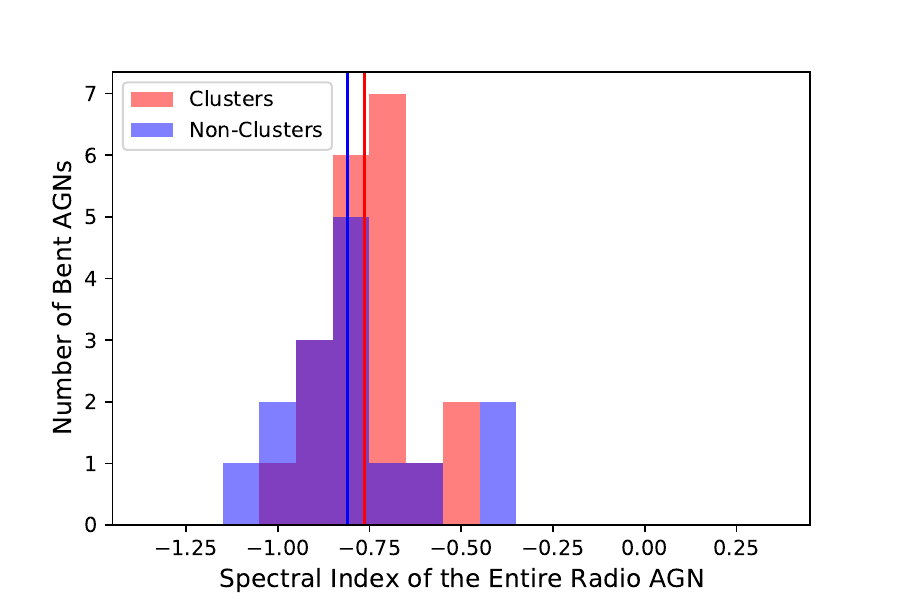}
\includegraphics[scale=0.55,trim={0.2in 0.in 0.4in 0.4in},clip=true]{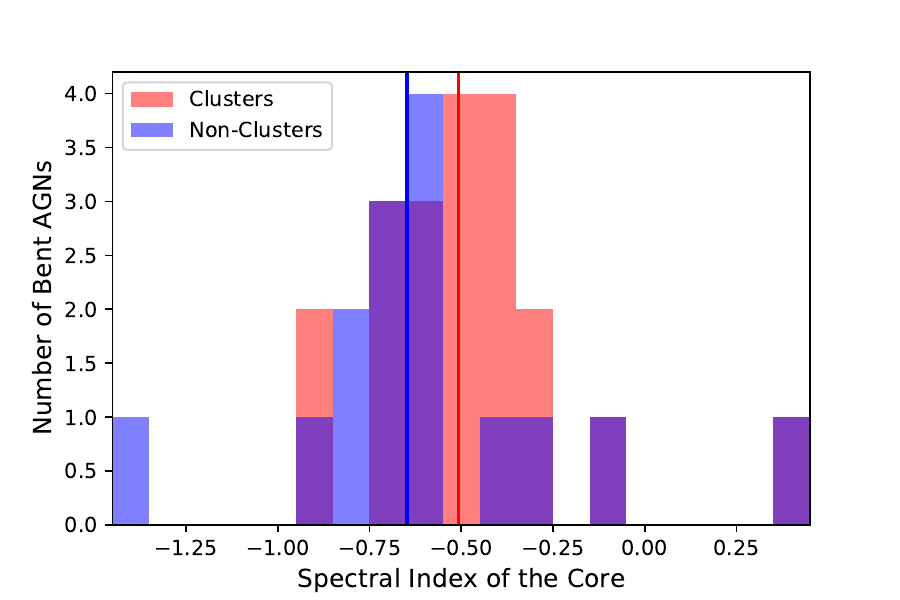}
\includegraphics[scale=0.55,trim={0.2in 0.in 0.4in 0.4in},clip=true]{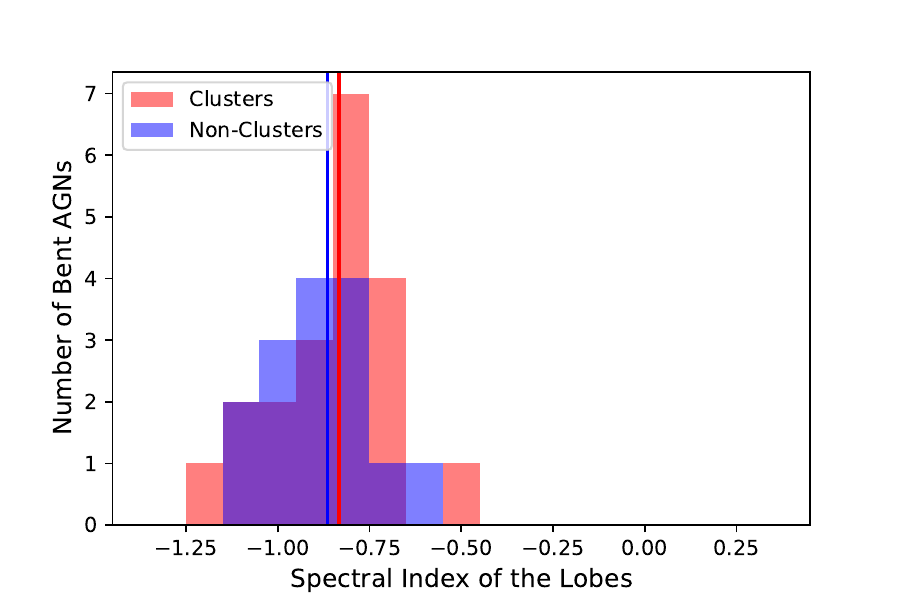}

\caption{Histograms of the spectral index values of the entire radio AGN, the radio core, and the radio lobes.  The radio AGNs in the cluster sample are shown in red and the radio AGNs in the non-cluster sample are shown in blue.  The median value of each distribution is shown as the vertical red and blue lines.  While the total number of radio AGNs in each sample is 20 and 15 respectively, there are sources in each sample where the core region is smaller than the beam size.  While we have included them here, they are flagged in Table\,\ref{tb:RadioProp-2}.}
\label{Fig:SI-Hist}
\end{center}
\end{figure}

As shown in the top panel of Figure~\ref{Fig:SI-Hist}, the total spectral indices of these sources span a similar range of values regardless of the environment.  However, we find a slight difference in the peak of the distribution of values, with the median of the cluster sample being -0.76 $\pm$ 0.01 and the median of the non-cluster sample being -0.81 $\pm$ 0.02.  Based on our error in the median, we find an $\approx$ 1.5$\sigma$ offset between the median values of the mean spectral index (see Table~\ref{tb:SI-median}).  This difference is further emphasized in Figure~\ref{Fig:SI-Hist}, where we find 10 of 20 bent radio AGNs in clusters with a flatter mean spectral index ($\geq$ -0.75) as compared to 4 of 15 in the non-cluster sample.  

Because quasars are more energetic than the typical bent radio AGNs in our sample, we also examine these populations separately.  Using the four cluster sample quasars and the six non-cluster sample quasars (see Table\,\ref{tb:RadioProp-2}), we find a median spectral index of -0.82 $\pm$ 0.05 (clusters) and -0.80 $\pm$ 0.07 (non-clusters), which are in agreement with one another.  In contrast, for the non-quasar sample, we find a median spectral index of -0.76 $\pm$ 0.01 for the cluster sample and -0.83 $\pm$ 0.02 for the non-cluster sample. This $>$ 2$\sigma$ difference between the non-quasars would appear to further highlight a potential environmental impact on bent AGNs.    

However, as the typical values of the spectral index of extended radio AGNs is between -0.7 and -0.8 \citep[e.g.,][]{Kellermann1988,Sarazin1988,Condon1992,Peterson1997,Lin2007,Miley2008}, and the mean value of the spectral index of all cross-matched sources from LoTSS DR2 discussed in \citet{Shimwell2022} is -0.792, all our bent radio AGNs appear to be ordinary extended radio sources.  While the overwhelming majority of sources can be described as having a spectral index with a flatter core and steeper lobes, even among sources where the core has a similar value to the lobes, we do have two sources, one in each sample (COBRA135838.1+384722 and COBRA170443.9+295246), where the core is steeper than the lobes (see Figures~\ref{Fig:SI-clusters} and \ref{Fig:SI-nonclusters}).

\begin{deluxetable*}{lccc}
    \tablecolumns{4}
    \tabletypesize{\small}
    \tablecaption{Median Values of the Mean Spectral Index\label{tb:SI-median}}
    \tablewidth{0pt}
    \tabletypesize{\footnotesize}

    \setlength{\tabcolsep}{0.05in}
    \tablehead{
    \colhead{Sample}&
    \colhead{Total Spectral Index}&
    \colhead{Core Spectral Index}&
    \colhead{Lobe Spectral Index}
    }
    
\startdata
Cluster Sample & -0.76 $\pm$ 0.01 & -0.51 $\pm$ 0.03 & -0.83 $\pm$ 0.02 \\
Non-Cluster Sample & -0.81 $\pm$ 0.02 & -0.65 $\pm$ 0.04 & -0.86 $\pm$ 0.03 \\
\enddata

\end{deluxetable*}
\subsubsection{The Spectral Index of the Core and Lobes}
To further probe the differences in the spectral index of bent AGNs as a function of environment, we measure the spectral index of the radio core.  For this analysis, we fiducially estimate the area of each core based on the spectral index maps and plot a histogram of the mean values in the middle panel of Figure~\ref{Fig:SI-Hist} (see Table\,\ref{tb:RadioProp-2}).  Although we report all values of the core spectral index, we flag measurements where either the core region is much smaller than the beam size, or in one case (COBRA104641.5+282028 - which has the most positive spectral index core of any source in the non-cluster sample) where the core is below the 10$\sigma$ LoTSS threshold, but detected at the 3$\sigma$ level in FIRST.  For the sources where the core is a distinct separate component and smaller than the beam, we may be underestimating the error in these values.  

We again find a difference between the populations of bent radio AGNs (see Table~\ref{tb:SI-median}).  The median core spectral index in the cluster sample is -0.51 $\pm$ 0.03 and in the non-cluster sample is -0.65 $\pm$ 0.04.  While both values are flatter than the median of the total spectral index, we see an $\approx$ 2$\sigma$ separation in these values.  This is further enhanced in examining the distribution of core spectral indices.  Specifically, 13 of 20 (65$\%$) of bent radio sources in the cluster sample have mean core spectral indices $\geq$ -0.6 (11 of 17 [64.7$\%$] if we remove those sources with a core that is poorly constrained), compared to 4 of 15 (26.7$\%$) of bent radio AGNs in the non-cluster sample (3 of 12 [25$\%$] bent radio AGNs in the non-cluster sample if we remove those with a poorly identified core).  Although the difference is not great enough to claim a distinct separate sample without additional observations, we note that while not all bent AGNs in clusters have flatter cores, it does appear that sources with a flatter spectral core are more commonly in clusters.  

Interestingly, of the SDSS-identified quasars in each sample (see Table\,\ref{tb:RadioProp-2}), three of four quasars in the cluster sample fall into the flat-core subsample, with a median value of -0.39 $\pm$ 0.15.  However, in the non-cluster sample, only three of six quasars fall within this sample, with a median value of -0.53 $\pm$ 0.06.  When probing the remaining non-quasar AGNs with well defined cores, we find an even greater difference, with the median of core spectral index of the cluster sample being -0.54 $\pm$ 0.03 and the median core spectral index of the non-cluster sample being -0.75 $\pm$ 0.03 which corresponds to a $>$ 3$\sigma$ difference.  Given the larger difference among the non-quasars and the small difference within the quasars, a flatter core spectral index in a non-quasar radio AGN could hint at further differences that might be used to identify radio AGNs in clusters (see Section~\ref{Sect:SI-Environment} for a more complete discussion on the differences in the spectral index). 

We further probe the differences in the populations of radio AGNs by examining the mean spectral index of the radio lobes (see the bottom panel of Figure~\ref{Fig:SI-Hist}).  The median values of the spectral index of the lobes slightly differ (-0.83 $\pm$ 0.02 for the cluster sample and -0.86 $\pm$ 0.03 for the non-cluster sample, an $\approx$ 0.6$\sigma$ offset).  In general, both values are in agreement that the spectral index steepens as a function of the distance from the radio core. 

\section{Discussion}\label{sect:discussion}
As highlighted in Section~\ref{sect:Asymmetry}, Section~\ref{sect:RadioPower-Comp}, and Section~\ref{sect:SI-Cluster-Comp}, we see slight differences in the properties of bent radio AGNs inside and outside of red sequence cluster candidates.  To contextualize what these differences might mean in terms of the environment, we examine how some characteristics scale with cluster richness and compare our results to similar studies of the impact of the cluster environment on radio AGNs.  We explore the impact of the cluster environment on radio source size in Section~\ref{sect:Size-Asymmetry-Environment}, the differences in radio luminosity in Section~\ref{sect:radioPower-environment}, and our spectral index of bent radio AGNs findings in Section~\ref{Sect:SI-Environment}.  

\subsection{Characterizing the Radio Lobes as a Function of Environment} \label{sect:Size-Asymmetry-Environment} 
\subsubsection{Radio Source Asymmetry}
There has been much work done to probe the impact of the ICM on radio lobe asymmetry.  \citet{Rodman2019} used a sample of low-$z$ ($z$ $<$ 0.3) FRI and FRII radio sources from the Radio Galaxy Zoo to probe the asymmetry of radio AGNs as a function of local environment and found an anti-correlation between the size of the radio lobe and the density of the surrounding environment.  \citet{Yates-Jones2021} similarly probed this result using simulations and found that the density of the environment can be a driving factor in the asymmetry of radio AGNs.  \citet{Garon2019}, using a large sample of low-$z$ ($z$ $<$ 0.7) bent radio AGNs found similar results using both the position of the bent radio AGN within the cluster and the pressure gradient across the cluster.

While a dearth of ICM studies of high-$z$ COBRA clusters (only five have pointed X-ray observations from either Chandra or XMM-Newton - see Blanton et al. in prep) prevents a detailed analysis of the density of the surrounding environment for most of our sample, Figure~\ref{Fig:Asymmetry-comparison} shows that the most asymmetric sources are not in clusters.  However, we do find nine bent sources in the cluster sample that are slightly more asymmetric (with a ratio of the area of the lobe $<$ 0.8).  We posit that this slight difference in radio lobe symmetry may result from the density of the surrounding gas.  

\begin{figure}
\begin{center}
\includegraphics[scale=0.6,trim={0.0in 0.0in 0.0in 0.6in},clip=true]{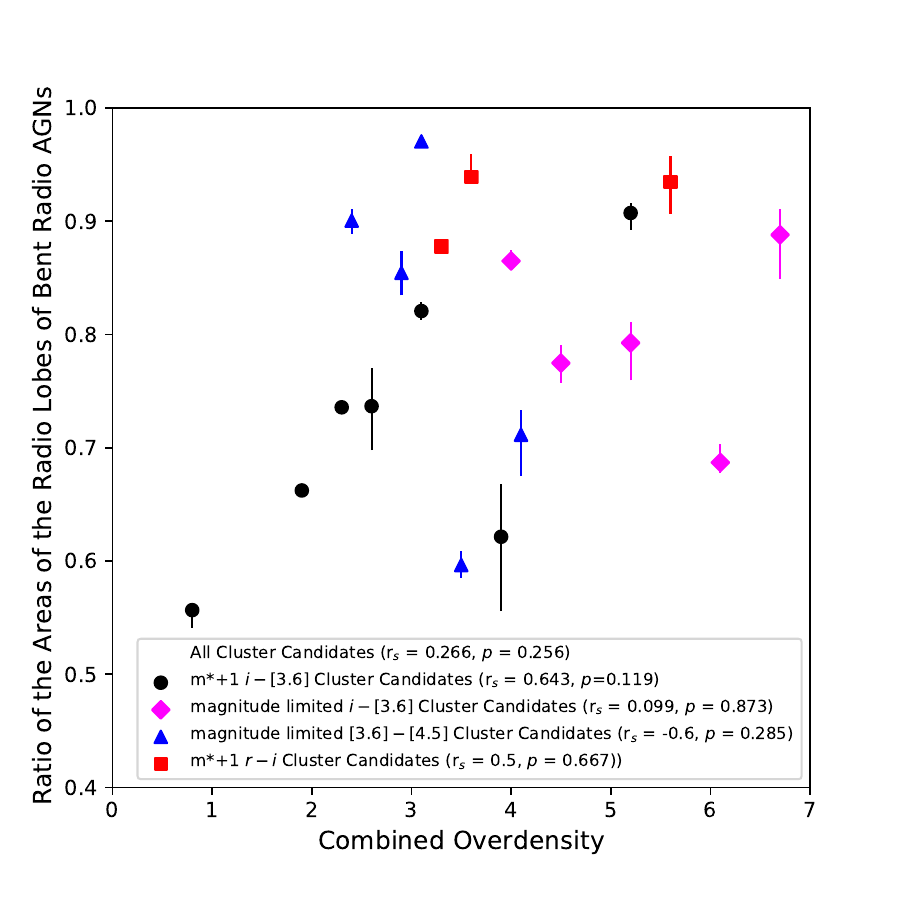}
\caption{The ratio of the projected physical areas of the radio lobes as a function of the significance of the combined overdensity from GM19, our cluster richness proxy, for the cluster candidates.  We plot the m*+1 $i - [3.6]$ cluster candidates (the most statistically robust sample in the initial analysis) in black circles, the magnitude limited $i - [3.6]$ cluster candidates in pink diamonds, the magnitude limited $[3.6] - [4.5]$ cluster candidates in blue triangles, and the m*+1 $r - i$ cluster candidates in red squares.  We find a weak correlation with little evidence to reject the null hypothesis for the entire sample (r$_{s}$ = 0.266, $p$ = 0.256).  While we find a stronger correlation in the m*+1 $i - [3.6]$ sample, we again find no evidence to reject the null hypothesis, making this trend statistically improbable (r$_s$ = 0.643, $p$ = 0.119).}

\label{Fig:Asymmetry-Overdensity}
\end{center}
\end{figure}

For our sample, we analyze the potential role of the ICM density on the radio lobe asymmetry using the combined overdensity measured in GM19 as our proxy for cluster richness (and ICM density; see Figure~\ref{Fig:Asymmetry-Overdensity}).  We find a weak correlation with no evidence to reject the null hypothesis (r$_s$ = 0.266, $p$ = 0.256).  While we see a stronger correlation between the degree of symmetry and overdensity if we focus on the more statistically robust m*+1 $i - [3.6]$ cluster candidate sample (r$_{s}$ = 0.643 and $p$ = 0.119; see Section~\ref{sect:COBRA} and GM19 and GM21 for a full discussion of the COBRA subsamples), we again find no evidence to reject the null hypothesis.  As some of our systems are in lower mass clusters/groups, these systems may be more impacted by cluster weather and instabilities, which would make cluster richness a poor proxy for ICM density and may explain the more scattered distribution.

We examine the symmetry of bent radio AGNs in the non-cluster sample as a function of the environment and find no trend between the degree of lobe asymmetry and overdensity.  Because the overdensity of many of these sources is poorly constrained or below the minimum three red sequence galaxy detection threshold in GM19, we do not show the correlation here.  

\begin{figure}
\begin{center}
\includegraphics[scale=0.6,trim={0.0in 0.2in 0.0in 0.6in},clip=true]{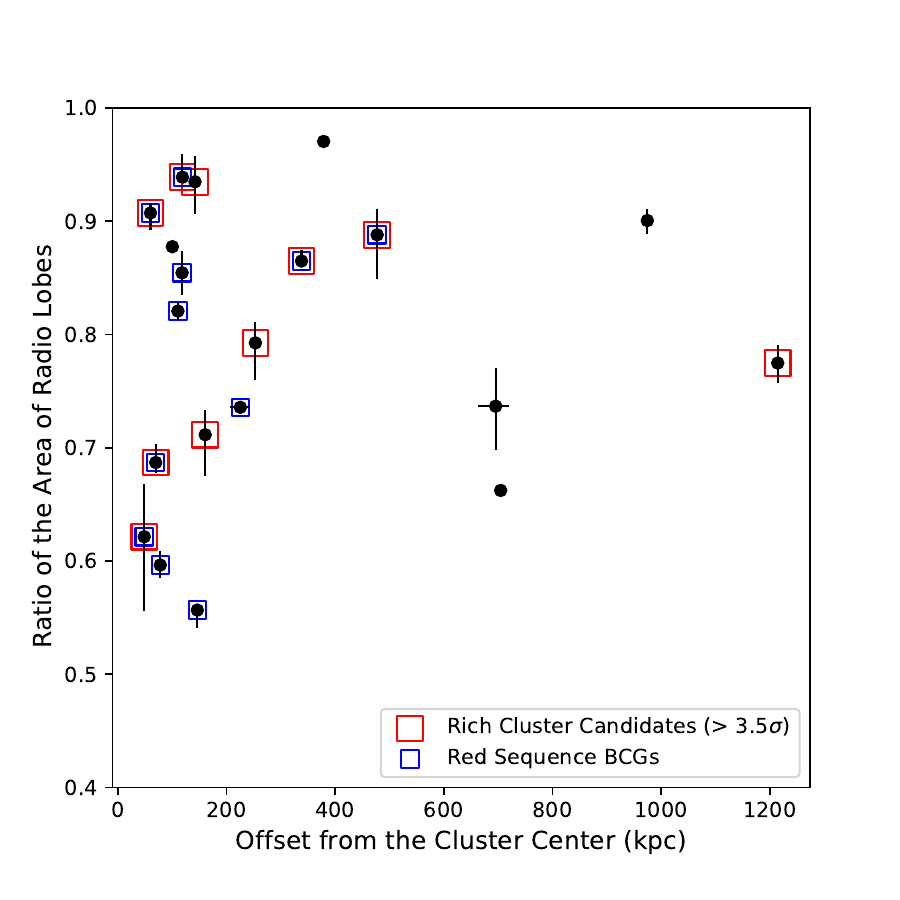}
\caption{The ratio of the projected physical areas of the radio lobes as a function of the offset from the cluster center.  The offset from the cluster centers is measured in GM19 and presented in GM21 based on the surface density of red sequence galaxies.  Richer cluster candidates, with a combined overdensity greater than 3.5$\sigma$, are shown in red boxes.  Bent radio AGNs hosted by BCGs are shown in blue boxes.  We find no overall trend between the symmetry of radio lobes and the offset from the cluster center.}
\label{Fig:Asymmetry-Offset}
\end{center}
\end{figure}
 
To further study the environmental drivers of radio lobe asymmetry, we plot the degree of asymmetry as a function of the offset from the cluster center for the cluster candidate sample in Figure~\ref{Fig:Asymmetry-Offset}.  Assuming a typical ICM density profile, for sources near the cluster center, any direction that the lobes expand into will encounter the same density gradient.  For bent sources at greater offsets from the cluster center, the direction the radio lobes expand may result in greater differences due to the surrounding ICM density, particularly if both lobes do not point radially toward or away from the cluster center.  Based on this simple model, differences in the surrounding ICM density distribution may cause some of the asymmetric lobes in our sample at large offsets ($>$ 600\,kpc), including for some of the BCGs found at larger offsets in our sample (see GM21 for a full discussion of the location of bent radio AGNs in COBRA clusters and \citealp{Gozaliasl2019} and \citealp{Zhang2019} for further discussion of BCGs offset from the cluster center).

A number of our bent radio AGNs are located near the cluster center, including some of the most asymmetric sources.  This result may be a function of the small sample size being probed, limiting our ability to identify highly symmetric sources at larger offsets from clusters.  However, because we do not know the velocity differences between the galaxies (i.e., how fast the host galaxy is moving relative to the parent cluster), we propose that the bent radio AGNs near the cluster center are likely not all hosted by infalling galaxies, but rather hosted by brightest cluster galaxies (BCGs).  As wide-angle-tail (WAT) radio sources are commonly hosted by BCGs \citep{Owen1976}, the ICM is set in motion from a large-scale cluster-cluster merger (both major and minor mergers), which results in the bending of these radio tails \citep[e.g.,][]{Blanton2011,Douglass2011,Paterno-Mahler2013,O'Dea2023}.  As shown in the radio contours presented in \citet{Paterno-Mahler2013}, bent radio AGNs associated with such systems can have asymmetries in their tails (highlighted by the width and extent of the tail in the bent radio AGN in Abell2029).  Since 11 of 20 bent radio radio AGNs in the cluster sample are hosted by BCGs (see Figure~\ref{Fig:Asymmetry-Offset}) and an additional four are quasars, which are not identified as BCGs in GM21 because they are not on the red sequence, it is possible that gas sloshing yields the range of radio lobe asymmetries observed among bent radio AGNs near the cluster center.  

Apart from gas sloshing, buoyant forces can also impact the bending of WATs \citep[e.g.,][]{Gull1973,Sakelliou1996,Smolcic2007}.  Since buoyancy is most impactful at larger radii \citep{Smolcic2007}, bending the sources towards regions of lower density, this may also explain the asymmetry within the lobes.  Furthermore, given the differences in the locations of the bent radio AGNs in clusters, the velocity of the host galaxy relative to the ICM may also impact radio lobe asymmetry, especially among the lower mass, potentially less dynamically relaxed systems.  

\subsubsection{Radio Source Size}
Beyond the radio source asymmetry, we also look at the difference in the size and location of these sources. We find a similar range of projected physical areas and lengths of bent AGNs in both samples, in agreement with \citet{Wing2011} at low redshift.  However, much work has been done to examine populations of rich clusters at $z$ $\approx$ 1 hosting radio AGNs from either the MaDCoWS sample - done using FIRST observations to identify the radio sources and high resolution VLA imaging in either L-band centered on 1.4\,GHz with an angular resolution of 1$\farcs$3 or C-band centered on 5.5\,GHz with an angular resolution of 1$\farcs$0 \citep[e.g.,][]{Moravec2019,Moravec2020} - and the ORELSE sample - done using 1.4\,GHz JVLA observations with a beam size of 5$\arcsec$ down to a $\approx$ 10\,$\mu$Jy 1$\sigma$ sensitivity \citep[e.g.,][]{Shen2017,Shen2019,Shen2020}.  These studies found that radio AGNs farther from the cluster center are larger.  If we assume spherically symmetric ICM distributions, these results imply that a less dense ICM at greater cluster-centric radii allows for the formation of larger radio AGNs. 

\begin{figure}
\begin{center}
\includegraphics[scale=0.6,trim={0.0in 0.2in 0.0in 0.5in},clip=true]{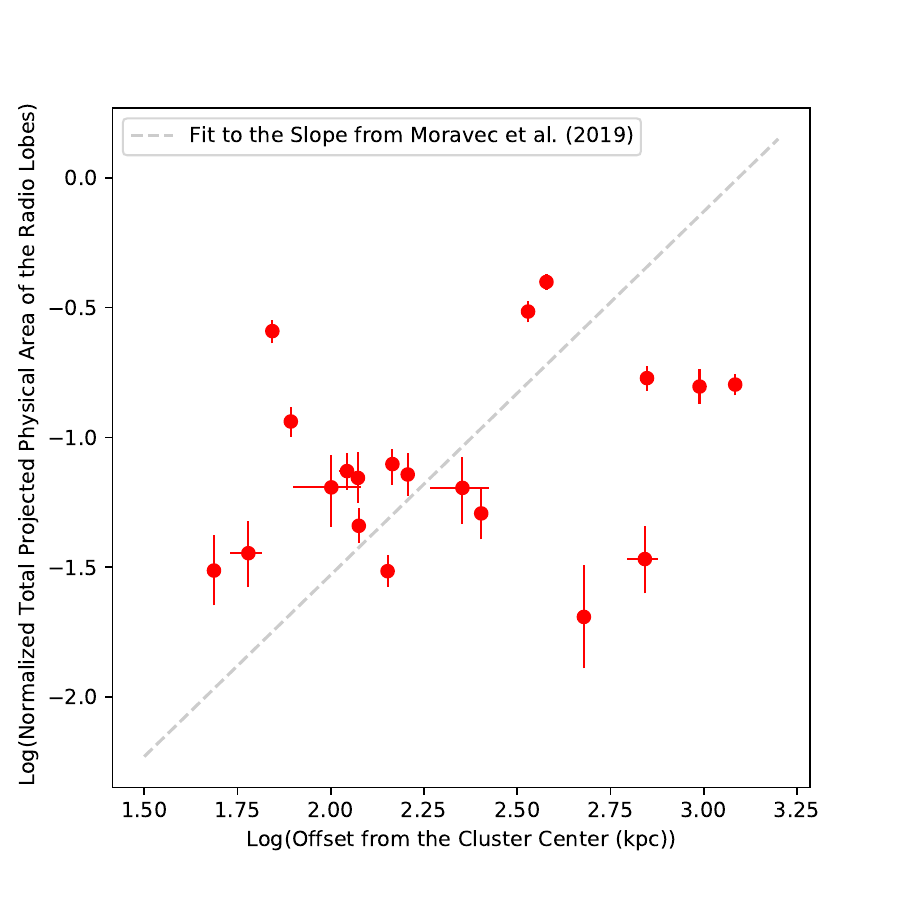}
\caption{The log of the normalized projected physical area as a function of the log of the offset from the cluster center.  The offsets from the cluster center are reported in GM21.  The normalization of the fit from \citet{Moravec2019} is shown in the grey dashed line.}
\label{Fig:NormalizedArea}
\end{center}
\end{figure}

Following the normalization of the projected physical size discussed in \citet{Moravec2019}, we do a similar analysis for our bent radio AGNs in clusters.  As our measurements of the projected physical size are less well constrained, we measure the normalized projected physical area as a function of the offset from the cluster center (see Figure~\ref{Fig:NormalizedArea}).  To determine whether our data follow the trend in \citet{Moravec2019,Moravec2020}, we plot the slope of that trend re-normalized to our data (the grey line in Figure~\ref{Fig:NormalizedArea}).  As in GM21, we see no agreement with this trend and a large degree of scatter among our own sample. 

While the sample of 49 radio AGNs in \citet{Moravec2020} includes some bent radio AGNs, the majority of the sources are straight (four are explicitly classified as bent tails and one source that is in this sample is identified as an FRII).  As such, it is possible that the more complicated geometries and projection effects associated with bent radio AGNs induce noise in our measurements and limit the effectiveness of this trend between AGN offset and AGN size.  Additionally, although the cluster samples used in \citet{Moravec2020} and \citet{Shen2020} (both of which show agreement with the trend) each have one cluster that overlaps with this COBRA sample, the typical cluster mass of both a MaDCoWS cluster and ORELSE cluster is much higher than the expected cluster mass of a high-$z$ COBRA target.  The median MaDCoWS M$_{500}$ is $\approx$ 10$^{14.20}$\,M$_{\odot}$ \citep{Gonzalez2019} and the median ORELSE cluster virial mass in \citet{Shen2020} is 10$^{14.5}$\,M$_{\odot}$.  However, for COBRA, only two of our clusters have mass estimates. COBRA113733.8+300010 ($z$ = 0.96), the only spectroscopically confirmed cluster \citep{Blanton2003}, has a virial mass of $\approx$ 10$^{14.1}$\,M$_{\odot}$ \citep{Lemaux2019} and COBRA164611.2+512915, the lowest redshift cluster in our sample ($z$ = 0.351), has a redMaPPer richness of 42.5, corresponding to M$_{200}$ $\approx$ 10$^{14.23}$\,M$_{\odot}$ \citep{Rykoff2014,Simet2017}. 

\subsubsection{Opening Angle}

\begin{figure}
\begin{center}
\includegraphics[scale=0.7,trim={0.in 0.0in 0.0in 0.5in},clip=true]{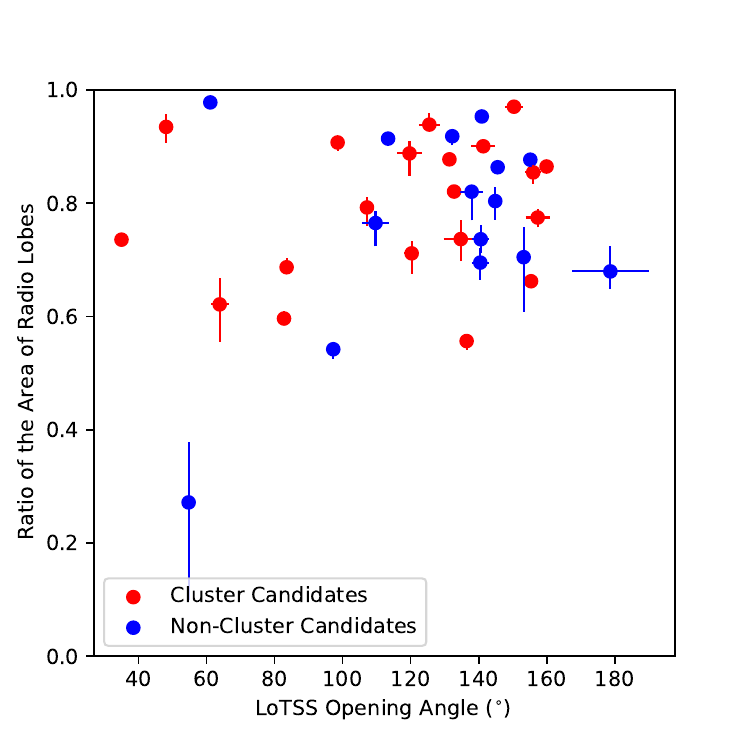}
\caption{The ratio of the projected physical area of the radio lobes as a function of the LoTSS opening angle.  While there is no clear distinction between the two populations, we find that narrower bent AGNs are generally more commonly in clusters, as expected.}
\label{Fig:Asymmetry-Angle}
\end{center}
\end{figure}

The opening angle is thought to depend on the local environment.  Thus, we plot the ratio of the areas of the radio lobes as a function of the size of the LoTSS opening angle of each bent radio AGN in both samples (Figure~\ref{Fig:Asymmetry-Angle}) to further explore the role of environment.  Although we find a larger fraction of narrower sources in clusters, we see no trend between opening angle and asymmetry and a strong similarity between the samples.  Moreover, Figure~\ref{Fig:Asymmetry-Angle} shows a large population of wide AGNs (opening angle $>$ 135$^{\circ}$) with moderate asymmetry ($<$ 0.8) in both samples. Thus, while we have evidence that the opening angle is dependent on the density of the ICM \citep[e.g.,][]{Garon2019}, we find no evidence to conclude that the narrowness of the opening angle and the degree of lobe asymmetry are correlated. 

It is interesting that we find that the widest sources in both environments in Figure~\ref{Fig:Asymmetry-Angle} show a range of lobe asymmetries.  In the cluster sample, this likely results from bent sources being far from the cluster center, while others are near the cluster center, creating discrepancies in the local environment (see Figure~\ref{Fig:Asymmetry-Offset}).  Similarly, we find a range of asymmetries among the narrowest sources in both samples (opening angles $<$ 80$^{o}$), including the most asymmetric source.  As we only have five narrow sources, we cannot determine whether this is driven by the environment, or the inclusion of both central and non-central AGNs (for the three in the cluster sample).  Of the cluster sample, one is at the cluster center (COBRA113733.8+30010, offset $<$ 50\,kpc, asymmetry $\approx$ 0.62), while the other two (COBRA164611.2+512915 and COBRA164951.6+310818) are infalling near the cluster center.  Given that the infalling sources are in clusters of different richness and differing asymmetries, we are unable to determine whether the narrow opening angle, the richness of the cluster, or a combination of the two results in the asymmetry.  Ultimately, because our sample only contains a few classical narrow angle tail (NAT) radio sources (opening angle $<$ 45$^{o}$; \citealp[e.g.,][]{Rudnick1976,ODea1985}), a larger sample including more NATs, like the one in \citet{deVos2021}, is needed to determine the role a narrow opening angle plays in the asymmetry of the radio lobes. 

\subsubsection{Differences in Radio AGN Environment?}
Despite the differences between the opening angles and spectral indices of bent radio AGNs inside and outside of clusters, we find much similarity between these two populations.  It is possible that beyond the opening angle, the local environment has little impact on the remaining aspects of morphology.  However, there are a few possibilities specifically, with respect to our definition of the non-cluster sample.  While the majority of non-cluster sources are in fields where GM19 was either unable to characterize a red sequence overdensity or had negative values, some systems are in weaker galaxy groups (overdensities of the order 1.5$\sigma$), which may imply that these systems are more like galaxy clusters than the field (i.e., the bent radio AGN is found either centrally within or on the outskirts of a quasi-symmetric intragroup medium).  This hypothesis is strengthened if we look at the 3.6\,$\mu$m overdensities from \citet{Paterno-Mahler2017}.  For the fields in the non-cluster sample, \citet{Paterno-Mahler2017} found 4 of 15 were in 2$\sigma$ overdensities within a 1$\arcmin$ search radius centered on the AGN and 9 of 15 were in 2$\sigma$ overdensities within a 2$\arcmin$ search radius.  These larger scale overdensities could point to structures offset from the AGNs.  Additionally, 12 of 15 non-cluster sources showed positive 3.6\,$\mu$m overdensities, which could further suggest that these sources reside in smaller structures, albeit ones not-identified as clusters in GM19.  Moreover, despite the challenges in identifying high-$z$ group candidates, there are numerous examples of bent radio AGNs found in galaxy groups \citep[e.g.,][]{Ekers1978,Venkatesan1994,Doe1995,Blanton2001,Freeland2008,Vardoulaki2021,Morris2022}, which support this hypothesis.

Bent radio AGNs are not always found near the cluster center and in some cases can be found at large offsets.  \citet{Sakelliou2000} found one bent source offset by $\approx$ 1.4\,Mpc from the cluster center in their sample of X-ray detected low-$z$ clusters hosting bent AGNs.  \citet{Edwards2010} identified the first known bent radio AGN in a filament offset from Abell1763 by 3.4\,Mpc.  \citet{Garon2019}, in their study of the large scale environment of bent radio AGNs at $z$ $<$ 0.7, found that more than 50$\%$ of their most bent sources are at offsets greater than 1.5\,R$_{500}$ ($\approx$ 1\,Mpc).  While \citet{Garon2019} note that bent sources at greater offsets are beyond the typical sphere of influence of a cluster, they found that for sources as offset as 2\,Mpc, these bent sources still are in slightly overdense environments, implying that they reside in infalling groups or filaments.  Similarly, \citet{deVos2021} study a sample of 208 NATs at 0.02 $<$ $z$ $<$ 0.8 from and used photometric redshifts from \citet{Wen2015} to associate each NAT with its host cluster.  Although most bent radio AGNs are closer to the cluster center, they also identify sources out to 10\,R$_{500}$ (which, following \citet{Garon2019}, is $\approx$ 6.7\,Mpc).

Given the 5$\arcmin$ $\times$ 5$\arcmin$ FOV of $Spitzer$ IRAC and the fact that all bent AGNs are at the center of the COBRA $Spitzer$ imaging, GM19 could only probe between $\approx$ 0.74\,Mpc (2$\farcm$5 at $z$ = 0.351) and $\approx$ 1.27\,Mpc (2$\farcm$5 at $z$ = 1.818, the highest redshift source in our cluster sample) from each AGN.  For all of the clusters in GM19, the IRAC FOV limits our ability to identify bent AGNs at the largest offsets from the cluster center reported \citep[e.g.,][]{Sakelliou2000,Edwards2010,Garon2019,deVos2021}.  Thus, it is possible our sample of non-cluster bent radio AGNs is not a representative field sample, but rather a sample of bent radio AGNs infalling into clusters at distances beyond the scope of the analysis in GM19.  This would imply that the differences seen in overdense environments (e.g., opening angle and asymmetry) are real, but could also account for the similarities as these sources are associated with galaxy clusters just at varying distances.  

\subsection{The Impact of the Cluster Environment on Radio Luminosity }\label{sect:radioPower-environment}

As shown in Figure~\ref{Fig:Area-RadioPower}, we find a strong correlation between the projected physical area and the radio luminosity of bent radio AGNs in COBRA cluster candidates.  We see agreement between our result and that of \citet{Moravec2020}, which found a similar trend between the projected physical extent and radio luminosity of radio AGNs in MaDCoWS clusters.  Similarly, GM21 reported a strong correlation between the projected physical size and the radio luminosity of the larger high-$z$ COBRA cluster candidate sample using FIRST observations. 

\begin{figure}
\begin{center}
\includegraphics[scale=0.6,trim={0.0in 0.2in 0.0in 0.5in},clip=true]{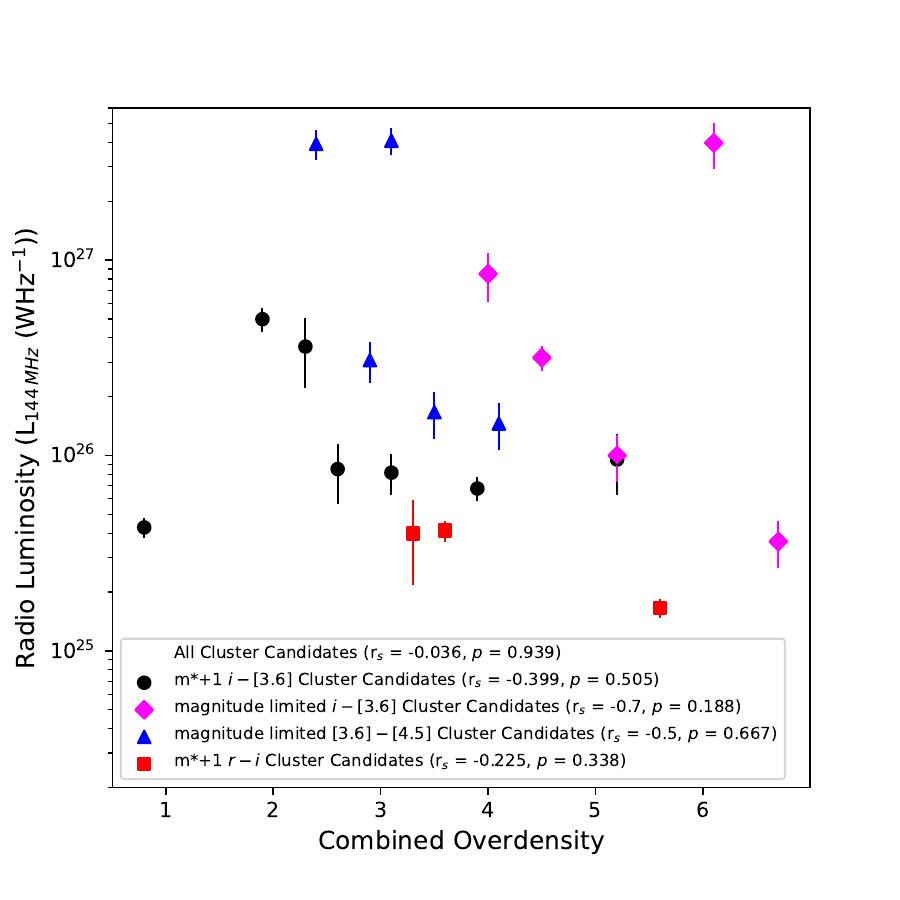}
\includegraphics[scale=0.6,trim={0.0in 0.2in 0.0in 0.5in},clip=true]{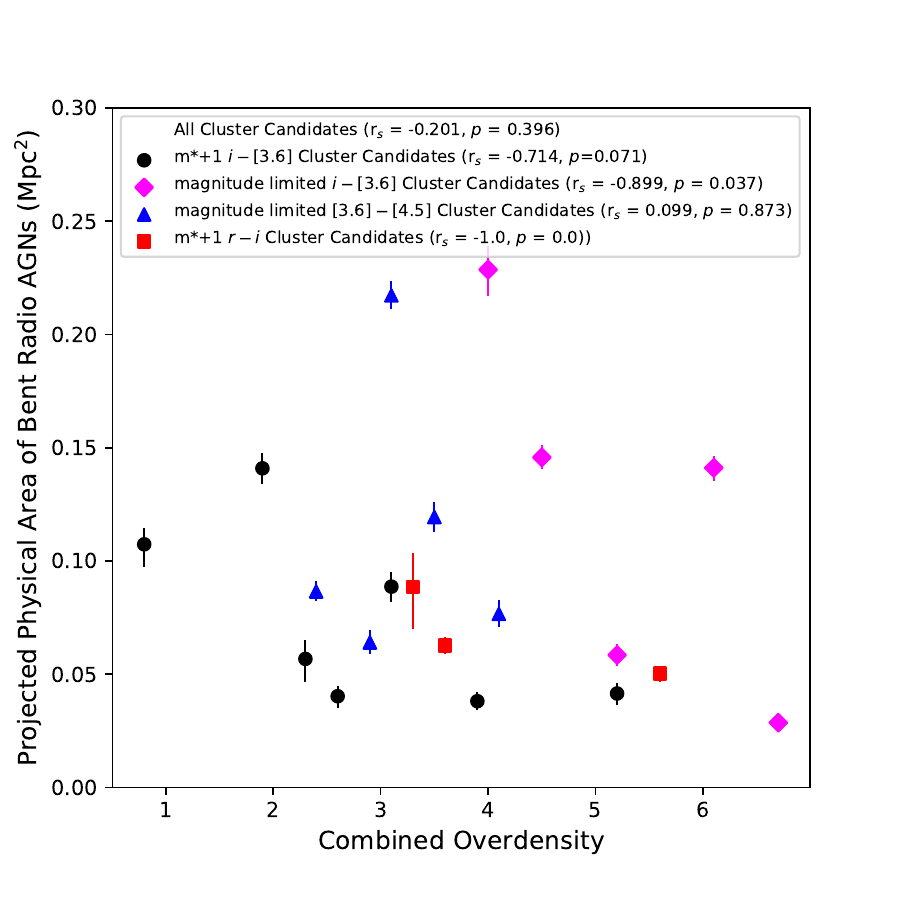}

\caption{The radio luminosity (L$_{144\,MHz}$) as a function of the combined overdensity for the cluster candidate sample in the top panel and the total projected physical area (Mpc$^{2}$) as a function of the combined overdensity in the bottom panel.  Each subsample is identified following the same legend as Figure~\ref{Fig:Asymmetry-Overdensity}.  We find no evidence of any correlations between the combined overdensity and the radio luminosity.  Although we see no trend between the projected physical size and combined overdensity for the entire sample, we do find a strong trends with weak to moderate evidence to reject the null hypothesis for the m*+1 $i - [3.6]$, magnitude-limited $i - [3.6]$, and m*+1 $r -i$ cluster samples.}
\label{Fig:Overdensity-RP}
\end{center}
\end{figure}

We find differences between the cluster and non-cluster sample in terms of the correlation between radio luminosity and the projected physical area, which leads to the question of whether environment impacts either property for bent radio AGNs.  At low redshift, the literature has not reached a consensus. \citet{Croston2019} studied a large sample of low-$z$ radio AGNs from LoTSS DR1 in clusters identified in redMaPPer, and found some evidence, with a large degree of scatter, that more powerful radio AGNs ($>$ 10$^{24.5}$\,WHz$^{-1}$) tend to reside in richer groups/clusters ($\approx$ 35 $<$ $\lambda$ $<$ 45; M$_{200}$ $\approx$ 10$^{14.14}$ - 10$^{14.25}$M$_{\odot}$).  In contrast, \citet{Wing2011} found no trends between richness and radio source size both inside and outside of clusters when looking at the parent sample of low-$z$ bent AGNs used to create COBRA (although their overdensity measurements differ from redMaPPer's richness, many have been identified in redMaPPer and have masses between 10$^{13.5}$ - 10$^{14.5}$\,M$_{\odot}$).  

We plot the radio luminosity as a function of the combined overdensity to determine the impact of the cluster environment for our LoTSS objects (see the top panel of Figure~\ref{Fig:Overdensity-RP}).  While we find some moderate trends between radio luminosity and richness within the individual subsamples, there is no statistical evidence to reject the null hypothesis.  Furthermore, for the entire sample, we find no statistical evidence for trends between the richness of the cluster environment and the radio luminosity of our bent radio AGNs.  Given that the trend identified by \citet{Croston2019} is strongest among systems likely more massive than many of our COBRA clusters and that they only probe a very narrow range of cluster masses, it is possible that the difference is due to the lower mass cluster candidates hosting bent radio AGNs in our sample and in the \citet{Wing2011} sample or that the effect strongly depends on the radio luminosity of sources in similar cluster masses and is not an overarching trend.

To further determine if environment plays any role in the size/energetics of the bent radio source, we plot the
total projected physical area as a function of the combined overdensity (lower panel of Figure~\ref{Fig:Overdensity-RP}).  We find a weak trend with no evidence to reject the null hypothesis among the entire sample (r$_{s}$ = -0.201, $p$ = 0.396).  Among the subsamples, we find moderate to strong anti-correlations with weak to moderate evidence to reject the null hypothesis among three of our cluster subsamples (m*+1 $i - [3.6]$ - r$_{s}$ = -0.714, $p$ = 0.071; magnitude limited $i - [3.6]$ - r$_{s}$ = -0.899, $p$ = 0.037; m*+1 $r - i$ - r$_{s}$ = -1.0, $p$ = 0.0). 

Given that our total sample agrees with the results from \citet{Wing2011}, it is difficult to determine whether the trend of larger radio AGNs being found in poorer clusters is real, or due to small number statistics.  However, if true, it would agree with some of the inferences from \citet{Moravec2019} and \citet{Moravec2020}, where the size of the radio AGNs was a function of local environment (assuming weaker overdensities have a weaker ICM density).  Additionally, given the difference in redshift between our sample and \citet{Wing2011}, this could also represent redshift evolution.  As shown in \citet{Vardoulaki2021}, the ICM gets denser at lower redshift, which can result in narrower bent AGNs because of the additional space for jet interactions.  Thus, a similar scenario could occur where at high redshift, the necessary ICM/gas density to constrict AGNs and bend the radio lobes only exists in richer clusters, while at low-$z$, clusters, groups, filaments, and/or fossil groups could have the minimum gas density to not only bend the radio lobes, but severely constrict the size.  Additionally, due to the 120$\arcsec$ selection criteria of the parent sample of COBRA from \citet{Wing2011}, our sample is sensitive to physically larger radio AGNs at high-$z$, which may account for the correlations we identify.    

Ultimately, as with our analysis of the asymmetry of bent radio AGN lobes, our analysis of the radio luminosity of bent AGNs finds little evidence of environmental dependence.  While it is possible that the projected physical area and luminosity of radio AGNs are not strongly dependent on environment, it is also possible that the environments of the cluster and non-cluster sample are similar.  As we currently lack X-ray observations for the majority of these sources and require a larger sample to better determine if the trends we identify between cluster richness and projected physical area are real, we plan to address these questions in the future using new data from eRosita and continuing observations from LOFAR. 

\subsection{A Comparison of Spectral Indices}\label{Sect:SI-Environment}
As discussed in Section~\ref{sect:SI-Cluster-Comp}, we measure the spectral index for the entire radio source, the spectral index of the radio core, and the spectral index of the radio lobes for the bent radio AGNs in our sample. We find that the median total spectral index of bent radio AGNs in clusters is slightly flatter than those not in clusters (-0.76 $\pm$ 0.01 vs -0.81 $\pm$ 0.02) and that this is especially true among the non-quasar populations (-0.76 $\pm$ 0.01 vs -0.83 $\pm$ 0.02). Past studies of bent AGNs or AGNs in clusters in general do not use identical frequencies, nor do they have the same sensitivity, so we can only make a qualitative comparison in terms of the similarities of the properties of bent radio AGNs and the environments of these sources.

In comparing our samples to other observations of bent AGNs (either head-tails or WATs), we find similar results to sources in low-$z$ clusters \citep[e.g.,][]{O'Donoghue1990,O'Dea2023,DiGennaro-o,DiGennaro2018a,Patnaik1986}. \citet{O'Donoghue1990} presented one of the largest samples of low-$z$ WATs with uniform observations that have spectral index measurements. Using VLA observations at 6\,cm ($\approx$ 4.996\,GHz) and 20\,cm ($\approx$ 1.444\,GHz), \citet{O'Donoghue1990} found that the core of these bent sources tend to have spectral indices of -0.5 $<$ $\alpha$ $<$ -0.6 and that the lobes are much steeper, with $\alpha$ $\approx$ -1.5, in agreement with our measurements.  Using individual targets, both \citet{Gendron-Marsolais2020}, who measured the spectral index between 344\,MHz and 610\,MHz, and \citet{Patnaik1986}, who measured the spectral index between 1.444\,GHz and 4.996\,GHz, saw bent radio AGNs with a flatter radio core, typically of the order $\alpha$ $>$ -0.5 and steeper radio lobes ($\alpha$ $<$ -1).  Similarly, \citet{DiGennaro-o} analyzed the spectral index in a similar fashion to our method (but between 235\,MHz and 8.4\,GHz), and also found flatter radio cores and steeper diffuse regions, which are akin to our radio lobes, for their two tailed radio AGNs.

At high redshift (1.5 $<$ $z$ $<$ 3.2), \citet{Barthel1988} identified a sample of 30 bent radio quasars (opening angles $<$ 160$^{\circ}$) among their sample of 80 high-$z$ quasars with steeper spectral indices ($\alpha$ $<$ -0.6 measured using VLA observations at 1.4\,GHz and 5\,GHz).  While we only study 10 quasars and our redshift range is 0.79 $<$ $z$ $<$ 2.346, we find similar results, with all but the lowest redshift quasar having a mean spectral index $<$ -0.6.  Although \citet{Barthel1988} did not identify which sources might be in protocluster/cluster environments, the similarities of our samples may be indicative of these quasars being found in regions of increased overdensity. 

Although not bent radio AGNs, \citet{Mahony2016} and \citet{deGasparin2018} studied large samples of radio AGNs out to high-$z$.  \citet{Mahony2016} used LOFAR observations (150\,GMz) and 1.4\,GHz observations to observe radio AGNs in the Lockman Hole and found that steep spectrum sources (-1.2 $<$ $\alpha$ $<$ -0.5) make up 82.1$\%$ of the sources in their Lockman-Wide sample, while flat sources ($\alpha$ $>$ -0.5) make up 5.7$\%$ of sources.  Similarly, we find all but two of our sources would be characterized as steep spectrum sources based on their criteria.  \citet{deGasparin2018} probed 147\,MHz - 1.4\,GHz across 80$\%$ of the radio sky and found a weighted mean spectral index of -0.7870 $\pm$ 0.0003.  Given the similar spectral index measurements between our samples and \citet{Mahony2016} and \citet{deGasparin2018}, it appears that the spectral index of bent radio AGNs is not a uniquely defining characteristic and that these sources have spectral indices of typical radio AGNs.  

In regards to the total spectral index and the spectral index of the radio core, we find an environmental difference between those in clusters and those not in clusters.  While the spectral index of the cores in both samples is generally flatter than the lobes, the spectral index of cores in the cluster sample are flatter than those in the non-cluster sample with a 2$\sigma$ significance.  Similar to our study, \citet{Dabhade2020} measured the mean spectral index of giant radio galaxies in both clusters and the field.  Although they used different surveys, they span a similar frequency range (NVSS at 1400\,MHz and LoTSS DR1), which makes the results comparable.  \citet{Dabhade2020} found a mean spectral index of -0.79 for the giant radio quasars and -0.78 for the giant radio galaxies.  Although not the focus of their analysis, \citet{Dabhade2020} identified 20 giant radio galaxies in low-$z$ ($z$ $<$ 0.55) clusters and report that 14 have spectral indices flatter than or equal to their reported mean values (two of the sources have no measurement), which indicates a lower mean spectral index for radio AGNs in clusters, in agreement with our findings.

To interpret the difference in the mean spectral index of the cores of bent radio AGNs, we look to \citet{deGasparin2018}, who fully modeled the spectral index as a function of the life cycle of the AGN.  \cite{deGasparin2018} found that younger, more core-dominated AGNs are both brighter and have flatter cores.  As the AGN ages, the lobes become more dominant as the AGN jets shut down.  Following this model for our sources would imply that the sources in clusters are more likely to be either younger or more recently turned on.  Given the obvious environmental difference, we argue that the bent radio AGNs in clusters have flatter cores because material was more recently accreted onto them, which may be due to the denser environment.  This could be particularly true of the quasars in the non-cluster sample, which have flatter cores, though not as flat as those in the cluster sample, indicating that they might be slightly older than their cluster counterparts.  Based on this interpretation, having a flatter mean spectral index, particularly in the core, would appear to be the strongest signpost for determining which future detected bent radio AGNs are in clusters without optical/IR follow-up (although not having a strong flat core does not rule out a bent radio AGN from being in a cluster).

\section{Conclusion}\label{sect:Conclusions}
This paper is the fourth paper in the series probing high-$z$ bent radio AGNs identified as part of the high-$z$ COBRA survey. This is the first to fully probe the impact of the cluster environment on the morphology of bent radio AGNs to determine differences between bent radio AGNs inside and outside of clusters.  Using new publicly available observations from LoTSS DR2, we identify a sample of 20 high-$z$ bent radio AGNs in clusters and 15 bent radio AGNs not in clusters.  Using this sample, we examine differences between the projected physical size, projected physical area, radio source asymmetry, opening angle, radio luminosity, and for the first time for a large sample of high-$z$ bent radio AGNs, the spectral index of these sources.  Below is a summary of our findings.
\begin{itemize}
    \item \textit{Differences in the Projected Physical Size, Area, and the Asymmetry of Bent Radio AGNs}:  For each of the 35 radio AGNs in our sample, we estimate the projected physical size and projected physical area of the bent radio AGNs within 10$\sigma$ contours. Within the cluster sample, we find tentative evidence of moderate to strong anti-correlations between the projected physical area of each radio AGN and the richness of the host cluster.  However, we find no distinct differences between the sources in clusters and the sources not in clusters.  
    
    \item \textit{Measurements of the Opening Angle}: We measure the opening angle of each bent AGN observed with LoTSS and find agreement with previous measurements from FIRST.  For the cluster sample, we do not see a statistically robust trend between the size of the opening angle and cluster richness.  However, we do find more narrow sources in clusters, implying some environmental differences in the populations of bent radio AGNs. 
    
    \item \textit{Probing Radio Luminosity}: We estimate the radio luminosity and examine it as a function of projected physical area and environment.  We find a moderate positive correlation between radio luminosity and the projected physical area of the AGN for the cluster sample and a weak correlation with no evidence to reject the null hypothesis for the non-cluster sample.  Despite these differences, a KS test implies that both samples are drawn from the same sample, further highlighting their similarities.  
    
    \item \textit{New Measurements of the Spectral Index of Bent Radio AGNs}: We measure the spectral index for a large sample of bent radio AGNs at high-$z$.  Using observations from LoTSS DR2 and FIRST, we find the median values of the spectral index differ by $\approx$ 1.6$\sigma$ (-0.76 $\pm$ 0.01 and -0.81 $\pm$ 0.02 respectively) and the core spectral index differ by $\approx$ 2$\sigma$ for the total sample (-0.51 $\pm$ 0.03 and -0.65 $\pm$ 0.04) and $>$ 3$\sigma$ for the non-quasars with a well defined core (-0.54 $\pm$ 0.03 and -0.75 $\pm$ 0.03).  We find that 13 of 20 cluster bent radio AGNs have a flatter core ($\geq$ -0.6) as compared to 4 of 15 in the non-cluster sample.  As the flatness of the core indicates the age of the radio emission, this suggests that these sources are indeed in richer cluster environments with more recent accretion onto the supermassive black hole.  It is possible that future cluster surveys using bent radio AGNs should target AGNs with a core with a flatter spectral index.
\end{itemize}

The goal of this study was to identify differences between the populations of bent radio AGNs based on their large scale environment to improve the identification of high-$z$ cluster candidates.  Although we found that sources with narrower opening angles and flatter spectral cores are indeed more commonly found in clusters, we find a surprising degree of similarity between the samples.  Ultimately, our populations of bent radio AGNs are similar in regards to their projected physical size, area, and radio luminosity, which may point to further similarities between the environments of these radio AGNs.  

While the non-cluster sample is not located in rich cluster overdensities of red galaxies as shown in GM19, it is possible that these sources reside in bent radio AGNs that are outside the cluster center, possibly beyond the area probed in GM19, either in the outer cluster regions or falling along filaments.  As bent radio AGNs require a gaseous medium to bend the radio lobes, the differences we find may be due to the differing affects of the local and large scale environment \citep[e.g.,][]{Rodman2019,Yates-Jones2021}.  This interpretation would explain why the radio source properties are similar (e.g., the projected physical area and the radio luminosity), despite the obvious outliers (narrow opening angles and flatter radio cores) that are indicative of richer cluster environments.  This stands to further strengthen the use of all bent radio AGNs as signposts for structure in large sky surveys. 

Given the similarities between these two populations, we aim to continue to study populations of bent radio AGNs to better characterize the host environments of our non-cluster sample.  Although X-ray observations of distant clusters and filaments are observationally expensive, with more data becoming available from telescopes like eRosita, we will continue to better quantify the gaseous environments to improve constraints on the role of the local gas density on radio morphology, especially in regards to the asymmetry.  In particular, we plan to better constrain the role of the opening angle on asymmetry by studying samples of NATS and straight radio AGNs.  Given the abundance of data available at low-$z$, including measurements of the ambient density, we will further probe the environments of low-$z$ bent radio AGNs to improve constraints on the mass of the host clusters and further characterize populations of bent radio AGNs that are potentially in filaments or fossil group galaxies.  

Additionally, we plan to further constrain the energetics of radio AGNs by probing larger samples at lower redshifts, where we can more accurately map differences over the entire radio source.  With new radio observations, from surveys like those being conducted with Meerkat and SKAO, we plan to better map the entirety of the spectral index of the radio source, allowing us to probe the asymmetry of the lobes of bent radio AGNs. 

Lastly, we will continue to work to identify new bent radio AGNs out to the highest redshifts with future LoTSS data releases.  Although none of the sources in our sample are characterized as ultra steep sources ($\alpha$ $<$ -1.4) \citep[e.g.,][]{Mahony2016,deGasparin2018}, these radio AGNs tend to be associated with high-$z$ ($z$ $>$ 2) protoclusters \citep[e.g.,][]{Wylezalek2013,Wylezalek2014} and include quasars with bent radio AGN morphology \citep{Barthel1988}.  As we continue to search for bent radio AGNs at progressively higher redshifts, it will be interesting to see if we identify any such sources in clusters as they could be the signposts for some of the earliest forming dense ICMs.

\section{acknowledgments}
EGM would like to thank the anonymous referee for their insightful comments that helped to shape and strengthen this paper.  EGM would also like to thank the organizers of the Early Stages of Galaxy Cluster Formation 2021 Virtual Conference for fostering stimulating discussions that led to the ideas addressed in this paper.  EGM would like to thank Brian C. Lemaux for connecting the core group of early career scientists studying radio AGNs in high-$z$ clusters, which ultimately led to this paper.  Additionally, EGM would like to thank George Miley for useful discussions regarding the environments of high-$z$ bent quasars and protoclusters.

EGM, RG, and HT acknowledge the support of this work by the Tsinghua Shui Mu Scholarship.  Additionally, EGM and ZC acknowledge that this work was funded by the National Key R\&D Program of China (grant no.\ 2018YFA0404503), the National Science Foundation of China (grant no.\ 12073014). The science research grants from the China Manned Space Project with No. CMS-CSST2021-A05, and Tsinghua University Initiative Scientific Research Program (No. 20223080023)

LOFAR data products were provided by the LOFAR Surveys Key Science project (LSKSP; https://lofar-surveys.org/) and were derived from observations with the International LOFAR Telescope (ILT). LOFAR (van Haarlem et al. 2013) is the Low Frequency Array designed and constructed by ASTRON. It has observing, data processing, and data storage facilities in several countries, which are owned by various parties (each with their own funding sources), and which are collectively operated by the ILT foundation under a joint scientific policy. The efforts of the LSKSP have benefited from funding from the European Research Council, NOVA, NWO, CNRS-INSU, the SURF Co-operative, the UK Science and Technology Funding Council and the Jülich Supercomputing Centre.

These results made use of the Lowell Discovery Telescope at Lowell Observatory. Lowell is a private, non-profit institution dedicated to astrophysical research and public appreciation of astronomy and operates the LDT in partnership with Boston University, the University of Maryland, the University of Toledo, Northern Arizona University, and Yale University. LMI construction was supported by a grant AST-1005313 from the National Science Foundation.

This work is based in part on observations made with the  $\sl{Spitzer}$ Space Telescope, which is operated by the Jet Propulsion Laboratory, California Institute of Technology under a contract with NASA. Support for this work was provided by NASA through an award issued by JPL/Caltech (NASA award RSA No. 1440385).

Funding for SDSS-III has been provided by the Alfred P. Sloan Foundation, the Participating Institutions, the National Science Foundation, and the U.S. Department of Energy Office of Science. The SDSS-III web site is http://www.sdss3.org/.

SDSS-III is managed by the Astrophysical Research Consortium for the Participating Institutions of the SDSS-III Collaboration including the University of Arizona, the Brazilian Participation Group, Brookhaven National Laboratory, Carnegie Mellon University, University of Florida, the French Participation Group, the German Participation Group, Harvard University, the Instituto de Astrofisica de Canarias, the Michigan State/Notre Dame/JINA Participation Group, Johns Hopkins University, Lawrence Berkeley National Laboratory, Max Planck Institute for Astrophysics, Max Planck Institute for Extraterrestrial Physics, New Mexico State University, New York University, Ohio State University, Pennsylvania State University, University of Portsmouth, Princeton University, the Spanish Participation Group, University of Tokyo, University of Utah, Vanderbilt University, University of Virginia, University of Washington, and Yale University.

This research made use of Astropy,\footnote{http://www.astropy.org} a community-developed core Python package for Astronomy \citep{astropy:2013, astropy:2018}.

\vspace{5mm}
\facilities{LDT, $Spitzer$, Sloan, LOFAR, FIRST}

\software{astropy \citep{astropy:2013}, 
          CASA \citep{CASA}}

\begin{deluxetable*}{lccclll}
    \tablecolumns{7}
    \tabletypesize{\footnotesize}
    \tablecaption{Radio Source Properties \label{tb:RadioProp}}
    \tablewidth{0pt}

    \setlength{\tabcolsep}{0.03in}
    \tablehead{
    \colhead{Field}&
    \colhead{Redshift}&
    \colhead{Opening Angle}&
    \colhead{Total Projected}&
    \colhead{Total Projected} &
    \multicolumn{2}{c}{Radio Lobe Lengths}
    \cr
    \colhead{}&
    \colhead{(z)}&
    \colhead{($^{\circ}$)}&
    \colhead{Physical Size}&
    \colhead{Physical Area}&
    \colhead{Lobe 1}&
    \colhead{Lobe 2}
    \cr
    \colhead{}&
    \colhead{}&
    \colhead{}&
    \colhead{(kpc)}&
    \colhead{(Mpc$^{2}$)}&
    \colhead{($\arcsec$)}&
    \colhead{($\arcsec$)}
    }
    
\startdata
\cutinhead{Cluster Candidate Sample}
COBRA072805.2+312857\tablenotemark{a} & 1.75 $\pm$ 0.1\tablenotemark{b}\tablenotemark{c} & 156.0 $\pm$ 2.3 & 471 $^{+26}_{-26}$ & 0.064 $^{+0.005}_{-0.005}$ & 28.5$^{+3.0}_{-3.0}$ & 27.2$^{+3.0}_{-3.0}$ \\
COBRA074410.9+274011\tablenotemark{a} & 1.30 $\pm$ 0.1\tablenotemark{b}\tablenotemark{d} & 119.7 $\pm$ 3.6 & 256 $^{+25}_{-25}$ & 0.029 $^{+0.004}_{-0.004}$ & 14.6$^{+3.0}_{-3.0}$ & 15.9$^{+3.0}_{-3.0}$ \\
COBRA100745.5+580713\tablenotemark{a}& 0.656 $\pm$ 0.0003\tablenotemark{b}\tablenotemark{c}\tablenotemark{d}\tablenotemark{e}\tablenotemark{f}  & 125.5 $\pm$ 3.1 & 271 $^{+21}_{-21}$ & 0.063 $^{+0.004}_{-0.003}$ & 19.6$^{+3.0}_{3.0}$ & 19.5$^{+3.0}_{-3.0}$ \\
COBRA100841.7+372513 & 1.20 $\pm$ 0.1\tablenotemark{c}\tablenotemark{d}& 120.3 $\pm$ 2.1 & 412 $^{+34}_{-48}$ & 0.077 $^{+0.006}_{-0.006}$ & 26.9$^{+3.5}_{-5.7}$ & 22.7$^{+3.1}_{-3.1}$ \\
COBRA103434.2+310352\tablenotemark{a} & 1.20 $\pm$ 0.1\tablenotemark{b}\tablenotemark{c} & 83.6 $\pm$ 0.2 & 675 $^{+26}_{-27}$ & 0.141 $^{+0.005}_{-0.006}$ & 41.1$^{+3.0}_{-3.0}$ & 40.4$^{+3.0}_{-3.0}$ \\
COBRA104254.8+290719\tablenotemark{a} & 1.35 $\pm$ 0.1\tablenotemark{b}& 82.8 $\pm$ 0.2 & 520 $^{+26}_{-26}$ & 0.119 $^{+0.006}_{-0.006}$ & 31.5$^{+3.0}_{-3.0}$ & 31.2$^{+3.0}_{-3.0}$ \\
COBRA113733.8+300010\tablenotemark{a} & 0.96 $\pm$ 0.005\tablenotemark{b}\tablenotemark{c}\tablenotemark{e}\tablenotemark{g}& 63.9 $\pm$ 2.6 & 335 $^{+25}_{-25}$ & 0.038 $^{+0.004}_{-0.004}$ & 19.3$^{+3.0}_{-3.0}$ & 23.0$^{+3.1}_{-3.1}$ \\
COBRA121128.5+505253\tablenotemark{h} & 1.364 $\pm$ 0.00042\tablenotemark{f} & 150.4 $\pm$ 2.4 & 526 $^{+25}_{-25}$ & 0.217 $^{+0.006}_{-0.006}$ & 32.5$^{+3.0}_{-3.0}$ & 30.1$^{+3.0}_{-3.0}$ \\
COBRA123940.7+280828\tablenotemark{a} & 0.92 $\pm$ 0.0682 \tablenotemark{b}\tablenotemark{e} & 132.8 $\pm$ 1.0 & 581 $^{+26}_{-27}$ & 0.089 $^{+0.007}_{-0.007}$ & 41.1$^{+3.0}_{-3.0}$ & 33.1$^{+3.0}_{-3.0}$ \\
COBRA130729.2+274659\tablenotemark{h} & 1.144 $\pm$ 0.006\tablenotemark{f} & 157.3 $\pm$ 3.6 & 724 $^{+25}_{-25}$ & 0.146 $^{+0.006}_{-0.005}$ & 52.1$^{+3.0}_{-3.0}$ & 35.9$^{+3.0}_{-3.0}$ \\
COBRA135136.2+543955 & 0.55 $\pm$ 0.1\tablenotemark{b}\tablenotemark{d}& 131.4 $\pm$ 1.8 & 479 $^{+43}_{-52}$ & 0.089 $^{+0.015}_{-0.018}$ & 39.6$^{+3.0}_{-3.0}$ & 35.2$^{+3.0}_{-3.0}$ \\
COBRA135838.1+384722 & 0.81 $\pm$ 0.1\tablenotemark{b}\tablenotemark{c}& 134.7 $\pm$ 5.1 & 257 $^{+25}_{-26}$ & 0.040 $^{+0.005}_{-0.005}$ & 16.0$^{+3.0}_{-3.0}$ & 18.1$^{+3.0}_{-3.0}$ \\
COBRA141155.2+341510\tablenotemark{h} & 1.818 $\pm$ 0.00019\tablenotemark{f}& 141.3 $\pm$ 3.6 & 431 $^{+26}_{-26}$ & 0.087 $^{+0.004}_{-0.004}$ & 28.7$^{+3.0}_{-3.0}$ & 22.5$^{+3.0}_{-3.0}$ \\
COBRA145023.3+340123 & 1.20 $\pm$ 0.1\tablenotemark{b}\tablenotemark{d}& 107.2 $\pm$ 1.0 & 390 $^{+26}_{-26}$ & 0.059 $^{+0.005}_{-0.005}$ & 23.0$^{+3.0}_{-3.0}$ & 24.1$^{+3.0}_{-3.0}$ \\
COBRA152647.5+554859\tablenotemark{a} & 1.10 $\pm$ 0.1\tablenotemark{b}\tablenotemark{c}& 159.9 $\pm$ 1.1 & 694 $^{+27}_{-28}$ & 0.229 $^{+0.011}_{-0.012}$ & 37.4$^{+3.0}_{-3.0}$ & 47.6$^{+3.0}_{-3.0}$ \\
COBRA154638.3+364420\tablenotemark{h} & 0.939 $\pm$ 0.00036\tablenotemark{f}& 155.4 $\pm$ 1.0 & 643 $^{+25}_{-25}$ & 0.141 $^{+0.007}_{-0.007}$ & 44.6$^{+3.1}_{-3.1}$ & 37.0$^{+3.0}_{-3.0}$ \\
COBRA164611.2+512915 & 0.351 $\pm$ 0.00008\tablenotemark{b}\tablenotemark{c}\tablenotemark{d}\tablenotemark{f}& 48.2 $\pm$ 1.0 & 554 $^{+17}_{-16}$ & 0.050 $^{+0.004}_{-0.003}$ & 53.5$^{+3.0}_{-3.0}$ & 58.5$^{+3.2}_{-3.1}$ \\
COBRA164951.6+310818 & 0.52 $\pm$ 0.07875\tablenotemark{b}\tablenotemark{e} & 35.0 $\pm$ 0.4 & 412 $^{+35}_{-40}$ & 0.057 $^{+0.009}_{-0.010}$ & 39.4$^{+3.0}_{-3.0}$ & 26.8$^{+3.0}_{-3.0}$ \\
COBRA170105.4+360958\tablenotemark{a} & 0.80 $\pm$ 0.1\tablenotemark{b}& 98.5 $\pm$ 0.6 & 290 $^{+25}_{-27}$ & 0.042 $^{+0.005}_{-0.005}$ & 22.1$^{+3.0}_{-3.0}$ & 26.8$^{+3.0}_{-3.0}$ \\
COBRA171330.9+423502\tablenotemark{a} & 0.698 $\pm$ 0.00018\tablenotemark{b}\tablenotemark{f} & 136.4 $\pm$ 1.7 & 423 $^{+22}_{-22}$ & 0.107 $^{+0.007}_{-0.010}$ & 30.7$^{+3.1}_{-3.1}$ & 32.0$^{+3.0}_{-3.1}$ \\
\cutinhead{Non-Cluster Sample}\\
COBRA090102.7+420746\tablenotemark{h} & 1.621 $\pm$ 0.00112\tablenotemark{f} & 144.8 $\pm$ 0.1 & 294 $^{+26}_{-26}$ & 0.052 $^{+0.004}_{-0.004}$ & 16.6$^{+3.0}_{-3.0}$ & 18.2$^{+3.0}_{-3.0}$\\
COBRA090745.5+382740\tablenotemark{h} & 1.743 $\pm$ 0.00017\tablenotemark{f} & 145.6 $\pm$ 0.5 & 406 $^{+33}_{-26}$ & 0.112 $^{+0.005}_{-0.005}$ & 21.2$^{+3.0}_{-3.0}$ & 26.7$^{+3.7}_{-3.0}$\\  
COBRA093726.6+365550 & 1.40 $\pm$ 0.1\tablenotemark{b} & 140.6 $\pm$ 2.5 & 770 $^{+27}_{-27}$ & 0.139 $^{+0.008}_{-0.008}$  & 51.3$^{+3.0}_{-3.0}$ & 40.1$^{+3.1}_{-3.1}$\\  
COBRA104641.5+282028 & 1.75 $\pm$ 0.1\tablenotemark{b} & 54.8 $\pm$ 0.7 & 473 $^{+27}_{-27}$ & 0.027 $^{+0.006}_{-0.005}$ & 33.1$^{+3.1}_{-3.1}$ & 22.9$^{+3.0}_{-3.0}$\\
COBRA111707.3+305307 & 0.33 $\pm$ 0.0514\tablenotemark{e} & 61.1 $\pm$ 0.33 & 317 $^{+34}_{-38}$ & 0.029 $^{+0.007}_{-0.006}$ & 32.9$^{+3.0}_{-3.0}$ & 33.9$^{+3.0}_{-3.0}$\\  
COBRA120654.6+290742 & 0.853 $\pm$ 0.1139\tablenotemark{b}\tablenotemark{c}\tablenotemark{d}\tablenotemark{e} & 178.7 $\pm$ 11.2 & 309 $^{+26}_{-28}$ & 0.053 $^{+0.005}_{-0.006}$ & 22.4$^{+3.0}_{-5.6}$ & 17.9$^{+3.0}_{-5.2}$\\   
COBRA123347.0+354133 & 0.87 $\pm$ 0.1\tablenotemark{b} & 132.2 $\pm$ 0.5 & 304 $^{+26}_{-26}$ & 0.047 $^{+0.005}_{-0.005}$ & 20.6$^{+3.0}_{-3.0}$ & 18.7$^{+3.0}_{-3.0}$\\ 
COBRA131854.0+231153 & 1.45 $\pm$ 0.1\tablenotemark{b} & 153.2 $\pm$ 1.8 & 475 $^{+26}_{-26}$ & 0.027 $^{+0.005}_{-0.005}$ & 35.4$^{+3.0}_{-3.0}$ & 20.8$^{+3.0}_{-3.0}$\\ 
COBRA132903.2+253110\tablenotemark{h} & 0.987 $\pm$ 0.00019\tablenotemark{f} & 97.3 $\pm$ 1.2 & 543 $^{+27}_{-28}$ & 0.106 $^{+0.006}_{-0.005}$ & 45.9$^{+3.2}_{-3.4}$ & 22.1$^{+3.0}_{-3.0}$\\  
COBRA143817.6+491233\tablenotemark{h} & 1.358 $\pm$ 0.00094\tablenotemark{f} & 113.4 $\pm$ 0.7 & 702 $^{+26}_{-26}$ & 0.157 $^{+0.012}_{-0.009}$ & 41.4$^{+3.0}_{-3.0}$ & 42.1$^{+3.0}_{-3.0}$\\  
COBRA144207.1+562522 & 1.80 $\pm$ 0.1\tablenotemark{b} & 137.9 $\pm$ 3.4 & 275 $^{+26}_{-26}$ & 0.067 $^{+0.005}_{-0.005}$ & 16.0$^{+3.0}_{-3.0}$ & 16.5$^{+3.0}_{-3.0}$\\   
COBRA145656.0+501748 & 0.88 $\pm$ 0.1\tablenotemark{b}\tablenotemark{c}& 140.9 $\pm$ 1.1 & 357 $^{+26}_{-27}$ & 0.062 $^{+0.005}_{-0.006}$ & 23.1$^{+3.0}_{-3.0}$ & 23.0$^{+3.0}_{-3.0}$\\ 
COBRA153317.4+391804\tablenotemark{h} & 0.789 $\pm$ 0.00016\tablenotemark{f} & 155.2 $\pm$ 1.2 & 838 $^{+23}_{-23}$ & 0.143 $^{+0.009}_{-0.008}$ & 50.8$^{+3.0}_{-3.0}$ & 61.3$^{+3.0}_{-3.0}$\\  
COBRA155000.5+294953\tablenotemark{h} & 2.328 $\pm$ 0.00062\tablenotemark{f} & 109.7 $\pm$ 4.0 & 308 $^{+25}_{-25}$ & 0.049 $^{+0.004}_{-0.004}$ & 17.7$^{+3.0}_{-3.0}$ & 19.9$^{+3.0}_{-3.0}$\\  
COBRA170443.9+295246 & 1.25 $\pm$ 0.1\tablenotemark{f} & 140.5 $\pm$ 2.6 & 426 $^{+26}_{-26}$ & 0.071 $^{+0.005}_{-0.005}$ & 29.4$^{+3.0}_{-3.0}$ & 21.7$^{+3.0}_{-3.0}$\\ 
\enddata
\tablenotetext{a}{Bent Radio AGNs hosted by BCGs identified in GM21}
\tablenotetext{b}{Photometric redshift estimate from comparing the $i - [3.6]$ color to EzGal models}
\tablenotetext{c}{Photometric redshift estimates from comparing the $[3.6] - [4.5]$ color to EzGal models}
\tablenotetext{d}{Photometric redshift estimates from comparing the $r - i$ color to EzGal models}
\tablenotetext{e}{Photometric redshift estimates from SDSS}
\tablenotetext{f}{Spectroscopic redshift from SDSS}
\tablenotetext{g}{Spectroscopic redshift from \citet{Blanton2003}}
\tablenotetext{h}{Bent Radio AGNs that are SDSS identified quasars.}

\end{deluxetable*}

\begin{deluxetable*}{lccclllllccc}
    \tablecolumns{9}
    \tabletypesize{\footnotesize}
    \tablecaption{Radio Source Properties \label{tb:RadioProp-2}}
    \tablewidth{0pt}

    \setlength{\tabcolsep}{0.03in}
    \tablehead{
    \colhead{Field}&
    \multicolumn{2}{c}{RMS}&
    \multicolumn{3}{c}{Flux Density}&
    \colhead{Radio Luminosity} &
    \multicolumn{3}{c}{Spectral Index}
    \cr
    \colhead{}&
    \colhead{LoTSS}&
    \colhead{FIRST}&
    \colhead{LoTSS 10$\sigma$}&
    \colhead{LoTSS 3$\sigma$} &
    \colhead{FIRST} &
    \colhead{} &
    \colhead{Total} &
    \colhead{Core} &
    \colhead{Lobes} 
    \cr
    \colhead{}&
    \colhead{(mJy/beam)}&
    \colhead{(mJy/beam)}&
    \colhead{(mJy)}&
    \colhead{(mJy)}& 
    \colhead{(mJy)} &
    \colhead{(10$^{25}$ WHz$^{-1}$)} &
    \colhead{} &
    \colhead{} &
    \colhead{}
    }
    
\startdata
\cutinhead{Cluster Candidate Sample}
COBRA072805.2+312857 & 0.20 & 0.15 & 115.8 $\pm$ 11.6 & 80.2 $\pm$ 8.0 & 17.0 $\pm$ 1.7 & 30.8 $^{+7.2}_{-7.1}$ & -0.67$\pm$ 0.14 & -0.10 $\pm$ 0.14  & -0.96 $\pm$ 0.15 \\
COBRA074410.9+274011 & 0.10 & 0.14 & 36.9 $\pm$ 3.7 & 18.8 $\pm$ 1.9 & 2.0 $\pm$ 0.3 & 3.6 $^{+1.0}_{-1.0}$ & -0.98 $\pm$ 0.16 & -0.71 $\pm$ 0.18\tablenotemark{a} & -1.07 $\pm$ 0.17 \\
COBRA100745.5+580713 & 0.06 & 0.15 & 67.6 $\pm$ 6.8 & 52.6 $\pm$ 5.3 & 11.2 $\pm$ 1.2 & 4.1 $^{+0.5}_{-0.5}$ & -0.67 $\pm$ 0.14 & -0.46 $\pm$ 0.19\tablenotemark{a} & -0.68 $\pm$ 0.15 \\
COBRA100841.7+372513 & 0.07 & 0.17 & 96.5 $\pm$ 9.6 & 65.2 $\pm$ 0.65 & 14.6 $\pm$ 1.5 & 14.6 $^{+3.9}_{-3.9}$ & -0.65 $\pm$ 0.14 & -0.45 $\pm$ 0.14 & -0.84 $\pm$ 0.15 \\
COBRA103434.2+310352 & 0.08 & 0.14 & 2089.9 $\pm$ 209.0 & 2010.1 $\pm$ 201.0 & 315.1 $\pm$ 31.5 & 397.7 $^{+107.1}_{-105.8}$ & -0.80 $\pm$ 0.14 & -0.31 $\pm$ 0.15 & -0.81 $\pm$ 0.14 \\
COBRA104254.8+290719 & 0.10 & 0.14 & 137.6 $\pm$ 13.8 & 83.9 $\pm$ 8.4 & 13.3 $\pm$ 1.4 & 16.7 $^{+4.4}_{-4.4}$ & -0.80 $\pm$ 0.14 & -0.62 $\pm$ 0.15 & -0.84 $\pm$ 0.14 \\
COBRA113733.8+300010 & 0.08 & 0.14 & 52.3 $\pm$ 5.2 & 46.3 $\pm$ 4.6 & 8.3 $\pm$ 0.9 & 6.7 $^{+0.9}_{-0.9}$ & -0.74 $\pm$ 0.14 & -0.60 $\pm$ 0.16 & -0.76 $\pm$ 0.14 \\
COBRA121128.5+505253\tablenotemark{b} & 0.09 & 0.12 & 1972.3 $\pm$ 197.3 & 1850.2 $\pm$ 185.0 & 234.5 $\pm$ 23.4 & 407.0 $^{+64.2}_{-64.2}$ & -0.90 $\pm$ 0.14 & -0.89 $\pm$ 0.14 & -0.89 $\pm$ 0.14 \\
COBRA123940.7+280828 & 0.10 & 0.15 & 120.6 $\pm$ 12.1 & 61.4 $\pm$ 6.1 & 10.1 $\pm$ 1.1 & 8.2 $^{+2.0}_{-1.9}$ & -0.78 $\pm$ 0.14 & -0.54 $\pm$ 0.14 & -0.79 $\pm$ 0.15 \\
COBRA130729.2+274659\tablenotemark{b} & 0.08 & 0.16 & 333.6 $\pm$ 33.3 & 187.7 $\pm$ 18.8 & 22.6 $\pm$ 2.3 & 31.7 $^{+4.7}_{-4.7}$ & -0.92 $\pm$ 0.14 & -0.42 $\pm$ 0.14 & -1.2 $\pm$ 0.15 \\
COBRA135136.2+543955 & 0.05 & 0.16 & 111.9 $\pm$ 11.2 & 70.6 $\pm$ 7.1 & 13.0 $\pm$ 1.3 & 4.0 $^{+1.9}_{-1.8}$ & -0.73 $\pm$ 0.14 & -0.63 $\pm$ 0.19\tablenotemark{a} & -0.74 $\pm$ 0.14 \\
COBRA135838.1+384722 & 0.08 & 0.11 & 89.7 $\pm$ 9.0 & 80.6 $\pm$ 8.1 & 12.3 $\pm$ 1.2 & 8.5 $^{+3.0}_{-2.9}$ & -0.82 $\pm$ 0.14 & -0.86 $\pm$ 0.15 & -0.81 $\pm$ 0.14\\
COBRA141155.2+341510\tablenotemark{b} & 0.13 & 0.14 & 1074.0 $\pm$ 107.4 & 1036.8 $\pm$ 103.7 & 190.6 $\pm$ 19.1 & 393.0 $^{+69.7}_{-69.7}$ & -0.73 $\pm$ 0.14 & -0.36 $\pm$ 0.14 & -0.97 $\pm$ 0.14 \\
COBRA145023.3+340123 & 0.07 & 0.14 & 77.8 $\pm$ 7.8 & 52.7 $\pm$ 5.3 & 7.3 $\pm$ 0.8 & 10.0 $^{+2.7}_{-2.7}$ & -0.86 $\pm$ 0.15 & -0.68 $\pm$ 0.15 & -0.94 $\pm$ 0.15 \\
COBRA152647.5+554859 & 0.09 & 0.16 & 563.6 $\pm$ 56.4 & 484.2 $\pm$ 48.4 & 78.8 $\pm$ 7.9 & 85.0 $^{+24.0}_{-23.7}$ & -0.79 $\pm$ 0.14 & -0.28 $\pm$ 0.14 & -0.83 $\pm$ 0.14 \\
COBRA154638.3+364420\tablenotemark{b} & 0.11 & 0.14 & 394.1 $\pm$ 39.4 & 308.0 $\pm$ 30.8 & 89.7 $\pm$ 9.0 & 49.8 $^{+6.8}_{-6.8}$ & -0.54 $\pm$ 0.14 & 0.35 $\pm$ 0.14 & -0.83 $\pm$ 0.14 \\
COBRA164611.2+512915 & 0.07 & 0.15 & 96.1 $\pm$ 9.6 & 66.6 $\pm$ 6.7 & 13.2 $\pm$ 1.4 & 1.7 $^{+0.2}_{-0.2}$ & -0.70 $\pm$ 0.14 & -0.48 $\pm$ 0.15 & -0.74 $\pm$ 0.15 \\
COBRA164951.6+310818 & 0.11 & 0.13 & 79.1 $\pm$ 7.9 & 728.6 $\pm$ 72.9 & 113.3 $\pm$ 11.3 & 36.1 $^{+14.4}_{-13.9}$ & -0.81 $\pm$ 0.14 & -0.54 $\pm$ 0.14 & -0.86 $\pm$ 0.14 \\
COBRA170105.4+360958 & 0.09 & 0.14 & 98.9 $\pm$ 9.9 & 87.5 $\pm$ 8.8 & 16.2 $\pm$ 1.6 & 9.5 $^{+3.5}_{-3.3}$ & -0.73 $\pm$ 0.14 & -0.70 $\pm$ 0.14 & -0.74 $\pm$ 0.14 \\
COBRA171330.9+423502 & 0.09 & 0.13 & 101.9 $\pm$ 10.2 & 57.0 $\pm$ 5.7 & 17.1 $\pm$ 1.7 & 4.3 $^{+0.5}_{-0.5}$ & -0.52 $\pm$ 0.14 & -0.44 $\pm$ 0.17\tablenotemark{a} & -0.53 $\pm$ 0.14  \\
\cutinhead{Non-Cluster Sample}\\
COBRA090102.7+420746\tablenotemark{b} & 0.09 & 0.14 & 86.2 $\pm$ 8.6 & 76.9 $\pm$ 7.7 & 18.3 $\pm$ 1.9 & 27.9 $^{+4.9}_{-4.7}$ & -0.62 $\pm$ 0.14 & -0.44 $\pm$ 0.14 & -0.86 $\pm$ 0.15 \\
COBRA090745.5+382740\tablenotemark{b} & 0.18 & 0.14 & 1047.9 $\pm$ 104.8 & 971.4 $\pm$ 97.1 & 160.3 $\pm$ 16.0 & 331.1 $^{+51.1}_{-57.6}$ & -0.78 $\pm$ 0.14 & -0.61 $\pm$ 0.14 & -0.87 $\pm$ 0.14 \\   
COBRA093726.6+365550 & 0.10 & 0.12 & 149.8 $\pm$ 15.0 & 54.8 $\pm$ 5.5 & 7.3 $\pm$ 0.8 & 12.8 $^{+2.9}_{-3.2}$ & -0.87 $\pm$ 0.14 & -0.83 $\pm$ 0.17 & -0.88 $\pm$ 0.15 \\  
COBRA104641.5+282028 & 0.11 & 0.13 & 18.4 $\pm$ 1.8 & 8.3 $\pm$ 0.8 & 3.3 $\pm$ 0.4 & 4.2 $^{+0.9}_{-1.0}$ & -0.41 $\pm$ 0.15 & 0.36 $\pm$ 0.22\tablenotemark{c} & -0.61 $\pm$ 0.16 \\
COBRA111707.3+305307 & 0.07 & 0.16 & 70.0 $\pm$ 7.0 & 35.6 $\pm$ 3.6 & 6.4 $\pm$ 0.7 & 0.8 $^{+0.3}_{-0.3}$ & -0.75 $\pm$ 0.15 & -0.65 $\pm$ 0.16 & -0.77 $\pm$ 0.15 \\  
COBRA120654.6+290742 & 0.11 & 0.14 & 152.9 $\pm$ 15.3 & 127.1 $\pm$ 12.7 & 16.9 $\pm$ 1.7 & 14.1 $^{+5.4}_{-5.2}$ & -0.88 $\pm$ 0.14 & -0.68 $\pm$ 0.14 & -0.97 $\pm$ 0.14 \\   
COBRA123347.0+354133 & 0.09 & 0.13 & 120.6 $\pm$ 12.1 & 97.6 $\pm$ 9.8 & 10.6 $\pm$ 1.1 & 10.6 $^{+3.6}_{-3.5}$ & -0.96 $\pm$ 0.14 & -0.79 $\pm$ 0.14 & -1.08 $\pm$ 0.15 \\ 
COBRA131854.0+231153 & 0.20 & 0.13 & 36.1 $\pm$ 3.6 & 29.2 $\pm$ 3.0 & 4.3 $\pm$ 0.5 & 7.4 $^{+1.8}_{-1.8}$ & -0.83 $\pm$ 0.15 & -0.71 $\pm$ 0.20\tablenotemark{a} & -0.85 $\pm$ 0.15 \\ 
COBRA132903.2+253110\tablenotemark{b} & 0.12 & 0.16 & 513.6 $\pm$ 51.4 & 437.2 $\pm$ 43.7 & 67.7 $\pm$ 6.8 & 63.9 $^{+8.7}_{-8.9}$ & -0.81 $\pm$ 0.14 & -0.61 $\pm$ 0.14 & -0.86 $\pm$ 0.14 \\  
COBRA143817.6+491233\tablenotemark{b} & 0.06 & 0.16 & 169.1 $\pm$ 16.9 & 114.0 $\pm$ 11.4 & 13.7 $\pm$ 1.4 & 24.5 $^{+3.6}_{-3.9}$ & -0.92 $\pm$ 0.14 & -0.34 $\pm$ 0.15 & -0.98 $\pm$ 0.14 \\ 
COBRA144207.1+562522 & 0.07 & 0.12 & 80.4 $\pm$ 8.0 & 72.3 $\pm$ 7.2 & 11.6 $\pm$ 1.2 & 25.4 $^{+4.9}_{-5.9}$ & -0.80 $\pm$ 0.14 & -0.71 $\pm$ 0.14 & -0.83 $\pm$ 0.14 \\   
COBRA145656.0+501748 & 0.06 & 0.11 & 62.2 $\pm$ 6.2 & 51.1 $\pm$ 5.1 & 8.7 $\pm$ 0.9 & 6.4 $^{+2.1}_{-2.0}$ & -0.77 $\pm$ 0.15 & -0.65 $\pm$ 0.16\tablenotemark{a} & -0.78 $\pm$ 0.14 \\ 
COBRA153317.4+391804\tablenotemark{b} & 0.08 & 0.15 & 114.9 $\pm$ 11.4 & 41.0 $\pm$ 4.1 & 16.5 $\pm$ 1.7 & 5.3 $^{+0.9}_{-0.7}$ & -0.39 $\pm$ 0.14 & -0.11 $\pm$ 0.14 & -0.68 $\pm$ 0.15  \\  
COBRA155000.5+294953\tablenotemark{b} & 0.12 & 0.12 & 228.2 $\pm$ 22.8 & 206.3 $\pm$ 20.6 & 19.8 $\pm$ 2.0 & 76.4 $^{+11.5}_{-15.1}$ & -1.02 $\pm$ 0.14 & -0.85 $\pm$ 0.14 & -1.05 $\pm$ 0.14 \\  
COBRA170443.9+295246 & 0.14 & 0.16 & 222.7 $\pm$ 22.2 & 124.9 $\pm$ 12.5 & 9.4 $\pm$ 1.0 & 20.3 $^{+5.4}_{-5.5}$ & -1.12 $\pm$ 0.14 & -1.35 $\pm$ 0.16 & -1.05 $\pm$ 0.15 \\ 
\enddata
\tablenotetext{a}{Cores where the spectral index is less well-constrained because it is much smaller than the beam size}
\tablenotetext{b}{Bent Radio AGNs that are SDSS identified quasars.}
\tablenotetext{c}{Cores where the spectral index is less well-constrained because the core was not detected at the 10$\sigma$ level in LoTSS DR2}

\end{deluxetable*}


\appendix
\label{sect:app}
\section{Identifying Host Galaxies}
\begin{figure*}
\begin{center}
\includegraphics[scale=0.14,trim={0.0in 0.0in 0.0in 0.0in},clip=true]{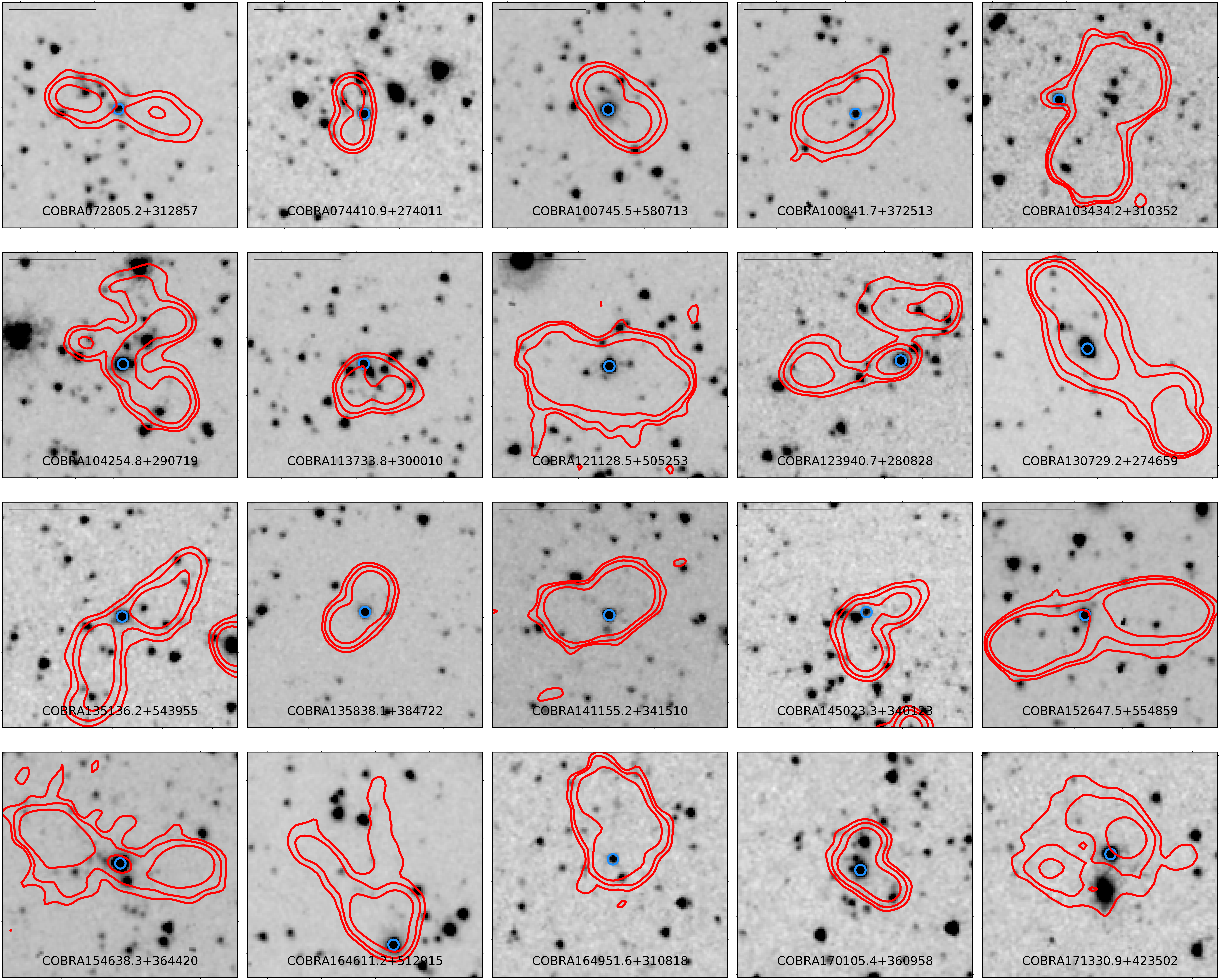}

\caption{3.6\,$\mu$m $Spitzer$ IRAC Cutouts of the 20 bent, double-lobed radio sources in the COBRA cluster candidate sample observed by LoTSS.  Each image shows the same FOV as in Figure~\ref{Fig:AllClusters} and the three red contours show the same 10$\sigma$, 20$\sigma$, and 50$\sigma$ contours based on the measurement of the LoTSS rms noise in each image.  Here, the blue circle identifies the host galaxy.  The black line shows 0$\farcm$5.}
\label{Fig:AllClustersIR}
\end{center}
\end{figure*}

\begin{figure*}
\begin{center}
\includegraphics[scale=0.14,trim={0.0in 0.0in 0.0in 0.0in},clip=true]{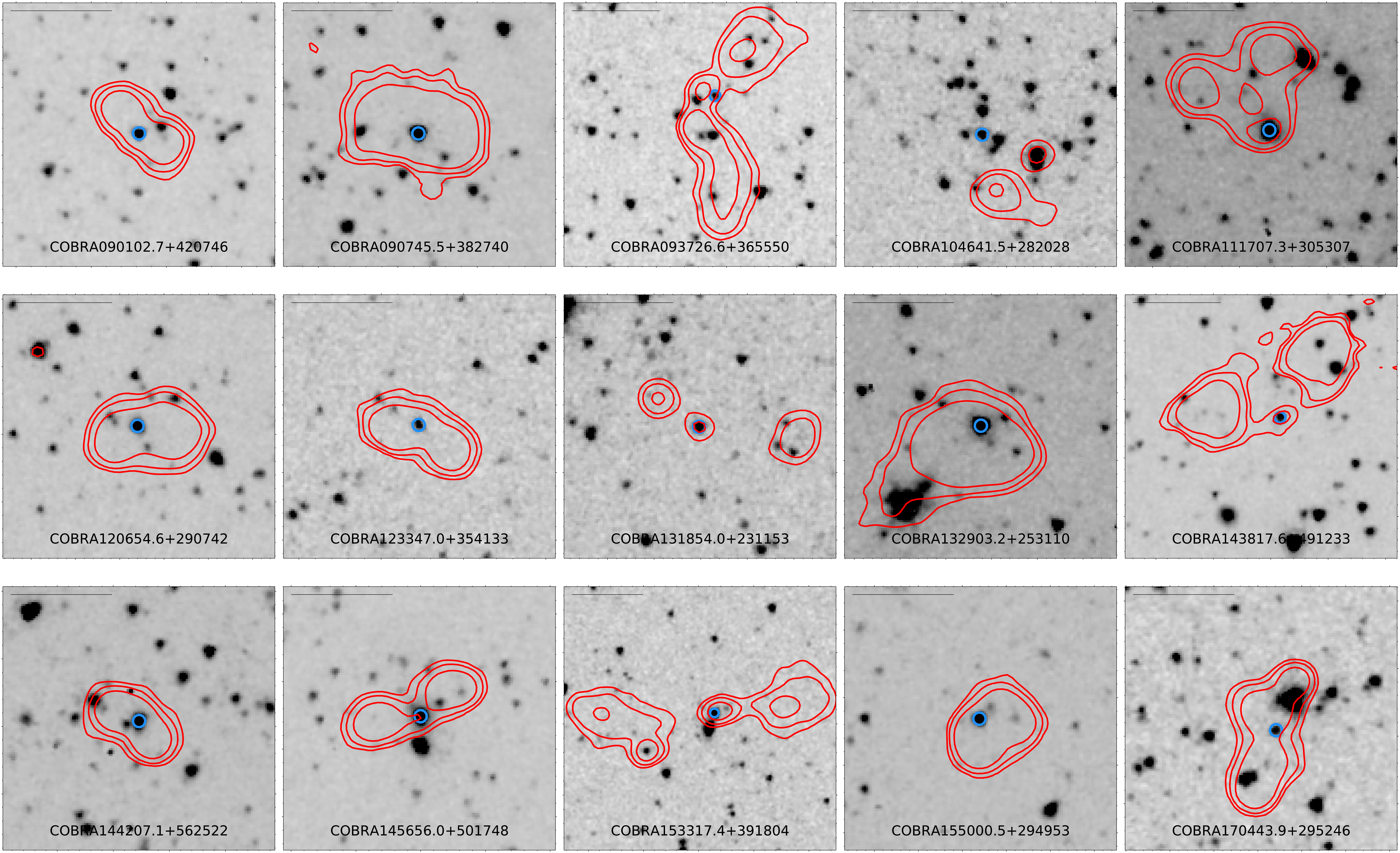}

\caption{3.6\,$\mu$m $Spitzer$ IRAC Cutouts of the 15 bent, double-lobed radio sources in the non-cluster sample observed by LoTSS.  Each image shows the same FOV as in Figure~\ref{Fig:AllNonClusters}.  The red contours show the same contours as in Figure~\ref{Fig:AllNonClusters}, while the blue circle identifies the host galaxy.}
\label{Fig:AllNonClustersIR}
\end{center}
\end{figure*}

As mentioned in Section~\ref{sect:COBRA}, \citet{Paterno-Mahler2017} used the 3.6$\mu$m $Spitzer$ IRAC images to identify host galaxies for each of the bent radio AGNs in the entire high-$z$ COBRA survey.  This was done by overlaying radio contours from FIRST on the $Spitzer$ images and cross-matching with SDSS to avoid accidentally identifying an incorrect low-$z$ host.  In Figures~\ref{Fig:AllClustersIR} and \ref{Fig:AllNonClustersIR}, we've overlaid the LoTSS contours (the same contours as in Figures~\ref{Fig:AllClusters} and \ref{Fig:AllNonClusters}) on the 3.6$\mu$m $Spitzer$ IRAC images to show the location of the host galaxy relative to the radio source. 

\section{Spectral Index Maps}
\begin{figure*}
\begin{center}
\includegraphics[scale=0.28,trim={0.8in 0.65in 1.0in 0.75in},clip=true]{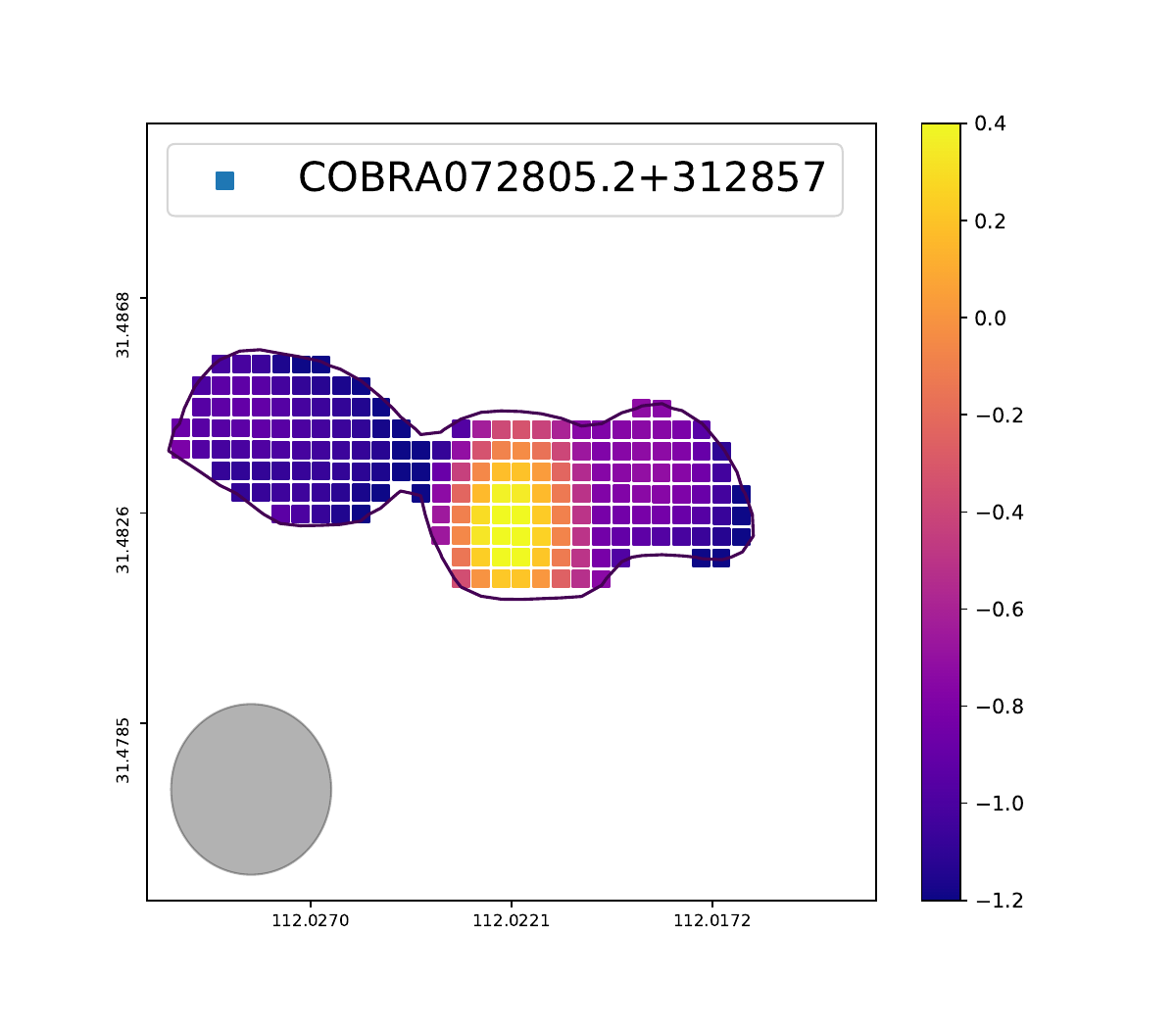}
\includegraphics[scale=0.28,trim={0.8in 0.65in 1.0in 0.75in},clip=true]{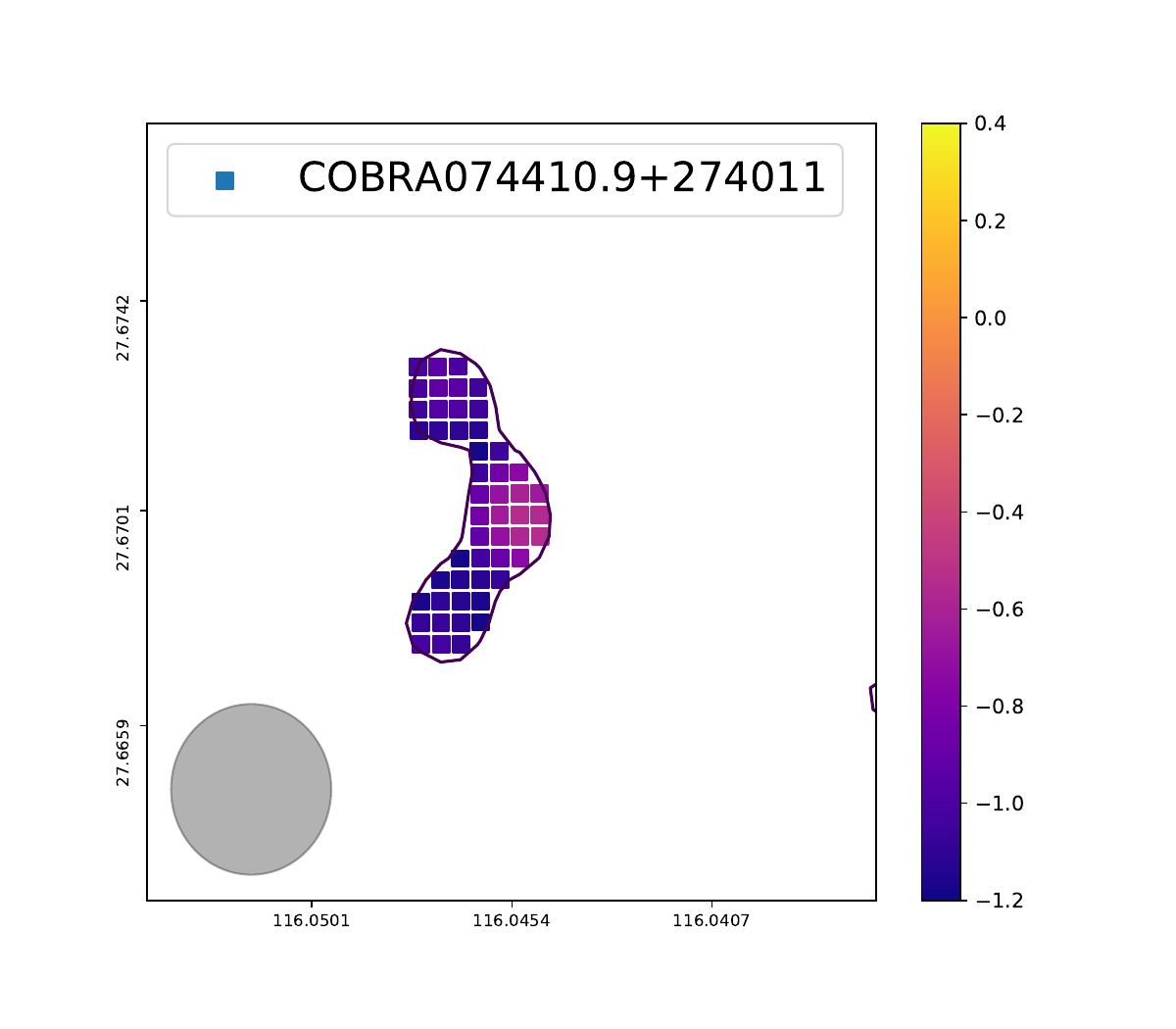}
\includegraphics[scale=0.28,trim={0.8in 0.65in 1.0in 0.75in},clip=true]{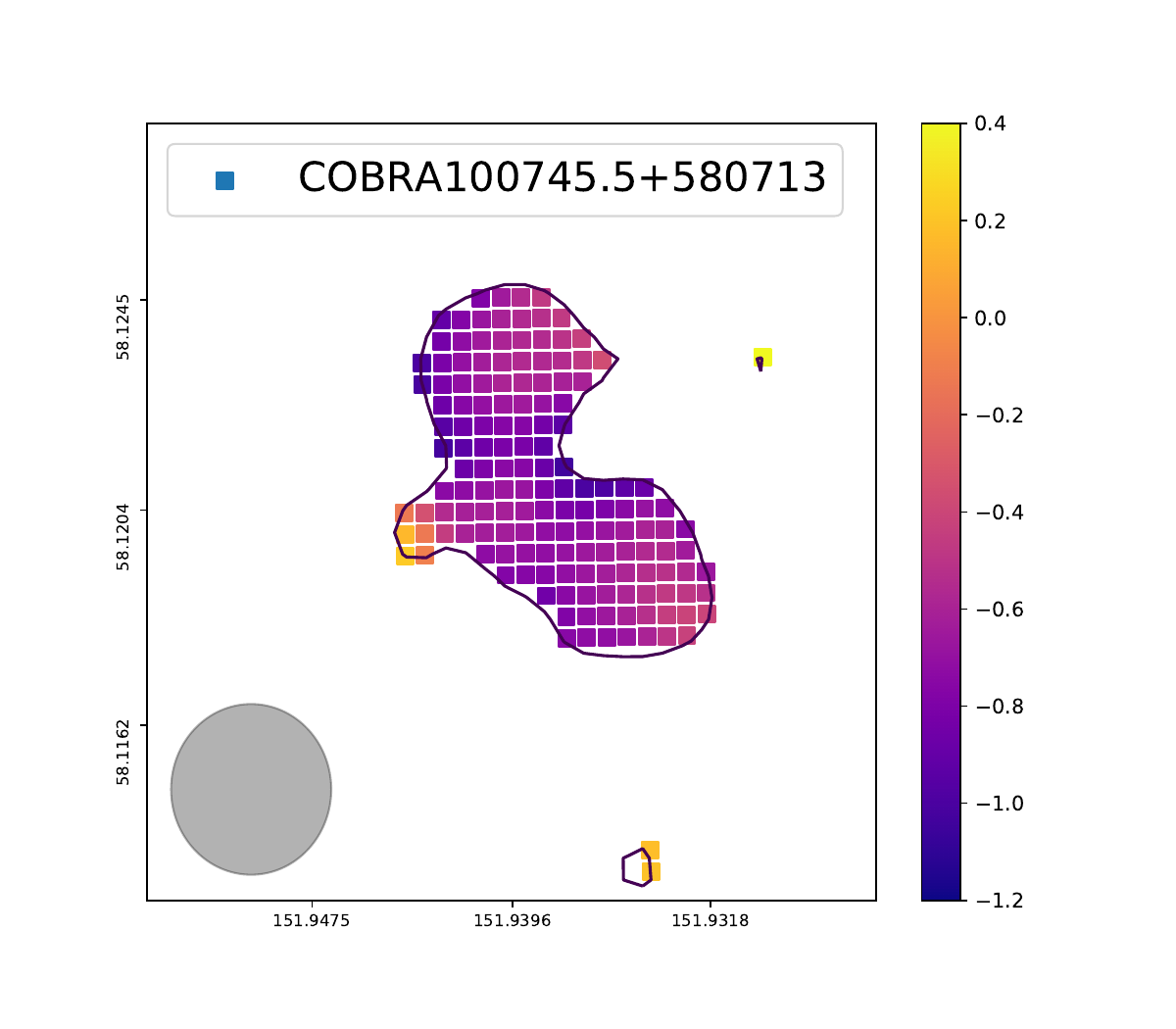}
\includegraphics[scale=0.28,trim={0.8in 0.65in 1.0in 0.75in},clip=true]{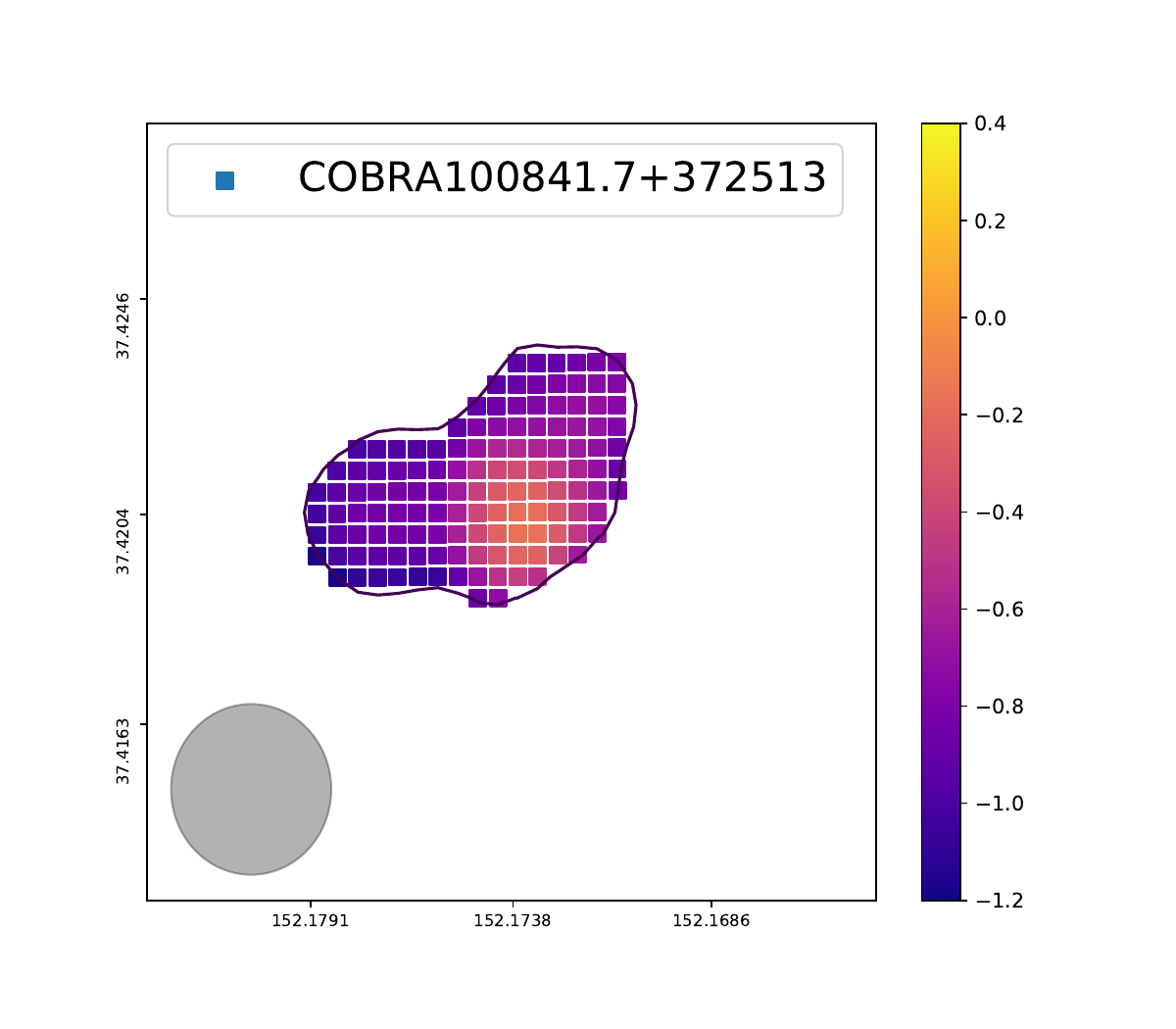}
\includegraphics[scale=0.28,trim={0.8in 0.65in 1.0in 0.75in},clip=true]{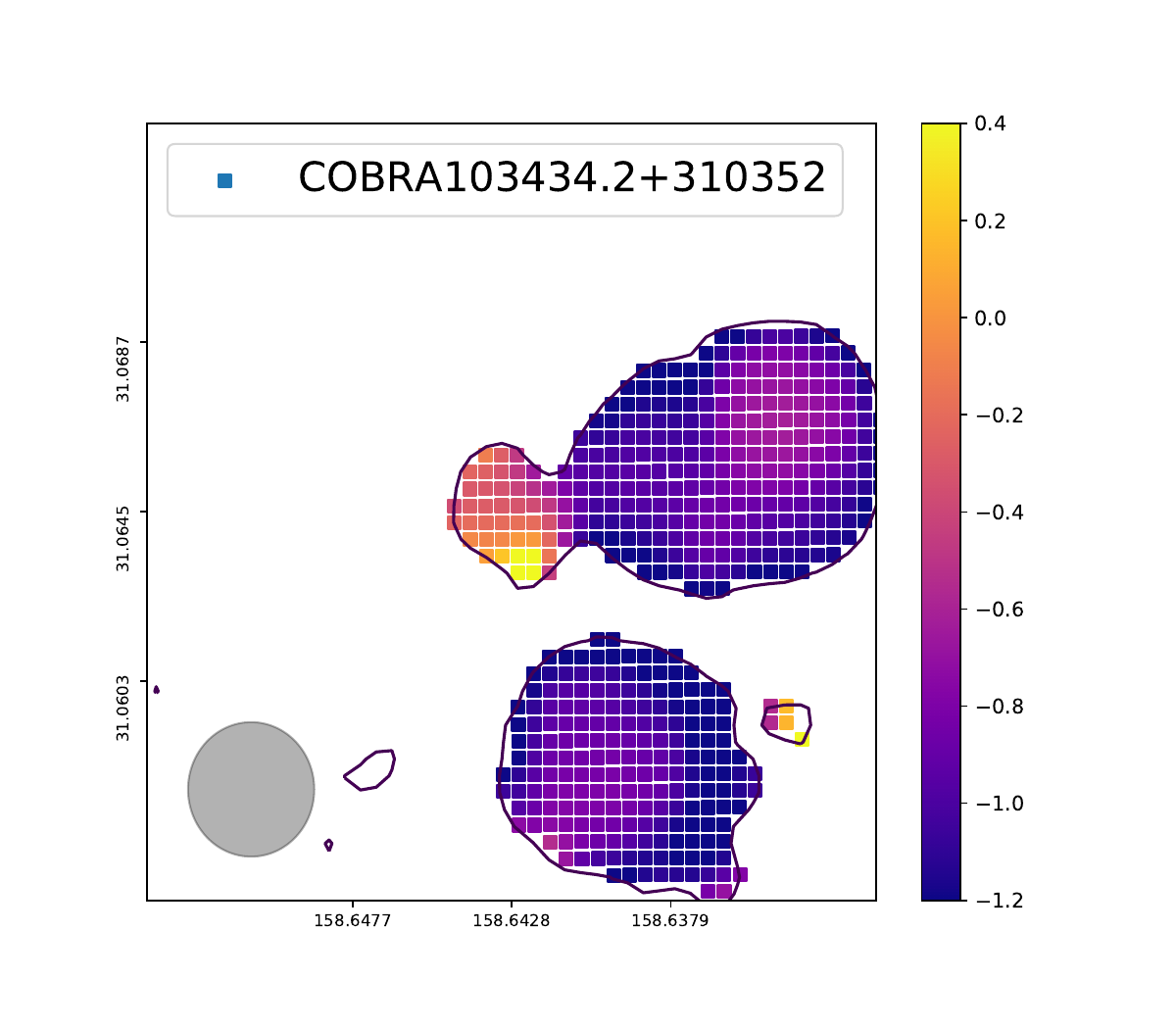}
\includegraphics[scale=0.28,trim={0.8in 0.65in 1.0in 0.75in},clip=true]{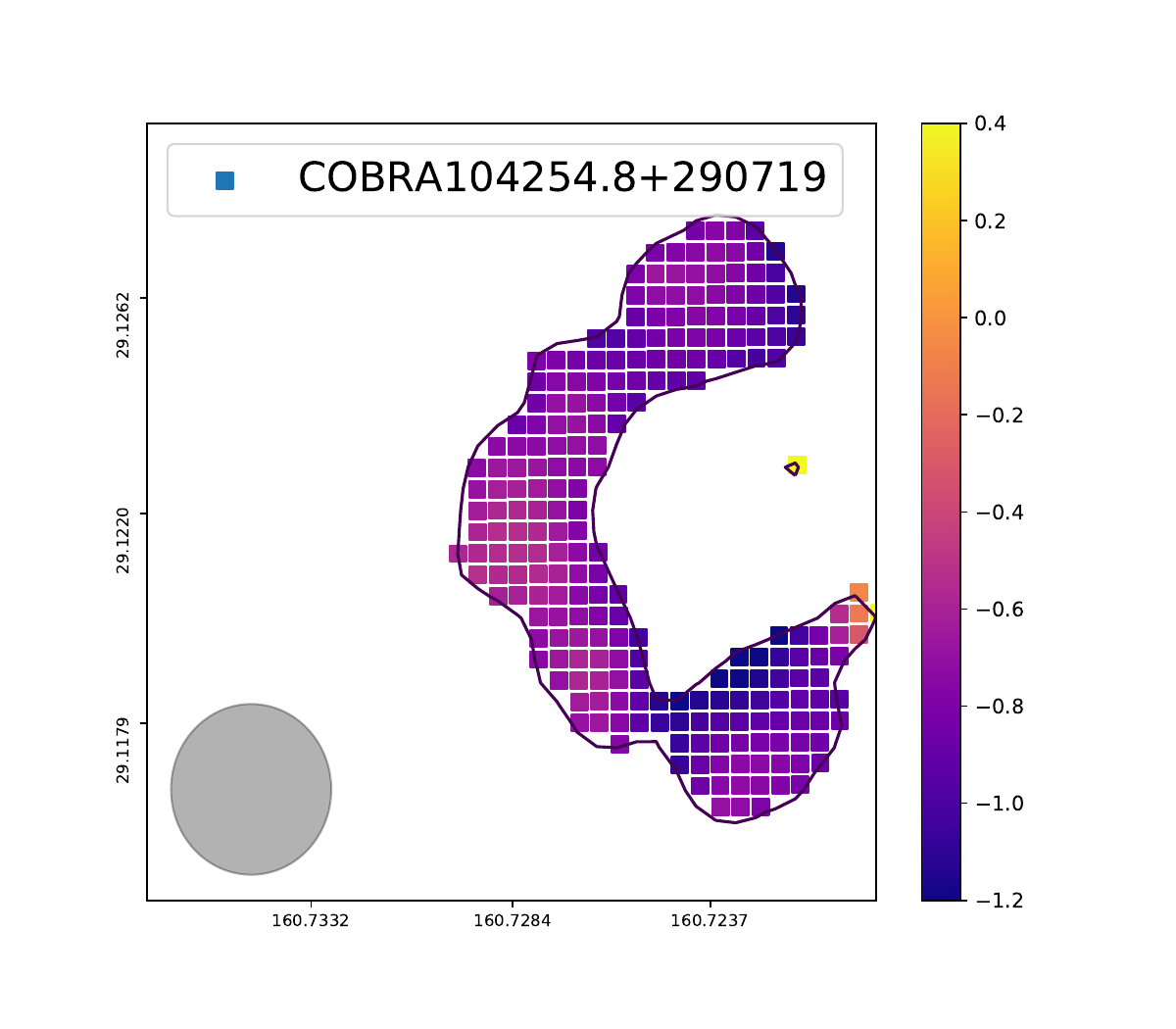}
\includegraphics[scale=0.28,trim={0.8in 0.65in 1.0in 0.75in},clip=true]{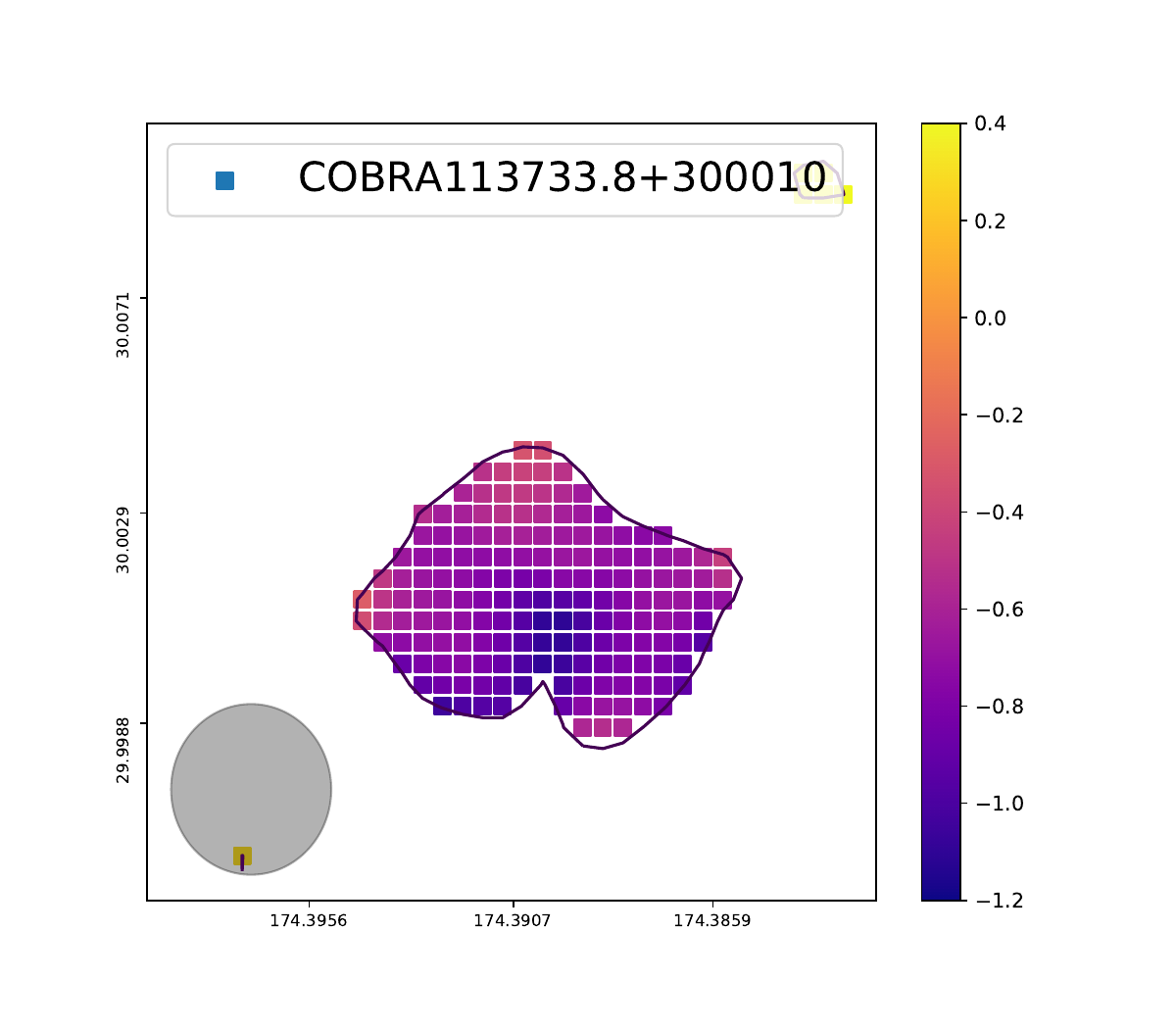}
\includegraphics[scale=0.28,trim={0.8in 0.65in 1.0in 0.75in},clip=true]{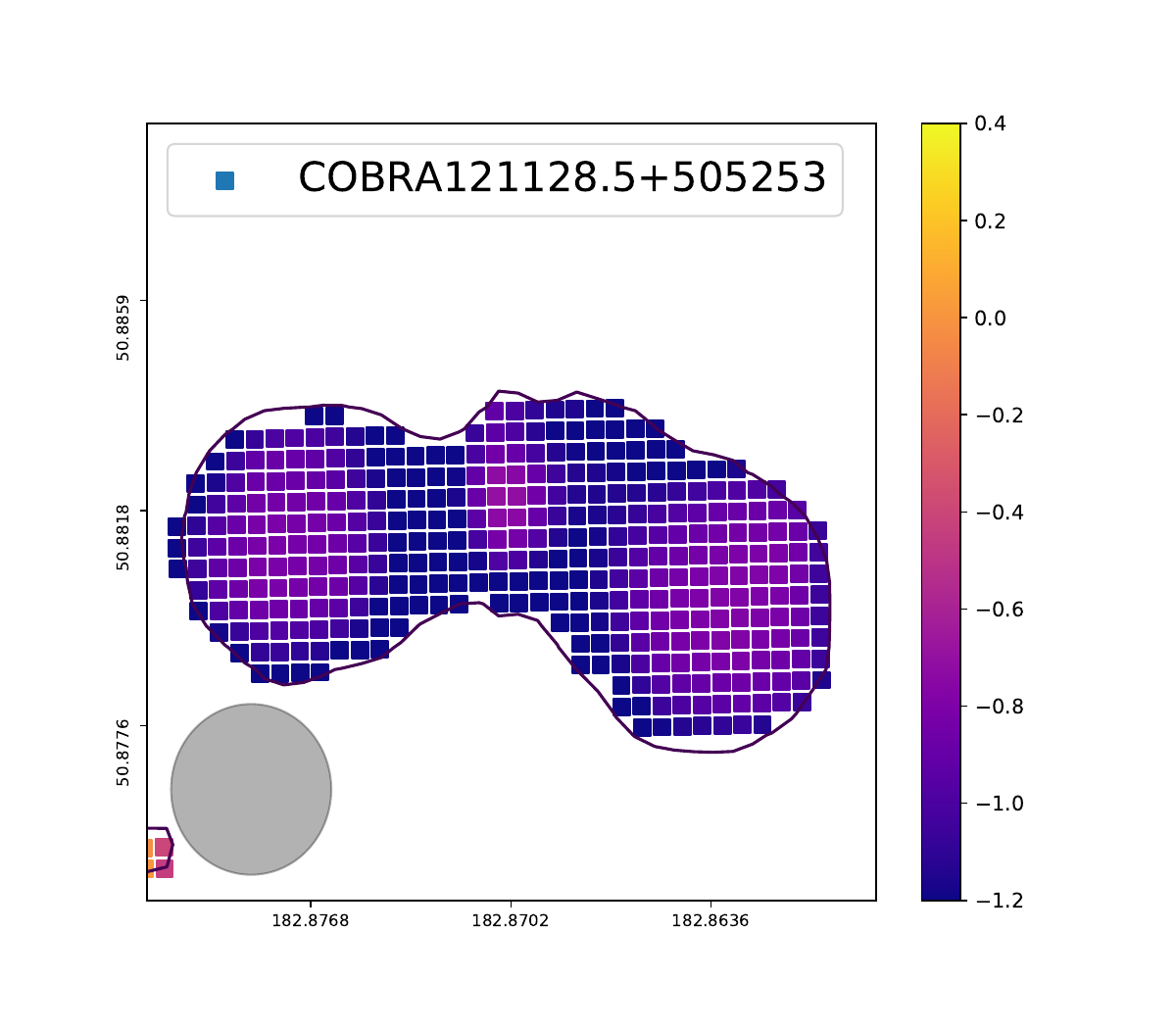}
\includegraphics[scale=0.28,trim={0.8in 0.65in 1.0in 0.75in},clip=true]{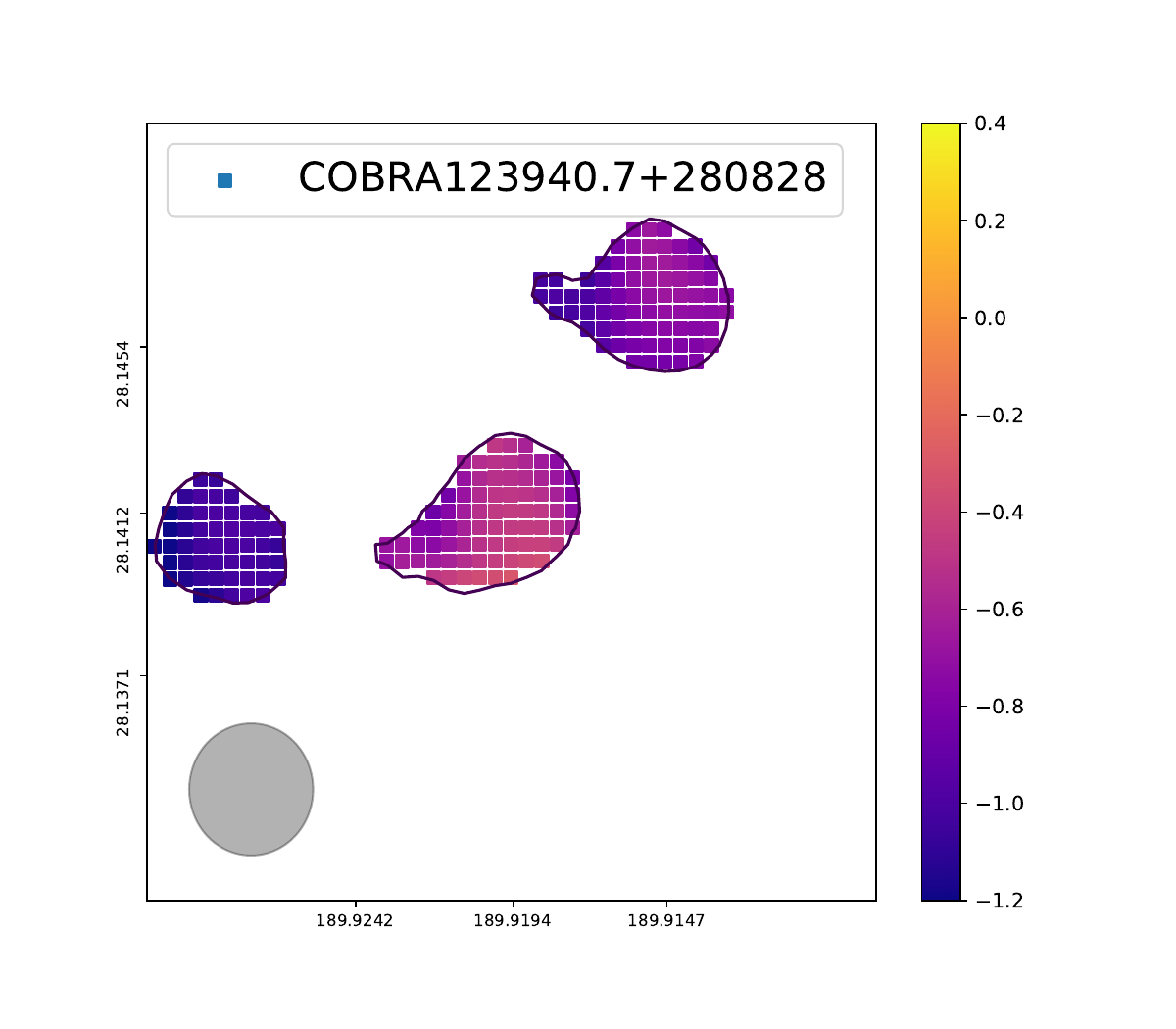}
\includegraphics[scale=0.28,trim={0.8in 0.65in 1.0in 0.75in},clip=true]{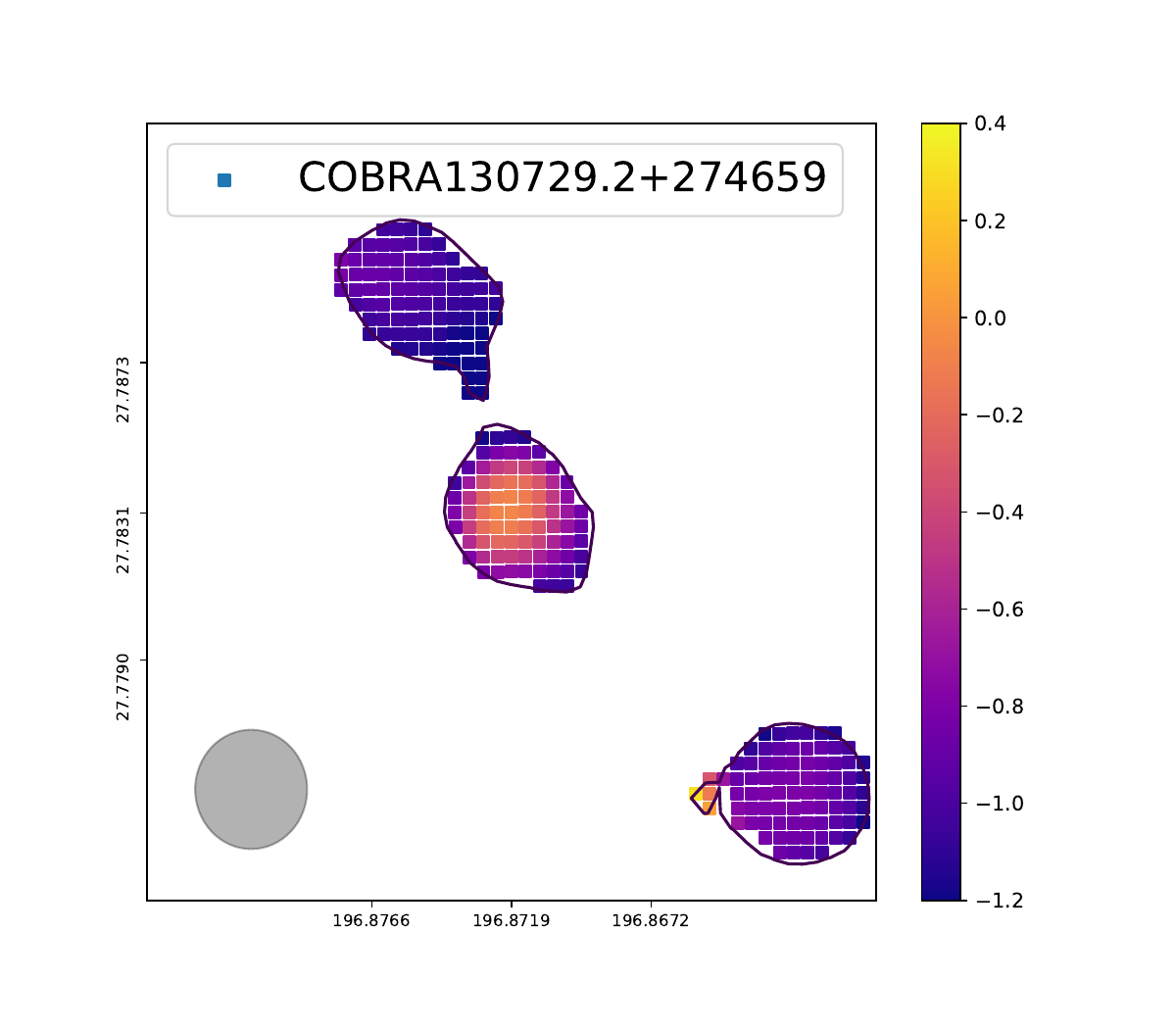}
\includegraphics[scale=0.28,trim={0.8in 0.65in 1.0in 0.75in},clip=true]{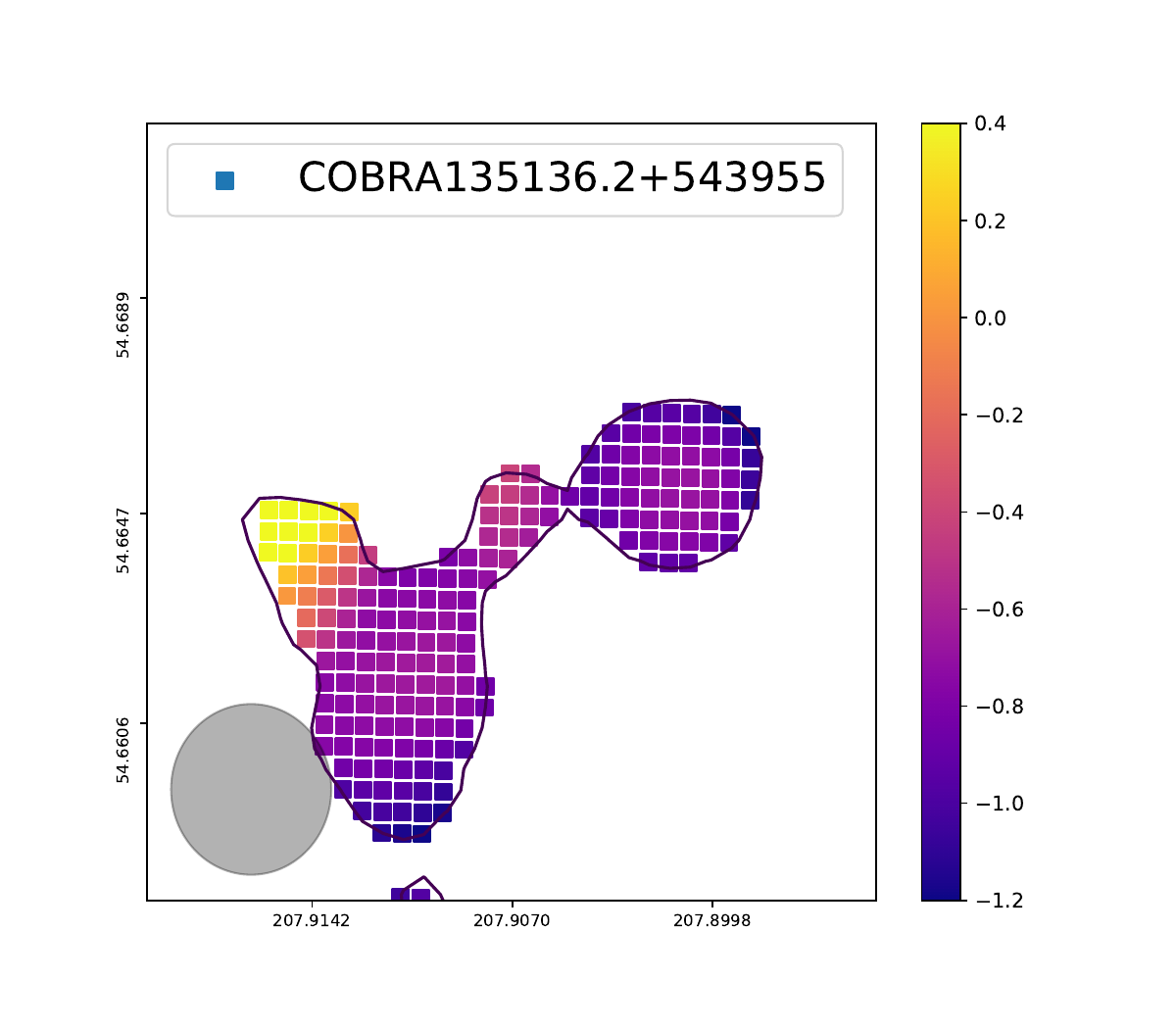}
\includegraphics[scale=0.28,trim={0.8in 0.65in 1.0in 0.75in},clip=true]{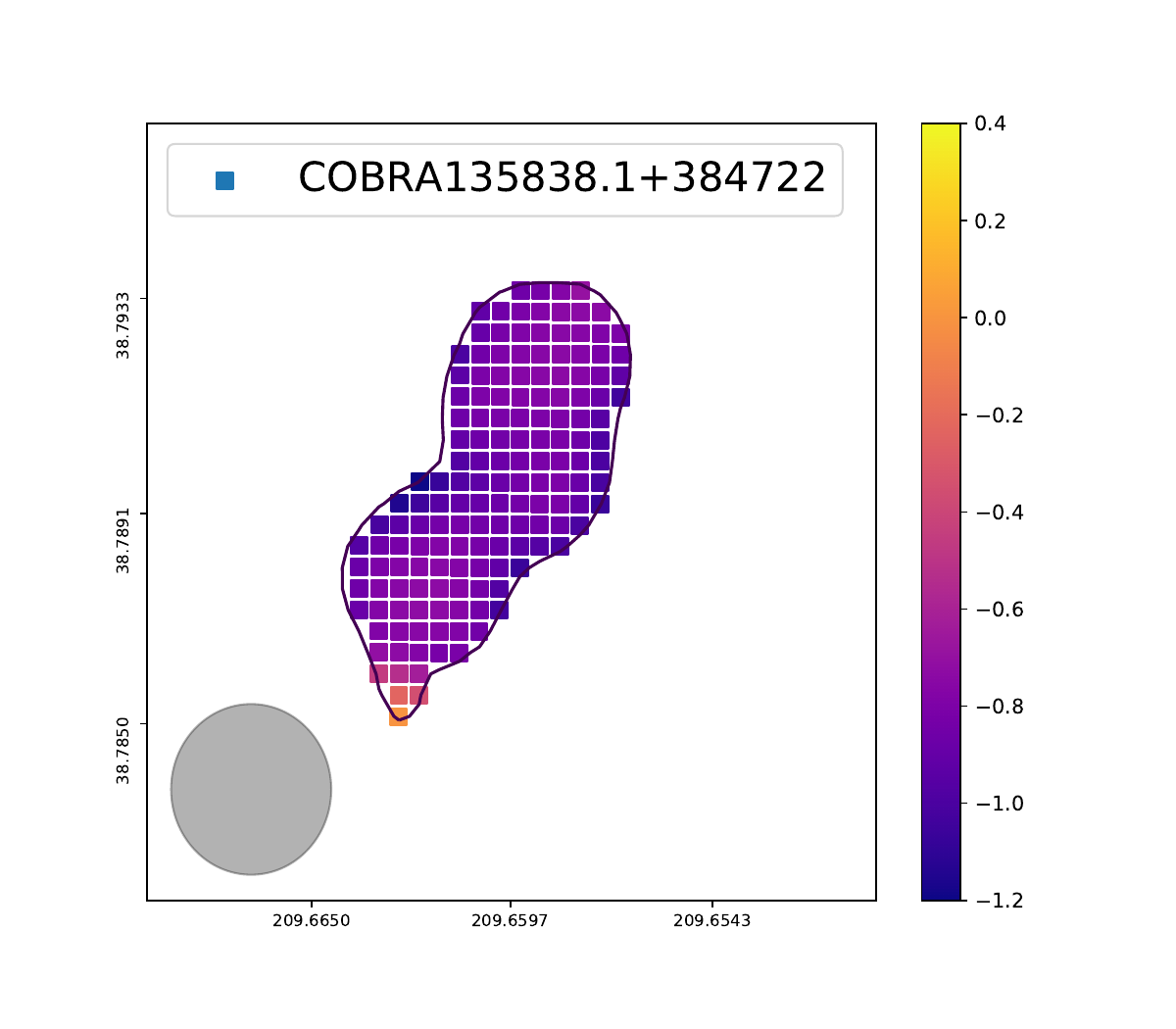}
\includegraphics[scale=0.28,trim={0.8in 0.65in 1.0in 0.75in},clip=true]{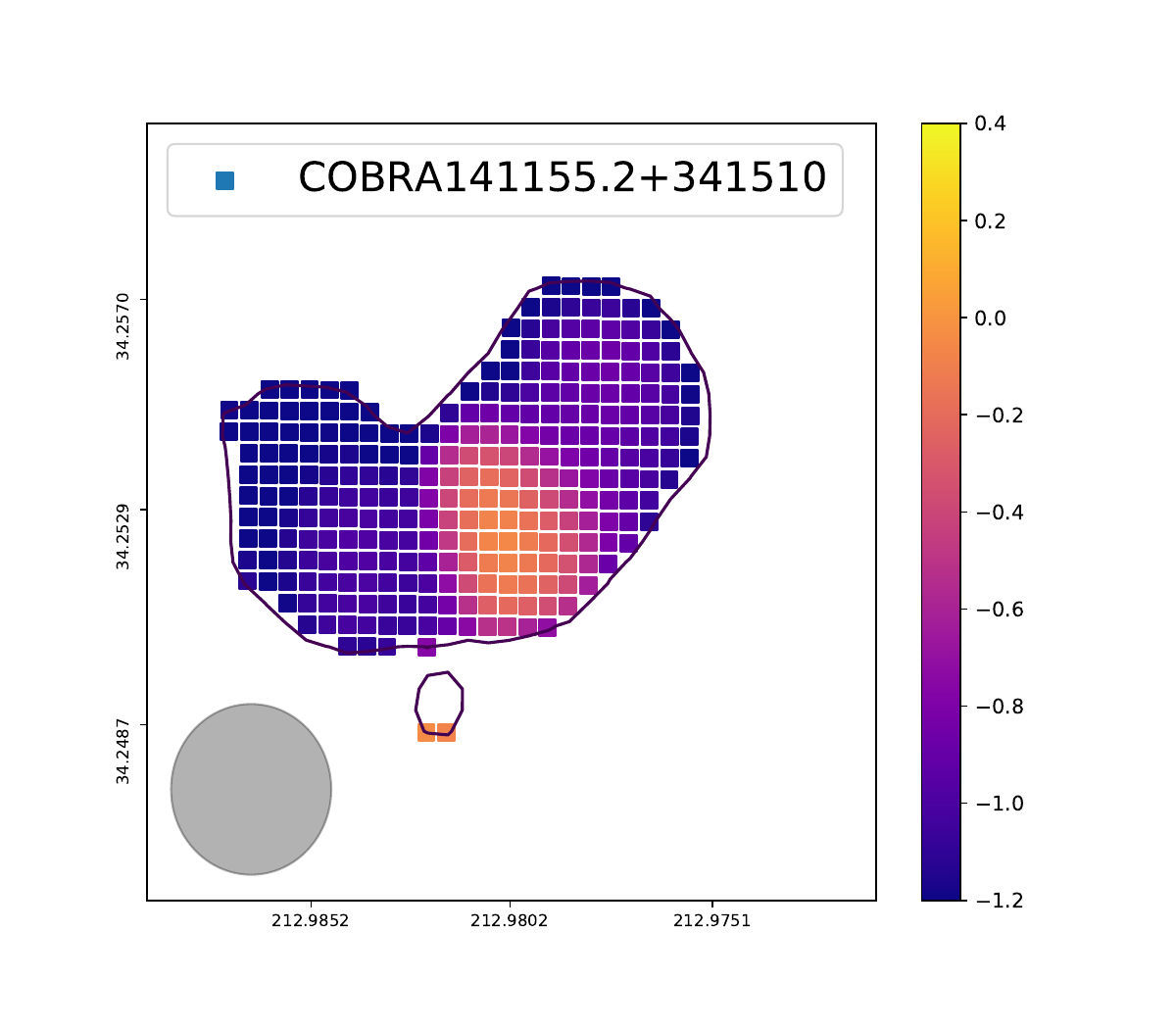}
\includegraphics[scale=0.28,trim={0.8in 0.65in 1.0in 0.75in},clip=true]{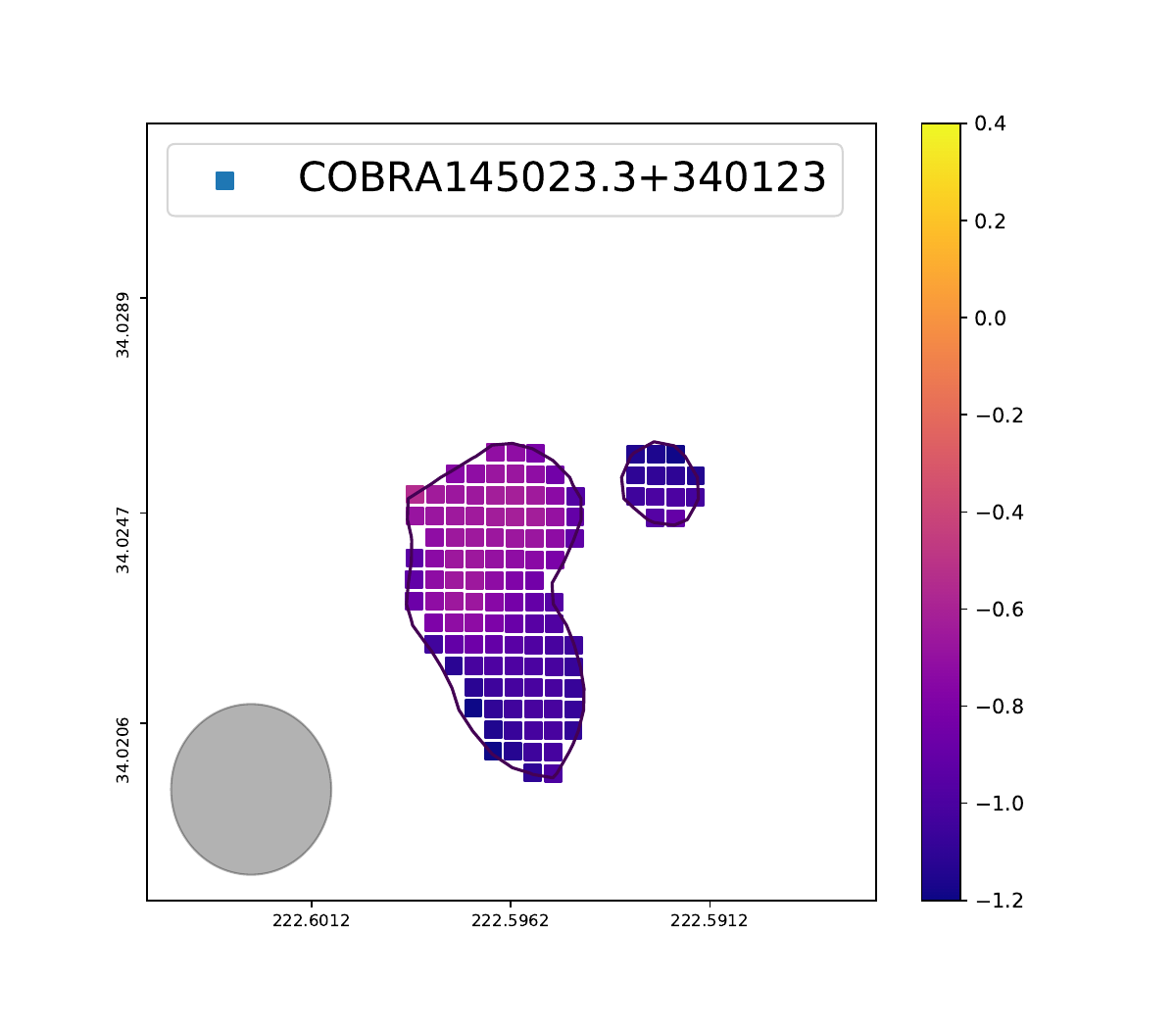}
\includegraphics[scale=0.28,trim={0.8in 0.65in 1.0in 0.75in},clip=true]{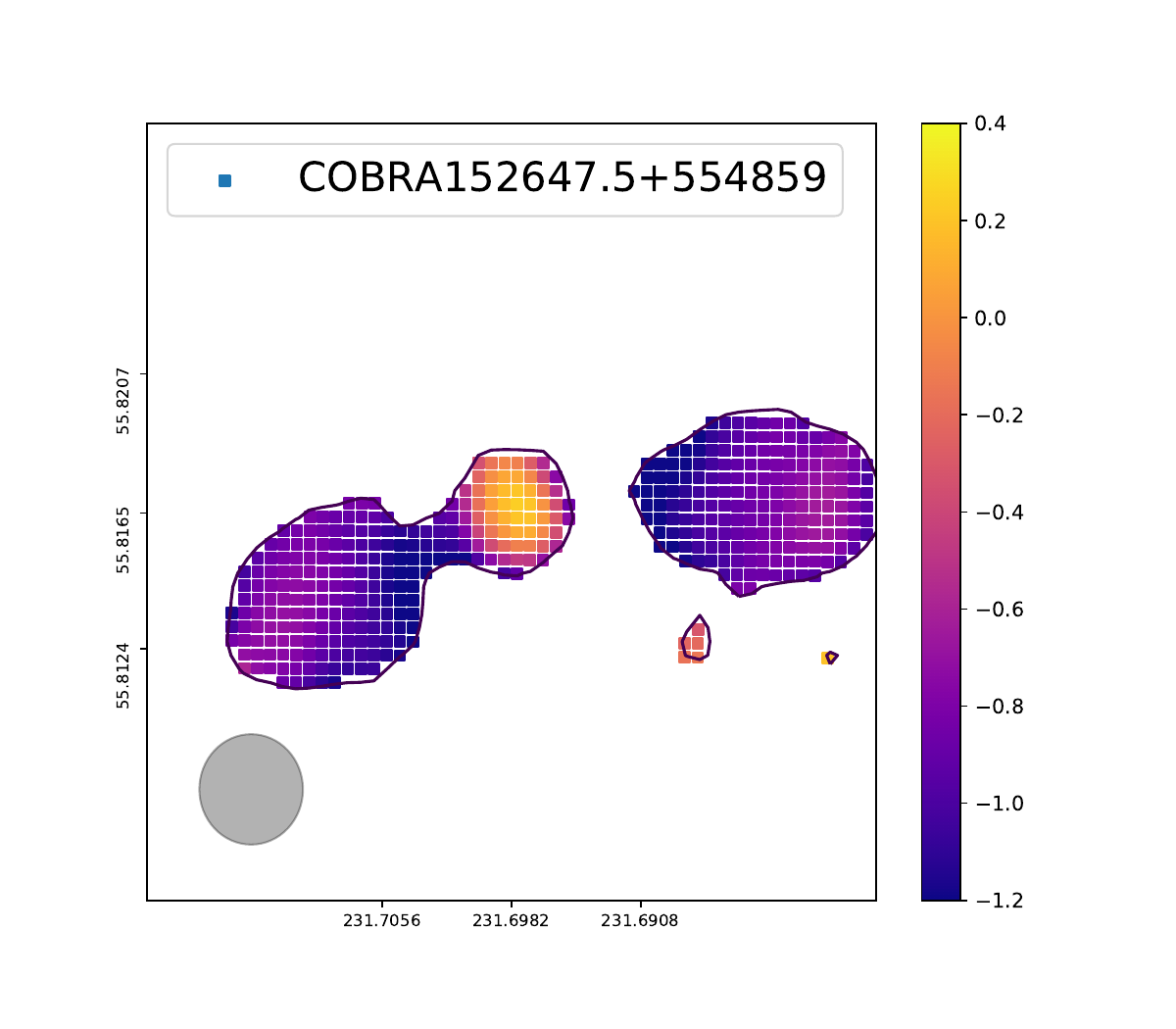}
\includegraphics[scale=0.28,trim={0.8in 0.65in 1.0in 0.75in},clip=true]{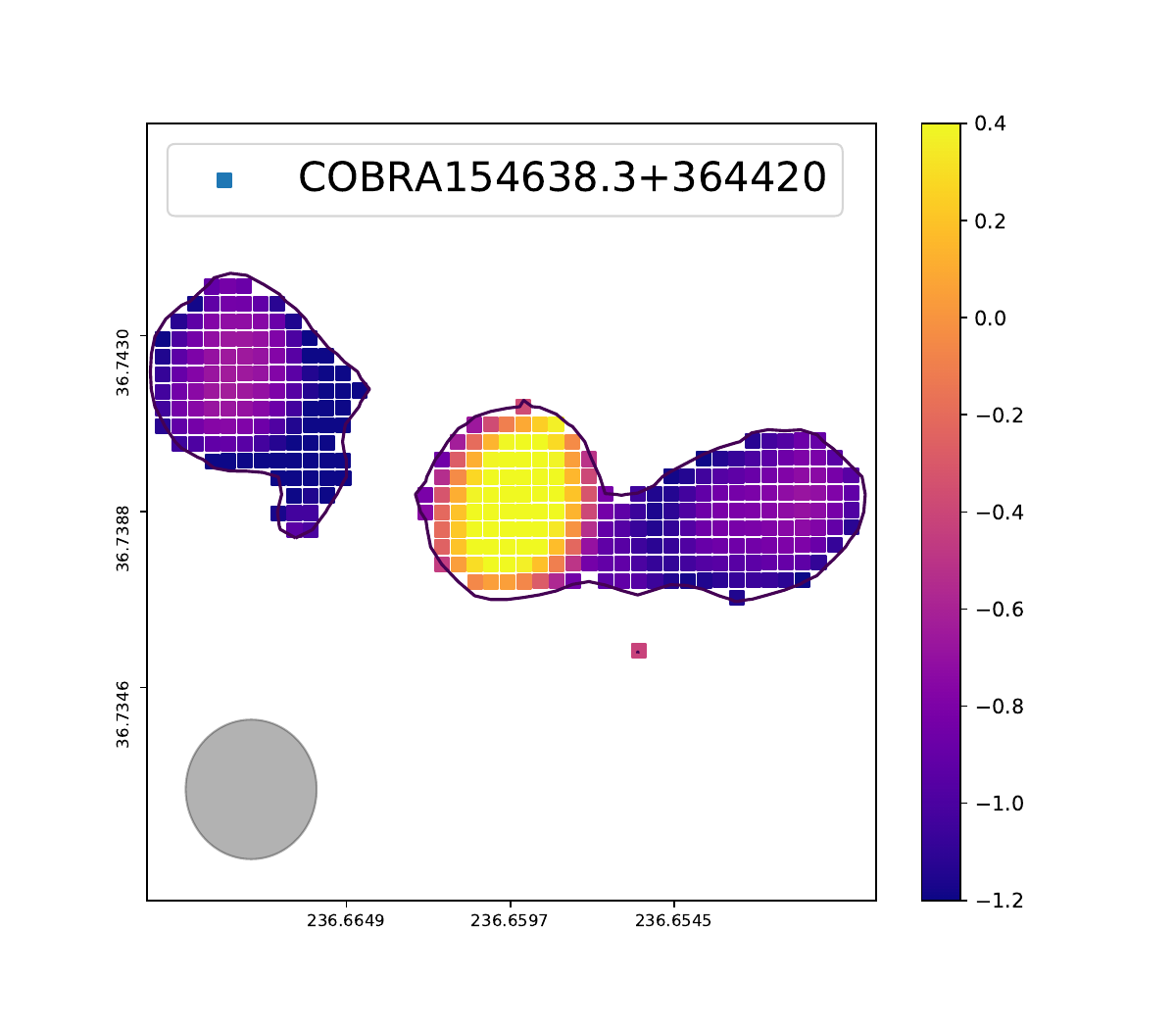}
\includegraphics[scale=0.28,trim={0.8in 0.65in 1.0in 0.75in},clip=true]{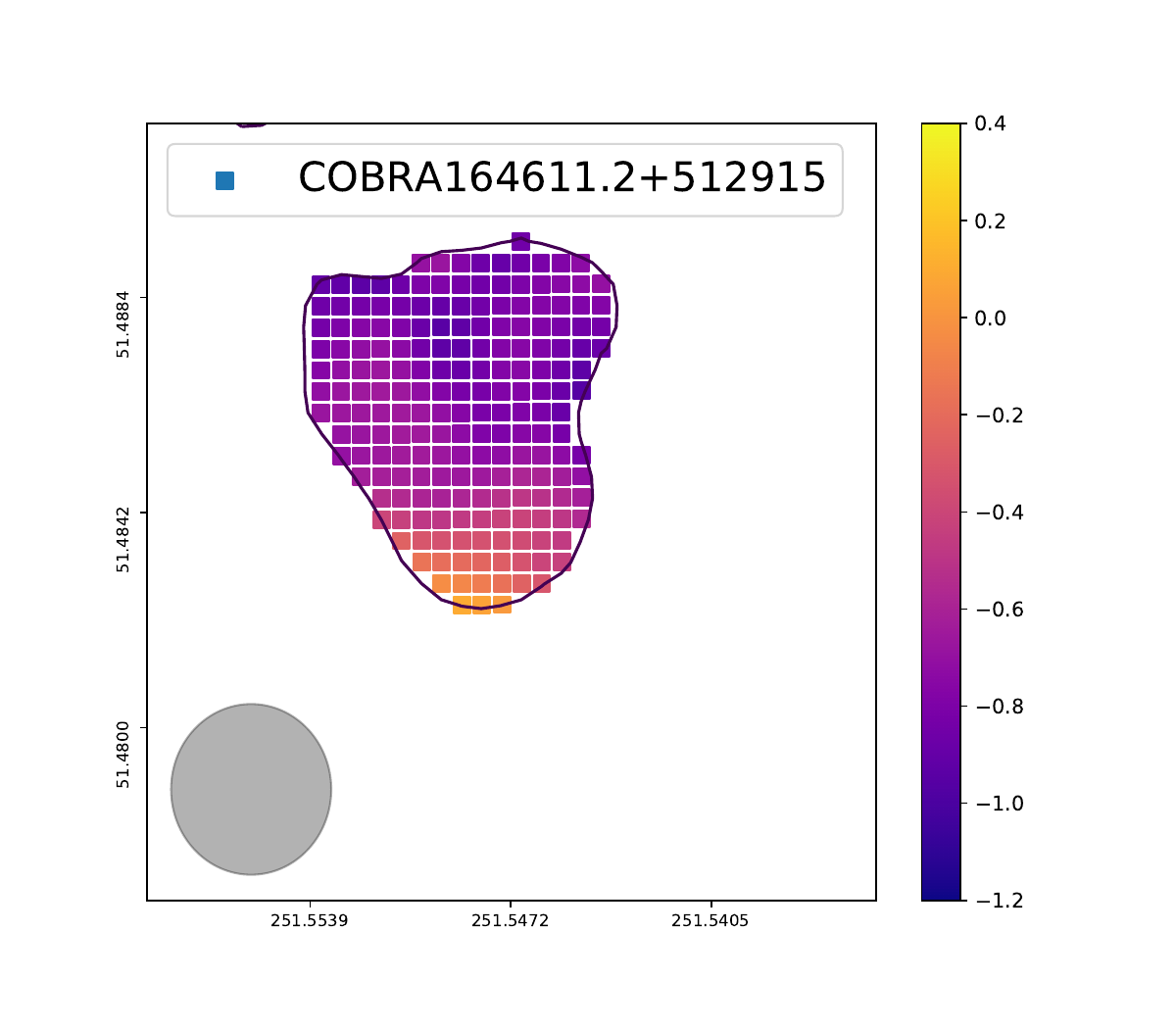}
\includegraphics[scale=0.28,trim={0.8in 0.65in 1.0in 0.75in},clip=true]{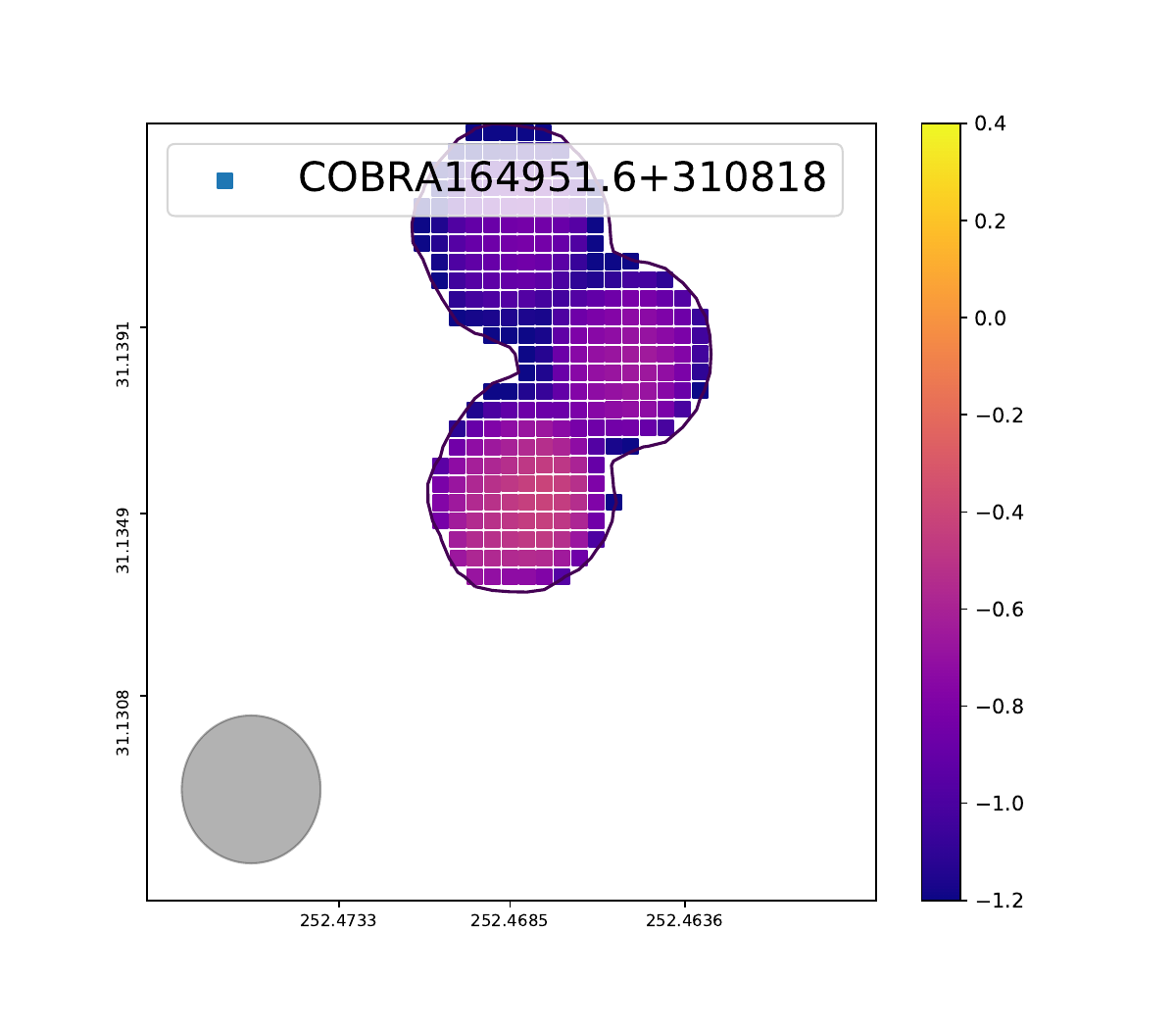}
\includegraphics[scale=0.28,trim={0.8in 0.65in 1.0in 0.75in},clip=true]{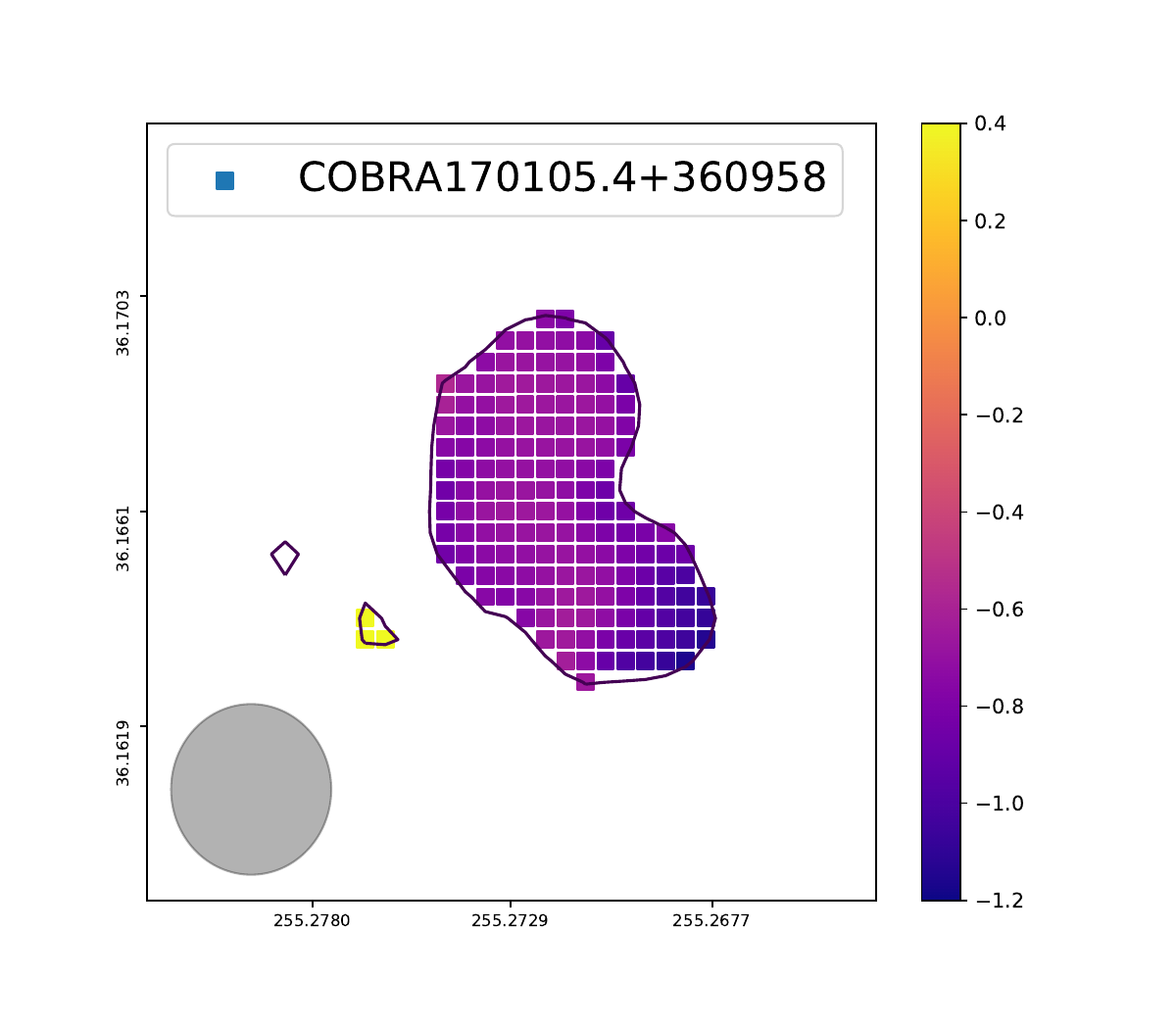}
\includegraphics[scale=0.28,trim={0.8in 0.65in 1.0in 0.75in},clip=true]{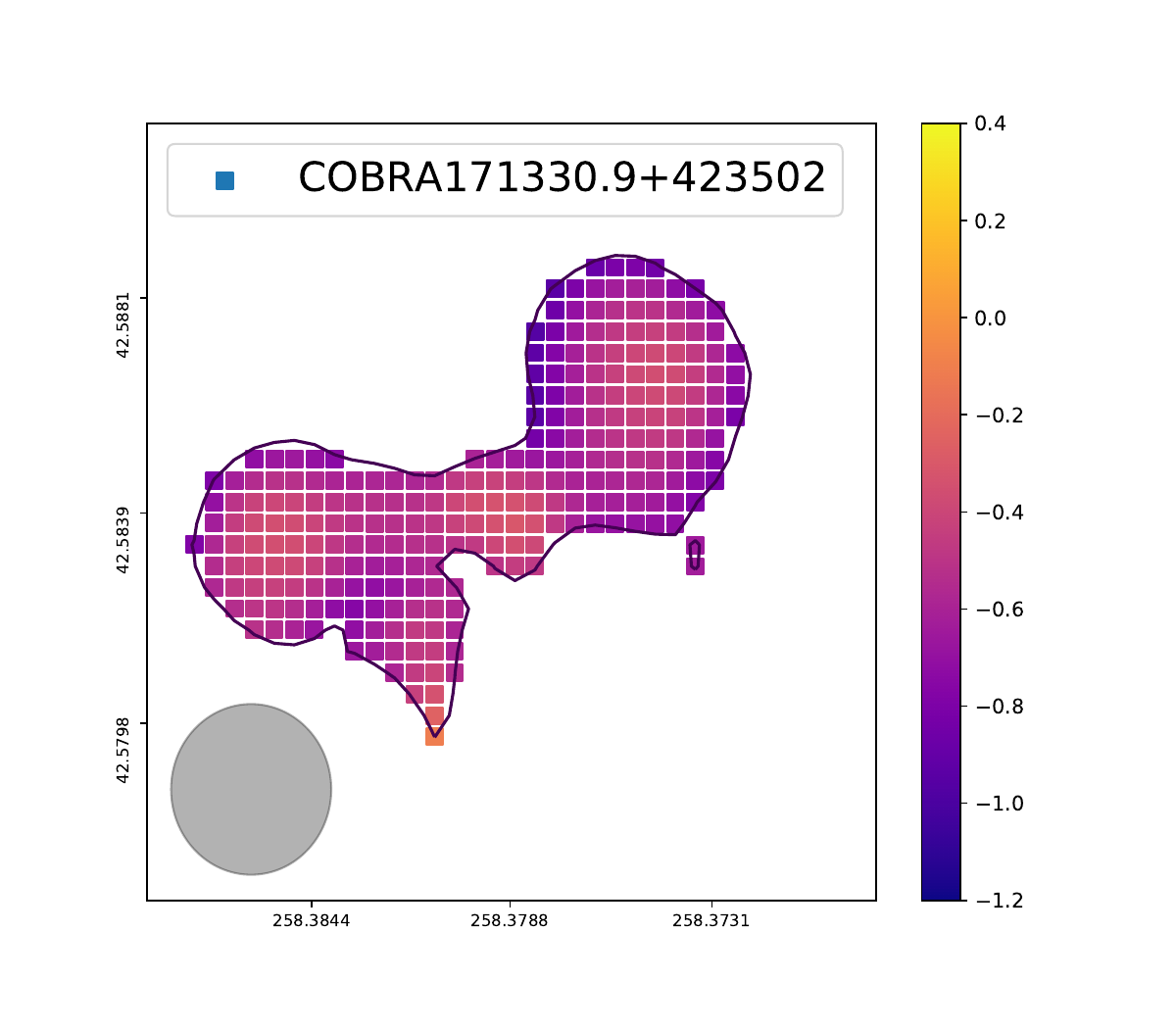}

\caption{Spectral index maps of bent radio AGNs in the cluster sample.  Each pixel corresponds to a 1$\farcs$5 $\times$ 1$\farcs$5 region of the sky.  The 6$\arcsec$ LoTSS beam is shown as a grey circle in the bottom right corner of each image. The 3$\sigma$ FIRST contours are shown.  Although most images are shown at the same size ($\approx$ 0$\farcm$45 $\times$ 0$\farcm$45 centered on the radio core), a few of the images are scaled differently to encompass the entire radio AGN.  We use a uniform color bar to show the value of the spectral index.  We see examples of both core-dominated and lobe-dominated sources within our sample.  Despite showing values of the spectral index for each individual pixel (the pixels are highly correlated because the pixel size is smaller than the beam size), we limit our analysis to larger structures and only identify the cores and lobes of our radio AGNs (see Table~\ref{tb:RadioProp-2}).}
\label{Fig:SI-clusters}
\end{center}
\end{figure*}

\begin{figure*}
\begin{center}
\includegraphics[scale=0.28,trim={0.8in 0.65in 1.0in 0.75in},clip=true]{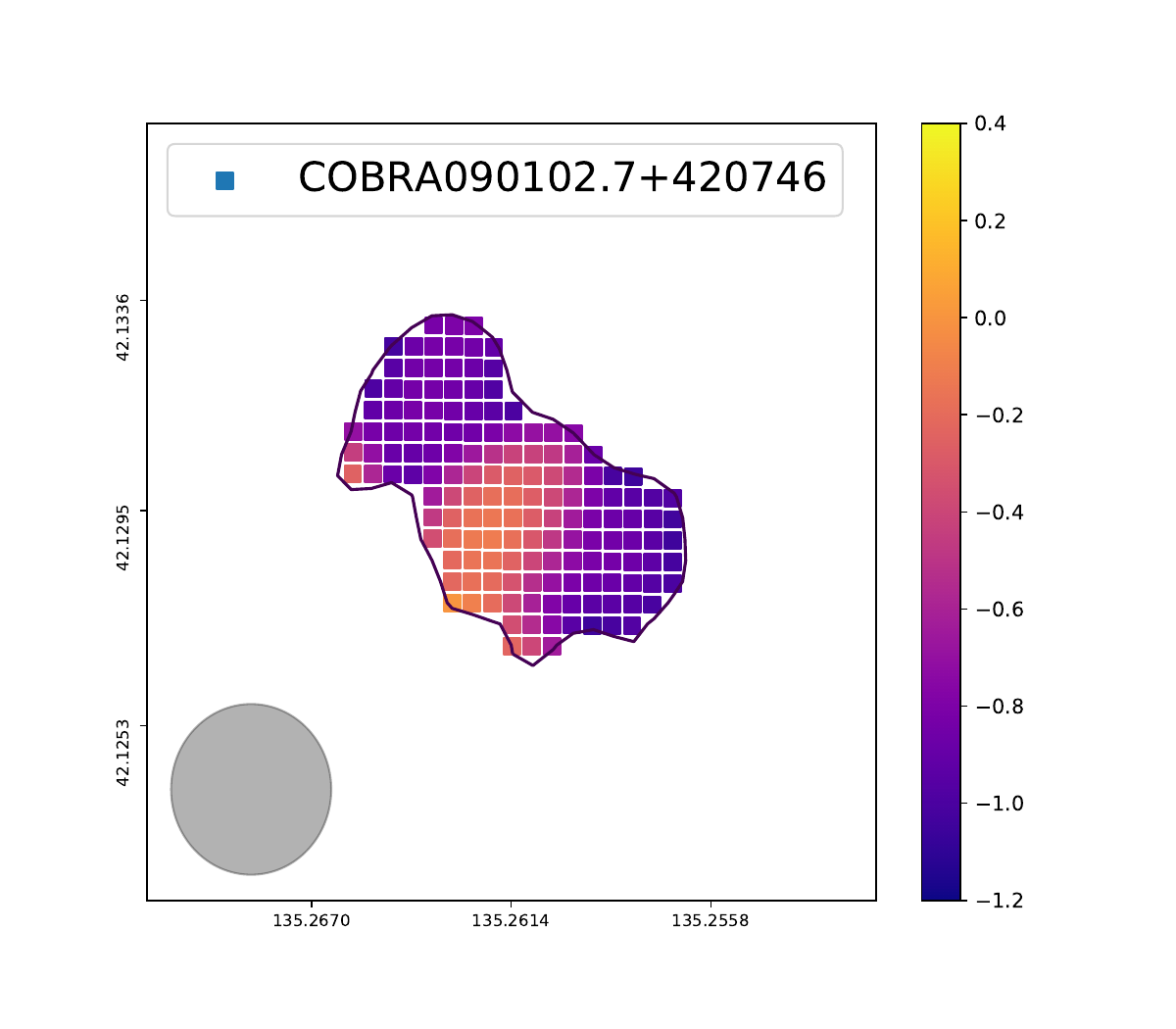}
\includegraphics[scale=0.28,trim={0.8in 0.65in 1.0in 0.75in},clip=true]{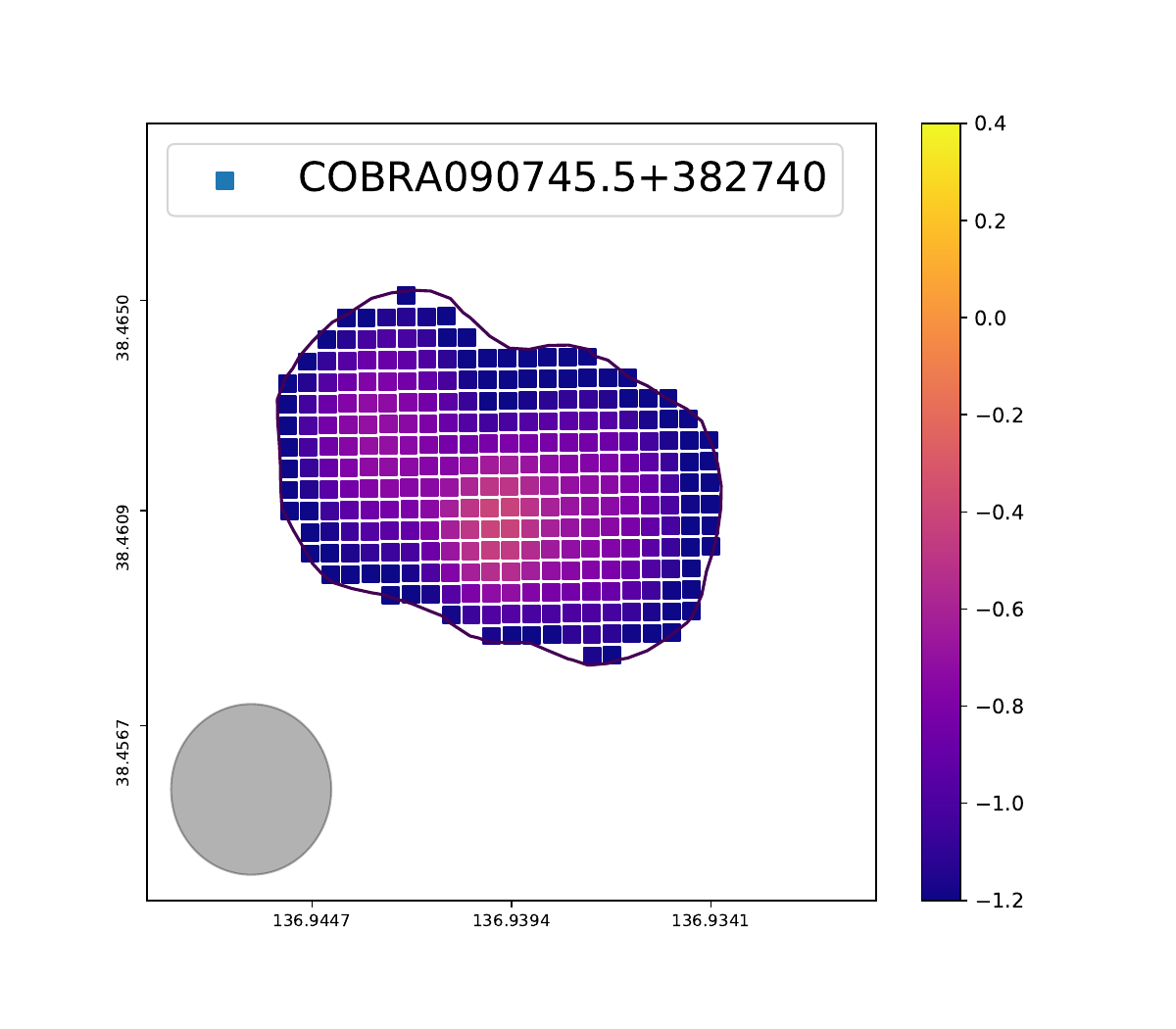}
\includegraphics[scale=0.28,trim={0.8in 0.65in 1.0in 0.75in},clip=true]{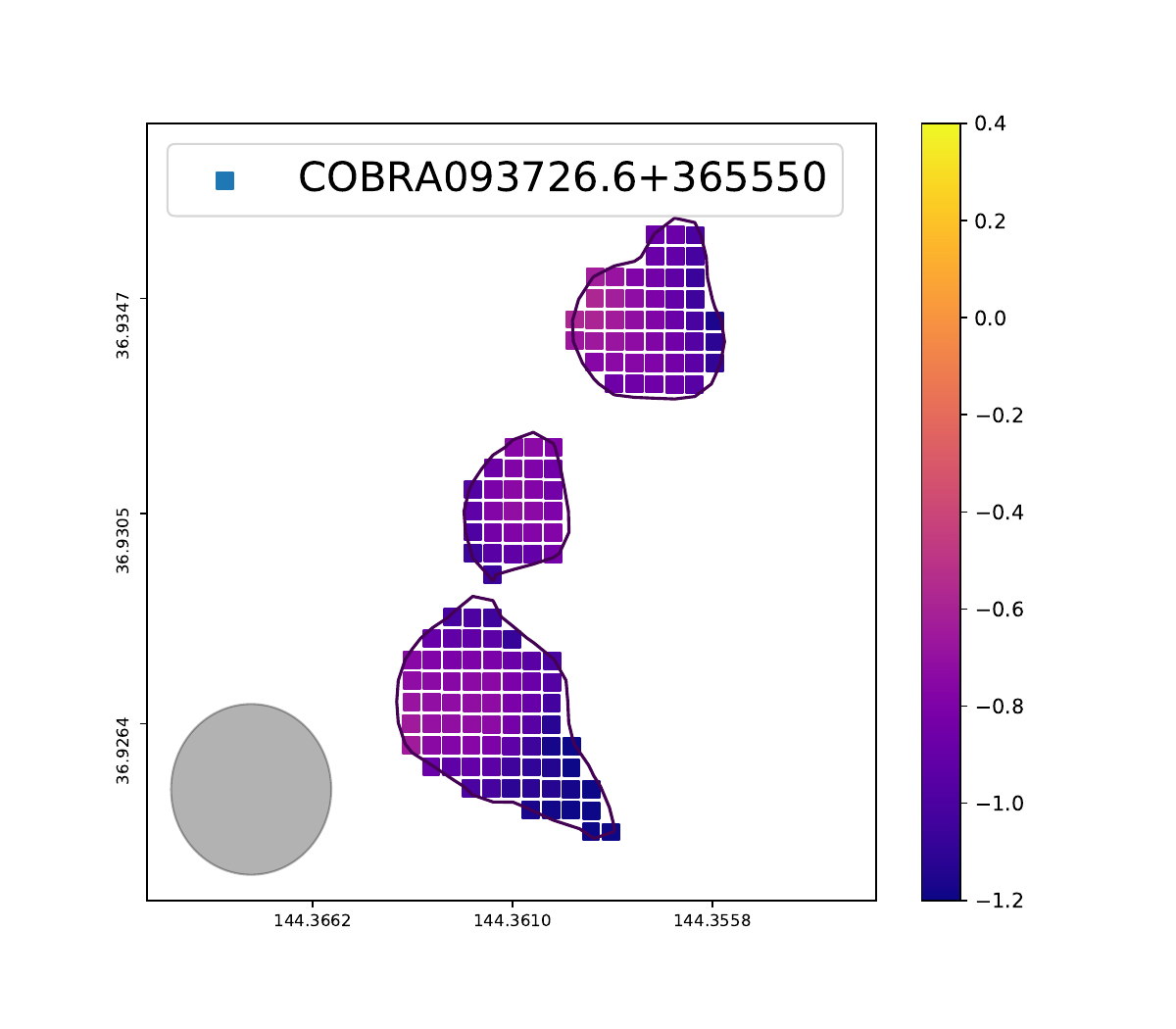}
\includegraphics[scale=0.28,trim={0.8in 0.65in 1.0in 0.75in},clip=true]{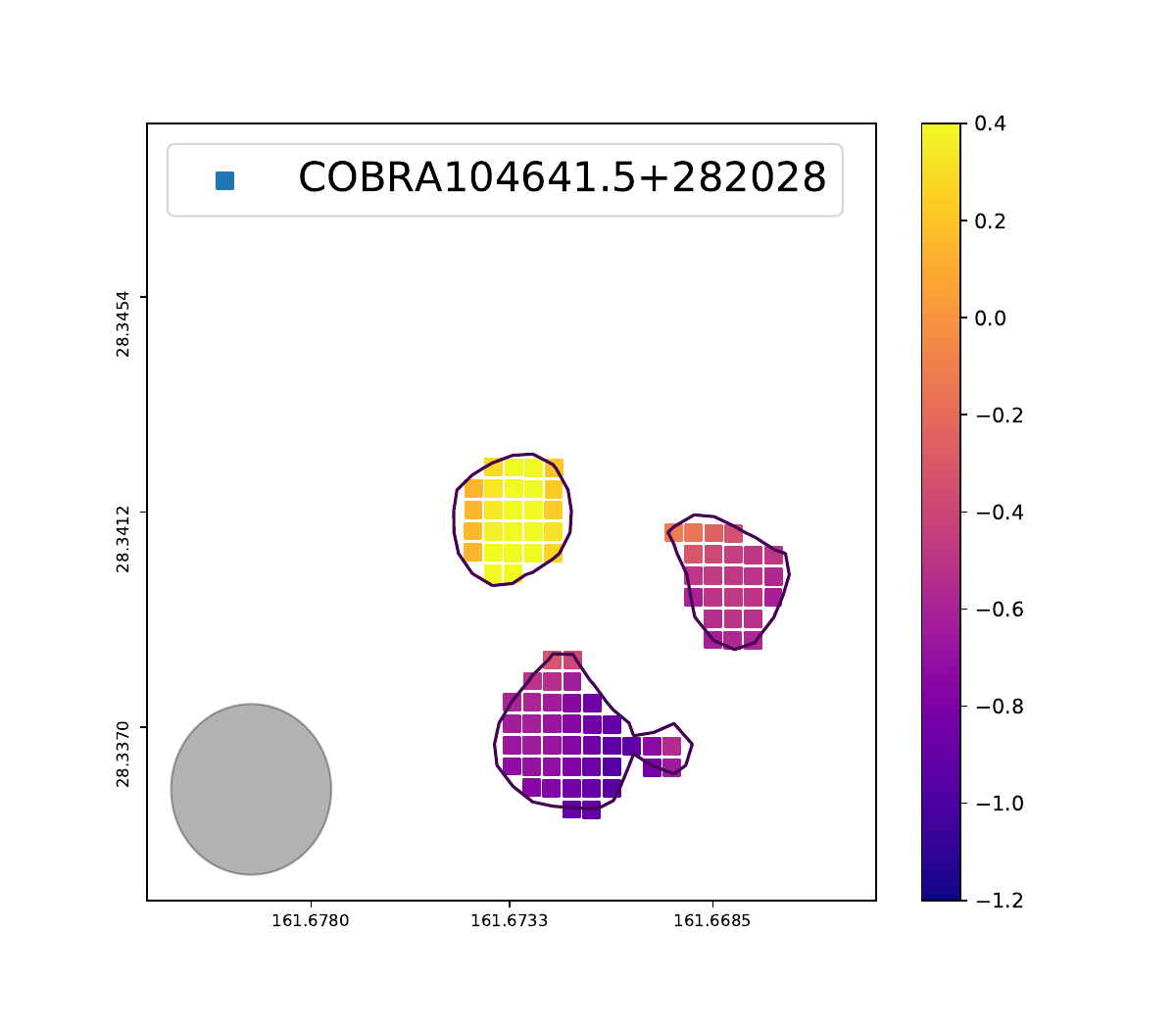}
\includegraphics[scale=0.28,trim={0.8in 0.65in 1.0in 0.75in},clip=true]{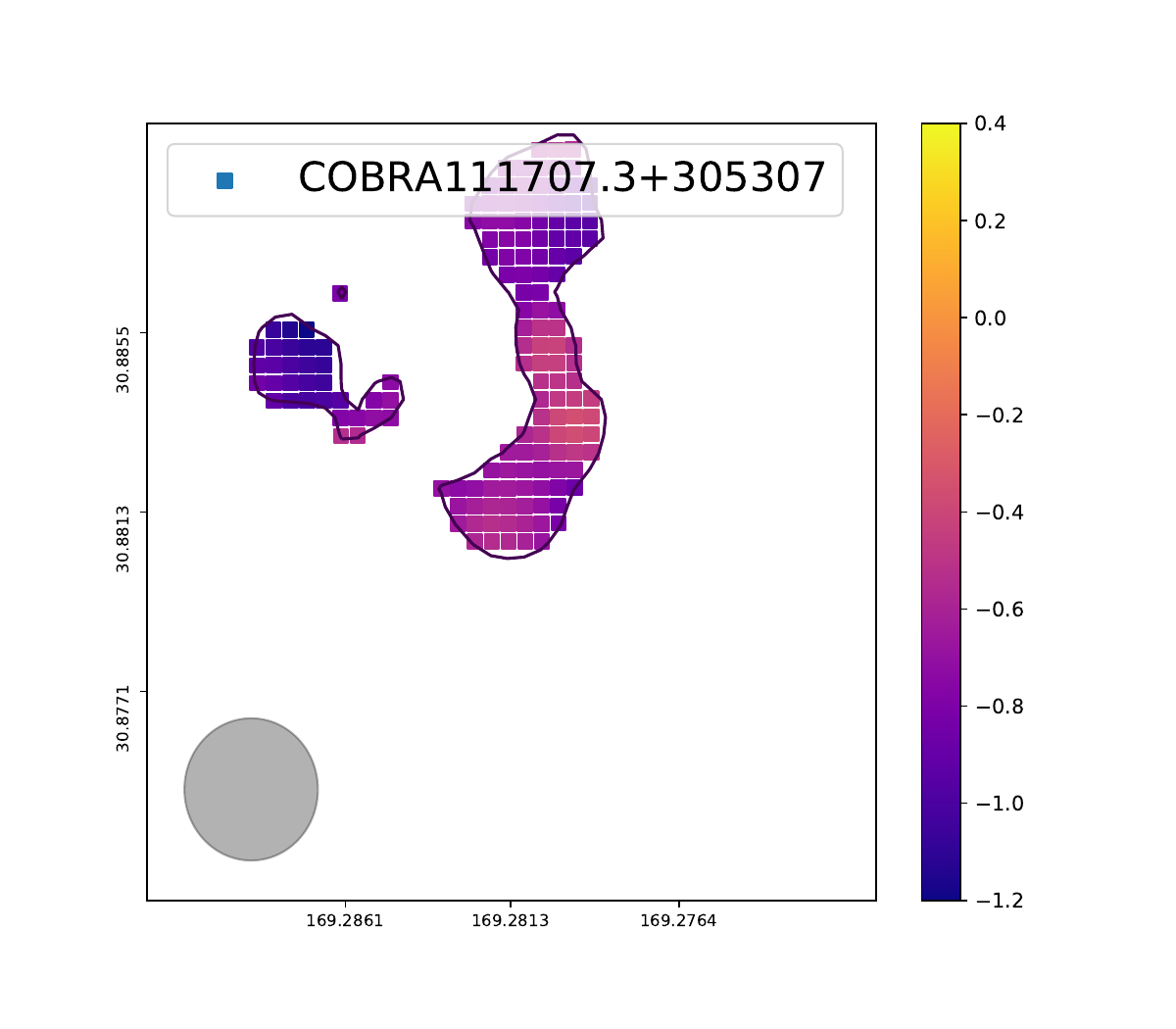}
\includegraphics[scale=0.28,trim={0.8in 0.65in 1.0in 0.75in},clip=true]{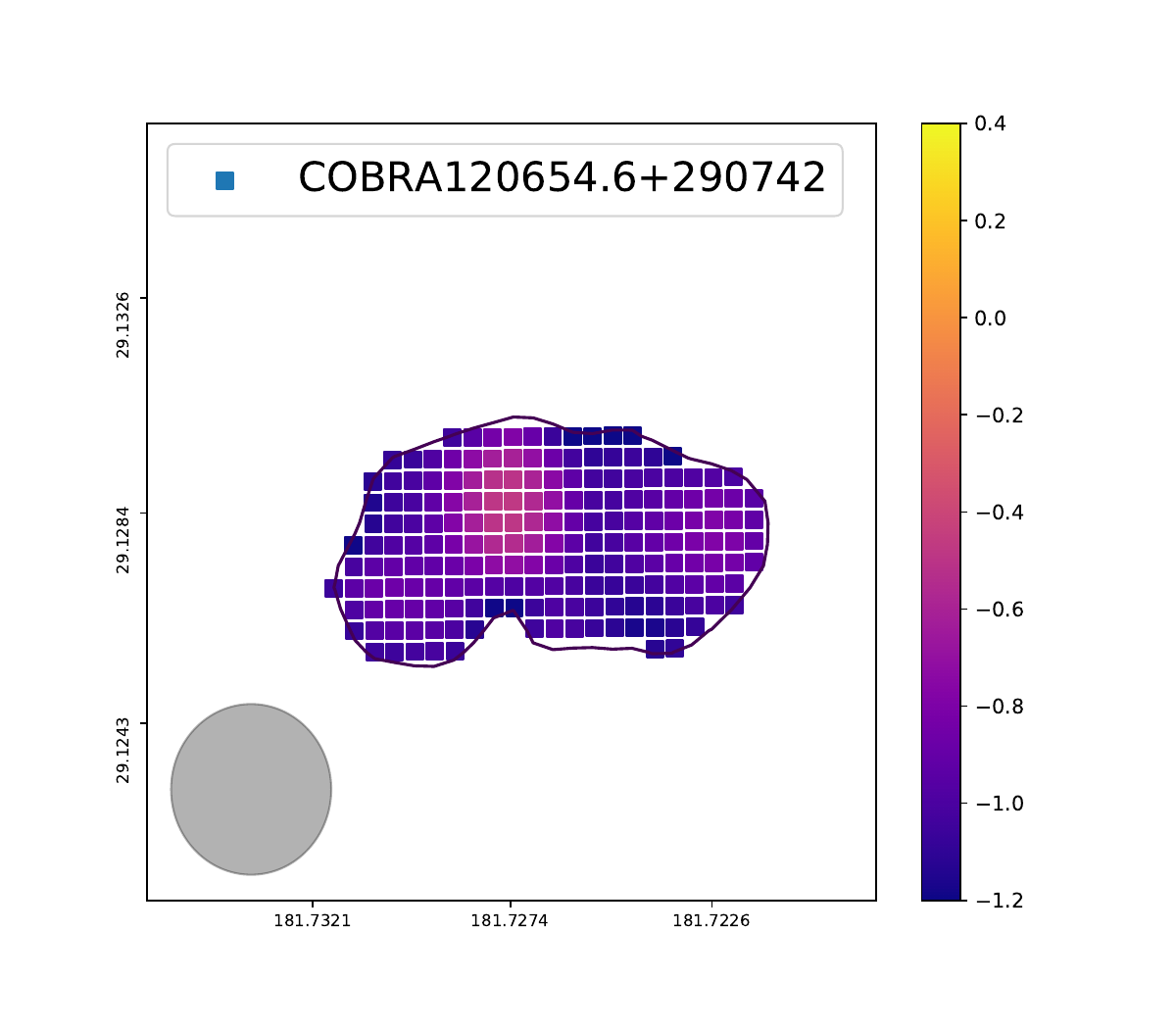}
\includegraphics[scale=0.28,trim={0.8in 0.65in 1.0in 0.75in},clip=true]{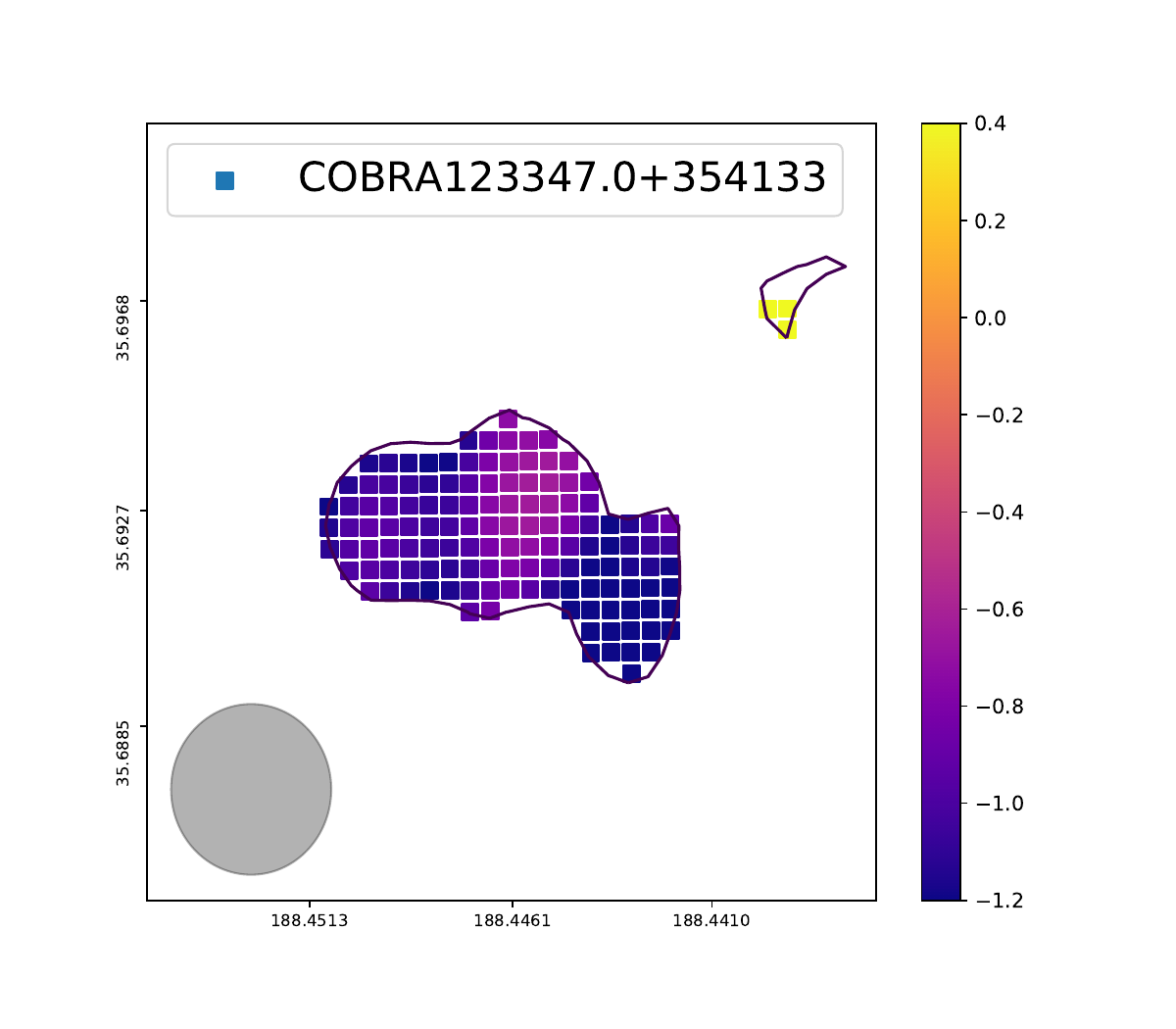}
\includegraphics[scale=0.28,trim={0.8in 0.65in 1.0in 0.75in},clip=true]{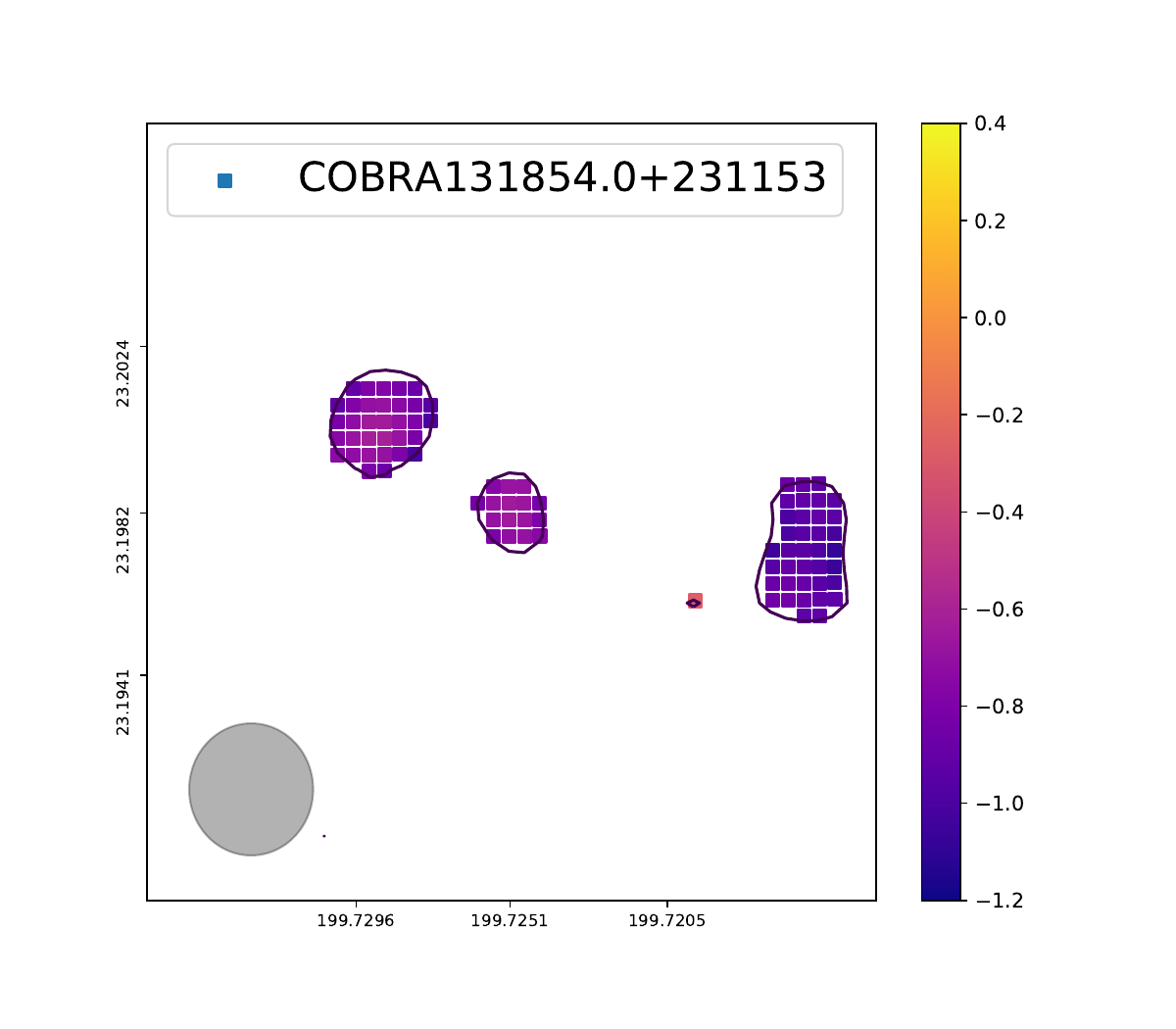}
\includegraphics[scale=0.28,trim={0.8in 0.65in 1.0in 0.75in},clip=true]{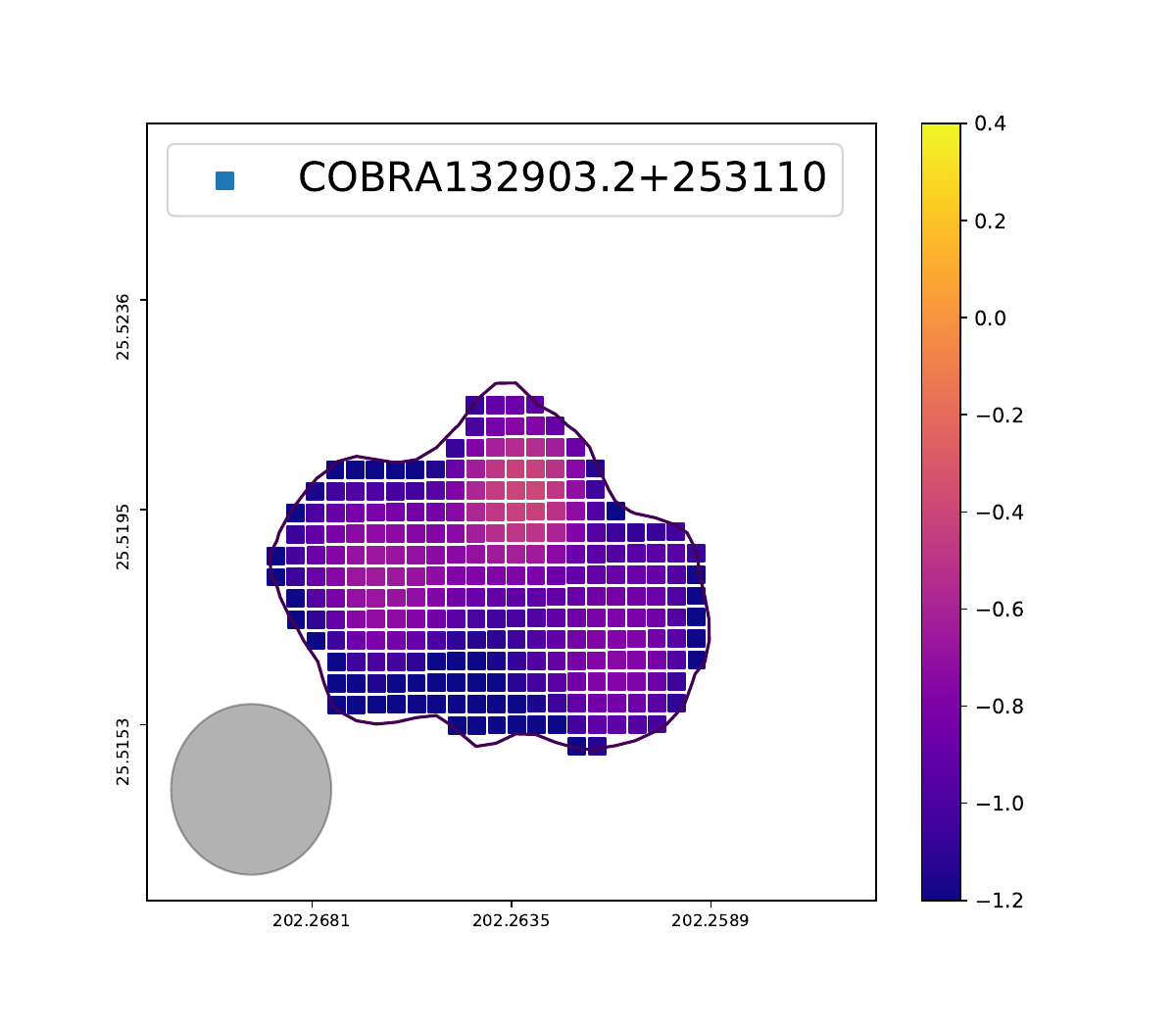}
\includegraphics[scale=0.28,trim={0.8in 0.65in 1.0in 0.75in},clip=true]{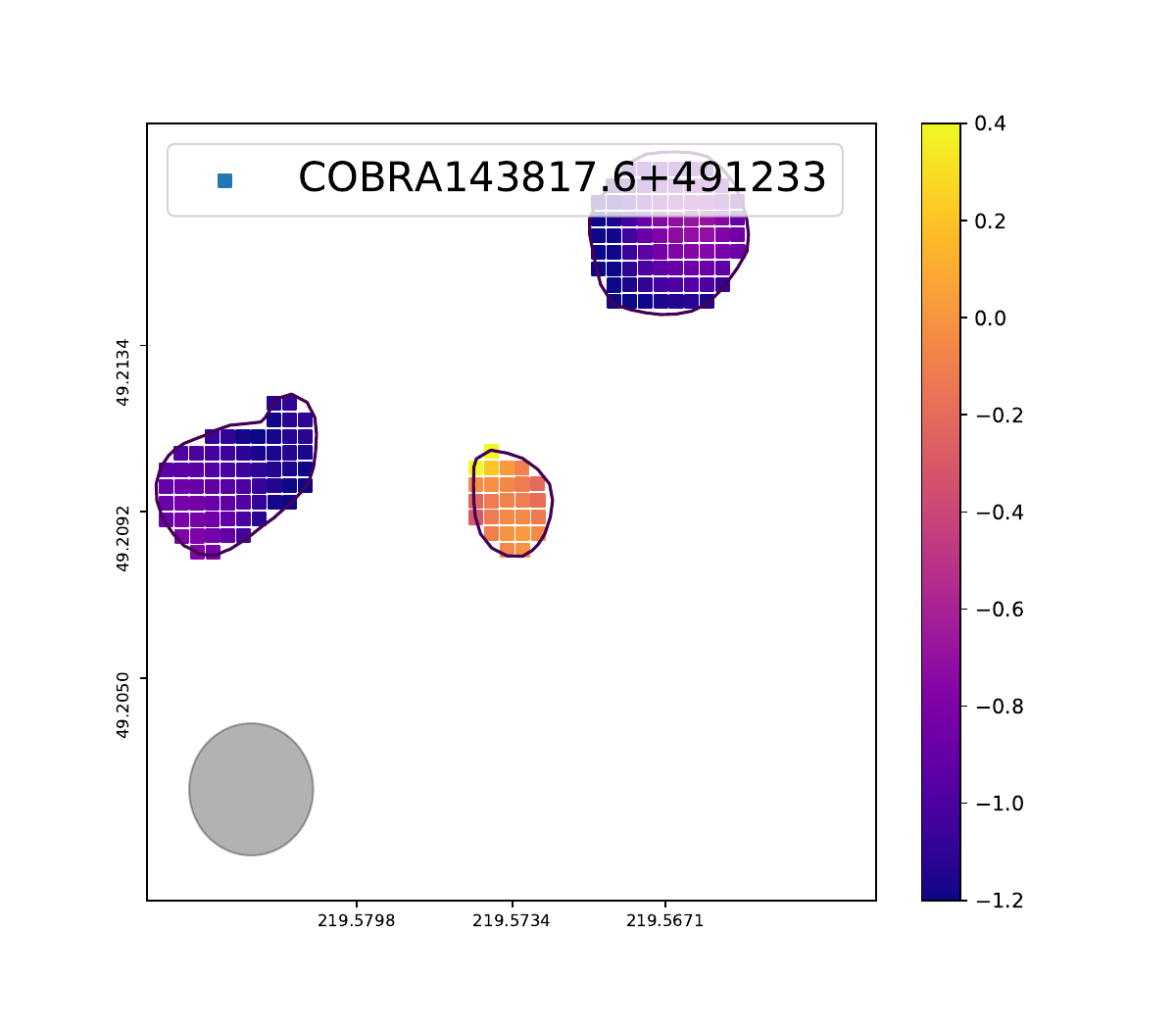}
\includegraphics[scale=0.28,trim={0.8in 0.65in 1.0in 0.75in},clip=true]{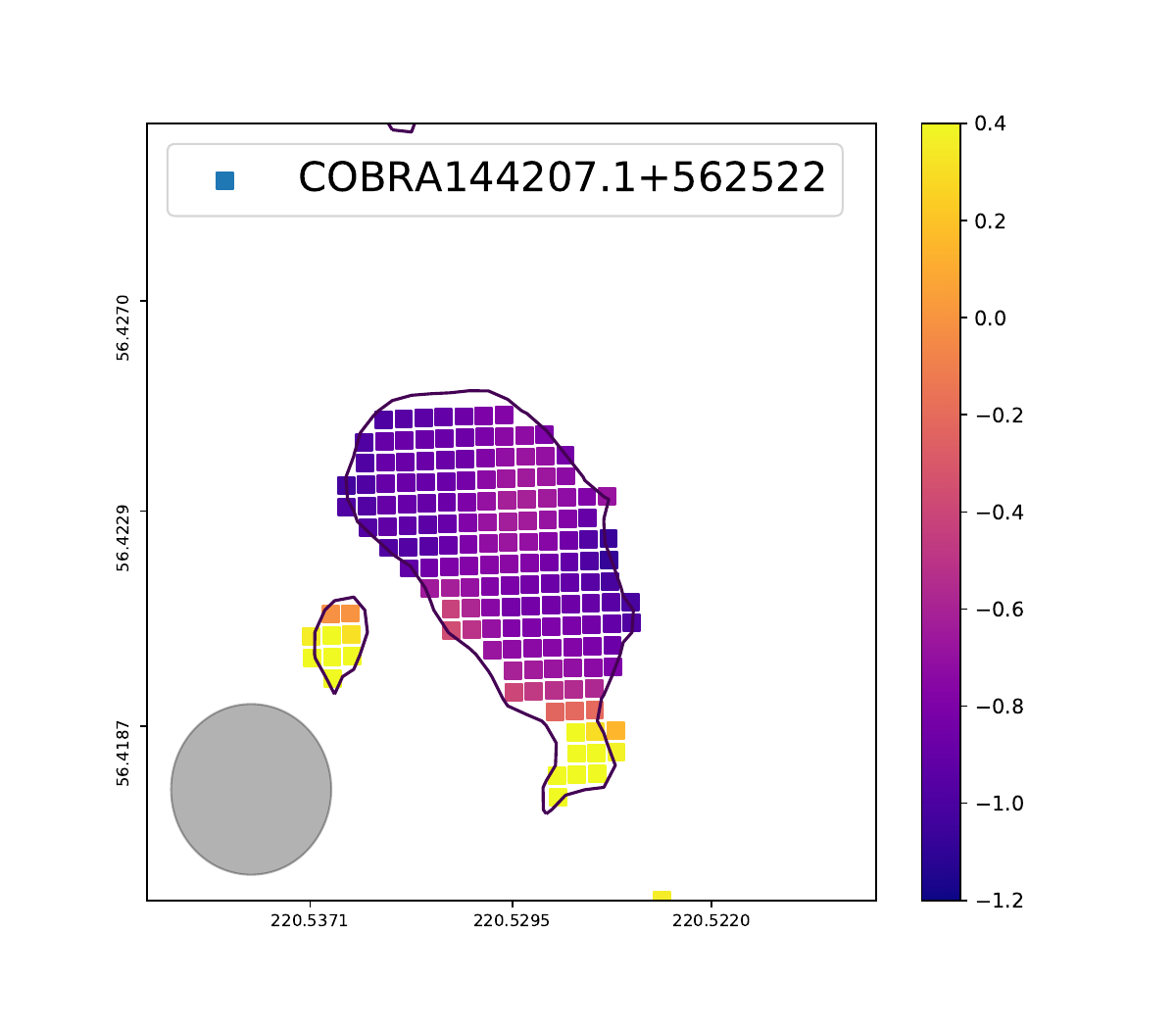}
\includegraphics[scale=0.28,trim={0.8in 0.65in 1.0in 0.75in},clip=true]{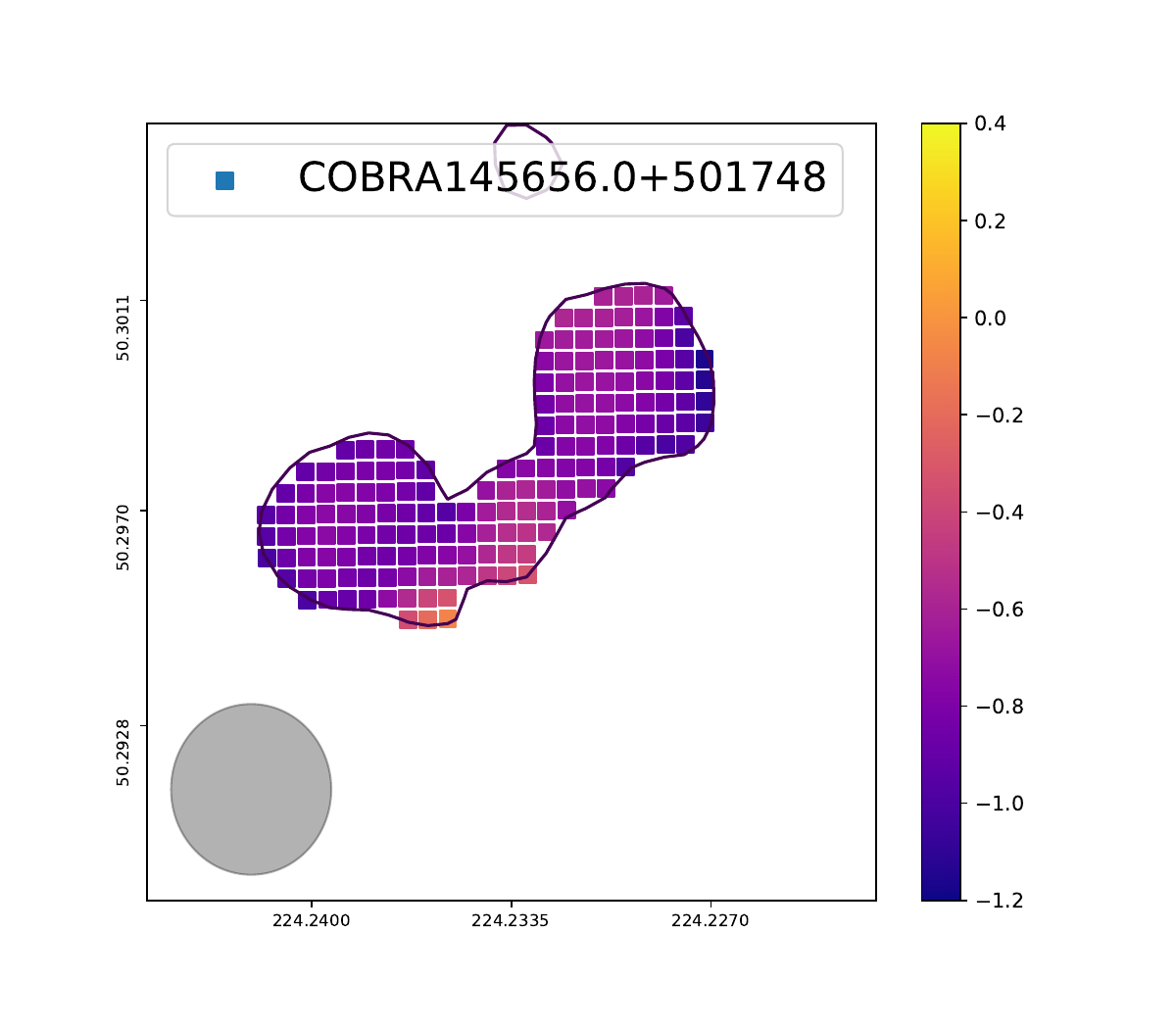}
\includegraphics[scale=0.28,trim={0.8in 0.65in 1.0in 0.75in},clip=true]{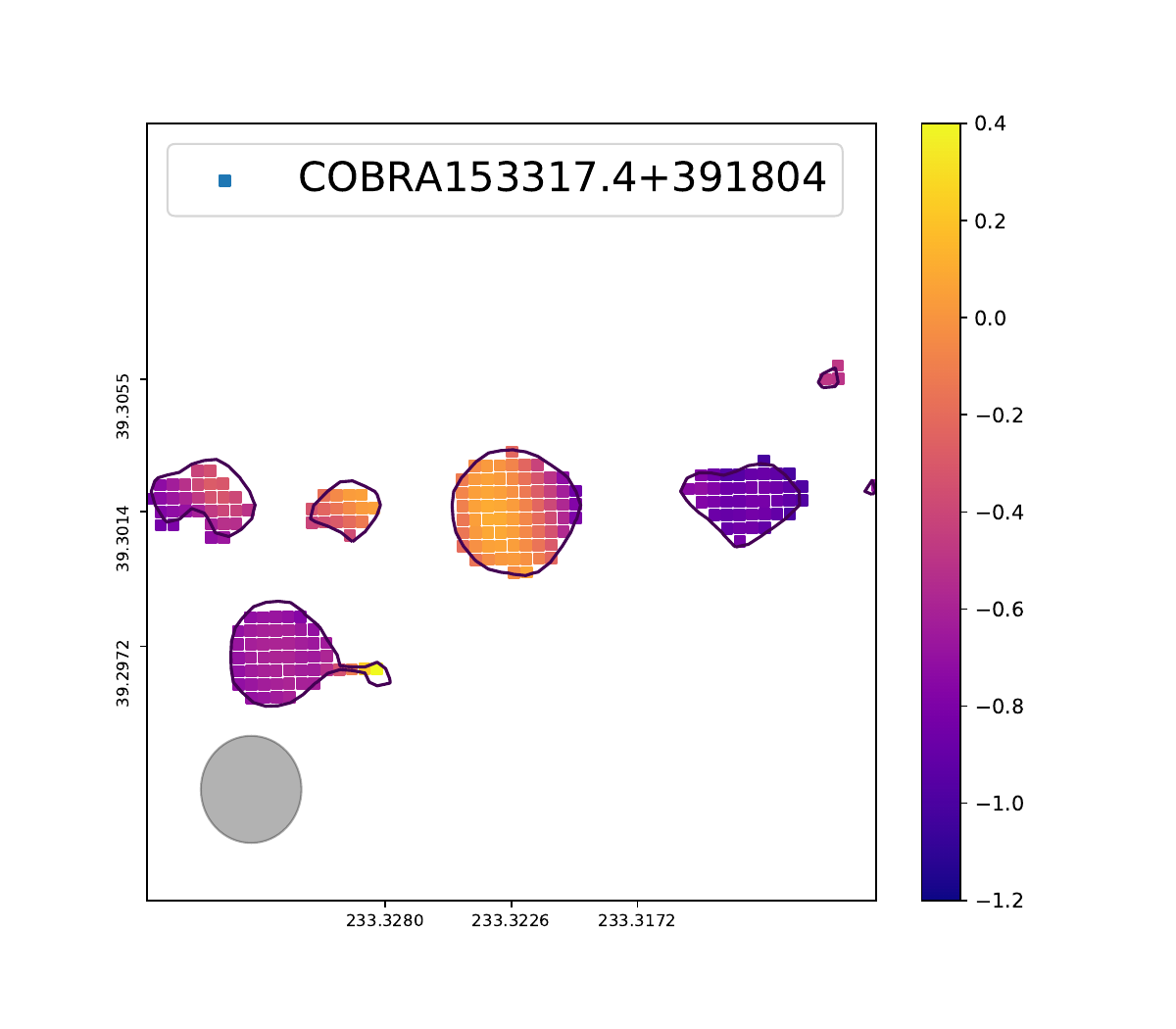}
\includegraphics[scale=0.28,trim={0.8in 0.65in 1.0in 0.75in},clip=true]{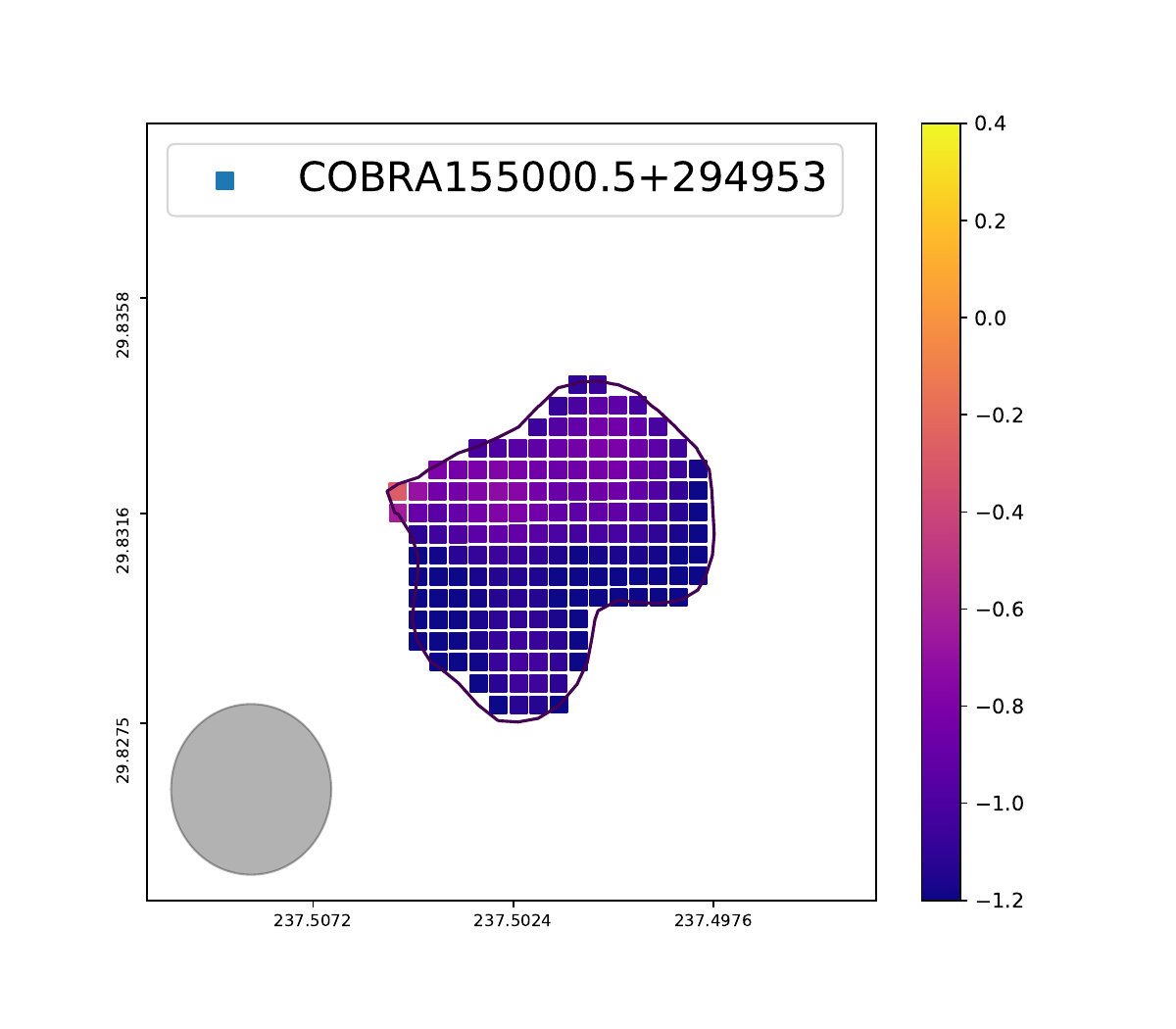}
\includegraphics[scale=0.28,trim={0.8in 0.65in 1.0in 0.75in},clip=true]{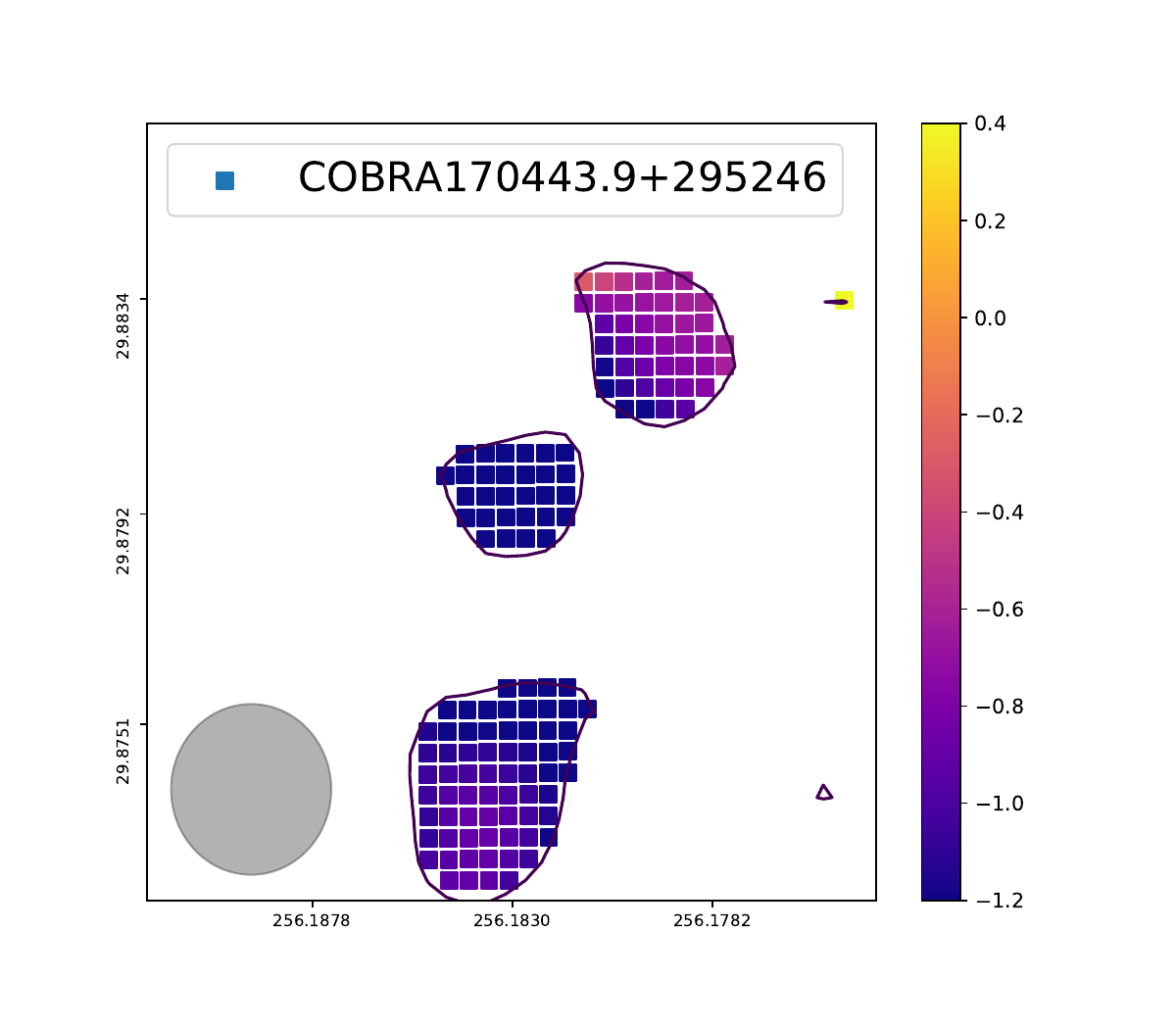}

\caption{Spectral index maps of bent radio AGNs in the non-cluster sample.  As in Figure~\ref{Fig:SI-clusters}, each pixel corresponds to a 1$\farcs$5 $\times$ 1$\farcs$5 region of the sky.  Although most images are scaled the same ($\approx$ 0$\farcm$45 $\times$ 0$\farcm$45 centered on the radio core), a few are scaled differently to encompass the entire radio AGN.  We use the same color bar as in Figure~\ref{Fig:SI-clusters}.}  \label{Fig:SI-nonclusters}
\end{center}
\end{figure*}

As mentioned in Section~\ref{sect:SpectralIndex}, because the LoTSS beam size is much larger than the pixel size, individual pixel measurements of the spectral index are highly correlated.  Thus, we limit our analysis of the spectral index to regions of approximately the beam size or larger (e.g., the entire radio source, the radio core, or the radio lobes).  However, to allow for an examination of the regions probed for each bent source, we do present our spectral index maps in Figures~\ref{Fig:SI-clusters} and \ref{Fig:SI-nonclusters}.  As can be seen, we have a number of sources in the cluster sample, as well as a few in the non-cluster sample, with very flat spectral index measurements in the core relative to the lobes.  As seen in our reported values, we see a steepening of the spectral index away from the core.  

\begin{figure*}
\begin{center}
\includegraphics[scale=0.28,trim={0.8in 0.65in 1.0in 0.75in},clip=true]{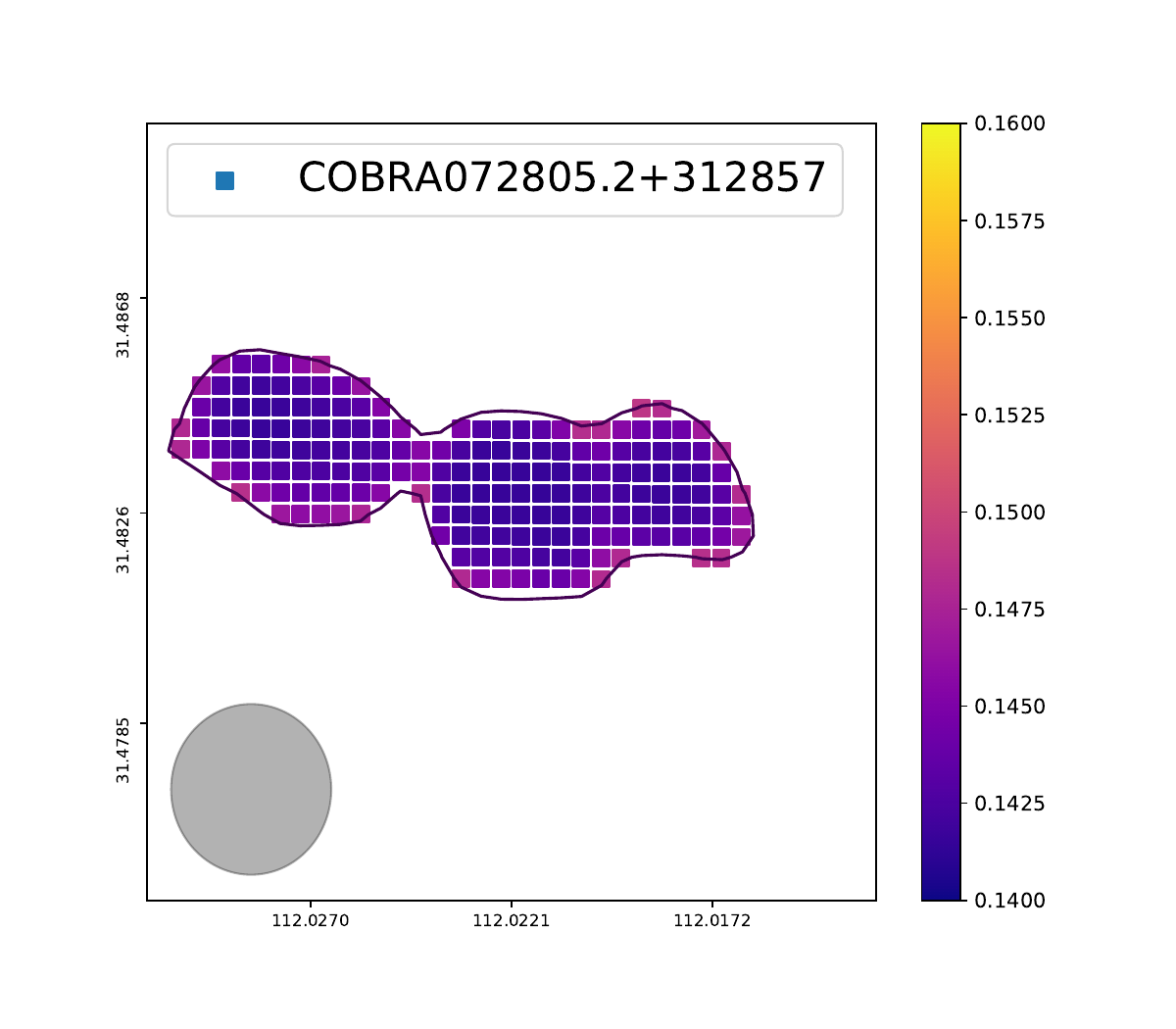}
\includegraphics[scale=0.28,trim={0.8in 0.65in 1.0in 0.75in},clip=true]{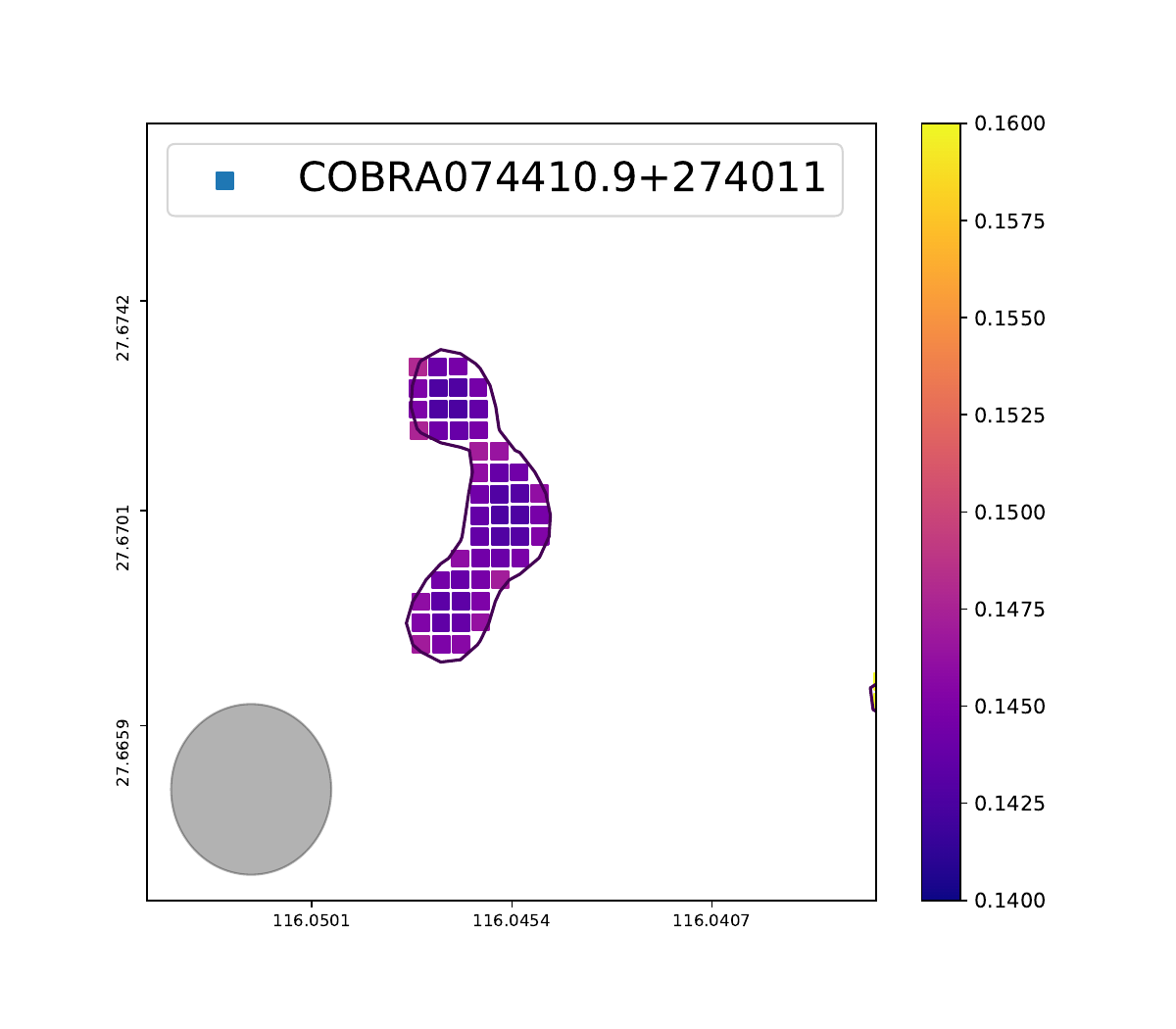}
\includegraphics[scale=0.28,trim={0.8in 0.65in 1.0in 0.75in},clip=true]{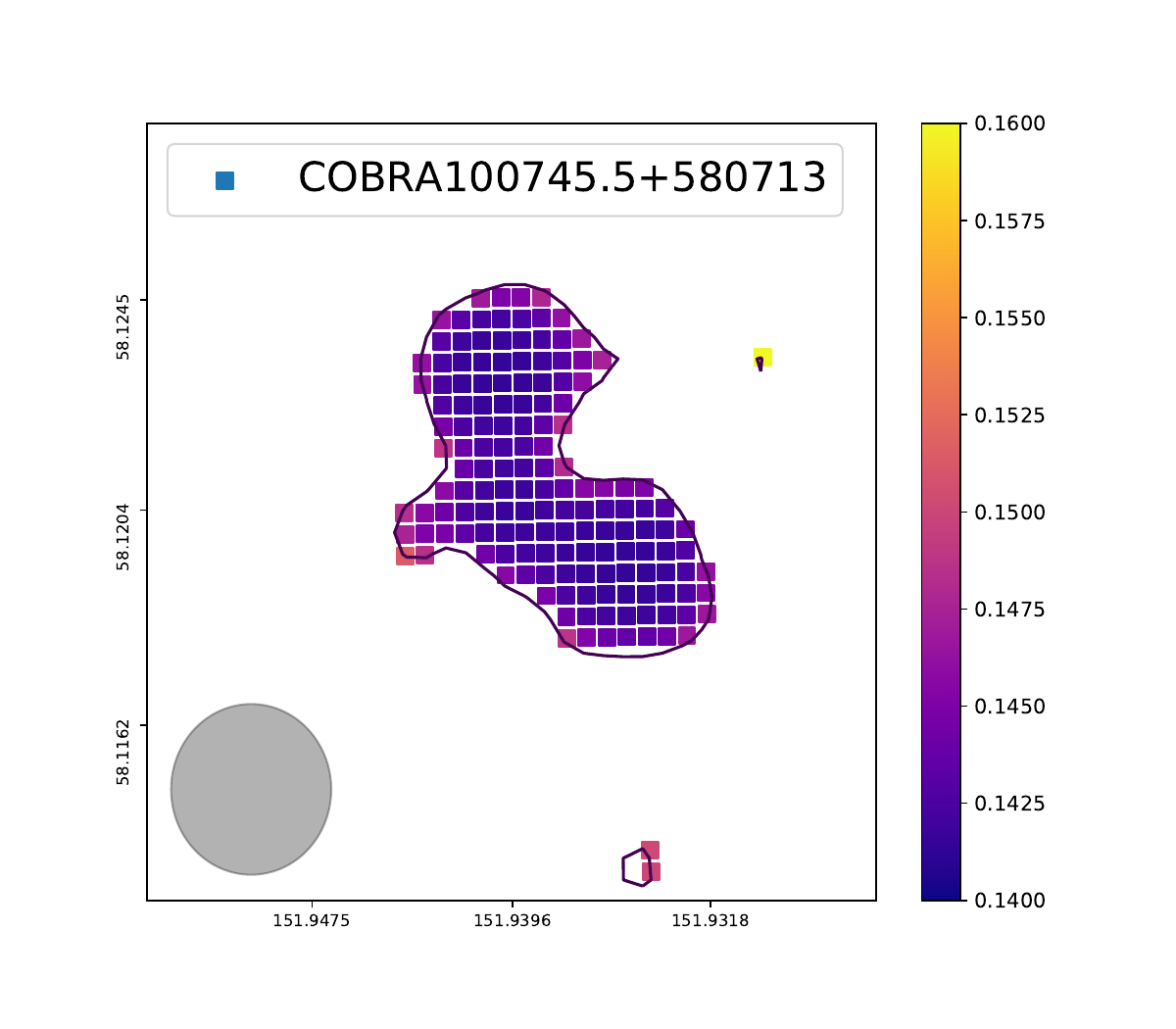}
\includegraphics[scale=0.28,trim={0.8in 0.65in 1.0in 0.75in},clip=true]{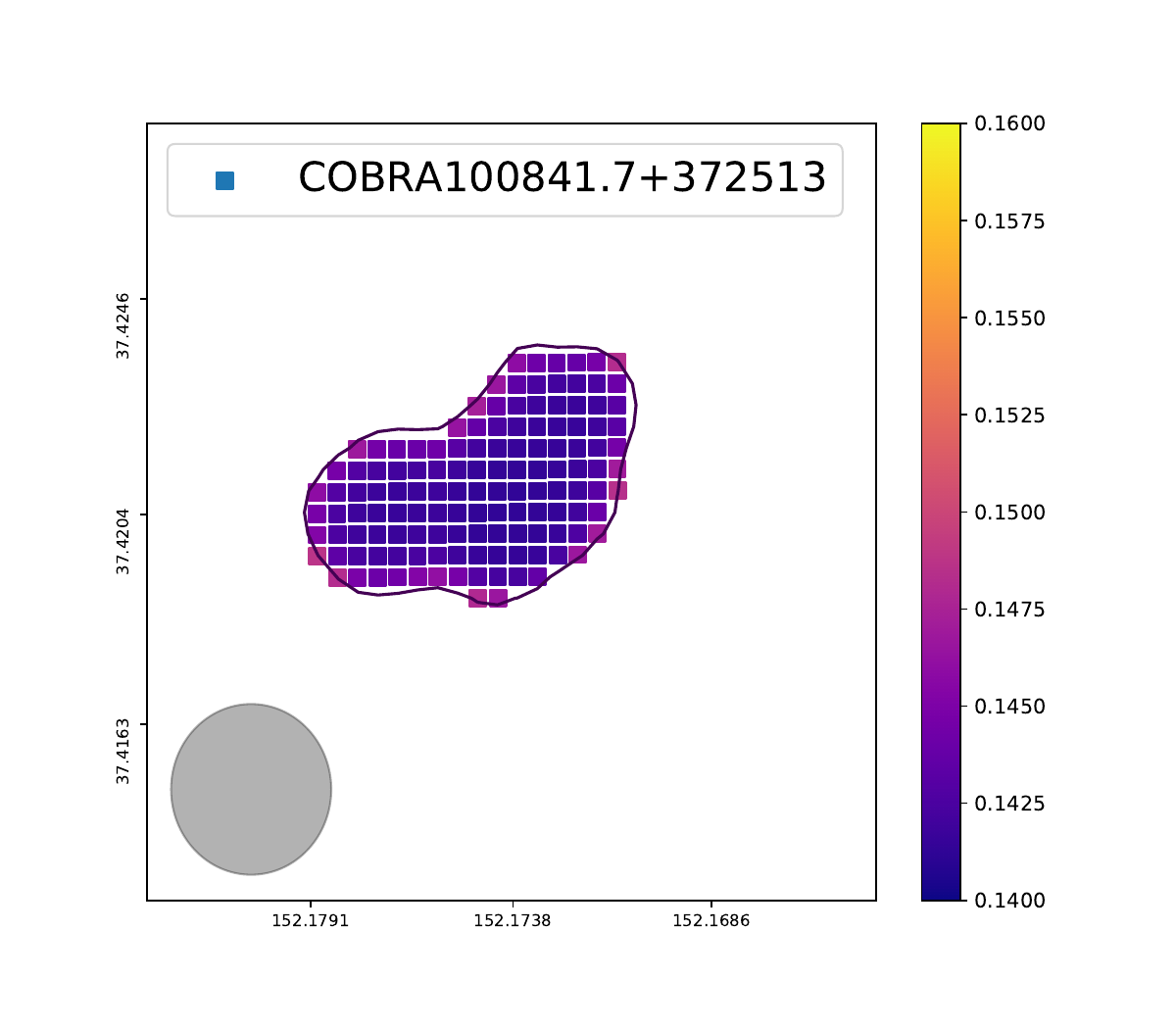}
\includegraphics[scale=0.28,trim={0.8in 0.65in 1.0in 0.75in},clip=true]{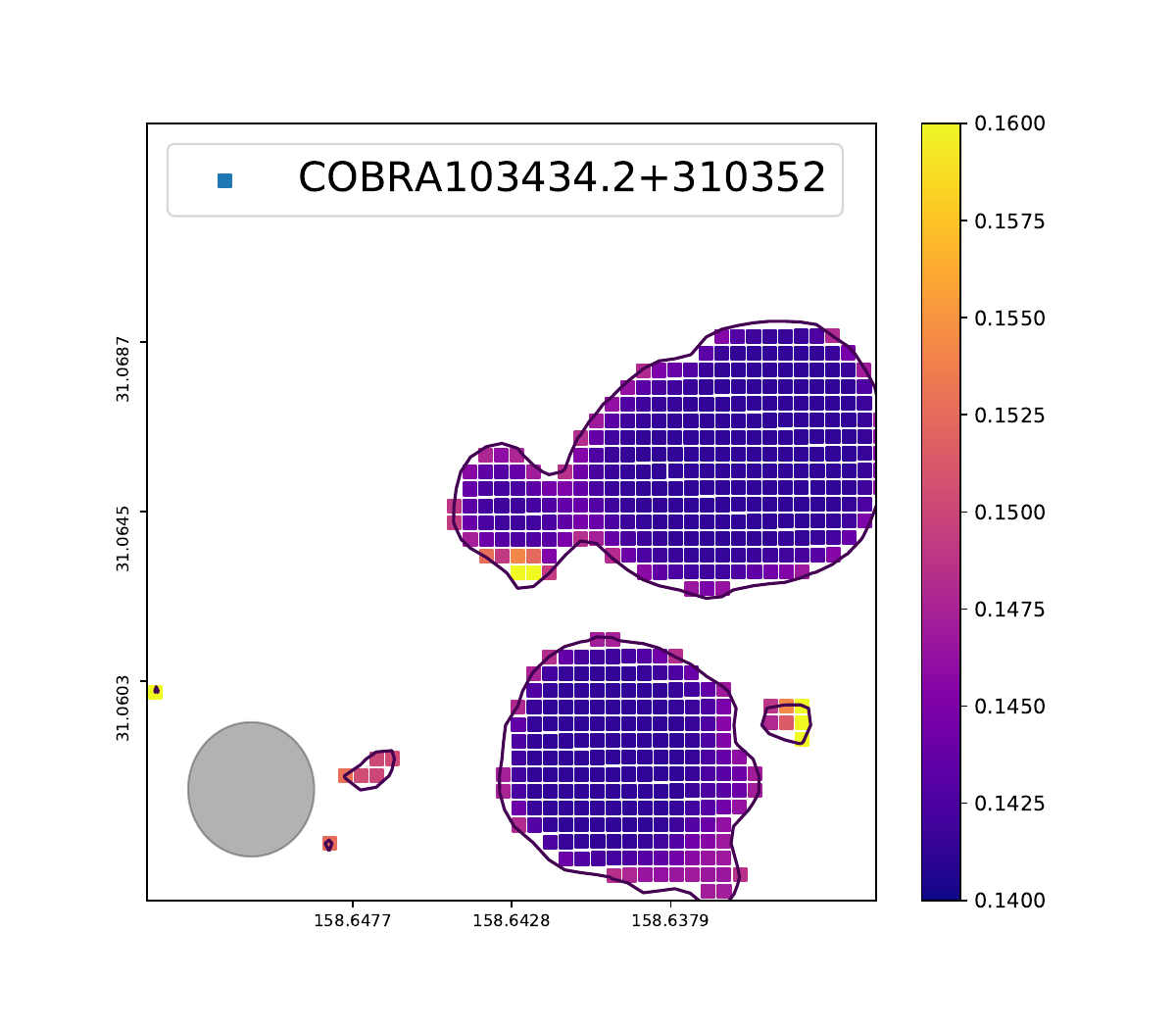}
\includegraphics[scale=0.28,trim={0.8in 0.65in 1.0in 0.75in},clip=true]{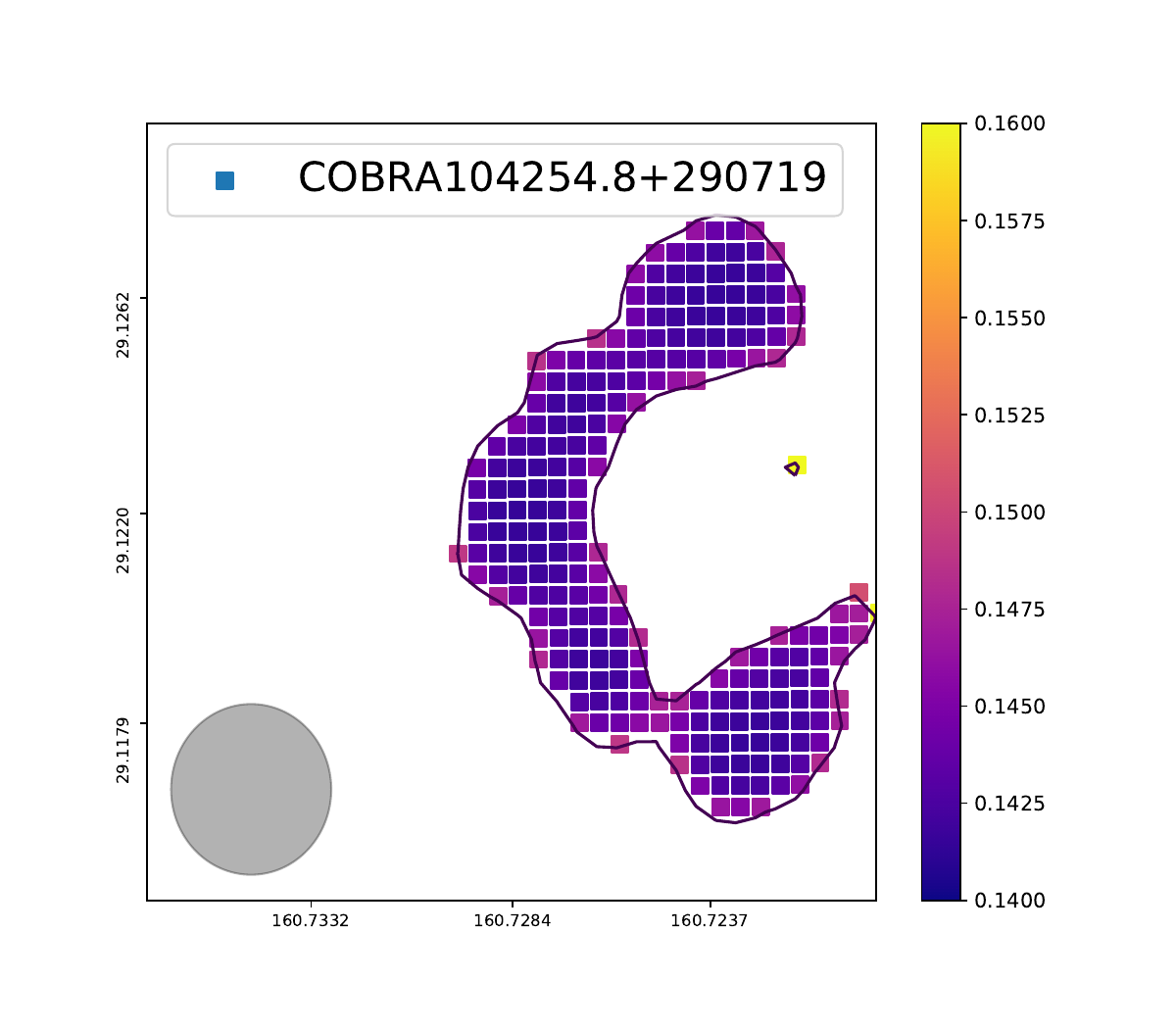}
\includegraphics[scale=0.28,trim={0.8in 0.65in 1.0in 0.75in},clip=true]{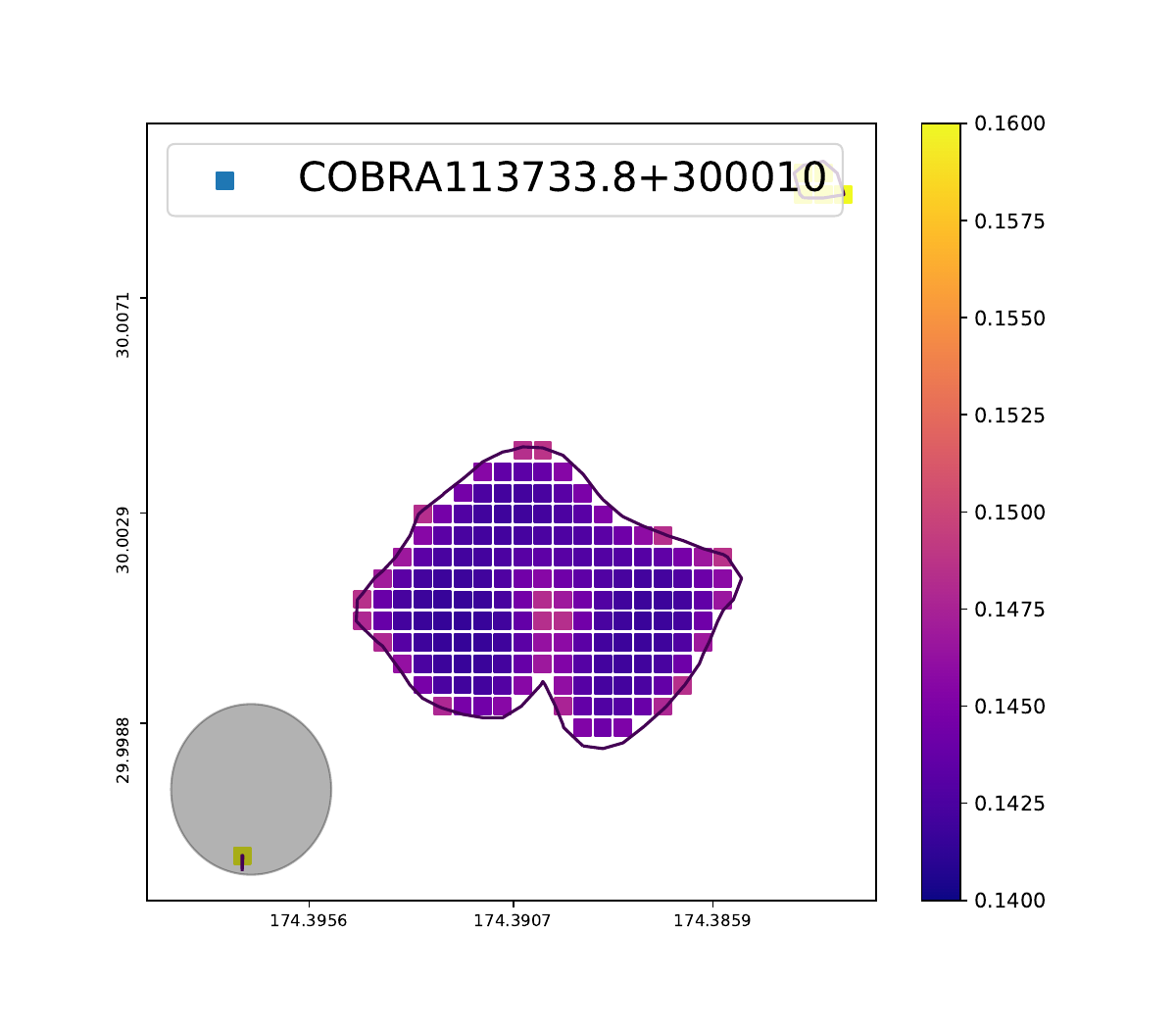}
\includegraphics[scale=0.28,trim={0.8in 0.65in 1.0in 0.75in},clip=true]{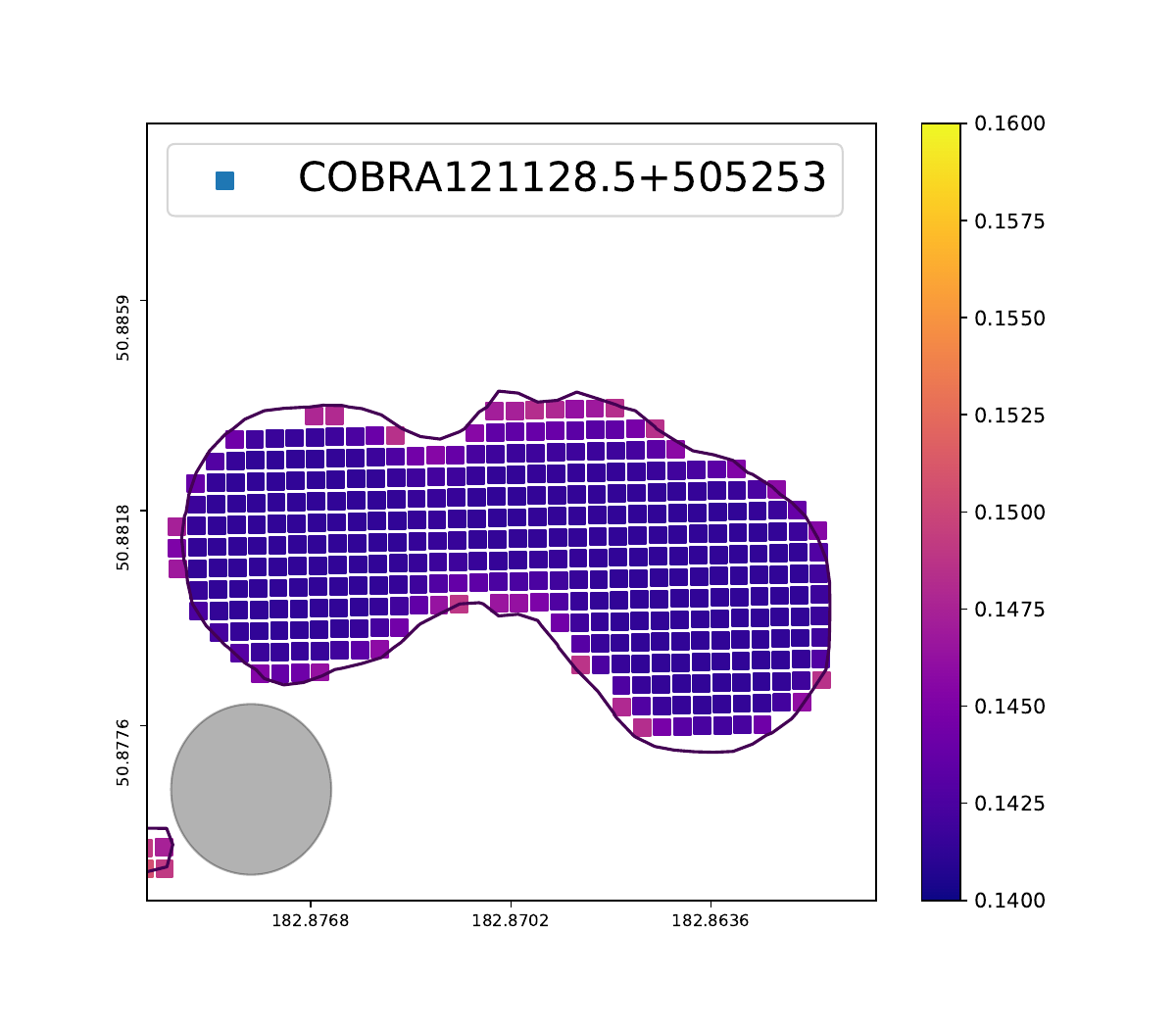}
\includegraphics[scale=0.28,trim={0.8in 0.65in 1.0in 0.75in},clip=true]{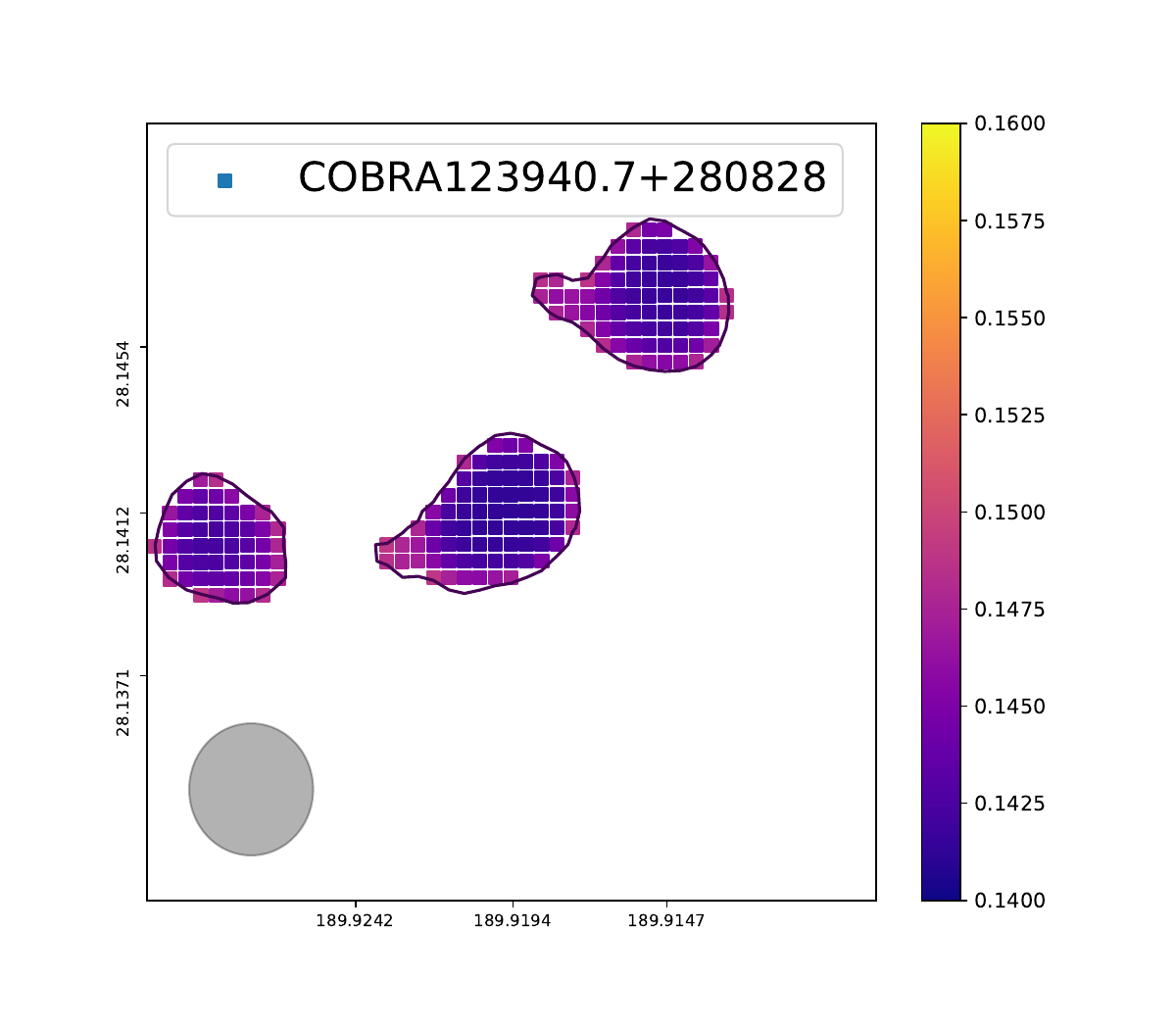}
\includegraphics[scale=0.28,trim={0.8in 0.65in 1.0in 0.75in},clip=true]{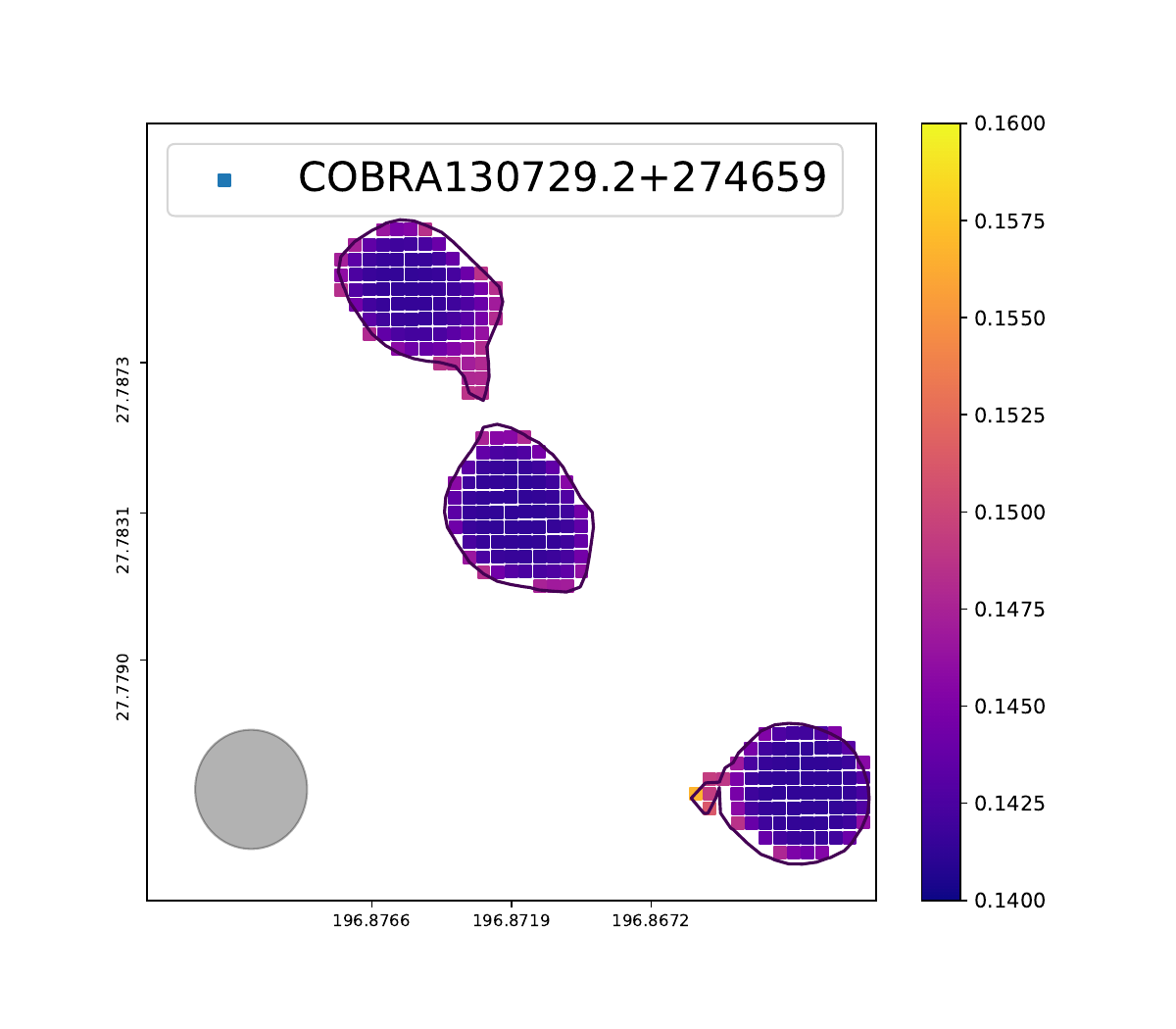}
\includegraphics[scale=0.28,trim={0.8in 0.65in 1.0in 0.75in},clip=true]{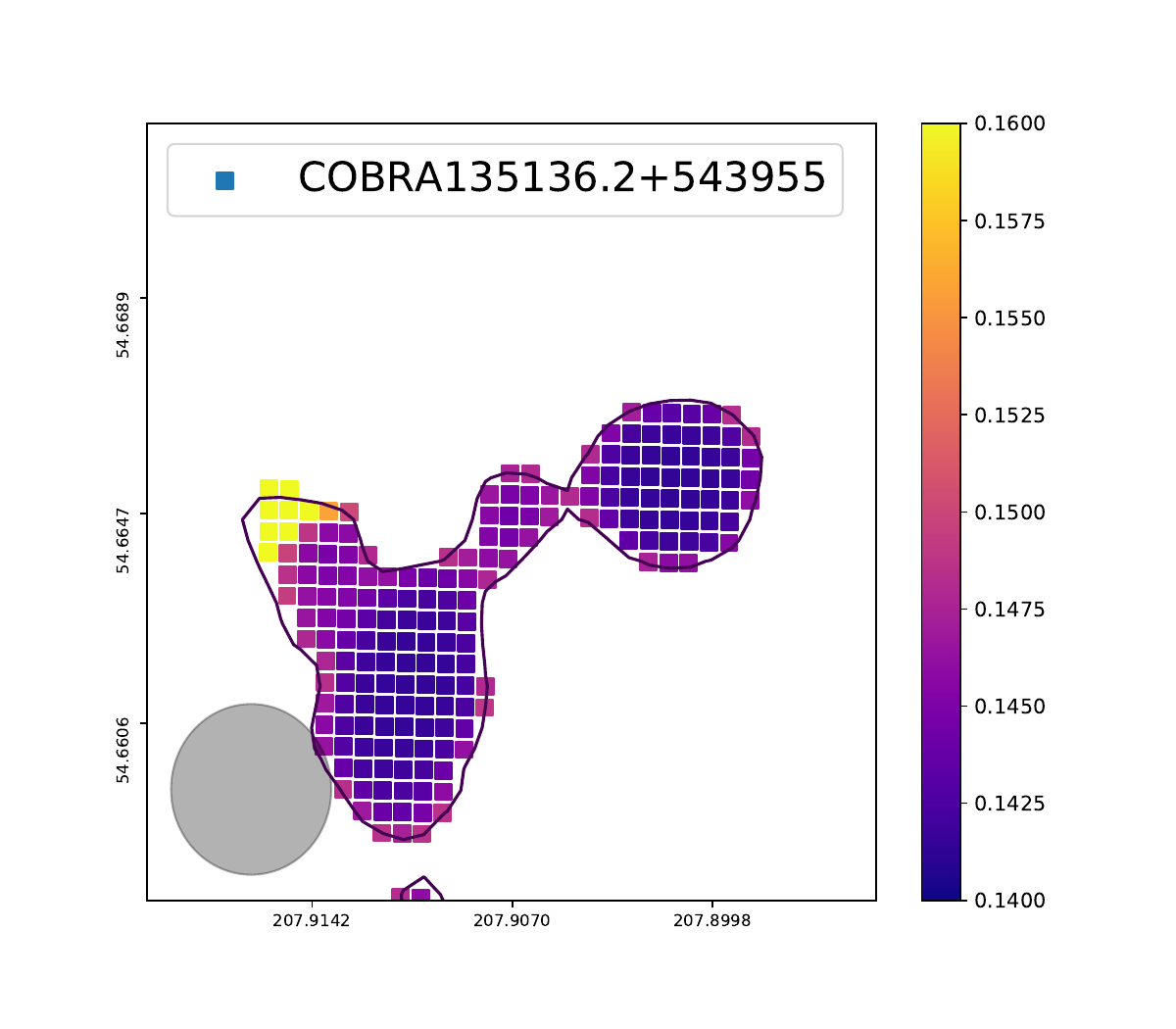}
\includegraphics[scale=0.28,trim={0.8in 0.65in 1.0in 0.75in},clip=true]{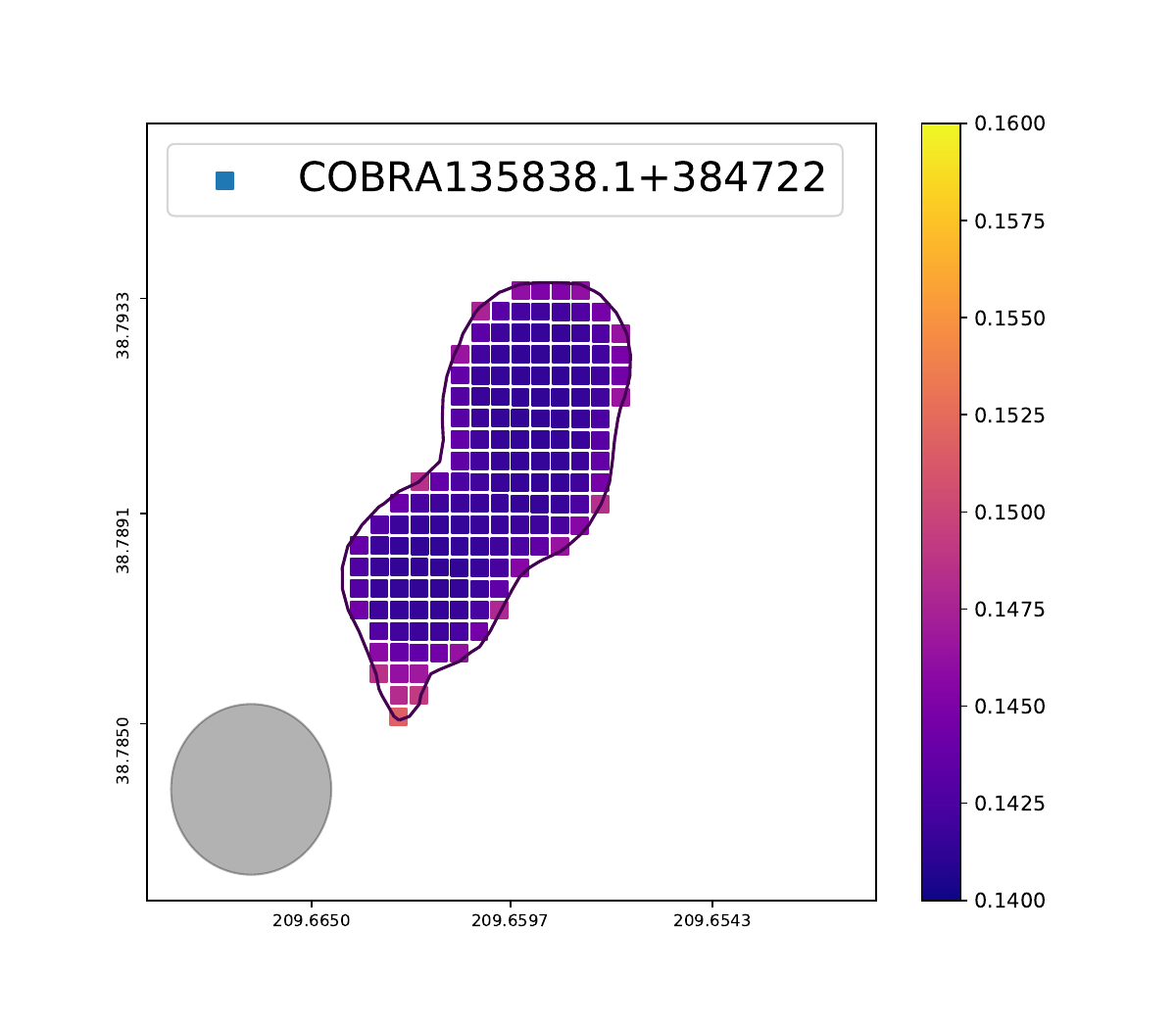}
\includegraphics[scale=0.28,trim={0.8in 0.65in 1.0in 0.75in},clip=true]{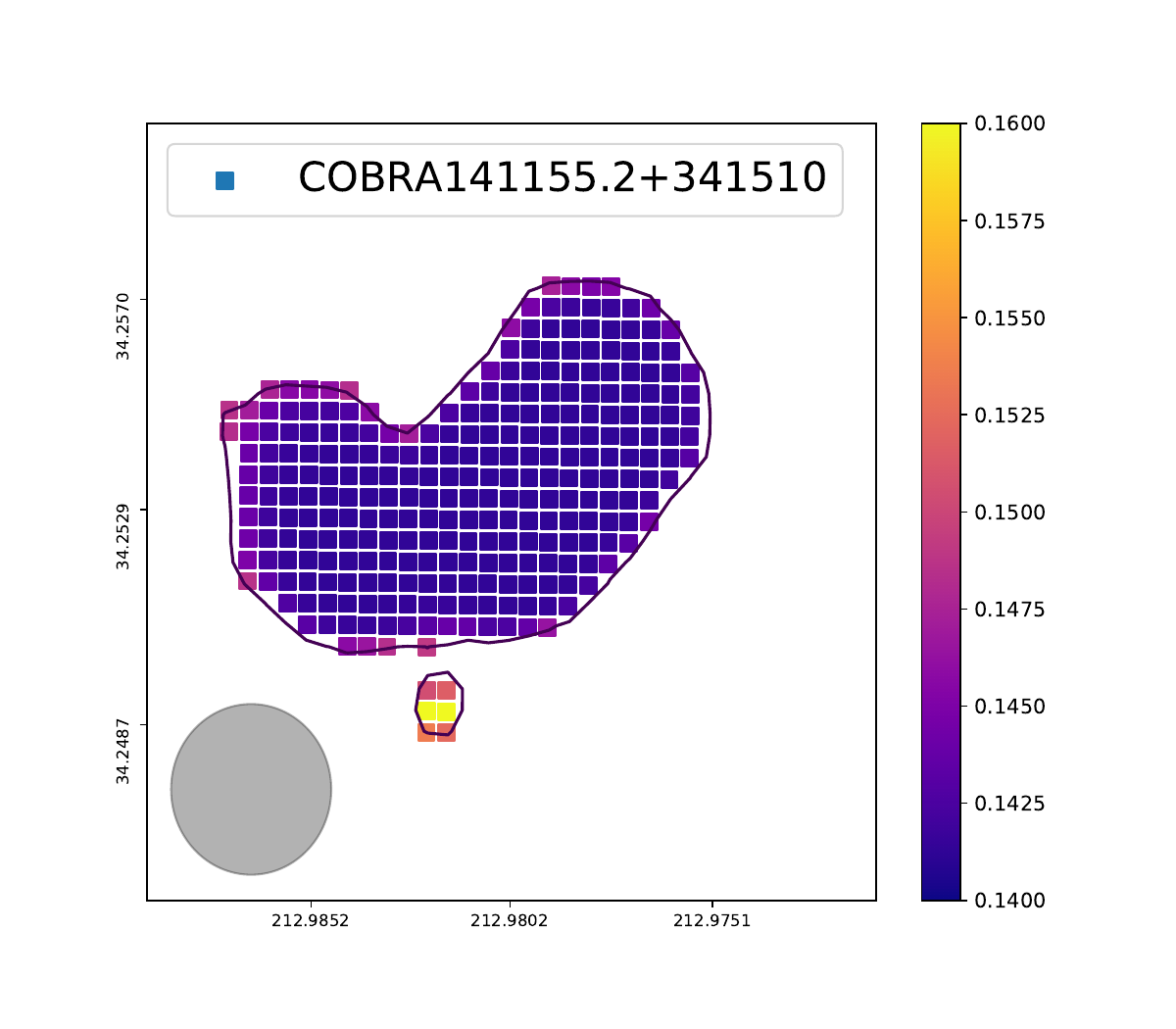}
\includegraphics[scale=0.28,trim={0.8in 0.65in 1.0in 0.75in},clip=true]{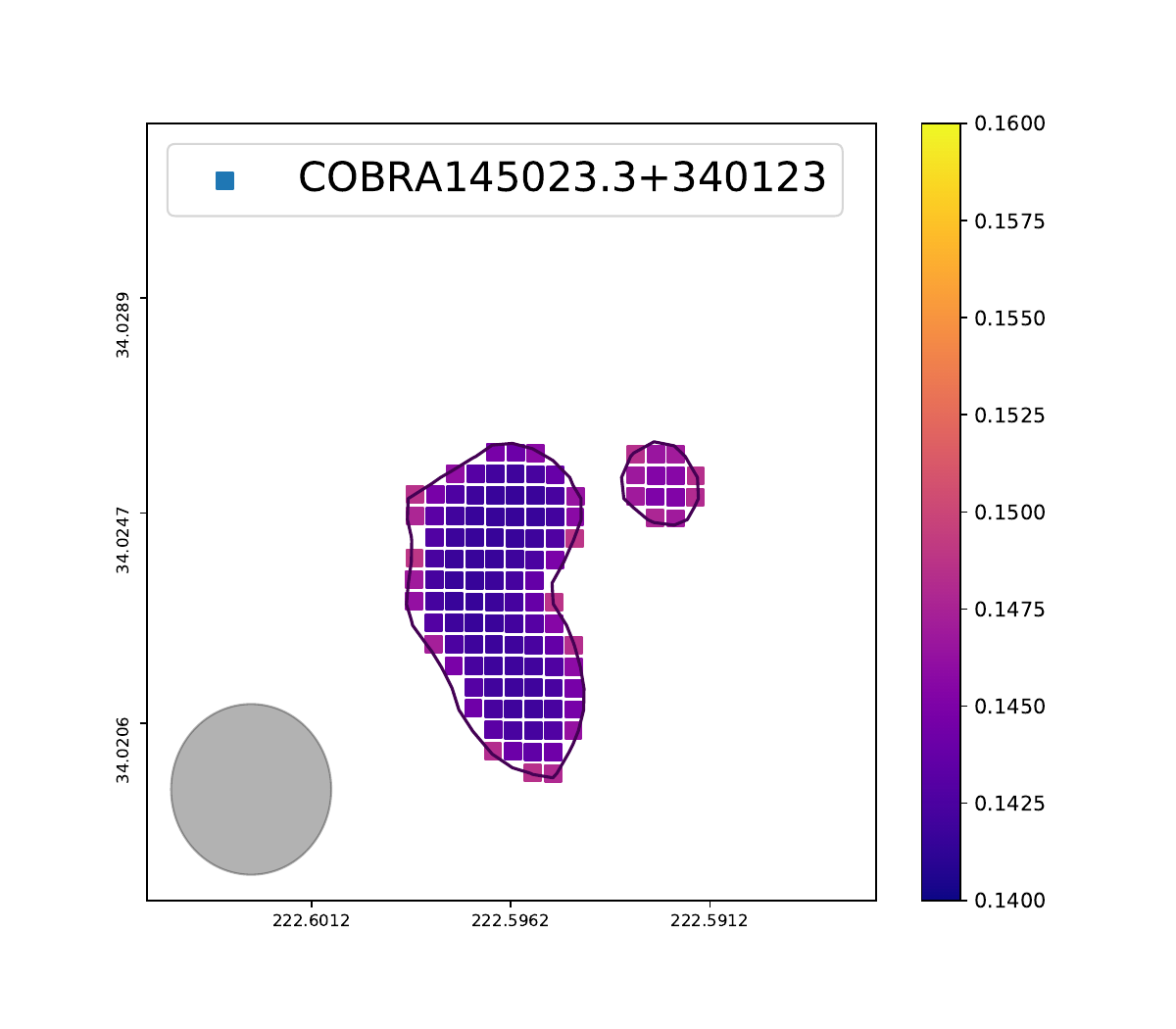}
\includegraphics[scale=0.28,trim={0.8in 0.65in 1.0in 0.75in},clip=true]{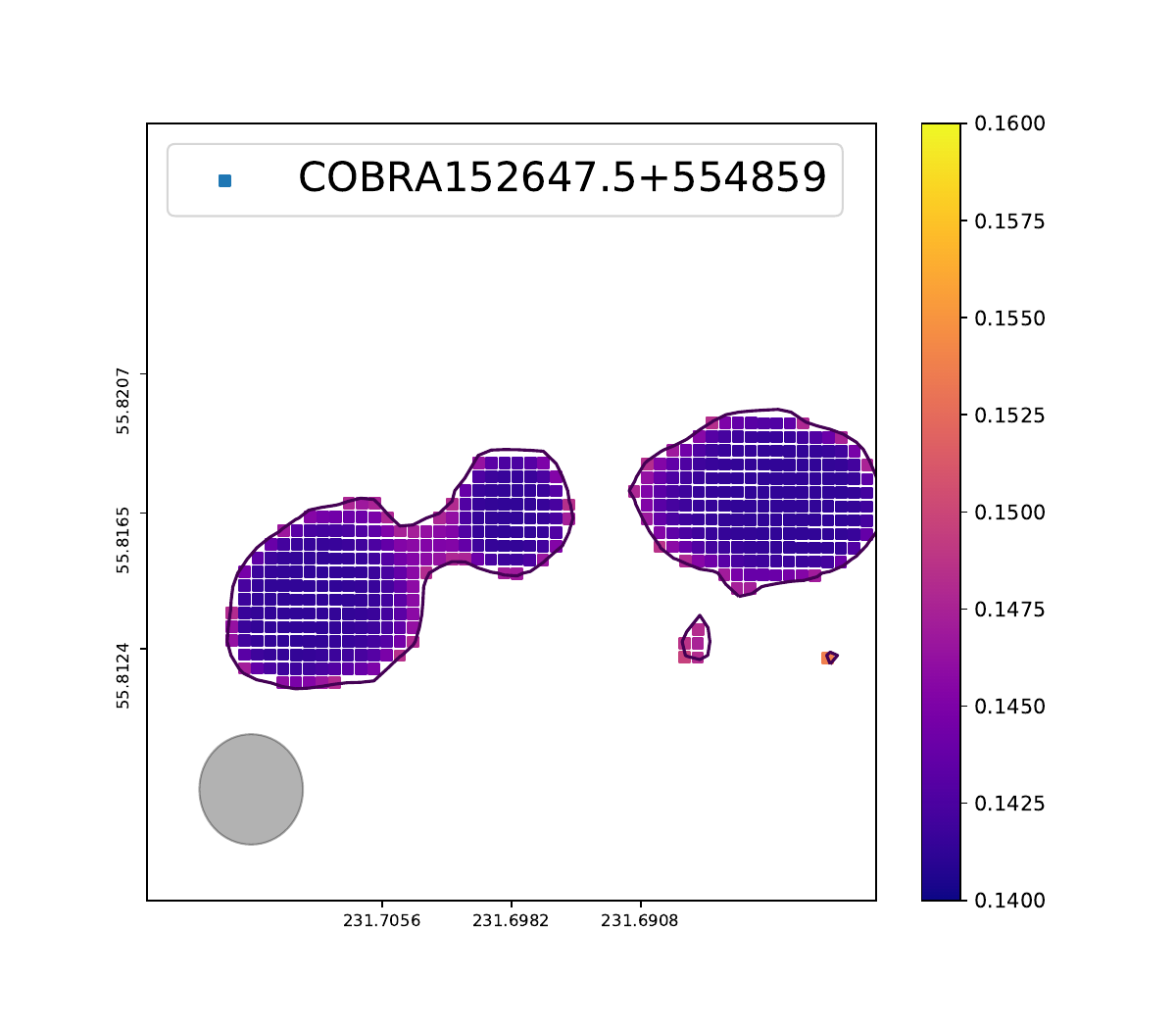}
\includegraphics[scale=0.28,trim={0.8in 0.65in 1.0in 0.75in},clip=true]{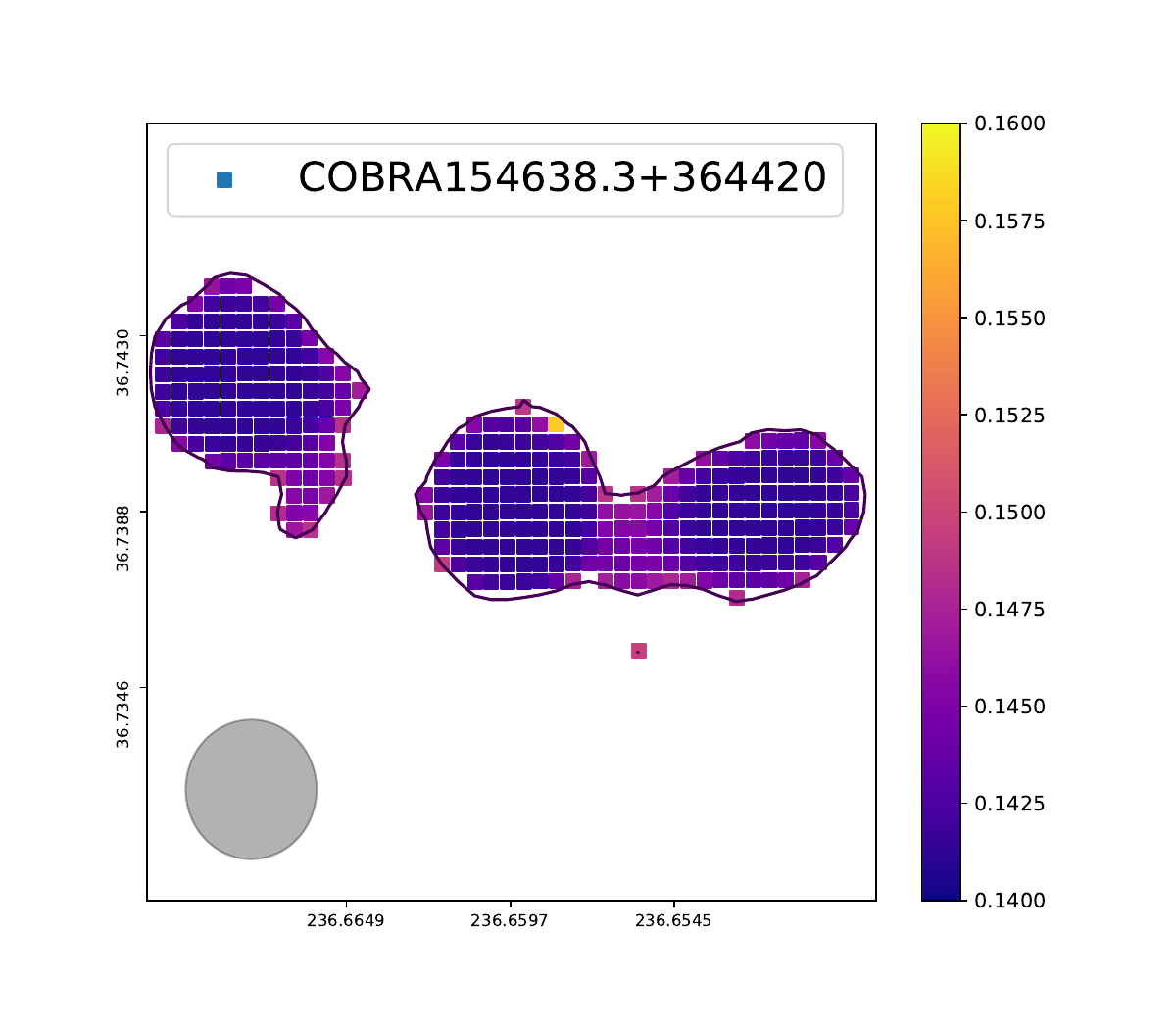}
\includegraphics[scale=0.28,trim={0.8in 0.65in 1.0in 0.75in},clip=true]{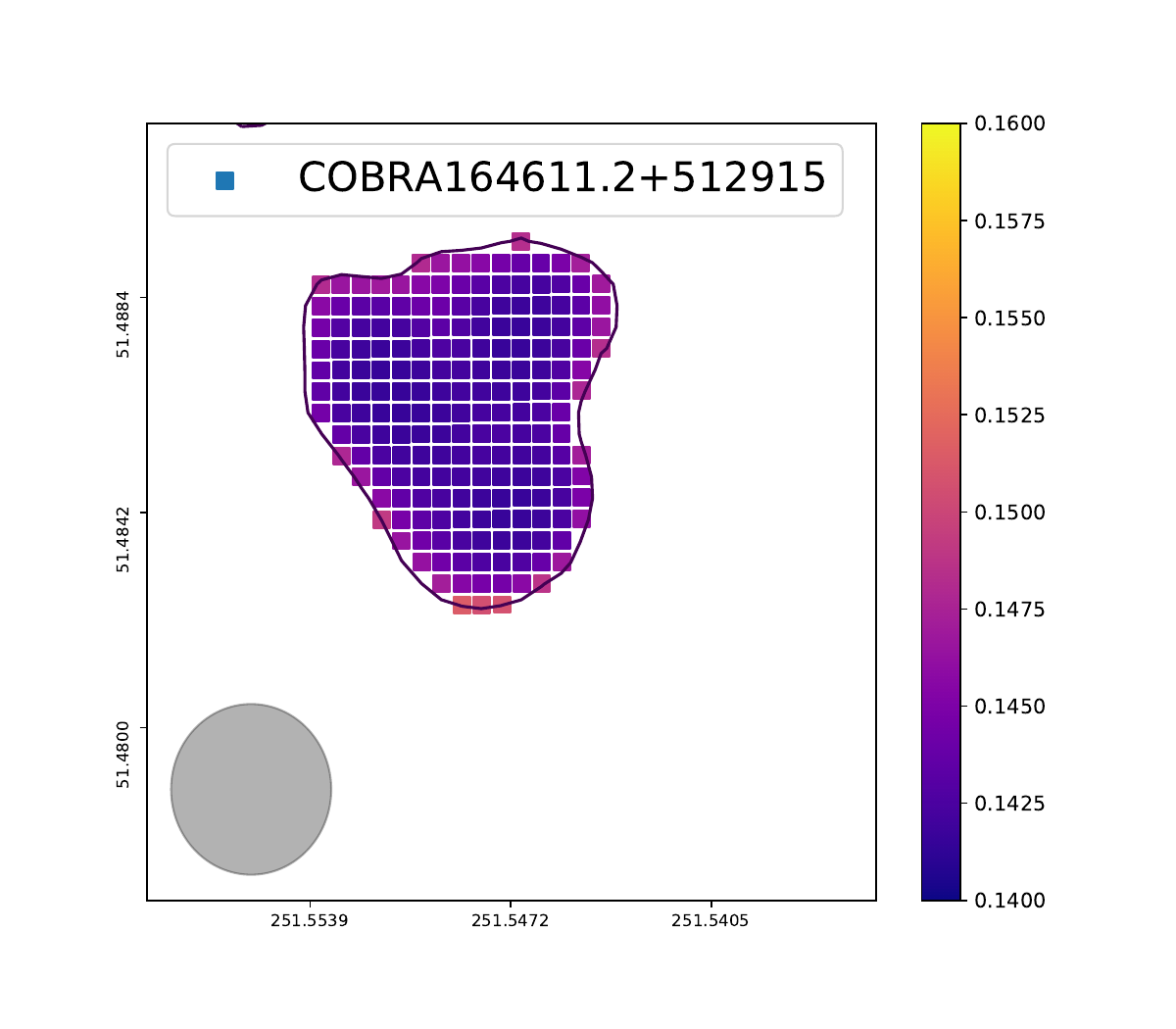}
\includegraphics[scale=0.28,trim={0.8in 0.65in 1.0in 0.75in},clip=true]{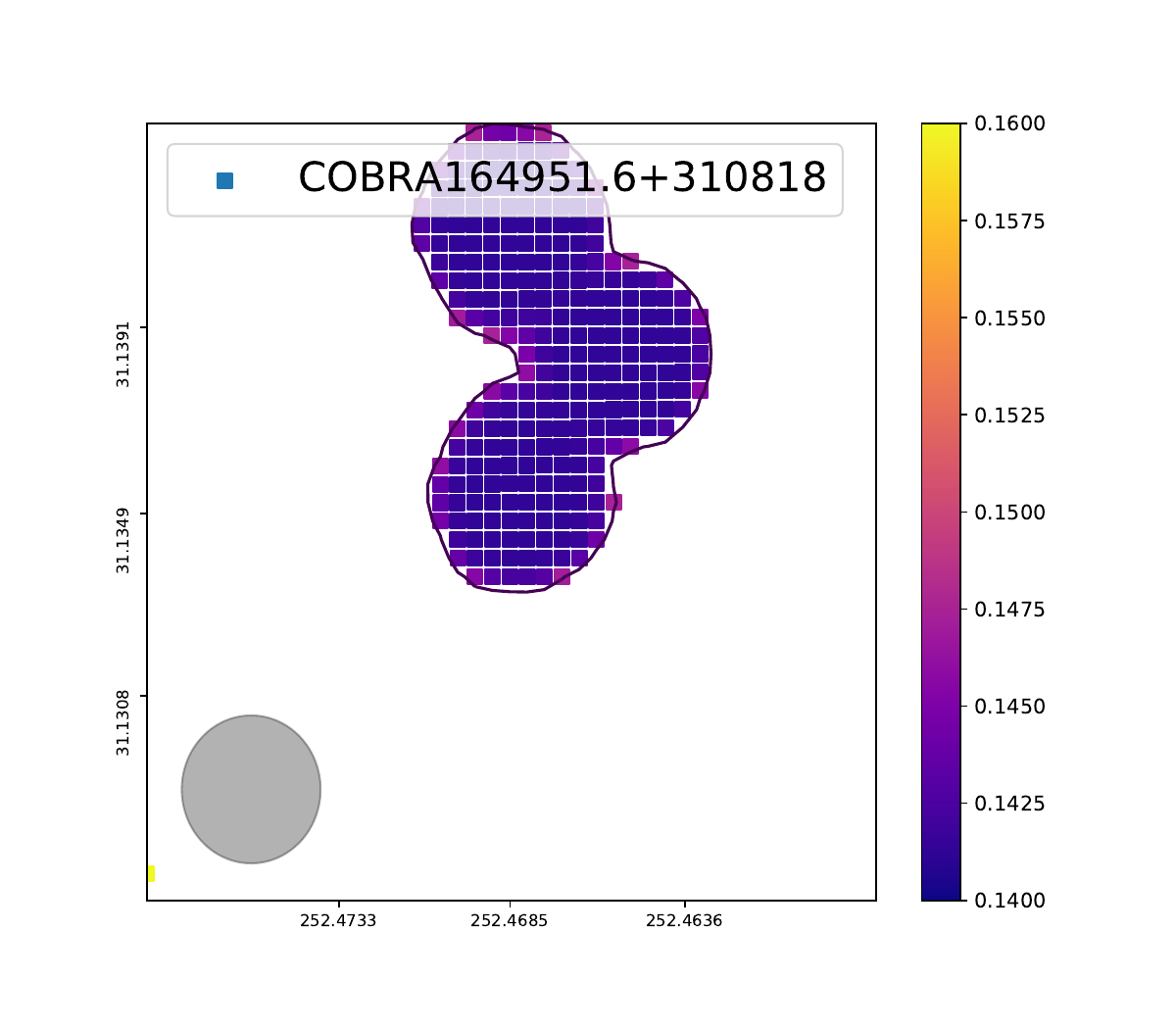}
\includegraphics[scale=0.28,trim={0.8in 0.65in 1.0in 0.75in},clip=true]{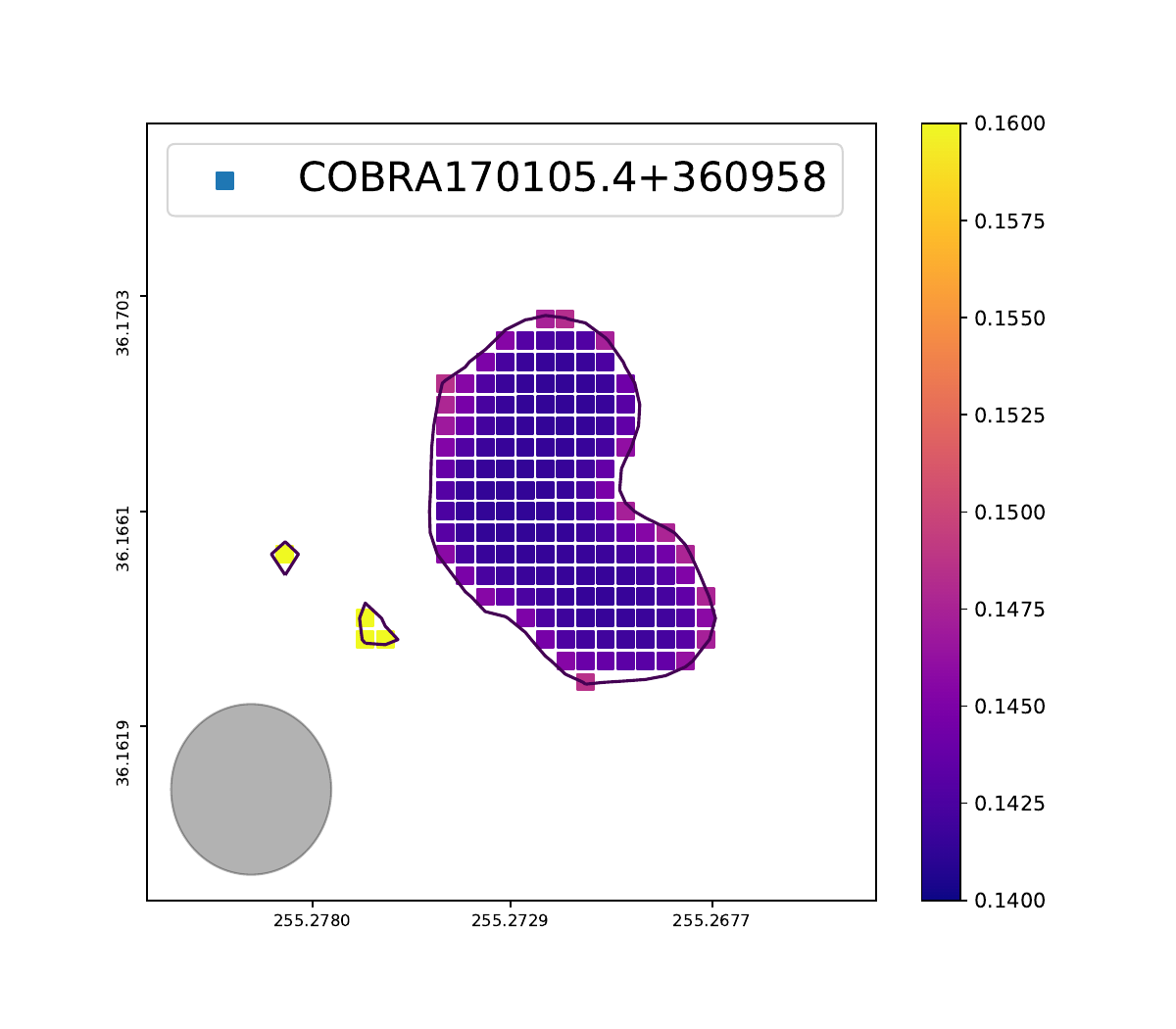}
\includegraphics[scale=0.28,trim={0.8in 0.65in 1.0in 0.75in},clip=true]{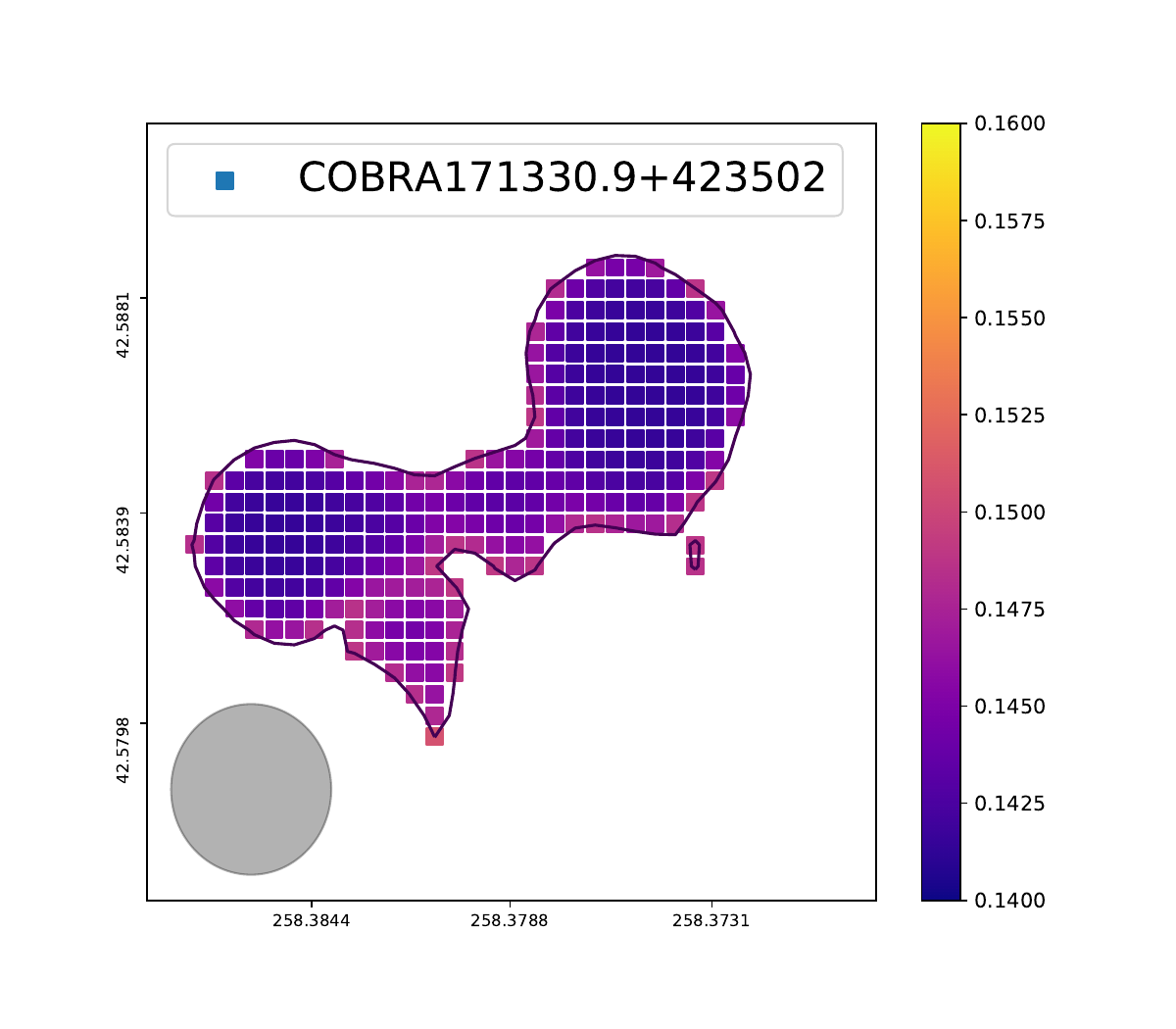}

\caption{Spectral index error maps of bent radio AGNs in the cluster sample.  Each pixel corresponds to a 1$\farcs$5 $\times$ 1$\farcs$5 region of the sky.  The 6$\arcsec$ LoTSS beam is shown as a grey circle in the bottom right corner of each image. The black contours show the 3$\sigma$ FIRST contours.  All images show the same region as in Figure~\ref{Fig:SI-clusters}.  While we see a roughly uniform error, we do use a uniform color bar to show our measure of the error in the value of the spectral index. As with the spectral index maps, we again note that the pixels are highly correlated because the pixel size is smaller than the beam size.}
\label{Fig:SI-clustersERROR}
\end{center}
\end{figure*}

\begin{figure*}
\begin{center}
\includegraphics[scale=0.28,trim={0.8in 0.65in 1.0in 0.75in},clip=true]{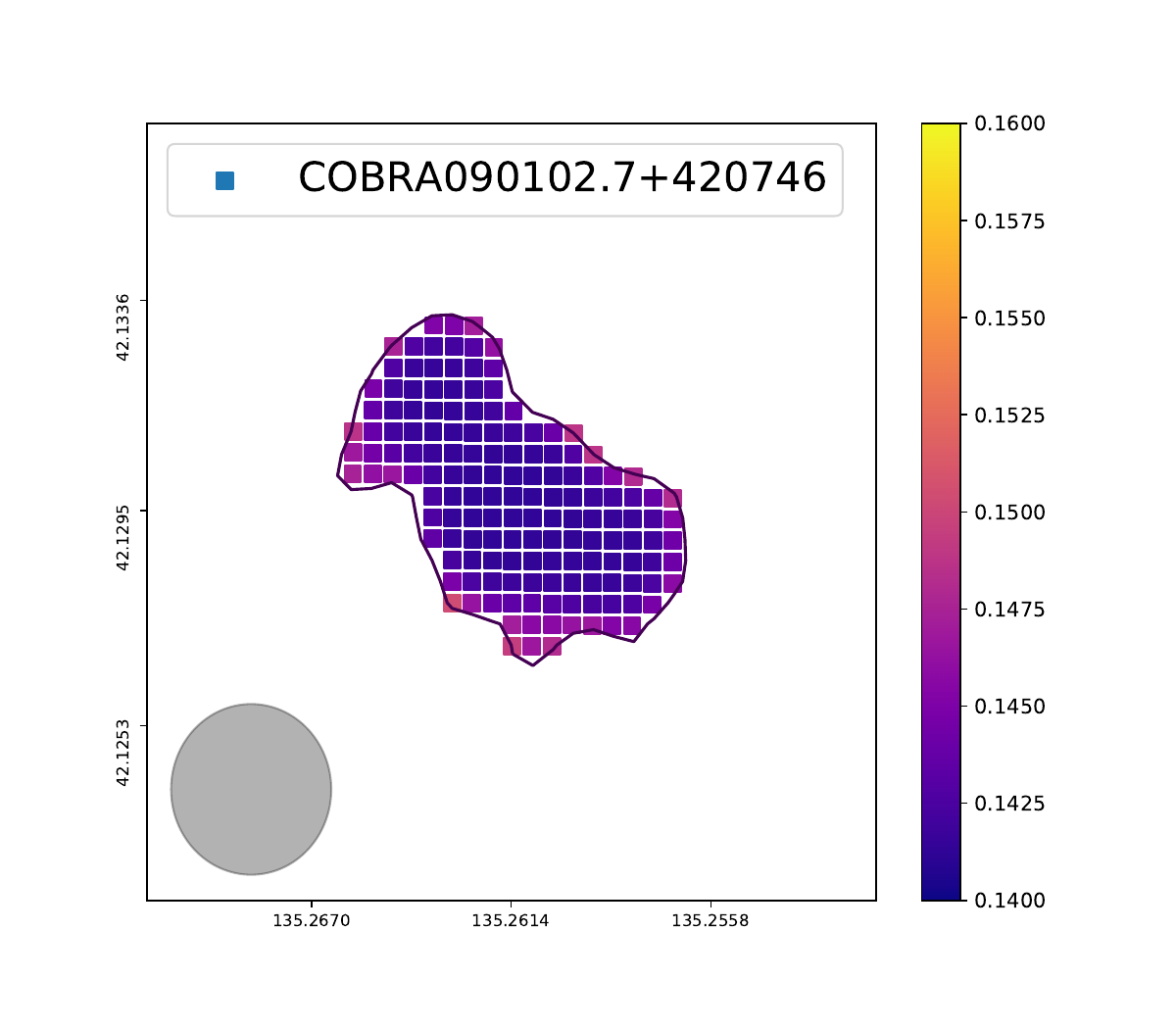}
\includegraphics[scale=0.28,trim={0.8in 0.65in 1.0in 0.75in},clip=true]{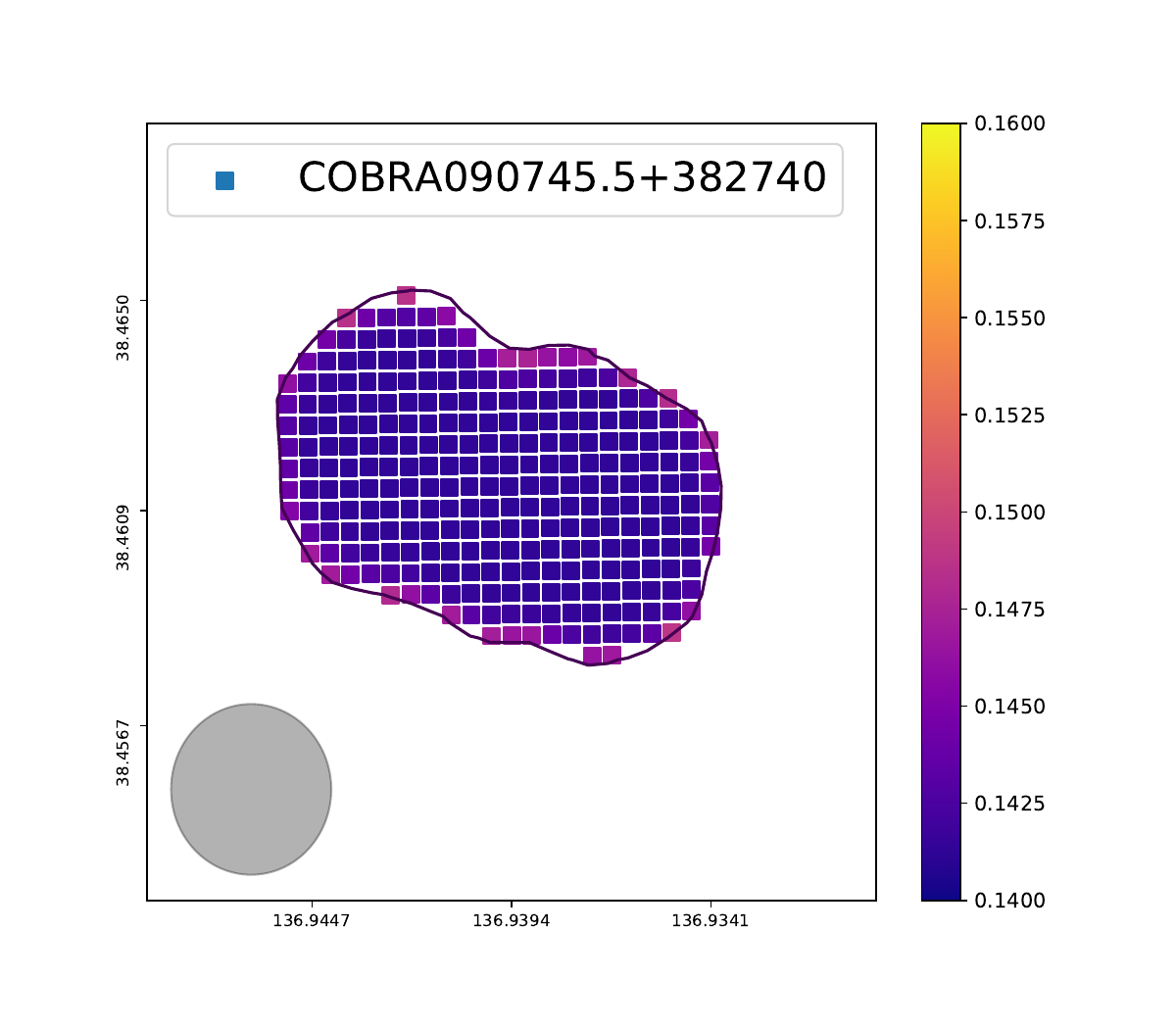}
\includegraphics[scale=0.28,trim={0.8in 0.65in 1.0in 0.75in},clip=true]{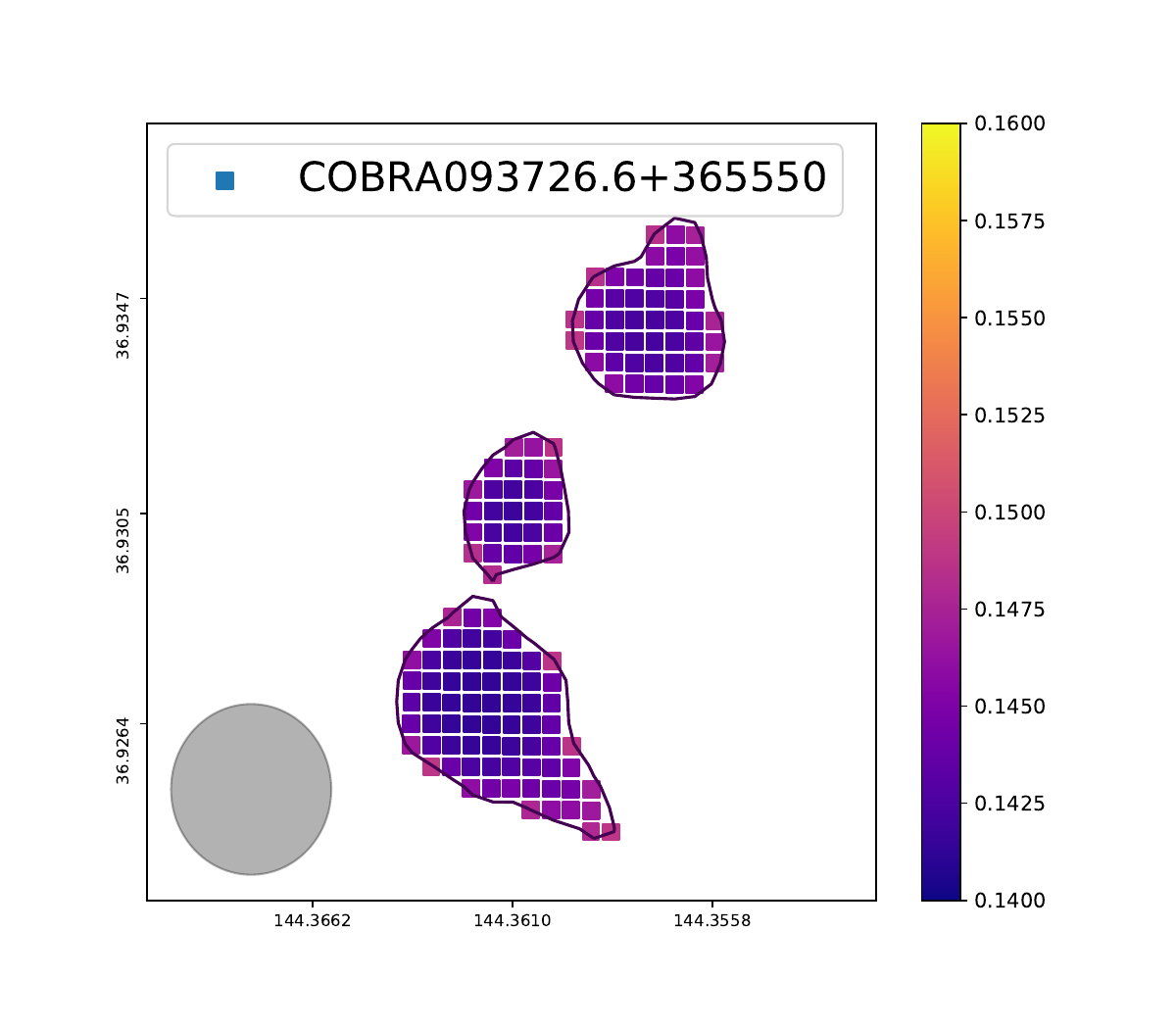}
\includegraphics[scale=0.28,trim={0.8in 0.65in 1.0in 0.75in},clip=true]{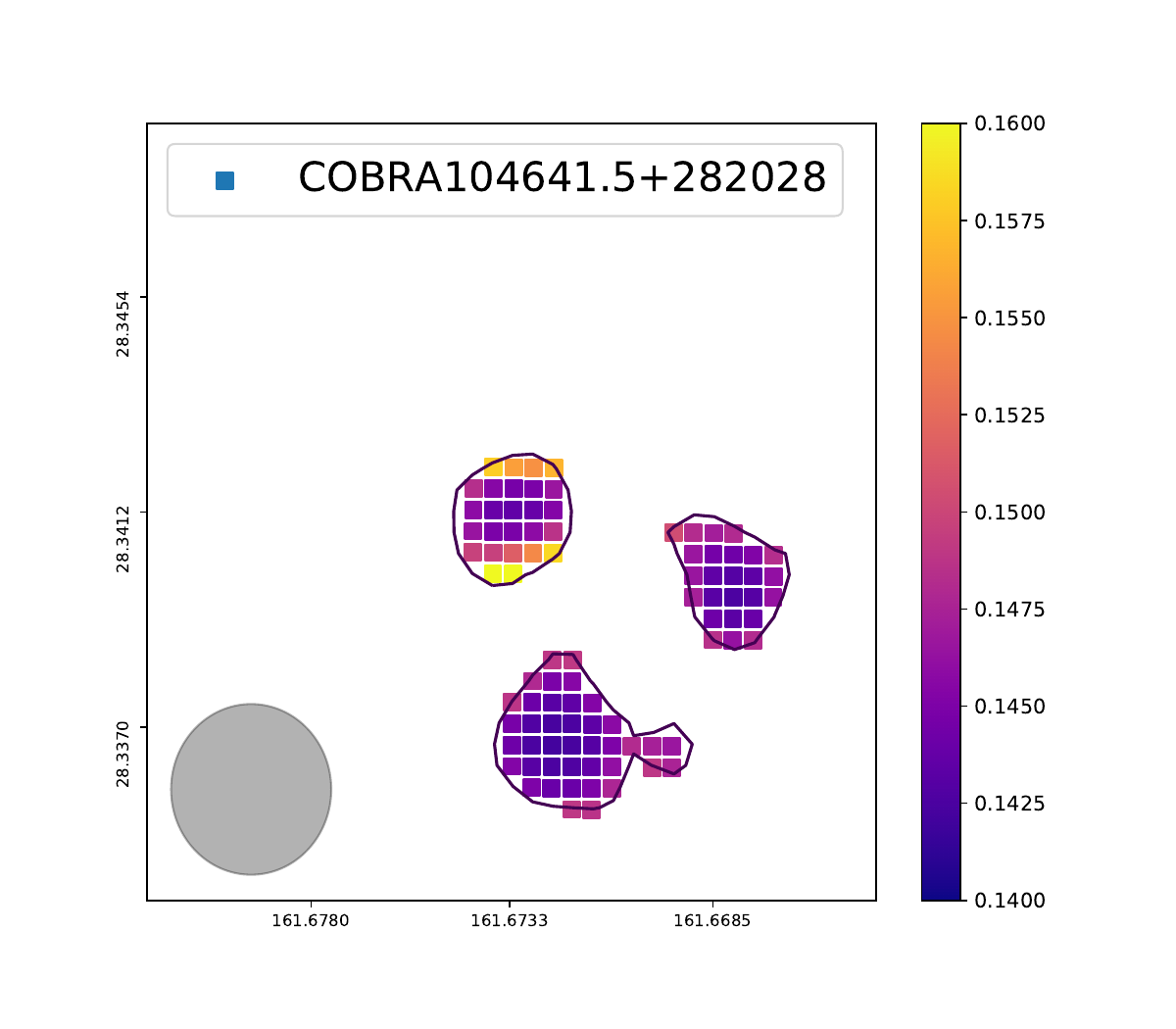}
\includegraphics[scale=0.28,trim={0.8in 0.65in 1.0in 0.75in},clip=true]{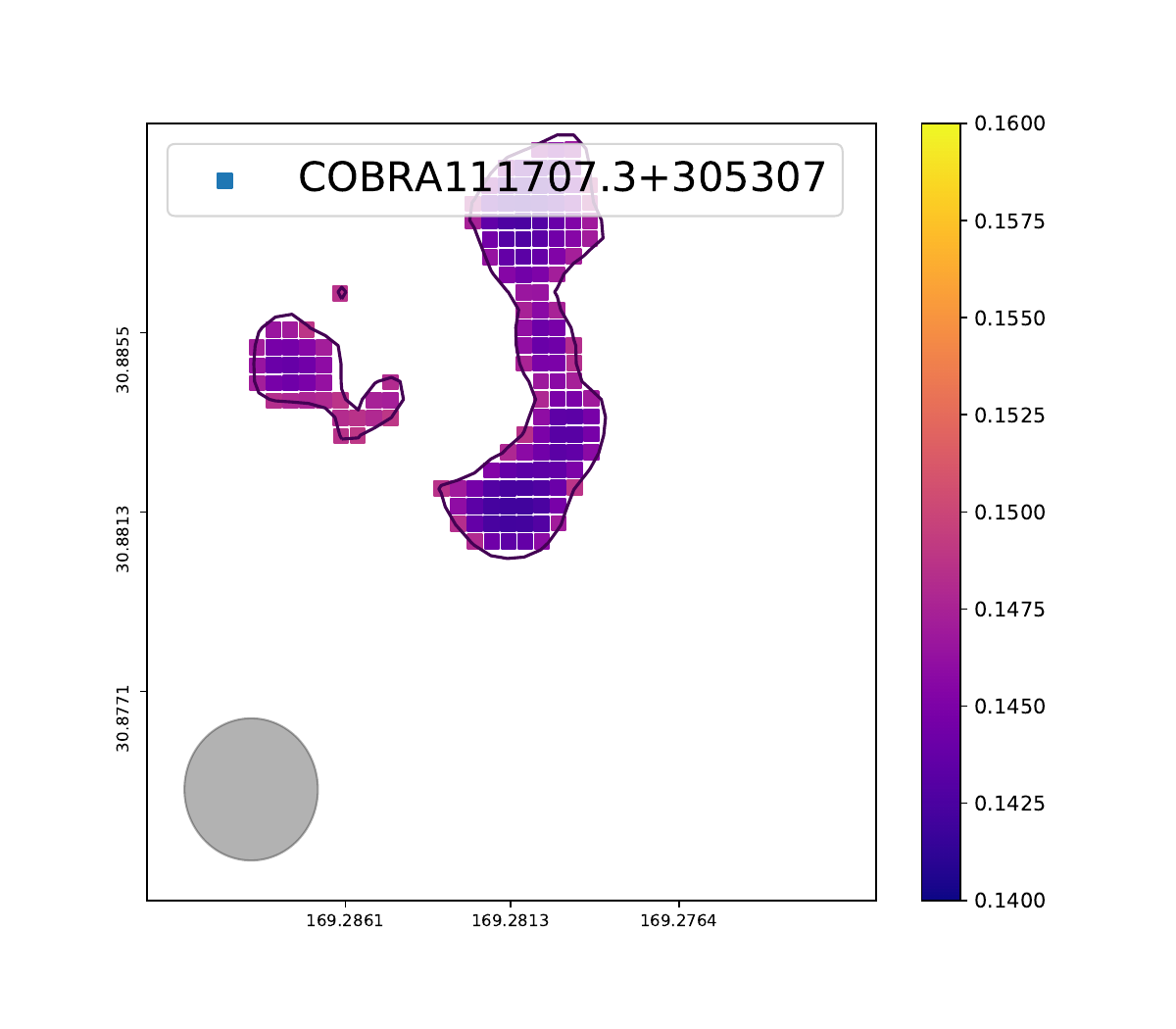}
\includegraphics[scale=0.28,trim={0.8in 0.65in 1.0in 0.75in},clip=true]{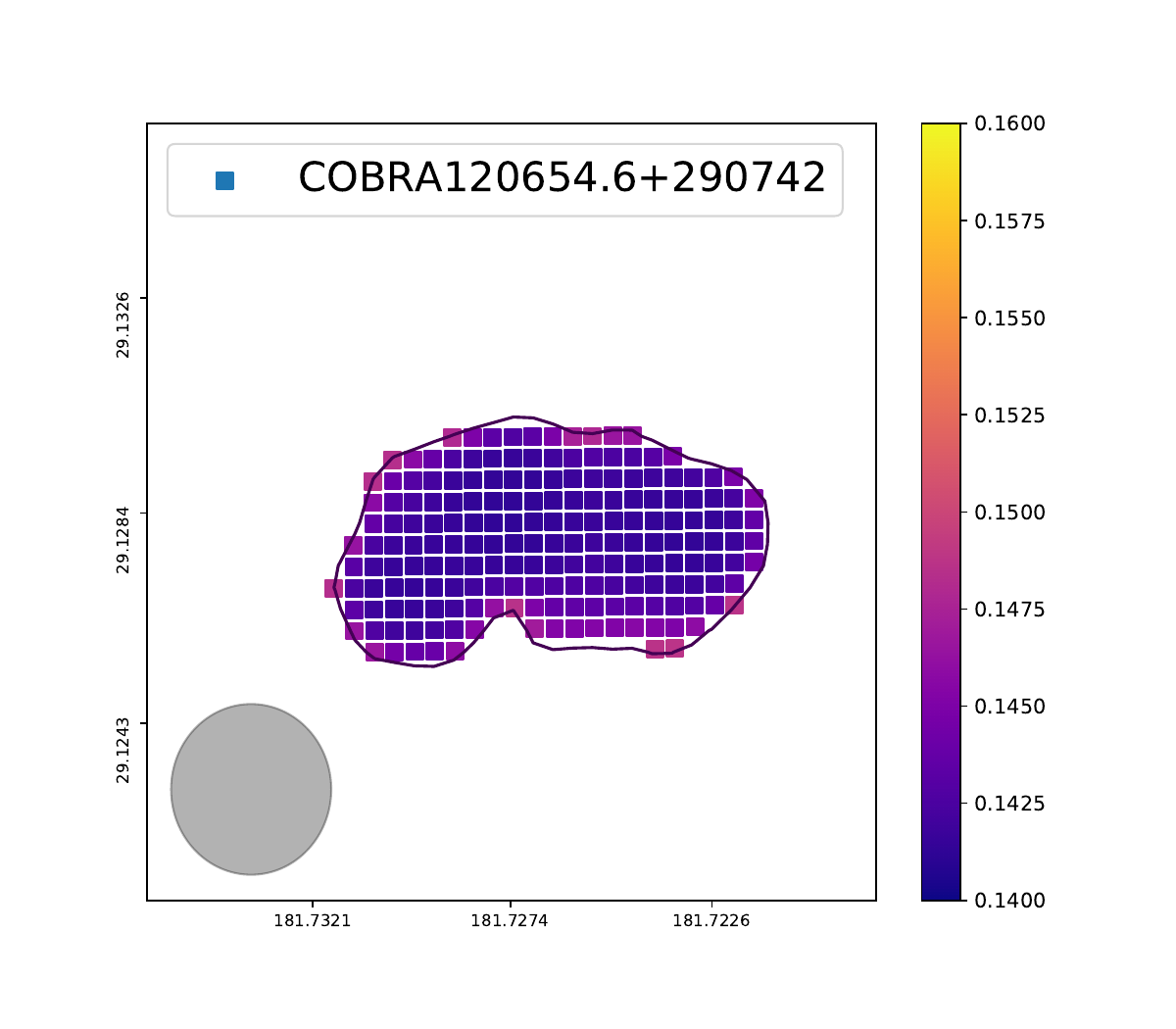}
\includegraphics[scale=0.28,trim={0.8in 0.65in 1.0in 0.75in},clip=true]{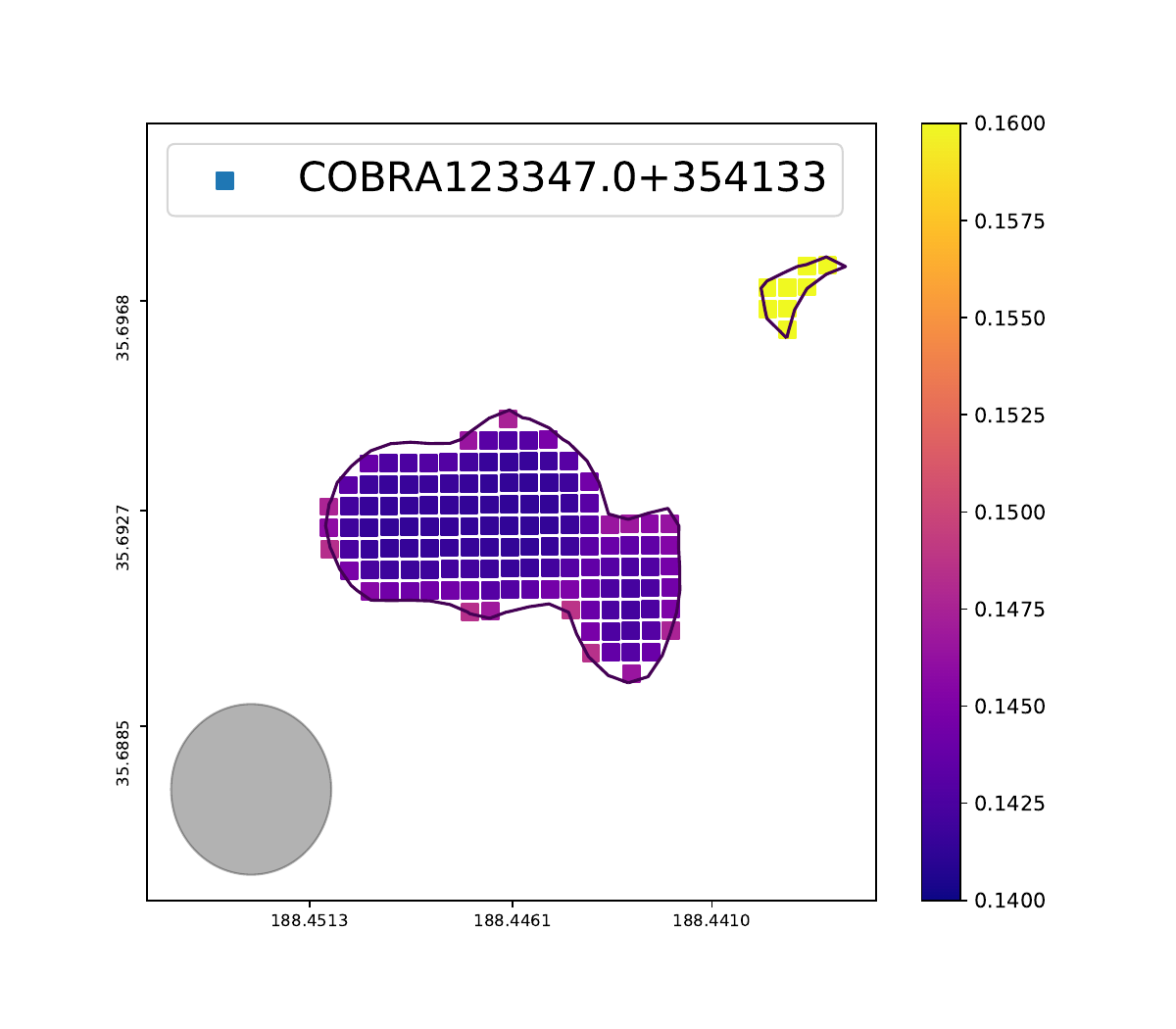}
\includegraphics[scale=0.28,trim={0.8in 0.65in 1.0in 0.75in},clip=true]{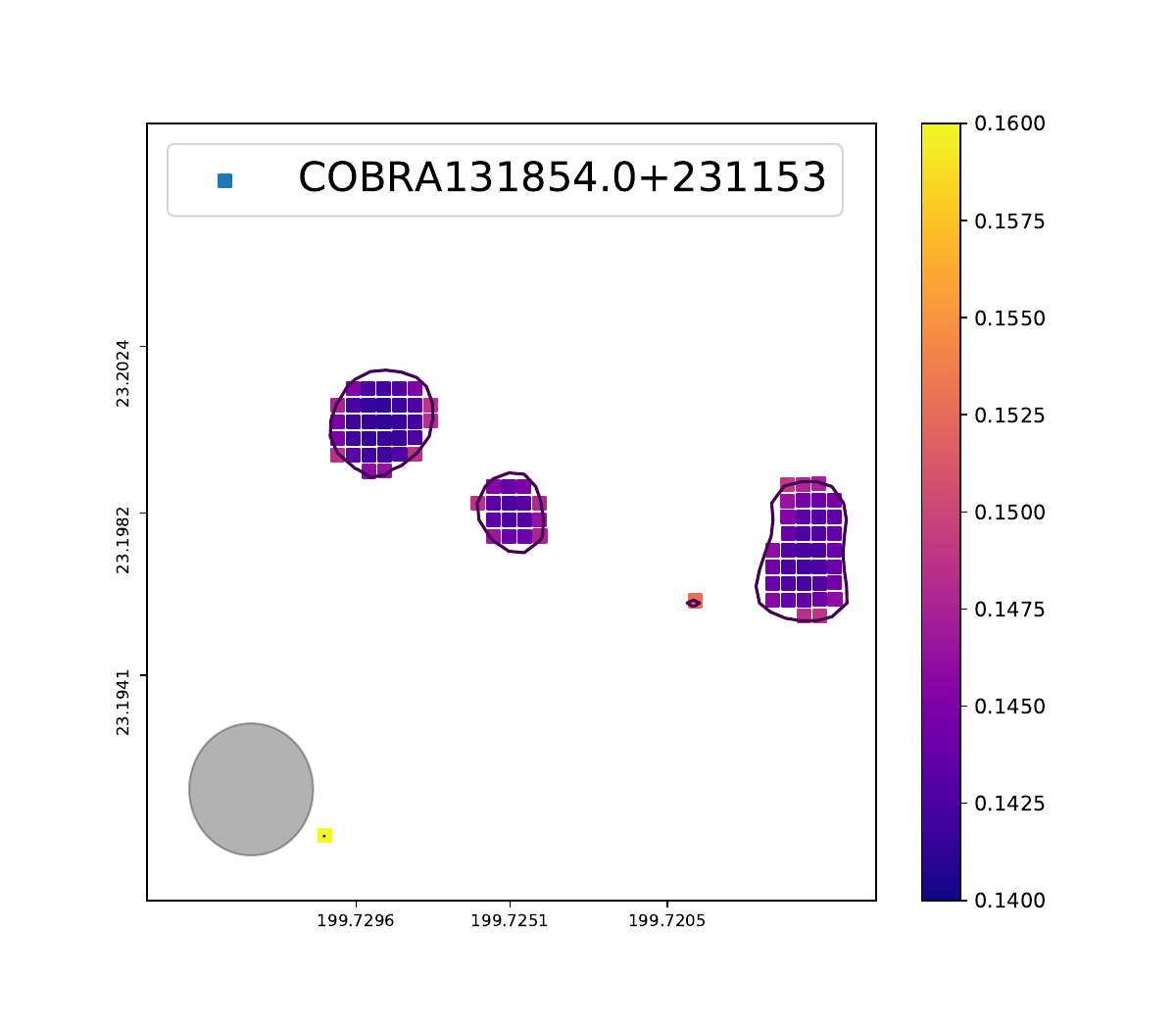}
\includegraphics[scale=0.28,trim={0.8in 0.65in 1.0in 0.75in},clip=true]{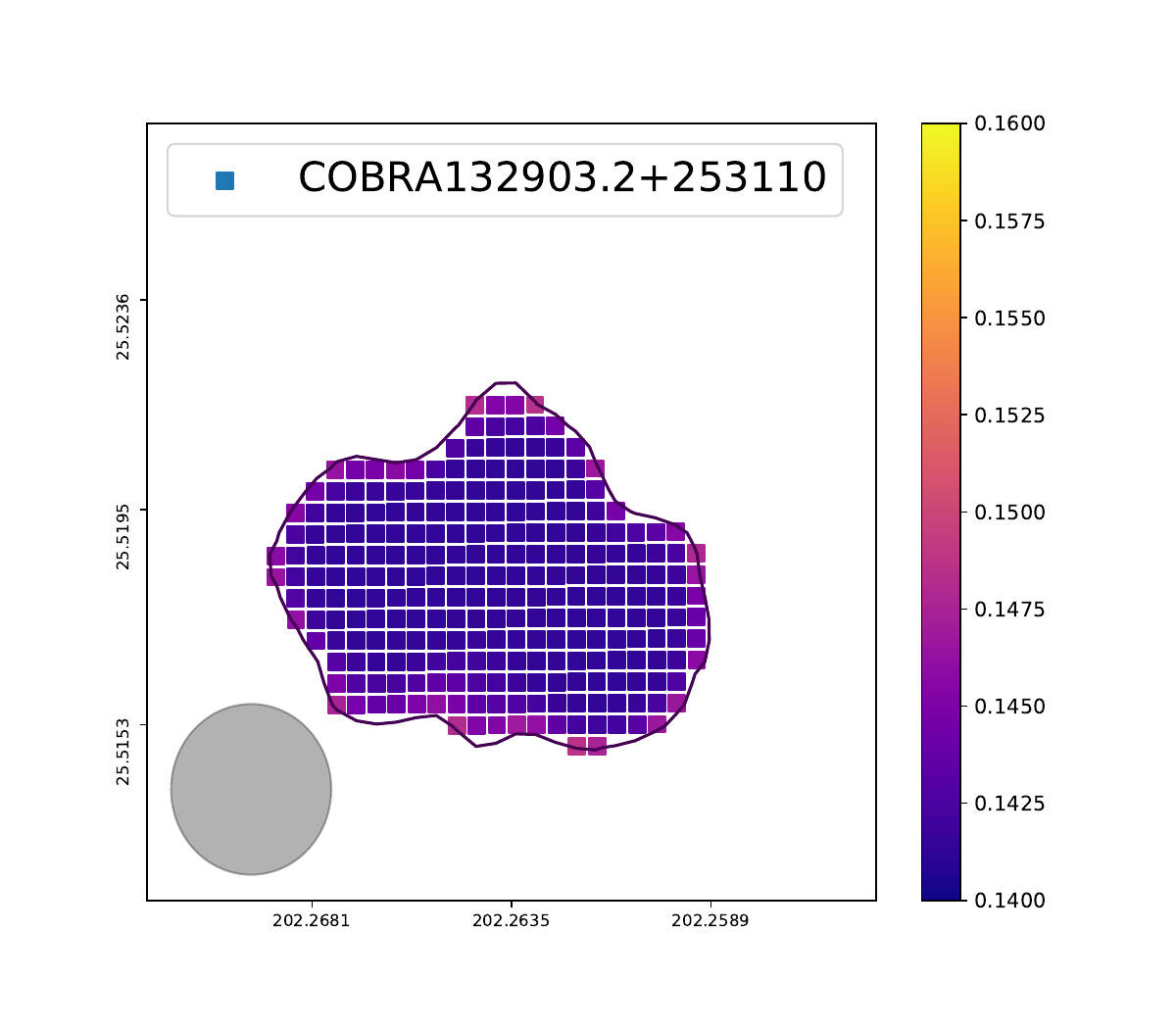}
\includegraphics[scale=0.28,trim={0.8in 0.65in 1.0in 0.75in},clip=true]{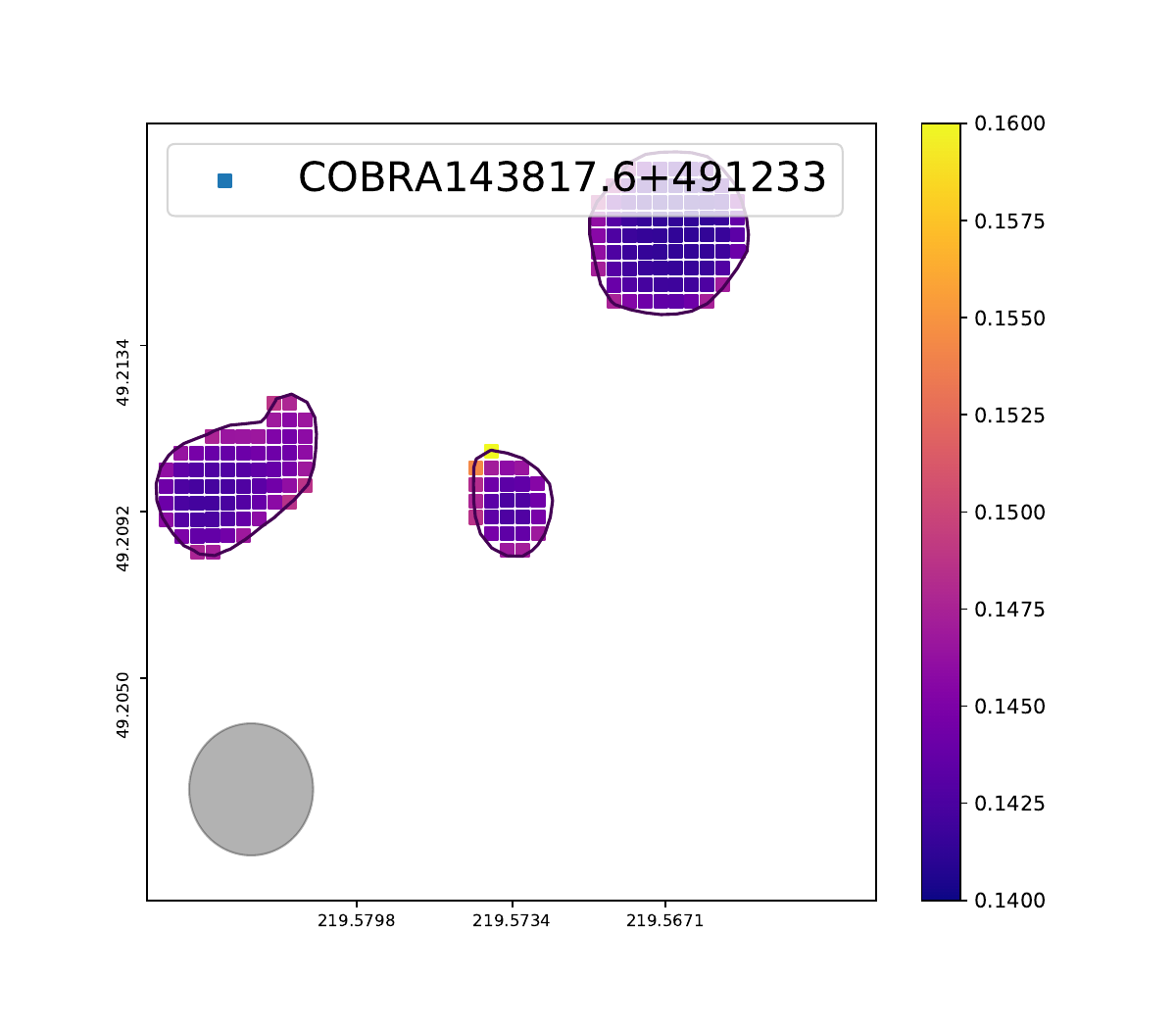}
\includegraphics[scale=0.28,trim={0.8in 0.65in 1.0in 0.75in},clip=true]{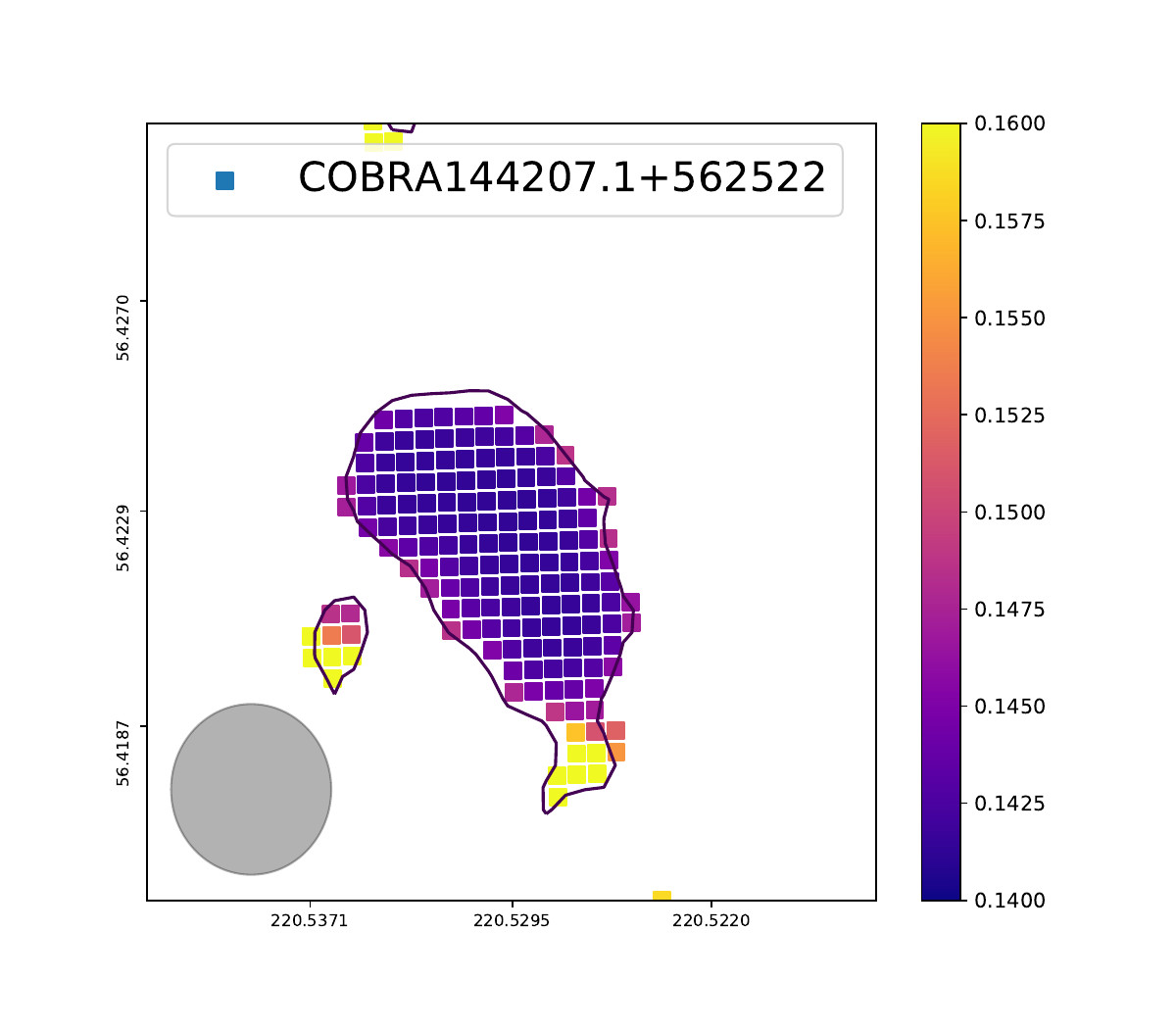}
\includegraphics[scale=0.28,trim={0.8in 0.65in 1.0in 0.75in},clip=true]{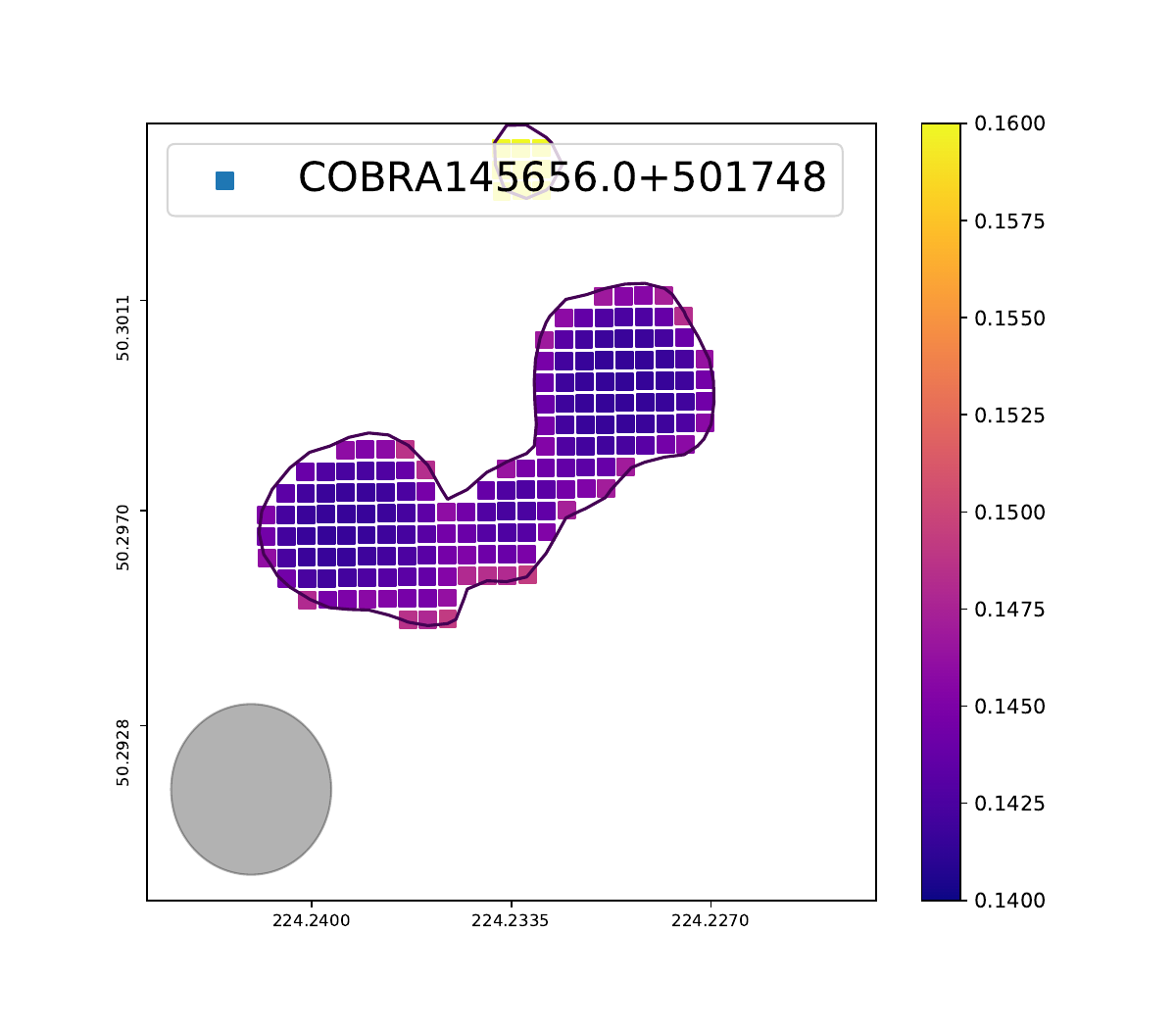}
\includegraphics[scale=0.28,trim={0.8in 0.65in 1.0in 0.75in},clip=true]{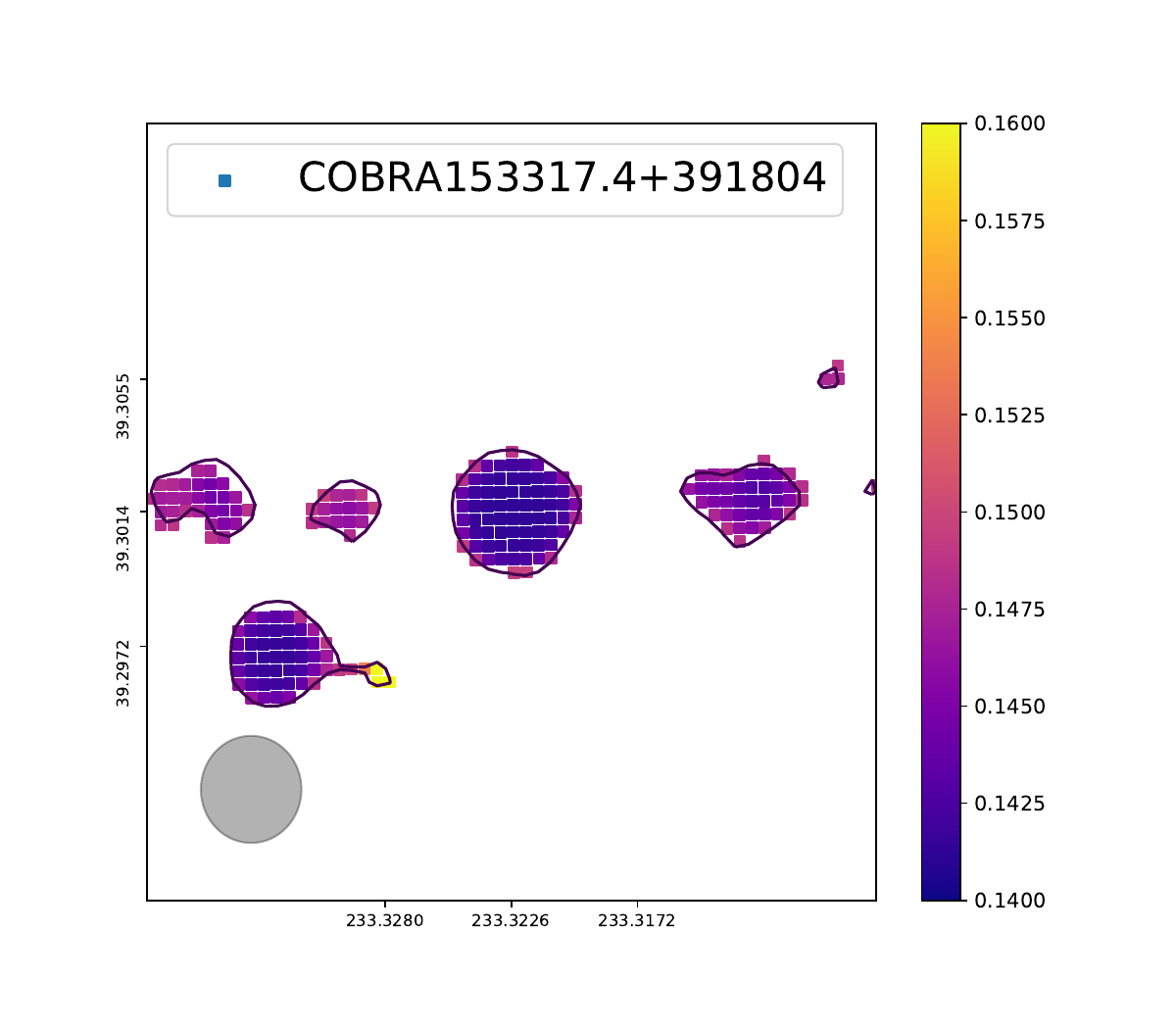}
\includegraphics[scale=0.28,trim={0.8in 0.65in 1.0in 0.75in},clip=true]{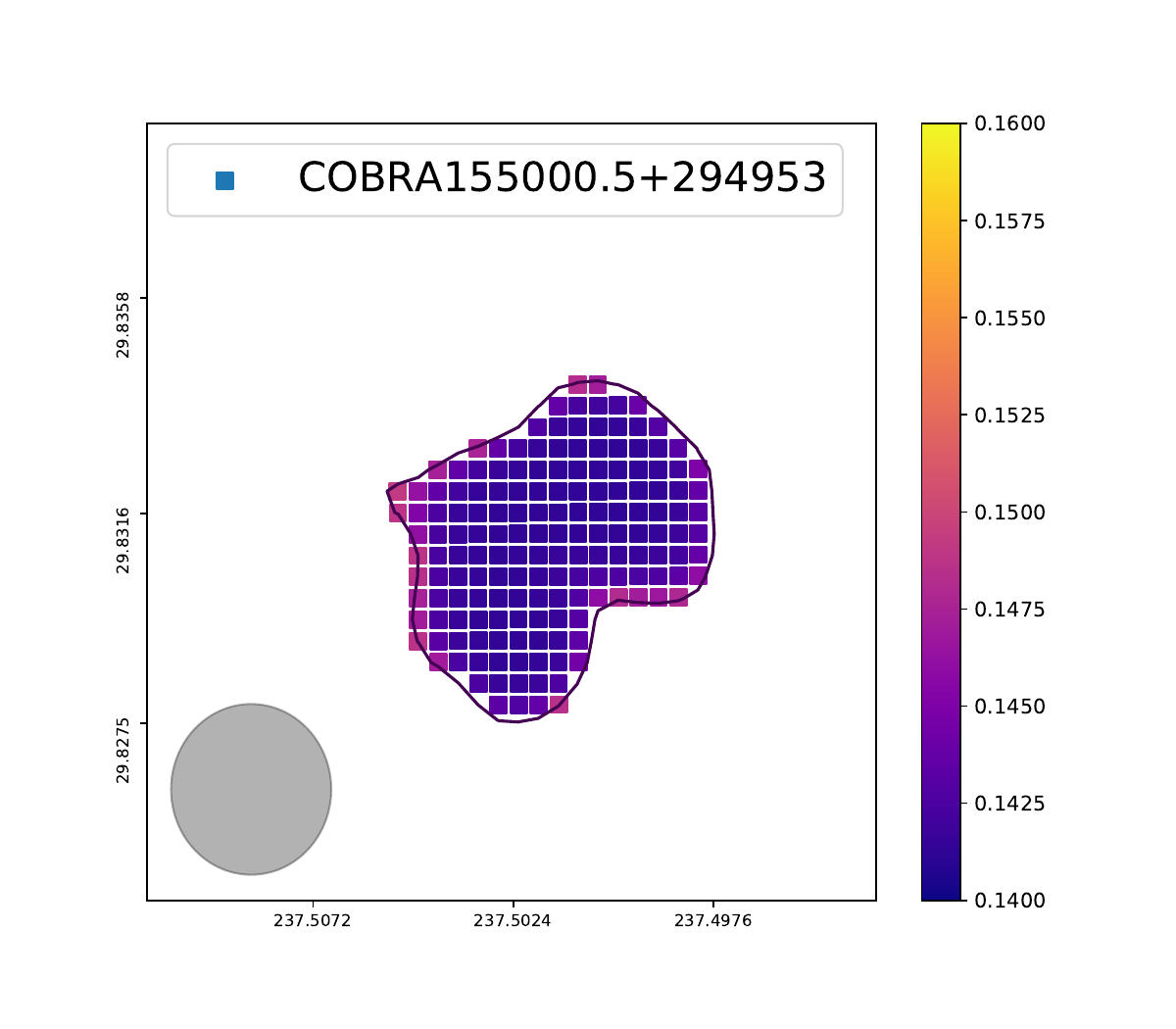}
\includegraphics[scale=0.28,trim={0.8in 0.65in 1.0in 0.75in},clip=true]{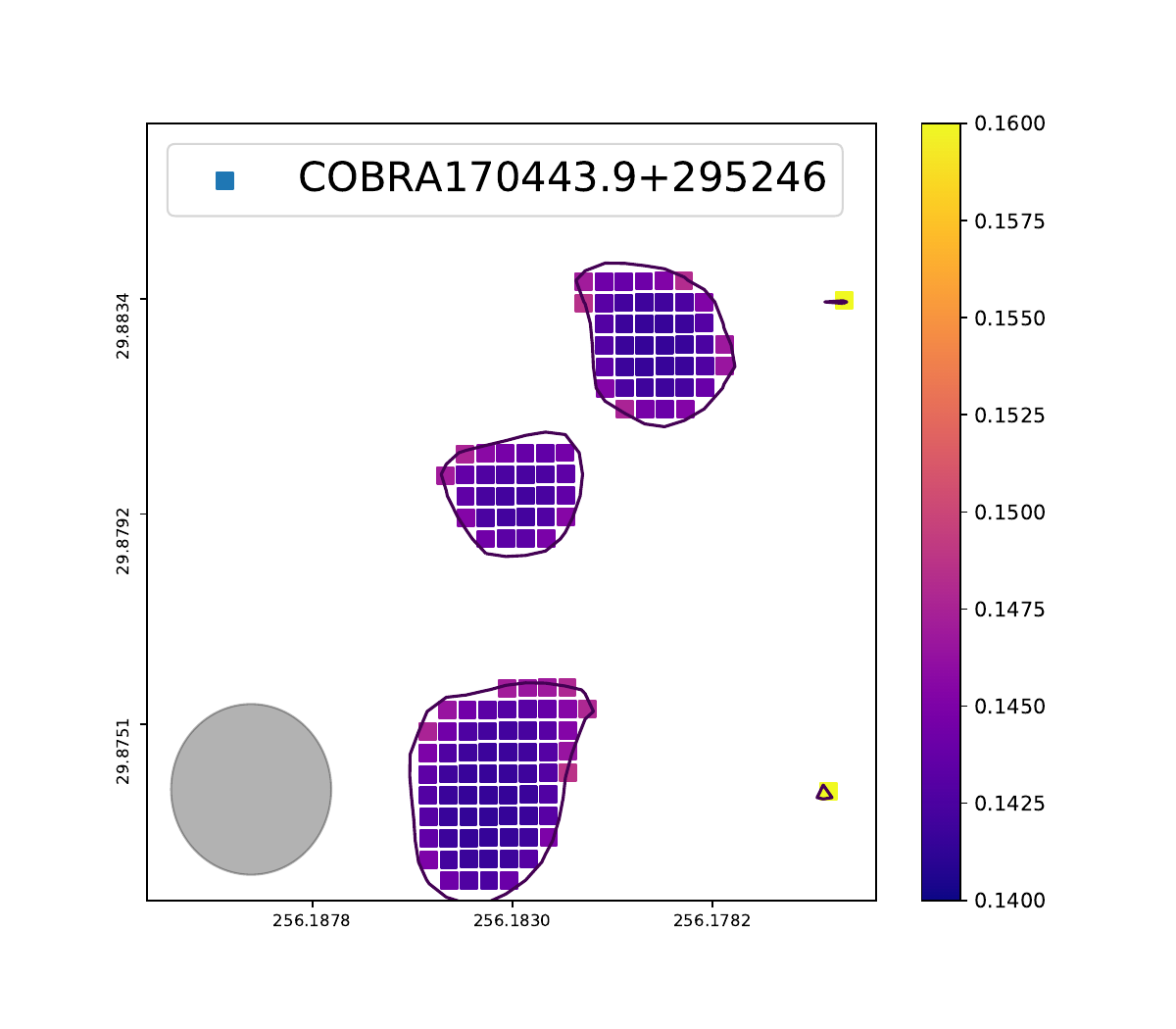}

\caption{Spectral index error maps of bent radio AGNs in the non-cluster sample.  As in Figure~\ref{Fig:SI-clustersERROR}, each pixel corresponds to a 1$\farcs$5 $\times$ 1$\farcs$5 region of the sky and the images cover the same region as those in Figure~\ref{Fig:SI-nonclusters}.  We use the same color bar as in Figure~\ref{Fig:SI-clustersERROR}.}
\label{Fig:SI-nonclustersERROR}
\end{center}
\end{figure*}

Additionally, we plot the error in the pixel measurements of the spectral index to highlight the regions that are subject to a slightly higher degree of error in Figures~\ref{Fig:SI-clustersERROR} and \ref{Fig:SI-nonclustersERROR}.  We measure the error following Equation 1 in \citet{DiGennaro-o}.  Unlike the spectral index maps shown in Figures~\ref{Fig:SI-clusters} and \ref{Fig:SI-nonclusters}, where the value does change rapidly across each AGN, we find a relatively uniform degree of error across the majority of each source, with only slightly higher errors at the edges of the sources.  This is in agreement with \citet{Gendron-Marsolais2020}, who show that the inner regions of their radio AGNs have roughly constant error values, similar to our own.  We do see a higher degree of error in the core of COBRA104641.5+282028, likely because the core region was not detected within the 10$\sigma$ LoTSS contours.

\end{document}